\documentclass[12pt]{book}

\usepackage[Lenny]{fncychap}

\usepackage[T1]{fontenc}

\usepackage{a4wide}
\usepackage{fancyhdr}
\usepackage{graphicx}
\usepackage{grffile}
\usepackage{placeins}
\usepackage[utf8]{inputenc}

\usepackage{verbatim}
\usepackage[normalem]{ulem}
\usepackage{amsfonts,amsmath,amssymb}
\usepackage{times}
\usepackage[italic]{mathastext}
\usepackage{enumitem}
\usepackage[hidelinks]{hyperref}
\usepackage{cleveref}
\usepackage{color}
\usepackage{upgreek}
\usepackage{flushend}
\usepackage{lineno}
\usepackage{siunitx}
\usepackage{mathrsfs}

\usepackage{placeins}
\usepackage{amsmath}
\usepackage{amssymb}
\usepackage{a4wide}
\usepackage{colortbl}
\usepackage[english]{babel}
\usepackage{subfig}

\usepackage{bm}

\usepackage{cite}
\usepackage[strict]{changepage}

\usepackage{setspace}
\usepackage{epigraph}
\usepackage{rotating}
\usepackage{enumitem}
\usepackage{emptypage}
\usepackage{todonotes}

\newcommand{\LMUTitle}[9]{
  \thispagestyle{empty}
  \vspace*{\stretch{1}}
  {\parindent0cm
   \rule{\linewidth}{.7ex}}
  \begin{flushright}

    \vspace*{\stretch{1}}
    \sffamily\bfseries\Large
    #1\\
    \vspace*{\stretch{1}}
    \sffamily\bfseries\large
    #2
    \vspace*{\stretch{1}}
  \end{flushright}
  \rule{\linewidth}{.7ex}
  \vspace*{\stretch{5}}
  \begin{center}
    \includegraphics[width=2in]{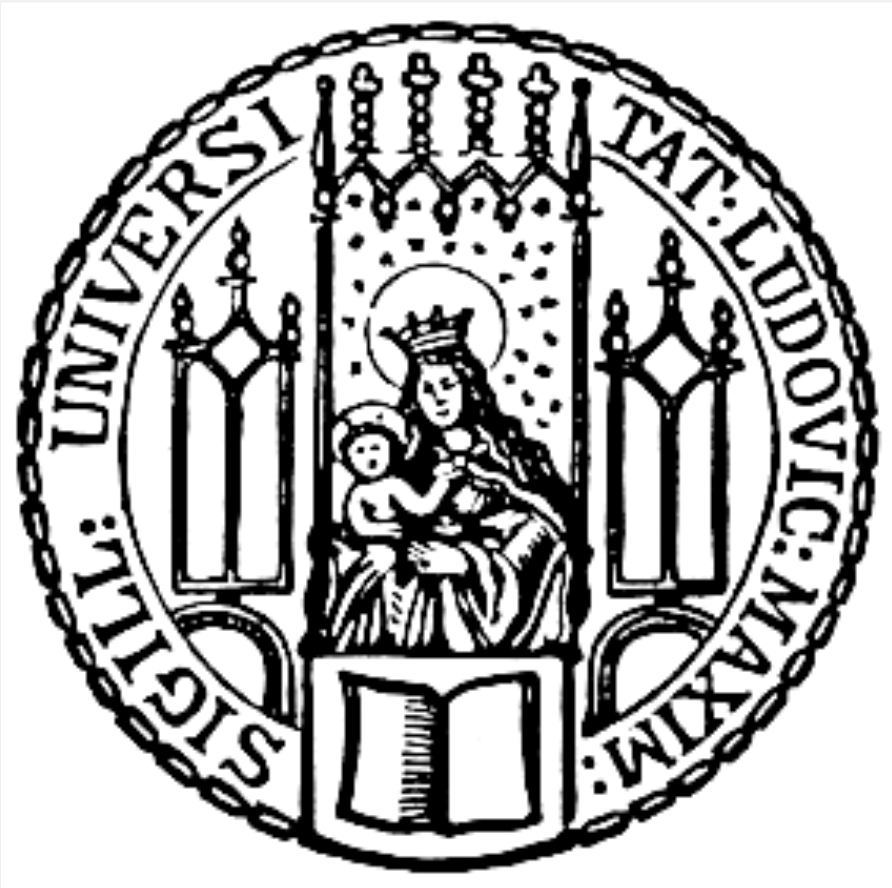}
  \end{center}
  \vspace*{\stretch{1}}
  \begin{center}\sffamily\large{#5}\end{center}
  \newpage
  \thispagestyle{empty}

  \cleardoublepage
  \thispagestyle{empty}

  \vspace*{\stretch{1}}
  {\parindent0cm
  \rule{\linewidth}{.7ex}}
  \begin{flushright}
    \vspace*{\stretch{1}}
    \sffamily\bfseries\Large
    #1\\
    \vspace*{\stretch{1}}
    \sffamily\bfseries\large
    #2
    \vspace*{\stretch{1}}
  \end{flushright}
  \rule{\linewidth}{.7ex}

  \vspace*{\stretch{3}}
  \begin{center}
    \large Dissertation\\
    \large an der #4\\
    \large der Ludwig--Maximilians--Universität\\
    \large München\\
    \vspace*{\stretch{1}}
    \large vorgelegt von\\
    \large #2\\
    \large aus #3\\
    \vspace*{\stretch{2}}
    \large München, den #6
  \end{center}

  \newpage
  \thispagestyle{empty}

  \vspace*{\stretch{1}}

  \begin{flushleft}
    \normalsize Erstgutachter:  #7 \\[1mm]
    \normalsize Zweitgutachter: #8 \\[1mm]
    \normalsize Tag der m\"undlichen Pr\"ufung: #9\\
  \end{flushleft}

  \cleardoublepage
}

 \renewcommand{\chaptermark}[1]%
 {\markboth{\thechapter.\ #1}{}}
 \renewcommand{\sectionmark}[1]%
 {\markright{\thesection\ #1}}
\begin{document}

  \frontmatter

  \LMUTitle
      {Simulating the inflationary Universe: \\from single-field to the axion-U(1) model} 
      {Angelo Caravano}                       
      {Neapel}                             
      {Fakultät für Physik}                         
      {München 2022}                          
      {21. Juli 2022}                            
      {Prof. Dr. Jochen Weller}                          
      {Prof. Dr. Eiichiro Komatsu}                         
      {15. September 2022}                         

  \tableofcontents
  \addcontentsline{toc}{chapter}{Table of Contents}
  \markboth{Table of Contents}{Table of Contents}

  \cleardoublepage
  \listoffigures
  \addcontentsline{toc}{chapter}{\listfigurename}
  \markboth{List of Figures}{List of Figures}

  \cleardoublepage

\chapter*{Zusammenfassung}
\small{

Die beobachtete Homogenität und räumliche Flachheit des Universums lassen vermuten, dass es unmittelbar nach dem Urknall eine Periode beschleunigter Expansion gab, die als Inflation bezeichnet wird. Generell wird angenommen, dass diese Expansion durch das Inflaton angetrieben wird, ein Skalarfeld jenseits des Standardmodells der Teilchenphysik. Wenn während dieser Epoche andere Felder vorhanden sind, können sie deutliche Spuren in Observablen hinterlassen, die mit Hilfe von zukünftigen Experimenten beobachtet werden könnten. Die Untersuchung der Phänomenologie solcher Felder ist eine besondere Herausforderung. Aufgrund der nichtlinearen Physik, die in verschiedenen nicht-minimalen Inflationsszenarien involviert ist, ist es oft nötig, über die Störungstheorie hinauszugehen. 

Wir präsentieren eine nichtlineare Studie der inflationären Ära, die auf numerischen Gittersimulationen basiert. Gittersimulationen sind ein bekanntes Werkzeug in der primordialen Kosmologie, und sie wurden ausgiebig zur Untersuchung der Wiedererwärmungsepoche am Ende der Inflation verwendet. Wir verallgemeinern dieses Verfahren auf die inflationäre Ära selbst. Da dies die erste Simulation der inflationären Epoche lange vor dem Ende der Inflation ist, konzentriert sich der erste Teil der Arbeit auf das einfachste Modell der Inflation, getrieben von einem einzelnen Feld. Wir diskutieren die konzeptionellen und technischen Voraussetzungen für die Simulation von Inflation auf einem Gitter. Die Simulation wird verwendet, um das nahezu invariante Spektrum skalarer Störungen sowie die Oszillationen im Leistungsspektrum zu reproduzieren, die durch eine Stufe im Potential verursacht werden.
 
Im zweiten Teil konzentrieren wir uns auf das komplexere Axion-U(1)-Inflationsmodell und präsentieren die erste Gittersimulation dieser Theorie während der frühen Inflationsepoche. Im Axion-U(1)-Modell führt eine effiziente Produktion von Eichbosonen oft zu starken Rückkopplungen, so dass man über die Störungstheorie hinausgehen muss, um die interessanten Vorhersagen zu untersuchen. Dank der Simulation entdecken wir neue statistische Eigenschaften von primordialen Skalarstörungen in diesem Modell. Im linearen Bereich der Theorie stellen wir fest, dass Nicht-Gaußsche Statistiken höherer Ordnung (jenseits von Bispektrum und Trispektrum) der Schlüssel zur Beschreibung der statistischen Eigenschaften von skalaren Störungen sind. Umgekehrt stellen wir fest, dass die Störungen im nichtlinearen Bereich der Theorie nahezu gaußförmig sind. Dies lockert die bestehenden Einschränkungen im Parameterraum, die sich aus der Überproduktion primordialer schwarzer Löcher ergeben, und deutet auf ein Gravitationswellensignal hin, das im beobachtbaren Bereich künftiger Experimente wie LISA liegt. 

}

\cleardoublepage

\chapter*{Abstract}
\normalsize


The observed homogeneity and spatial flatness of the Universe suggest that there was a period of accelerated expansion just after the Big Bang, called inflation. In the standard picture, this expansion is driven by the inflaton, a scalar field beyond the standard model of particle physics. 
If other fields are present during this epoch, they can leave sizable traces on inflationary observables that might be revealed using upcoming experiments. Studying the phenomenological consequences of such fields often requires going beyond perturbation theory due to the nonlinear physics involved in several non-minimal inflationary scenarios.

We present a nonlinear study of the inflationary epoch based on numerical lattice simulations. Lattice simulations are a well-known tool in primordial cosmology, and they have been extensively used to study the reheating epoch after inflation. We generalize this known machinery to the inflationary epoch. Being this the first simulation of the inflationary epoch much before the end of inflation, the first part of the thesis focuses on the minimal single-field model of inflation. 
We discuss the conceptual and technical ingredients needed to simulate inflation on a lattice. The simulation is used to reproduce the nearly scale-invariant spectrum of scalar perturbations, as well as the oscillations in the power spectrum caused by a step in the potential.

In the second part, we focus on the more complicated axion-U(1) model of inflation, and present the first lattice simulation of this model during the deep inflationary epoch. 
We use the simulation to discover new properties of primordial scalar perturbations from this model. In the linear regime of the theory, we find high-order non-Gaussianity (beyond bispectrum and trispectrum) to be key to describing the statistical properties of scalar perturbations. Conversely, we find perturbations to be nearly Gaussian in the nonlinear regime of the theory. This relaxes existing constraints from the overproduction of primordial black holes, allowing for a gravitational waves signal in the observable range of upcoming experiments such as LISA. Our results show that lattice simulations can be a powerful tool to study the inflationary epoch and its observational signatures.

\cleardoublepage
\chapter*{Acknowledgments}
\small
{
First of all, I would like to thank my advisors Jochen and Eiichiro. To Jochen, for giving me the chance of pursuing my research in his wonderful group, which has been like a second home for me. To Eiichiro, for teaching me to look where no one else was looking, and for patiently guiding me in my very first steps. I feel honored and humbled to have had such great supervisors, without whom I would have never made it this far.

Next, I would like to thank Kaloian Lozanov, who has been the best collaborator I could wish of. He has been like a third supervisor for me, teaching me the art of lattice simulations, and supporting me during the most challenging times.  

I thank also Sebastien Renaux-Petel, for being an excellent Master advisor and for the ongoing collaboration and interesting discussions. He introduced me to the exciting physics of inflation and to lattice simulations for the first time. Without him, my path into physics would have been very different.

I thank my colleagues at the USM who have been very close to me, both personally and professionally. Nico Hamaus, Giorgia Pollina, Nico Schuster, Steffen Hagstotz, Barbara Sartoris, Martin Kerscher, Kerstin Paech, and Sven Krippendorf. Thank you so much for everything. A special thank you to Marina Ricci for all the precious advice and for the useful comments on the manuscript. 

A special thanks goes to other fellow PhD students in Munich. Giordano Cintia, for being a great friend and collegue. Thank you very much for all the fun, and for the insightful scientific discussions. I am sure our friendship and professional relationship will continue in the future. Stefano De Nicola and Nazarena Tortorelli, for bringing me back home with all the food and magic songs. Marvin Lüben, for guiding me during the first and crucial months of PhD, where we shared good and bad moments. Micheal Zantedeschi, for being a colleague but more importantly a good climbing partner. 

I would like to thank Collegio Ghislieri and the alumni, with whom I shared the stimulating and beautiful years of my Bachelor's studies. Without this place, I certainly would not be here today.

I would also like to thank all those who inspired and passed on their passion for science. To Roberto Nesci, for giving me the opportunity of learning and doing astronomy before starting my path into physics. To Paolo Tini Brunozzi, for being the best high school teacher I could wish for.

Finally, I would like to thank my family. Giulia, who has always been close despite the distance. To my parents, for giving me my first telescope and always supporting my curiosity. 

Last but not least, I would like to thank Francesca with all my heart. She has been with me for all these years, being by my side in the hardest and most joyful moments.
}

\cleardoublepage
\normalsize

\mainmatter\setcounter{page}{1}
\markboth{Introduction}{Introduction}  

\chapter{Introduction}

\section{The inflationary paradigm}


Inflation, the accelerated expansion of the primordial Universe, was originally introduced to explain the homogeneity and spatial flatness of the Universe on very large scales \cite{PhysRevD.23.347,Sato:1980yn,Linde:1981mu,PhysRevLett.48.1220,STAROBINSKY198099}. Nowadays, this accelerated expansion is a very important piece of our understanding of the early Universe. 
The theory of inflation is powerful not only because it explains the observed homogeneity and flatness. It also provides a natural mechanism to generate primordial fluctuations, observed as small anisotropies in the Cosmic Microwave Background (CMB) and paving the way for the formation of large-scale structures. In the inflationary model, these are described as quantum vacuum fluctuations of the matter content present during this early phase \cite{Starobinsky:1979ty,Mukhanov:1981xt,HAWKING1982295,PhysRevLett.49.1110,STAROBINSKY1982175,Abbott:1984fp}. These fluctuations are generated on very small scales and then stretched to large cosmological scales thanks to the accelerated expansion. For this reason, inflation is a very interesting theoretical playground: it connects quantum physics to gravity, challenging our understanding of the most fundamental physical laws.

In the standard picture, this accelerated expansion of the early universe is driven by a scalar field, the so-called \textit{inflaton}. This field is assumed to be a degree of freedom beyond the standard model (SM) of particle physics. Most of the energy budget of the inflationary universe is contained in the inflaton field, which acts as a source for the accelerated expansion with an equation of state $p\simeq -\rho$. The quantum fluctuations of this scalar field fit very well the observed anisotropies in the CMB. In particular, the scalar field model of inflation predicts two important properties of primordial fluctuations: their Gaussian statistics \cite{Planck:2019kim} and the fact that they are nearly scale invariant \cite{Planck:2018jri} .

\section{Inflation as a high energy physics laboratory}

Although the simplest single-field model is compatible with all current observations, inflation provides a unique opportunity to test our most fundamental laws of nature and search for new physics. The predictions of inflationary cosmology are very sensitive to the particle content of the early Universe. If we modify the minimal single-field picture by adding other degrees of freedom during inflation, they can leave sizable traces on inflationary observables that might be observed using upcoming experiments. 
For example, if another massive scalar particle is present during inflation and interacts with the inflaton, it can leave a characteristic signature in the three-point function of primordial scalar perturbations \cite{Arkani-Hamed:2015bza}. Hunting for small signatures in inflationary observables could reveal new physics from inflation, and might give us crucial information about the high energy description of quantum gravity.

The recent discovery of gravitational waves with ground-based interferometers \cite{LIGOScientific:2016aoc}, together with future space missions such as LISA \cite{LISA1}, opens a new and unexplored window of cosmological signals and offers a unique opportunity in this direction. Several non-minimal models of inflation predict a sizable amount of gravitational waves in the form of a stochastic background. If observed, such a signal would give crucial information about the physics at play in the early Universe.

In this thesis, we will manly consider a particular family of extensions of the minimal scenario called axion-gauge models of inflation. In these models, a gauge field \textit{and} a pseudo-scalar field, often called \textit{axion}, are present during inflation. The axion could be the inflaton field, sourcing the accelerated expansion, or some other spectator field present during the inflationary epoch. The axion-gauge system gives rise to unique observational signatures, such as non-Gaussianity and parity-violating gravitational waves, which might be observed with next-generation experiments \cite{Komatsu:2022nvu,Campeti:2020xwn}. For this reason, these models have been extensively studied in the literature, both in the case where the gauge field is Abelian \cite{Anber_2006,Anber_2010,Barnaby_2011_Large, Barnaby_2011, Anber_2012} and non-Abelian \cite{Maleknejad_2011,maleknejad2013gaugeflation,Adshead_2012,Adshead_2013,maleknejad2021su2r}. We will consider the case of an U(1) Abelian field analogous to the electromagnetic field of the SM, which is coupled to the inflation field. This is usually called the axion-U(1) model of inflation.




\section{The need for simulations}

Computing precise theoretical predictions from non-minimal models of inflation is particularly challenging. The reason is twofold. First, the quasi-exponential expansion, translating into a large spacetime curvature, makes it necessary to include gravity in the quantum field theory description. This makes the computation of particle physics processes during inflation much more complicated than in a flat Minkowski space. Second, many models of inflation leading to sizable observational signatures are characterized by nonlinear physics, invalidating the perturbation theory approach typically used for computing predictions. There is plenty of examples where this occurs: from models of axion-gauge inflation mentioned above, to models with multiple scalar fields with a strong turn in the field-space trajectory \cite{Fumagalli:2020nvq}, or single-field models with a large step in the scalar field potential \cite{Inomata_2022}.  In these models, the computation of observational signatures, such as GW emission, often requires going beyond perturbation theory \cite{Ferreira:2015omg,Peloso_2016,Papageorgiou:2018rfx,Maleknejad_2019,Lozanov_2019,Mirzagholi:2019jeb,Papageorgiou:2019ecb,Fumagalli:2020nvq, Inomata_2022}. 

To address the first problem, i.e. dealing with the quasi-exponential expansion, several analytical techniques have been developed in the past two decades. Important examples are the well-established "in-in" formalism \cite{Weinberg:2005vy,Adshead:2009cb}, or the recently developed cosmological bootstrap method \cite{Arkani-Hamed:2015bza}, which has been shown to be an efficient analytical tool to compute observable quantities in a quasi-de Sitter spacetime \cite{Lee:2016vti,Arkani-Hamed:2018kmz,Baumann:2022jpr,Pimentel:2022fsc,Jazayeri:2022kjy}. In this thesis, we are going to develop an alternative and complementary tool to compute theoretical predictions from inflation based on numerical simulations. This will also tackle the second problem, allowing to study models of inflation beyond perturbation theory.

 Numerical simulations are becoming more and more useful in understanding physical systems, and are particularly important when it comes to cosmology. Due to the nonlinear physics characterizing many cosmological phenomena, simulations are nowadays an essential tool to test the fundamental theories behind the evolution of the Universe and the structures within it. An important example are N-body simulations, which are crucial in studying the nonlinear physics involved in the formation of large-scale structures. In this thesis, we are going to consider a particular kind of cosmological simulations called \textit{lattice simulations}. This kind of simulations have been extensively used to study the end of inflation and the reheating epoch after it, where the inflaton decays and transfers all its energy to the other degrees of freedom of the Universe. In this context, various lattice simulations have been developed in the last decades to study both scalar \cite{Khlebnikov_1996,Prokopec_1997,latticeeasy,Frolov_2008,hlattice,Sainio_2012,Child_2013,Easther_2010} and gauge fields \cite{Lozanov_2020,figueroa2021cosmolattice} models. This thesis aims at generalizing these lattice techniques to the inflationary epoch itself.

Our work represents the first lattice simulation of the deep inflationary epoch much before the end of inflation. For this reason, in the first part of the thesis we focus on simulating the simplest single-field model of inflation. We introduce the methodology and discuss the conceptual and technical aspects of simulating inflation on the lattice. 
 In the second part, we generalize this technique to the more complicated axion-U(1) model of inflation mentioned above. We use it to explore the axion-U(1) system beyond perturbation theory, which allows to discover new properties of the phenomenology of this model during inflation. We focus on studying the statistics of primordial scalar perturbations. 
 In the linear regime of the theory, we find non-Gaussianity to be quite unique: high-order statistical correlators, beyond bispectrum and trispectrum, are crucial to describe the statistical properties of scalar perturbations. On the contrary, non-Gaussianity is unexpectedly suppressed during the nonlinear dynamics, with major observational implications. The latter result invalidates an existing bound in the literature coming from overproduction of primordial black holes. This allows for a GW signal from the axion-U(1) system above the projected sensitivity of future experiments such as LISA. Our work shows that lattice simulations can be a powerful tool to investigate inflationary models and their theoretical predictions.

 
 
 


\section{Content of the thesis}
The thesis is organized into two parts. The first part is focused on simulating the minimal single-field model of inflation. It contains the following chapters:
\begin{itemize}
	\item In \cref{sec:inflation}, we give an introduction to the standard single-field model of inflation. This will also establish the notation used in the rest of the manuscript.
	
	\item In \cref{sec:inflationsim}, we introduce the lattice simulation for the single-field model of inflation and use it to study the inflationary Universe much before the end of inflation. 
	
\end{itemize}
In the second part we focus on the axion-U(1) model of inflation, extending the methodology developed in the first part. It is organized in the following chapters:
\begin{itemize}	
	\item  In \cref{sec:axioninf}, we give a brief review of the axion-U(1) model of inflation and summarize the known results regarding the phenomenology of this model.
	
	\item  In \cref{sec:axionsim} we generalize the technique developed in \cref{sec:inflationsim} to study the axion-U(1) model of inflation using a lattice simulation. This will allow to discover new properties of this model, and to confirm and improve upon previous results in the literature. 
	
	\item  In \cref{sec:conclusions}, we provide a summary of the results and discuss possible future applications of the work of this thesis.
\end{itemize}
The content of the thesis is based on the following publications:
\begin{itemize}
	\item \textit{Lattice Simulations of Inflation} \cite{Caravano_2021}\\ A. Caravano, E. Komatsu, K.D. Lozanov and J. Weller\\\hyperlink{https://iopscience.iop.org/article/10.1088/1475-7516/2021/12/010}{JCAP 12 (2021) 12, 010} [\hyperlink{https://arxiv.org/abs/2102.06378}{2102.0637}] 
	\item \textit{Lattice simulations of Abelian gauge fields coupled to axions during inflation } \cite{caravano2021lattice}\\ A. Caravano, E. Komatsu, K.D. Lozanov and J. Weller\\\hyperlink{https://journals.aps.org/prd/abstract/10.1103/PhysRevD.105.123530}{Phys.Rev.D 105 (2022) 12, 123530} [\hyperlink{https://arxiv.org/abs/2110.10695}{2110.10695}] 
	\item \textit{Lattice simulations of axion-U(1) inflation} \cite{Caravano:2022epk}\\ A. Caravano, E. Komatsu, K.D. Lozanov and J. Weller\\ \hyperlink{https://arxiv.org/abs/2204.12874}{2204.12874} 
\end{itemize}
During the doctoral studies, the candidate also took part in the following article, which is not included in the thesis:
\begin{itemize}
	\item \textit{Combining cosmological and local bounds on bimetric theory} \cite{Caravano:2021aum}\\ A. Caravano, M. L\"uben and J. Weller\\\hyperlink{https://iopscience.iop.org/article/10.1088/1475-7516/2021/09/035}{JCAP 09 (2021) 035} [\hyperlink{https://arxiv.org/abs/2101.08791}{2101.08791}] 
	\end{itemize}

\part{Single-field inflation}

\cleardoublepage

\chapter{Introduction to inflation}
\label{sec:inflation}
This chapter serves as an introduction to the standard paradigm of inflationary cosmology. We introduce the main equations that are needed in the rest of the manuscript and highlight the differences between standard computations and the lattice approach developed in this thesis. A more detailed and pedagogical introduction to the topic can be found, for example, in Ref. \cite{Baumann:2009ds}. Inflation is defined as an accelerated expansion of the early Universe. We start from the observational motivations for introducing this accelerated expansion. Then, we introduce the scalar field model of inflation and its predictions.

\section{Why inflation?}
\label{sec:inflationwhy}
\subsection{FLRW Universe}
On very large scales, the Universe appears to be homogeneous and isotropic. In general relativity, the most general metric that describes such a Universe is the so-called Friedmann-Lemaitre-Robinson-Walker (FLRW) metric, which can be written in spherical coordinates $(r,\theta,\phi)$ as:

\begin{equation}
\label{eq:FLRW}
	ds^2=-c^2dt^2+\frac{a^2(t)}{1-Ka^2}dr^2+a^2(t)d\Omega^2,
\end{equation}
where $d\Omega^2=d\theta^2+\sin^2(\theta)d\phi^2$. This metric is very simple, and it is fully determined by a constant $K$, describing the spatial curvature of 3-dimensional hypersurfaces, and by the scale factor $a(t)$, describing the expansion of the Universe as a function of time. $c$ is the speed of light, that we set to 1 throughout this work. Observations tell us that the curvature $K$ is very close to zero \cite{refId0}. Therefore, we assume $K=0$ for the rest of this work. This observed flatness is one of the main problems of the original Big Bang model. At the end of this section, we will see that this property of the Universe is a natural consequence of inflation.

The rate of expansion is described by the Hubble parameter $H(t)\equiv \dot{a}/a$. The evolution of the scale factor $a$ as a function of the matter content of the Universe is determined by the Einstein field equations, that in this case are called \textit{Friedman equations}\footnote{Here and throughout this work, we use the dot to indicate derivatives in cosmic time $dt=a\,d\tau$, i.e. $\dot f=\frac{df}{dt}$. We will also use the prime to indicate derivatives in conformal time $ f^\prime=\frac{df}{d\tau}$ (conformal time will be defined shortly).}:
\begin{align}
\label{eq:Friedmann}
\begin{split}
\Big(\frac{\dot{a}}{a}\Big)^2
&=\frac{1}{3M^2_{\rm Pl}}\rho\\
2\frac{\ddot{a}}{a}+\Big(\frac{\dot{a}}{a}\Big)^2
&=-\frac{1}{M^2_{\rm Pl}}p
\end{split}
\end{align}
where $\rho$ and $p$ are respectively the energy-density and the pressure of the matter content of the Universe, assumed to be a perfect fluid. $M_{Pl}$ is the reduced Planck mass $M_{Pl}=1/\sqrt{8\pi G}$, being $G$ Newton's gravitational constant. 

Before proceeding, let us introduce two important quantities in FLRW cosmology. The first is the \textit{conformal time} $\tau$, defined from the cosmological time $t$ as $dt=a(t)d\tau$. In this time coordinate, the metric is conformal to the Minkowski one\footnote{We are assuming $K=0$ and expressing the metric in Cartesian coordinates $\vec{x}$.}:
\begin{equation}
	ds^2=a^2(\tau)\left(-d\tau^2+d\vec{x}^2\right).
\end{equation}
The other quantity is the number of $e$-folds $N_e$, defined by the relation $dN_e=H(t)dt=d\log a(t)$. If we take two times $t_1$ and $t_2$, this quantity represents the logarithmic growth of the scale factor between these times $N^{(2)}_e-N^{(1)}_e=\log a(t_2)-\log a(t_1)$. $N_e$ is particularly useful in inflationary cosmology, as during inflation the scale factor $a$ grows by many orders of magnitude.


\subsection{The horizon problem}
\begin{figure}
	\centering
	\includegraphics[scale=0.15]{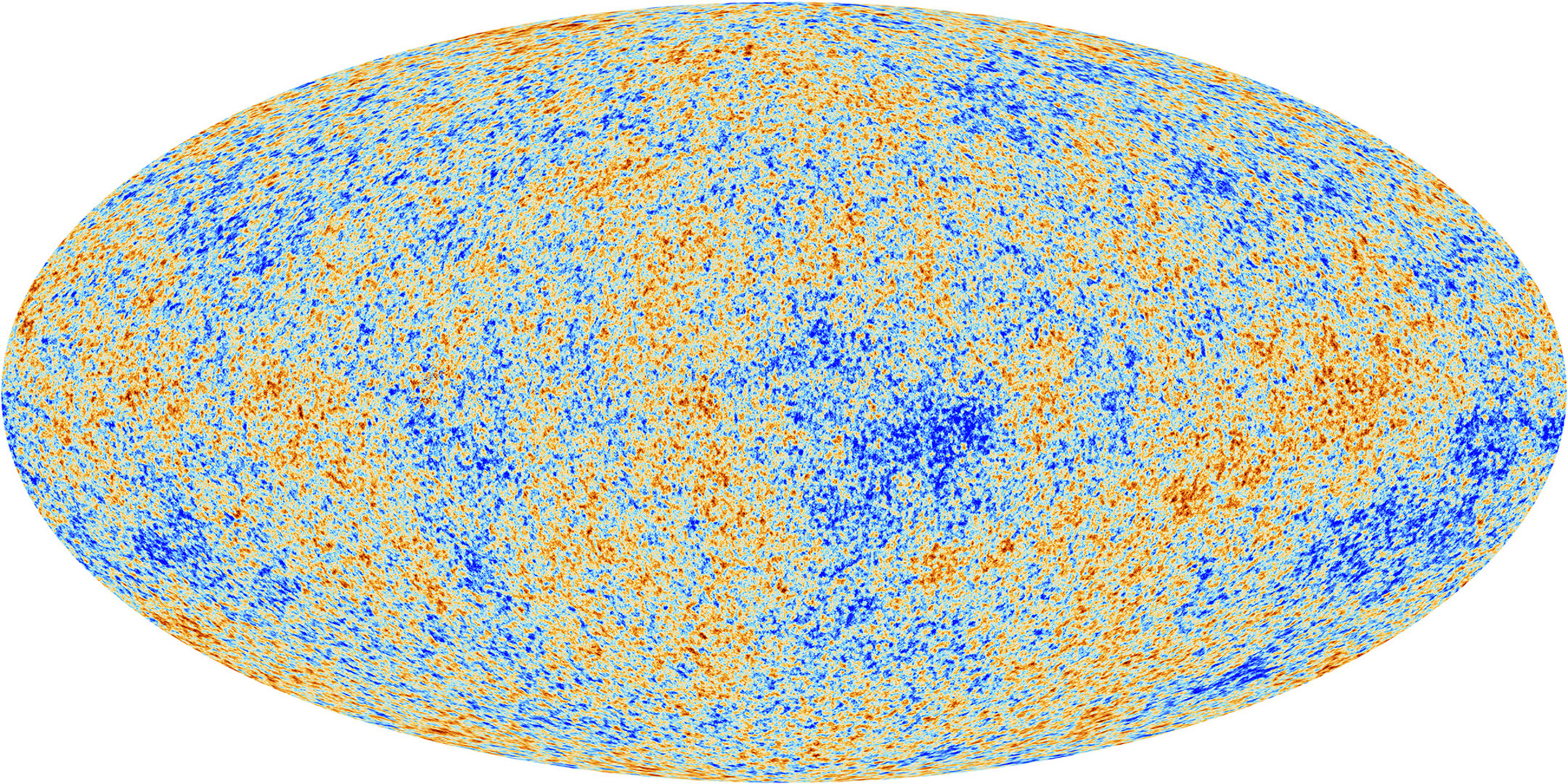}
	\caption{Cosmic Microwave Background as seen from the Planck satellite. The radiation has the same temperature $T\simeq 2.7\,$K across all sky. The fluctuations shown in the figure are very small, of order $\simeq10^{-5}\,$K.} \label{fig:CMB}
\end{figure}

Inflation was originally introduced to solve some observational problems of the original Big Bang model \cite{PhysRevD.23.347,Sato:1980yn,Linde:1981mu,PhysRevLett.48.1220,STAROBINSKY198099}. One of these is the so-called horizon problem, related to the homogeneity of the Universe on very large scales. Thanks to observations, we know that the Universe was already homogeneous at the epoch of recombination. This epoch, occurred roughly 370 thousand years after the Big Bang, is when the Universe cooled down enough to allow electrons and protons to form neutral hydrogen atoms. At this time, the Universe became transparent to electromagnetic radiation, which was emitted everywhere and is still observable today in the form of a background radiation permeating the Universe: the Cosmic Microwave Background (CMB). The CMB has a special property: its temperature does not depend on the particular direction we observe it. This can be seen in \cref{fig:CMB}, where we show the CMB radiations as seen from the Planck satellite. This radiation is homogeneous, and the fluctuations on top of it are very small (of order $10^{-5}$) and statistically independent from the direction.

This property of the CMB is a clear evidence that the Universe was already homogeneous during this early time. Unfortunately, this cannot be explained using the original Big Bang model. To see this, let us compute the physical distance that a photon travels between times $t_1$ and $t_2$. Setting $ds^2=0$ in the metric \eqref{eq:FLRW}, it is easy to obtain this quantity as:
\begin{equation}
\Delta r=\int_{t_1}^{t_2}\frac{dt}{a(t)}=\Delta\tau.
\end{equation}
If we take a time $t$, we can define the \textit{particle horizon} as the distance that a photon travels between the Big Bang $t=0$ and that time:
\begin{equation}
\label{eq:horizon}
h=\int_{0}^{t}\frac{dt^\prime}{a(t^\prime)}=\int(aH)^{-1}dN_e.
\end{equation}
At a given time, this quantity represents the maximum distance between two points such that they are causally connected. From the second equality of \eqref{eq:horizon} we can see that the horizon depends on the evolution in $e$-folds time $N_e$ of the so-called Hubble radius $r_H=(aH)^{-1}$. 

If we compute the particle horizon at the time of recombination using the old Big Bang model, the result is too small to explain the homogeneity of the CMB. Indeed, one can use this equation to compute that only $\sim1$ degree patches in the sky could be causally connected at the time of emission, which is in contrast with the fact that the CMB has the same temperature across all sky. This is known as the \textit{horizon problem}.

A solution to this problem is assuming that, for some time after the Big Bang and prior to CMB emission, there was a period in which the Hubble radius was shrinking:
\begin{equation}
\label{eq:sch}
\frac{d}{dt}(aH)^{-1}<0.
\end{equation}
If this happens, it is clear looking at \eqref{eq:horizon} that the lapse of conformal time between the Big Bang and CMB emission increases and can solve the horizon problem. This condition is equivalent to a positive acceleration $\ddot{a}>0$, and it is called inflation. In order the explain the homogeneity of the CMB, $N_e\simeq60$ $e$-folds of inflation are needed. 

To parametrize the accelerated expansion, it is useful to define the parameter $\varepsilon$:
\begin{equation}
\label{eq:condition}
\varepsilon=-\frac{\dot{H}}{H^2}.
\end{equation}
It is straightforward to prove that $\ddot a>0$ implies $\varepsilon<1$. To see how to achieve an accelerated expansion, let us assume that the Universe is filled by a perfect fluid with an equation of state $p=w\rho$, were $p$ and $\rho$ are pressure and energy density. Then, the second Friedmann \cref{eq:Friedmann} can be rewritten as:
\begin{equation}
\label{eq:friedH}
\dot{H}+H^2=-\frac{1}{6M_{\rm Pl}^2}(\rho+3p)=-\frac{H^2}{2}(1+3w).
\end{equation}
By looking at this equation, we see that $\varepsilon<1$ implies $w<-1/3$ for the equation of state. However, all familiar forms of matter in the Universe satisfy the so-called \textit{strong energy condition} $w>-1/3$. In \cref{sec:scfinf}, we will see how this problem is solved by introducing a scalar field as a source of inflation.
\subsection{Flatness explained}
Before proceeding, let us see how a period of accelerated expansion explains the spatial flatness of the Universe. Allowing for a nonzero spatial curvature $K$, the first Friedmann equation can be written as:
\begin{equation}
	H^2=\frac{\rho}{3 M^2_{\rm Pl}}-\frac{K}{a^2}.
\end{equation}
The spatial curvature of the Universe can be quantified as a deviation of the energy-density of the Universe from the critical density $\rho_c=3M^2_{Pl}H^2$:
\begin{equation}
\frac{\rho}{\rho_c}=1+\frac{K}{(aH)^2}.
\end{equation}
The critical value of the energy-density $\rho=\rho_c$ corresponds to $K=0$. During the 60 $e$-folds of inflation $aH$ drastically increases, and the energy density of the Universe converges to this critical value. Therefore, whatever the value of $K$ at the beginning of inflation, the residual curvature after inflation will be extremely small. This explains why the spatial curvature of the Universe is very small. 
\section{Scalar field inflation}
\label{sec:scfinf}
Let us assume that the matter content of the early Universe was dominated by a homogeneous scalar field $\phi(t)$, the so-called inflaton field. The action that describes a Universe filled with a scalar field, and its interaction with gravity, is the following:
\begin{equation}
\label{eq:action}
S=\int d^4x \text{ } \sqrt{-g}\biggl(\frac{1}{2}M^2_{\rm Pl}{R}-\frac{1}{2}g^{\mu\nu}\partial_\mu\phi\partial_\nu\phi-V(\phi)\biggr),
\end{equation} 
$V(\phi)$ is the inflaton potential and ${R}$ is the Ricci scalar. Varying the action \eqref{eq:action} with respect to the field, assuming $\phi=\phi(t)$, yields to the Klein-Gordon equation:
\begin{equation}
\label{eq:KG}
\partial_\tau^2\phi+2\mathcal{H}\partial_\tau{\phi}=-a^2\frac{\partial V(\phi)}{\partial{{\phi}}},
\end{equation}
where $\mathcal{H}=a^{-1}\partial_\tau a$. This equation determines the motion of the inflaton.

Computing the stress-energy tensor as a functional derivative of the action gives the energy density and pressure associated with the scalar field\footnote{We are assuming a spatially homogeneous inflaton field $\phi=\phi(t)$}:
\begin{equation}
T_{\mu\nu}=-\frac{2}{\sqrt{-g}}\frac{\delta S}{\delta g^{\mu\nu}}\quad\Longrightarrow\quad
\begin{aligned}
\rho&= -T^{0}_0=\frac{1}{2}\dot{\phi}^2+V(\phi) \\
p&= \frac{1}{3}\sum_iT^{i}_i=\frac{1}{2}\dot{\phi}^2-V(\phi).
\end{aligned}
\end{equation} 
In \cref{sec:inflationwhy}, and in particular from \cref{eq:friedH}, we have seen that the Universe can undergo an accelerated expansion if $w=p/\rho<-1/3$. From these equations, we can see that this can easily be achieved by the scalar field if the potential is flat enough, i.e. $\dot{\phi}\ll V(\phi)$. This is usually called slow-roll condition. In particular, one can see from the Friedmann equations that the limit $\dot{\phi}\rightarrow 0$, in which the scalar field is frozen, corresponds to a de Sitter Universe $a\propto e^{Ht}$, where $H$ is a constant and the Universe expands exponentially. 

\begin{figure}
	\centering
	\includegraphics[scale=.35]{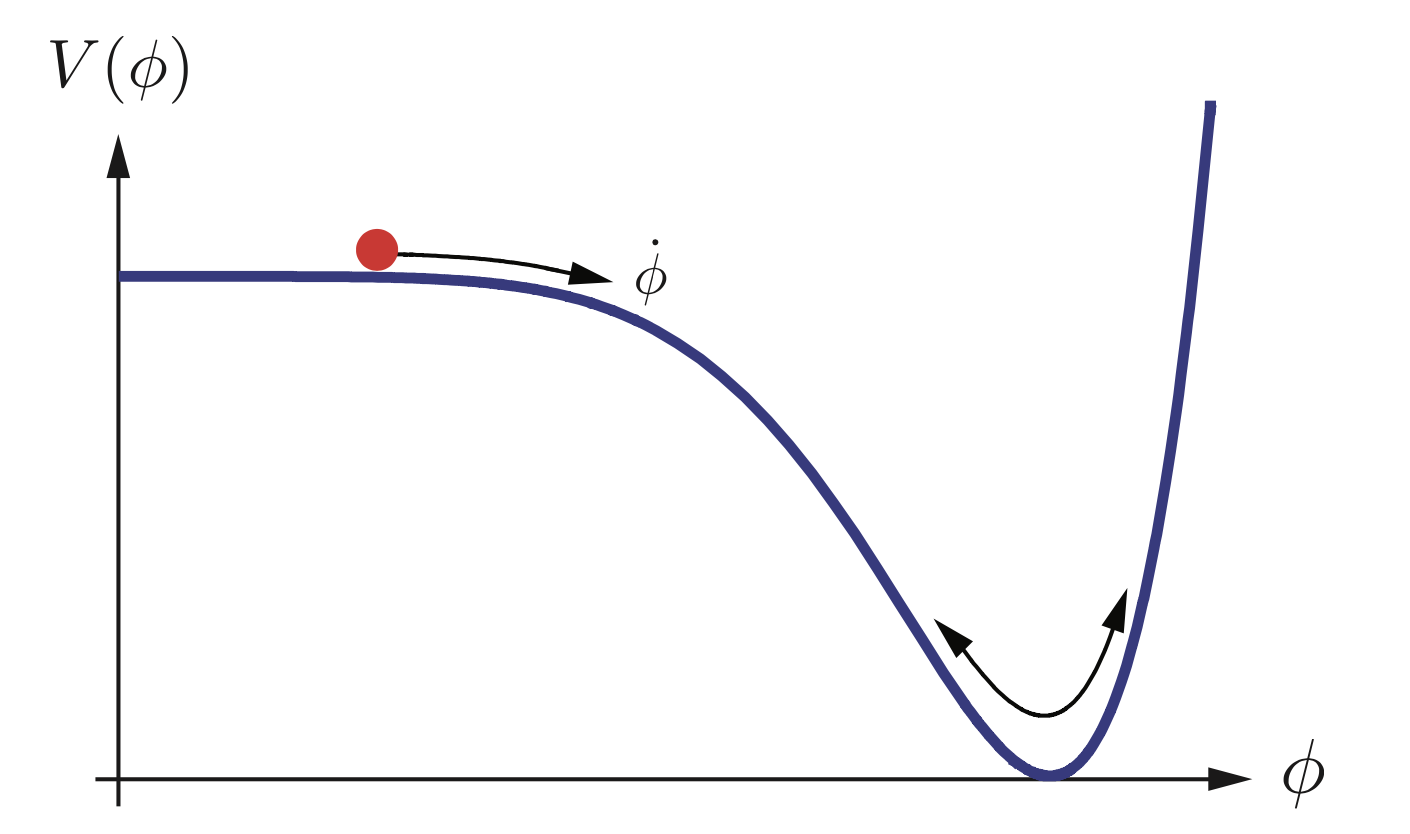}
	\caption{A schematic picture of a potential giving rise to slow roll inflation. This plot is taken from Ref. \cite{Baumann:2009ds}.} \label{fig:baumanninf}
\end{figure}

Although an exponential expansion is appealing to solve the horizon problem, de Sitter inflation is problematic because the acceleration goes on forever and one has to assume some other mechanism to end inflation. For this reason, one usually assumes that the inflaton $\phi$ is in a flat region of the potential for $N_e>60$ $e$-folds, resulting in a quasi-de Sitter expansion during this time. Later, the field slowly reaches a minimum of the potential, where it starts oscillating and inflation ends. This picture is well illustrated by \cref{fig:baumanninf}.

\section{Quantum origin of perturbations}
\label{sec:quantum}
In the last section, we introduced the inflaton as a function of time only $\phi=\phi(t)$. This was enough to explain the accelerated expansion of the Universe, needed to solve the horizon problem, and gave a very natural mechanism to explain why the Universe is spatially flat today. We shall now see how inflation provides also natural way to generate fluctuations in the matter content of the Universe, which are observed as small anisotropies in the CMB and are the seeds for formation of large-scale structures. This is the most valuable prediction of the inflationary model.

We start by allowing the inflaton to have a perturbation on top of its background value:
\begin{equation}
\phi(t)\rightarrow \phi(\vec{x},t)=\bar \phi(t)+\delta\phi(\vec{x},t),
\end{equation}
where we assume $\delta\phi\ll\phi$, i.e. that the perturbation is small. This ensures that the perturbations do not influence the accelerated expansion induced by $\bar \phi(t)$ discussed in the previous section. Moreover, this is physically well-motivated as CMB observations tell us that perturbations in the early Universe were very small, at least on large cosmological scales. 

We now review the standard procedure for analyzing inflationary perturbations, which can be summarized in the following two steps:

\begin{enumerate}
	\item We first see how $\delta\phi$ must be different from zero if we think of the inflaton as a quantum field. This will also determine the shape of inflationary perturbations at very small scales, corresponding to the asymptotic past of the inflationary Universe.
	
	\item We study the evolution of $\delta\phi$ during inflation using the well-established cosmological perturbation theory. 
\end{enumerate}
As we will see in \cref{sec:inflationsim}, the approach developed in this thesis is different from this standard picture. In our case, we only use step 1 and we substitute step 2 with the lattice simulation to evolve the quantum perturbations.

\subsection{Quantizing inflationary perturbation}
\label{sec:quantization}
Although we might not know the laws of physics during inflation, it is natural to assume that the inflaton is a quantum field described by relativistic quantum mechanics, just like the fields involved in the standard model of particle physics. In this framework, the inflaton is promoted to a quantum operator:
\begin{equation}
\label{eq:quantization}
\hat{\delta\phi}(\vec{x},\tau)=\int\frac{d^3\vec{k}}{(2\pi)^{3/2}}\Bigl[{\hat a}_{\vec{k}}{\delta\phi}(\vec{k},\tau)e^{i\vec{k}\cdot\vec{x}}+{\hat a}_{\vec{k}}^\dagger{\delta\phi}^\ast(\vec{k},\tau) e^{-i\vec{k}\cdot\vec{x}}\Bigr],
\end{equation}
where $\hat a$ and $\hat a^\dagger$ are the creation and annihilation operators satisfying $$[\hat a_{\vec{k}},\hat a^\dagger_{\vec{k}^\prime}]=\delta(\vec{k}-\vec{k}^\prime).$$ The fact that we are working with an expanding spacetime introduces an ambiguity in choosing the vacuum state of the theory and in identifying the corresponding mode function $\delta\phi(\vec{k},\tau)$. In the case of inflation, this ambiguity is solved with a physical input. On comoving length scales much smaller than the Hubble length $L\ll 1/(aH)$, the field should not feel any effect induced by the spacetime curvature. Therefore,
it is natural to assume that at these length scales the field looks like a massive free quantum scalar field in Minkowski spacetime, implying:
\begin{equation}
\label{eq:BD}
{\delta\phi}(\vec{k},\tau)=\frac{1}{a\sqrt{2\omega_k}}e^{- i \omega_{k}\tau},\quad\quad\quad \omega_{k}^2=k^2+m^2,\quad\quad k\gg aH,
\end{equation}
where $m=V^{\prime\prime}(\phi)$ is the mass of the inflaton. This is called the Bunch-Davies vacuum. The condition $k\gg aH$ is time-dependent, as $aH$ increases during inflation. Therefore, this condition is valid for every mode if we go far enough in time. Note that we have introduced a scale factor $a$ in the denominator of \cref{eq:BD}. This is because the kinetic term for $\phi$ in the action of \cref{eq:action} is not canonically normalized like in Minkowski space. A field redefinition $\phi\rightarrow a\phi$ makes the action canonically normalized. In other words, only the rescaled field $a\phi$ behaves like the canonically normalized scalar field in Minkowski spacetime in the asymptotic past. This will be evident in the next section.




\subsection{Cosmological perturbation theory}
\label{sec:cpt}
Now that we have seen that the inflaton must have some spatial fluctuations as a result of its quantum nature, we study the evolution of perturbations during inflation.
The main problem in studying the evolution of perturbations in general relativity is that fluctuations in the matter content of the Universe, such as $\delta\phi$, will also introduce perturbations in the metric:
\begin{equation}
	g_{\mu\nu}= \bar{g}_{\mu\nu}+\delta g_{\mu\nu},
\end{equation}
where $ \bar{g}_{\mu\nu}$ is the unperturbed FLRW metric. Therefore, studying the evolution of $\delta\phi$ is not enough.
\subsubsection*{SVT decomposition}
 The most general perturbed metric around a spatially flat FLRW background with conformal time $\tau$ can be written as:
\begin{equation}
\label{eq:pertmet}
ds^2=a^2(\tau)\biggl((1+2A)d\tau^2-2B_idx^id\tau-(\delta_{ij}+h_{ij})dx^idx^j\biggr).
\end{equation}
The perturbations $A$, $B_i$ and $h_{ij}$ can be decomposed in a clever way using the so-called Scalar-Vector-Tensor (SVT) decomposition. According to the SVT decomposition, one splits the three degrees of freedom of $B_i$ writing it in the following way: $B_i=\partial_iB+\hat{B}_i$ where $B$ is a scalar and $\hat{B}_i$ a transverse vector, such that $\partial^i\hat{B}_i=0$. The same is done for $h_{ij}$, which can be written as $h_{ij}=2\Psi\delta_{ij}+2(\partial_i\partial_j-\frac{1}{3}\delta_{ij}\nabla^2)E+\partial_i\hat{h}_j+\partial_j\hat{h}_i+\hat{h}_{ij}$, where $\Psi$ and $E$ are scalars, $\hat{h}_i$ a transverse vector and $\hat{h}_{ij}$ a traceless tensor. 

Thanks to this decomposition, one can separately describe the scalar, vector, and tensor perturbations of the metric. In this thesis, we mostly focus on scalar perturbations. We completely neglect vector perturbations, as one can show that they get suppressed very quickly during inflation. Tensor perturbations are important, but we momentarily neglect them as 
they do not play a role in the evolution of the scalar sector at linear order (i.e. if they are small). In the end, we can describe the perturbation of the metric with 4 scalar quantities $A$, $B$, $\Psi$ and $E$.
\subsubsection*{Gauge redundancy}

Perturbing the metric introduces a redundancy in the degrees of freedom that are used to describe the system, which is somewhat similar to what happens with gauge field theories (such as electromagnetism). 
The redundancy comes from the fact that if we do a coordinate transformation $x^{\mu}\rightarrow x^\mu + \xi^\mu$, where $\xi^\mu$ is an infinitesimal vector, perturbation quantities $\delta \phi$ and $\delta g_{\mu\nu}$ eventually change. In other words, the definition of perturbations depends on the particular coordinate choice. 

The gauge redundancy is a symptom that out of the 4 scalar degrees of freedom $A$, $B$, $\Psi$ and $E$ introduced above, only 2 are physical. There are two ways to deal with this problem. The first is to find gauge invariant quantities and work out their evolution. The second approach is to fix a gauge and then perform all computations in that given gauge. We will follow this second approach. In particular, throughout this thesis we will implicitly assume unless specified, that we work in the so-called spatially flat gauge $\Psi=E=0$. As we are focusing on scalar perturbations, this is equivalent to set $\delta g_{ij}=0$, and this is why it is called spatially flat gauge.

Although we work in this fixed gauge, it is still meaningful to introduce the following gauge invariant quantity \cite{Riotto:2002yw}:
\begin{equation}
\label{eq:R}
\mathcal{R} = \Psi+H\frac{\delta\phi}{\dot\phi}\,\,\overset{\text{spatially flat}}{=}\,\, H\frac{\delta\phi}{\dot\phi}.
\end{equation}
This is called \textit{comoving curvature perturbation} and is a combination of the inflaton perturbation $\delta\phi$ and the gravitational potential $\Psi$ (that we set to 0 by gauge choice). This quantity has a very important property: it freezes on super-horizon scales. Indeed, one can prove that $\dot{\mathcal{R}}_k=0$ for every mode $k\ll aH$ at all orders in perturbation theory \cite{Lyth:2004gb}. For this reason, once a given mode becomes super-horizon $k\ll aH$, its comoving curvature perturbation will remain frozen until it renters the horizon later after inflation. Another gauge invariant and physically meaningful quantity is the \textit{curvature perturbation on slices of uniform energy density} \cite{Riotto:2002yw}:
\begin{equation}
\label{eq:zeta}
\zeta = \Psi+H\frac{\delta\rho}{\dot{\rho}},
\end{equation}
that, by definition, coincides with the gravitational potential $\Psi$ in the gauge of uniform energy density $\delta\rho=0$. One can show that $\zeta\simeq\mathcal{R}$ at leading order in slow-roll expansion and on super-horizon scales \cite{Riotto:2002yw}, which makes this quantity approximately conserved for $k\ll aH$.

\subsection{Scalar perturbations from inflation}
\label{sec:scalarpert}
Studying the evolution of scalar perturbations at linear order is rather nontrivial. In order to do so, one needs to expand the action of \cref{eq:action} at second order in perturbations $\delta\phi$, $A$, and $B$ and derive the corresponding equations of motions. Then, the equations of motion for $A$ and $B$ have to be solved at linear order to eliminate these variables in favor of $\delta\phi$. The final result is the famous Mukhanov-Sasaki (MS) equation for scalar fluctuations:
\begin{equation}
\label{eq:MS}
\partial^2_\tau v +(k^2+m^2_{\rm eff}(\tau))v=0,
\end{equation}
where $v(\vec{k},\tau)=a\delta\phi(\vec{k},\tau)$ is the Mukhanov variable\footnote{The gauge-invariant expression for the Mukhanov variable is $v=a(\delta\phi-\dot{\bar{\phi}} \psi/H)$, but we are working in the spatially flat gauge $\psi=0$.} and $k=|\vec{k}|$. This equation describes an harmonic oscillator with a time dependent mass term:
\begin{equation}
m^2_{\rm eff}(\tau)=-\frac{H}{ \dot{{\bar\phi}}}\partial_\tau^2(\dot{{\bar\phi}}/H).
\end{equation}
The time dependence of the mass is strictly related to the problem of defining the vacuum state of the theory discussed in \cref{sec:quantization}. After identyifing the vacuum state of the theory, which corresponds to setting \cref{eq:BD} as the initial conditions of \cref{eq:MS}, one can numerically solve this equation together with \cref{eq:KG} to determine the evolution of $\delta\phi$ in time.

\subsubsection*{de Sitter limit}
In order to derive an analytical solution, let us assume an exact de Sitter space $a=e^{Ht}$. This corresponds to neglecting metric perturbations as well as the slow-roll corrections due to the quasi-de Sitter dynamics. In this case, the effective mass appearing in the MS equation simplifies to:
\begin{equation}
\label{eq:MSmassive}
m^2_{\rm eff}=m^2a^2-\frac{2}{\tau^2}.
\end{equation}
where $m=V^{\prime\prime}(\phi)$ is the mass of the inflaton. We can write an analytical solution to the MS equation with the initial conditions given by \cref{eq:BD} in the asymptotic past $\tau=-\infty$. The solution can be written as:
\begin{equation}
\label{eq:solution}
v(\vec{k},\tau)=a{\delta\phi}(\vec{k},\tau)=\frac{\sqrt{-\pi \tau}}{2}H_\mu^{(1)}(-k\tau),\quad\quad \mu^2=\frac{9}{4}-\frac{m^2}{H^2},
\end{equation}
where $H_\mu^{(1)}$ is the modified Hankel function of the first kind. We can use this solution to write the power spectrum of inflaton perturbations $P_{\phi}(k)=|{\delta\phi}(\vec{k},\tau)|^2$. Perturbations are usually described by the dimensionless power spectrum, defined as:
\begin{equation}
\label{eq:thprediction}
\mathcal{P}_{\phi}(k)=\frac{k^3}{2\pi^2}P_{\phi}(k)=\frac{H^2}{8\pi}(-k\tau)^3|H_\mu^{(1)}(-k\tau)|^2.
\end{equation}
Using the asymptotic behavior of the Hankel function, and assuming $m\ll H$, one can simplify this result on super-horizon scales:
\begin{equation}
\label{eq:thprediction_simple}
\mathcal{P}_{\phi}(k)=\frac{H^2}{(2\pi)^2},\quad\quad\quad k\ll aH.
\end{equation}
This is the famous scale invariant spectrum of primordial perturbations. Before proceeding with the lattice simulation, let us briefly discuss some observational constraints on inflationary perturbations.

\subsubsection*{Observational constraints}

Taking into account gravitational effects and quasi-de Sitter corrections to the background dynamics results in a weak momentum dependence of the power spectrum, that will depend on the particular shape of inflationary potential $V(\phi)$. This is usually parameterized by the following parameter:
\begin{equation}
	n_s=1+\frac{d\, \mathcal{P}_{\mathcal{\zeta}}(k)}{d\, \log k},
\end{equation}
that is typically defined from the power spectrum of the curvature perturbation $\mathcal{\zeta}$. This parameter can be constrained using the power spectrum measured from the CMB radiation. The latest results from the Planck satellite are:
\begin{equation}
	n_s=0.9649 \pm 0.0044
\end{equation}
at 68\% confidence level \cite{Planck:2018jri}. This is perfectly compatible with slow-roll inflation, which predicts a small deviation from $n_s=1$ as a consequence of the quasi-de Sitter dynamics.

Another feature of the scalar perturbations predicted from inflation is their Gaussian statistics. Indeed, one can go to higher order in perturbation theory and compute non-Gaussianity such as the three-point function of inflationary perturbation $\langle \delta\phi(k_1)\delta\phi(k_2)\delta\phi(k_3)\rangle$. The result happens to be undetectably small \cite{Maldacena:2002vr}, and this is compatible with all current observations, such as the ones from Planck \cite{Planck:2019kim}. 

These two predictions are the strongest evidence in favor of the scalar field model of inflation introduced in this chapter. But they are not the only ones. 
Single-field inflation also predicts perturbations in the tensor sector in the form of gravitational waves, which are in principle observable using the polarization of the CMB radiation. Their signature, however, remains undetected. Observing gravitational waves from inflation would tell us the energy scale $H$ at the time of emission, which is not possible using only the scalar power spectrum. Nevertheless, the observational constrains on the tensor power spectrum, together with the precise measurement of $n_s$, give already important information about the allowed shapes of inflationary potential $V(\phi)$. Some types of slow-roll potentials, like the simple quadratic potential $\frac{1}{2}m^2\phi^2$, are already ruled out by observations \cite{Planck:2018jri,BICEP:2021xfz}.

\cleardoublepage
\newpage\null
\chapter{Lattice simulations of inflation}
\label{sec:inflationsim}
We now introduce the lattice simulation as a numerical tool to study the inflationary universe. In this chapter, we focus on the single-field model of inflation presented in \cref{sec:inflation}. A nonlinear lattice simulation is not needed to understand the physics of this model, which lies well within the regime of validity of linear perturbation theory. However, recovering the well-known results is a necessary step if we want to use the simulation to understand more complicated inflationary models beyond perturbation theory, which will be the topic of \cref{sec:axionsim}.

We will mostly focus on the conceptual issues of simulating the inflationary universe on the lattice and neglect many details about the numerical implementation. Our code is inspired on LATTICEEASY \cite{latticeeasy}, a publicly available lattice code that has been developed to study the reheating phase of the universe. Similar to LATTICEEASY, our code is written in C++ and it is OpenMP parallelized. Moreover, we inherit various numerical routines from LATTICEEASY, like the computation of lattice Fourier transform \cite{ffteasy} and the way we deal with the periodic boundary conditions. Our code, however, is substantially different from LATTICEEASY. As we will see, we generate initial conditions in a different way, use a different numerical integrator for the equations of motion, and have different outputs routines. In the relevant parts of the text, we will highlight which are the techniques inherited from LATTICEEASY. 

The content of this chapter is based on Ref. \cite{Caravano_2021}, and constitutes original results from the doctoral studies.
In \cref{sec:lattice} we introduce the lattice approach for inflation and list the conceptual steps that are followed in the rest of the chapter. The results of the simulations are mostly contained in \cref{sec:results}.

\section{The lattice approach}
\label{sec:lattice}
The idea behind a lattice simulation is simple and it consists of simulating the dynamics of continuum fields on a finite cubic lattice. The lattice is defined as a collection of $N^3$ points separated by comoving lattice spacing $\Delta x=L/N$, where $L$ is the comoving physical size of the box. To any given field $f(x)$ in continuous space, we associate $N^3$ values to each point of the cubic lattice:
\begin{equation}
\label{eq:discretization}
f(\vec{x}), \quad \vec{x}\in \mathbb{R}^3\quad\quad \longrightarrow \quad\quad f(\vec{n}),\quad \vec{n}\in\mathbb{N}^3,\quad n_i\in\{1,\dots,N\}.
\end{equation}
We take the lattice to be periodic so that, for example, $f(N,n_2,n_3)=f(1,n_2,n_3)$.

Contrarily to what is done in perturbation theory, in the simulation we do not split in background and perturbation quantities. Indeed, the inflaton is evolved altogether using the classical Euler-Lagrange equations in real space. In the case of the single scalar field model of \cref{eq:action}, the equation of motion for the inflaton is the following:
\begin{equation}
\label{eq:fullcont}
\partial_\tau^2\phi+2\mathcal{H}\partial_\tau{\phi}-\nabla^2\phi+a^2\frac{\partial V}{\partial \phi} =0,
\end{equation}
To derive this equation we assumed a simple unperturbed FLRW metric, neglecting the curvature of the spacetime induced by the inhomogeneities. We will discuss later the reasons behind this assumption. 

To solve this equation, we associate $N^3$ values to the inflaton as in \cref{eq:discretization}. In this way, \cref{eq:fullcont} becomes:
\begin{equation}
\label{eq:disc}
\partial_\tau^2\phi(\vec{n})+2\mathcal{H}\partial_\tau{\phi}(\vec{n})-[\nabla^2\phi](\vec{n})+a^2\frac{\partial V}{\partial \phi}(\vec{n}) =0.
\end{equation}
Although they look similar, \cref{eq:fullcont} and \cref{eq:disc} are fundamentally different. 
While the former is a single partial differential equation (PDE) in the field $\phi(\vec{x})$, the latter constitutes a set of $N^3$ ordinary differential equations (ODE), one for each lattice point $\phi(\vec{n})$. These equations are coupled to each other through the discrete Laplacian $[\nabla\phi](\vec{n})$, which we will define below. 

This chapter is dedicated to numerically solving this set of equations. In order to do so, the following ingredients are required:

\begin{itemize}
	\item \textit{Discretization scheme}. After defining the lattice as a collection of $N^3$ points, we need to define how these points are connected. This corresponds to specifying how the set of equations in \cref{eq:disc} are coupled to each other, and it is given by the definition of the discrete Laplacian $[\nabla\phi](\vec{n})$. This will be the topic of \cref{sec:disc}.
	\item \textit{Spacetime evolution}. Spacetime is evolved assuming a FLRW metric and neglecting metric perturbations. This means that, to evolve the metric, we just study the evolution of the scale factor $a$ in \cref{eq:disc}.  We will justify this important assumption. The evolution of spacetime will be discussed in \cref{sec:gravity}
		\item \textit{Initial conditions}. One of the main ingredients in solving any differential equation is choosing the initial conditions. This is where the quantum nature of the fields involved in the simulation is relevant, and will be discussed in \cref{sec:IC}.
	\item \textit{Numerical integrator}. After fixing all the previous ingredients, we need to choose a numerical integrator to evolve this system of $2N^3+2$ values: $2N^3$ values for the field and their time derivatives, plus the scale factor and its time derivative. This is done in \cref{sec:numint}.

		\item \textit{Outputs}. Last but not least, we need to use the lattice simulation to compute observable quantities. To do so, we need to identify the physical properties of the lattice that are independent of the numerical implementation. In \cref{sec:output} we describe how outputs are computed in our code, and in \cref{sec:results} we show the results from the simulation.

\end{itemize}
Before proceeding to discuss these topics one by one, let us briefly discuss the justification of the semi-classical lattice approach.
\subsection{The semi-classical approximation}
In \cref{sec:quantum}, we saw that the small inhomogeneities in the early universe can be described as quantum fluctuations of the inflaton field. In the lattice approach, however, we use the classical equations of motion to evolve the inflaton field from an initial configuration $\phi_i(\vec{n})$ to a final one $\phi_f(\vec{n})$. These configurations are fully described by their numerical values across the $N^3$ points of the lattice (plus the values of its velocity $\phi^\prime$). This picture is very accurate when describing the inflationary universe on super-horizon scales, i.e. when the comoving lattice size is bigger than the Hubble horizon $L>aH$. In this case, the quantum properties of the inflaton are negligible, and the inflaton is fully determined by its configuration in real space. As we will see in \cref{sec:initialconditions}, we start the simulation when the size of the box is smaller than the horizon, $L<aH$. At these length scales, describing the inflaton and its velocity as a (deterministic) collection of $N^3$ real space values is not realistic. In order to mimic the uncertainty related to the quantum nature of the Universe at these scales, we take a statistical point of view and think of the $N^3$ values of the inflaton as random realizations of a stochastic process. This statistical sampling will be done at the initial time, and it is described in \cref{sec:initialconditions}. 

This approach represents a semi-classical approximation. As we will see, thinking of the inflaton as a stochastic classical field turns out to be a good approximation in predicting the statistical properties of the inflationary universe on large scales. This should not come as a surprise: the Mukhanov-Sasaki \cref{eq:MS} is a tree-level classical equation for the perturbations, which does not incorporate any quantum effect. This semi-classical approximation can be better understood using the path integral formulation of quantum mechanics. In this framework, the probability of having a given field configuration $\phi_f(\vec{x},t_f)$ at time $t=t_f$ starting from an initial one $\phi_i(\vec{x},t_i)$ at $t=t_i$ is written as:
\begin{equation}
	P\{ \phi_f(\vec{x}),t_f \rvert \phi_i(\vec{x}),t_i\}=\left|\int_{\phi(\vec{x},t_i)}^{\phi(\vec{x},t_f)}[\mathcal{D}\phi]e^{i\frac{\mathcal S[\phi(\vec{x},t)]}{\hbar}}\right|^2,
\end{equation}
where the path integral is the sum over all paths that bring the system from the initial configuration $\phi_i(\vec{x})$ to the final one $\phi_f(\vec{x})$. Out of all possible paths, the classical trajectory is only one, and it is the solution to \cref{eq:fullcont} with initial condition $\phi_i(\vec{x},t_i)$. Let us call this trajectory $\phi_{\rm cl}(\vec{x},t)$. Our semi-classical approximation represents the case in which the path integral is dominated by the classical contribution $\mathcal S[\phi_{\rm cl}(\vec{x},t)]\equiv \mathcal S_{\rm cl}\gg \hbar$. In this case, the path integral can be expanded around the saddle point corresponding to the classical trajectory. Expanding all trajectories around the classical one as $\phi(\vec{x},t)=\phi_{\rm cl}(\vec{x},t)+\xi(\vec{x},t)$, the path integral can be rewritten as \cite{Rattazzi2009ThePI,Celoria:2021vjw}:
\begin{equation}
\int [\mathcal{D}\phi]e^{i\frac{\mathcal S[\phi(\vec{x},t)]}{\hbar}}=e^{i\frac{\mathcal{S}_{\rm cl}}{\hbar}}\int[\mathcal{D}\phi] e^{\frac{i}{h}\left(\frac{\delta \mathcal{S}[\phi(\vec{x},t)]}{\delta \phi(\vec{x},t)}\rvert_{ \phi_{\rm cl}}\xi(\vec{x},t)+\frac{\delta^2 \mathcal{S}[\phi(\vec{x},t)]}{\delta \phi^2(\vec{x},t)}\rvert_{ \phi_{\rm cl}}\xi^2(\vec{x},t)+\dots \right)}\simeq e^{i\frac{\mathcal{S}_{\rm cl}}{\hbar}}.
\end{equation}
The last step, where we neglected the quantum contributions deviating from the classical trajectory, represents our semi-classical approach. The effects of quantum corrections on inflationary observables have been shown to be negligible \cite{Weinberg:2005vy,Weinberg:2006ac,Sloth:2006nu,Seery:2007wf,Gorbenko:2019rza,Senatore:2009cf}. Nevertheless, a full understanding of the quantum corrections due to gravitational interactions is still an open problem in inflationary cosmology. In fact, it has been shown that the quantum properties might cause a departure from the semi-classical trajectory if such corrections accumulate over a long time \cite{Dvali:2017eba}. These effects, however, are not relevant during the few $e$-folds of the lattice simulation. In this thesis, we are going to neglect all quantum effects in the evolution of inflationary perturbations, and we assume that the inflaton can be approximately described as a stochastic classical field during the $N_e\simeq 7$ e-folds of simulation. 






\section{Discretization scheme}
\label{sec:disc}
Out of all the ingredients involved in studying inflation on the lattice, the discretization scheme is probably the most important one. This should not come as a surprise: a field theory on a discrete structure behaves differently than a field theory in continuous space. A good discrete field theory will approach the continuous one in the limit where the separation of the points $dx=L/N$ goes to zero. On the lattice, however, we have a finite $dx$, which makes it important to understand the implications of the discretization. This is the purpose of this section.
\subsection{Discrete Fourier Transform}
Before proceeding, we need to define the Fourier transform on the lattice. First, we introduce the reciprocal lattice, defined by the following discrete momenta:
\begin{equation}
\label{eq:modes}
\vec{\kappa}_{\vec{m}}=\frac{2\pi}{L}\vec{m},\quad\quad m_i\in{1,...,N}.
\end{equation}
On the reciprocal lattice, the discrete Fourier transform (DFT) of a field $f(\vec{n})$ is defined as follows:
\begin{equation}
\label{eq:DFT}
\text{ DFT}[f](\vec{\kappa}_{\vec{m}})\equiv {f}(\vec{\kappa}_{\vec{m}})=\frac{dx^3}{N^{3}}\sum_{\vec{n}}f(\vec{n})\text{ }e^{-i\frac{2\pi}{N}\vec{n}\cdot\vec{m}}.
\end{equation}
We adopt the same convention of continuous space, for which we call a Fourier transform of a given field $f(\vec{n})$ using the same letter but with different argument $f(\vec{\kappa})$.

The Fourier transform in \cref{eq:DFT} is analogous to the continuous one except for the different normalization, which is just a convention. The prefactor $N^3$ is analogous to the $(2\pi)^3$ in the continuous transform. The prefactor $dx^3$ is slightly more subtle, and it is introduced to take into account the physical discretization of space. It comes from the $d^3x$ appearing in the integral inside the definition of the continuous Fourier transform:
\begin{equation}
{f}(\vec{k})=\frac{1}{(2\pi)^{-3/2}}\int d^3x f(\vec{x})e^{-i\vec{k}\cdot\vec{x}},
\end{equation}
and it is introduced to make the DFT dimensionally analogous to the continuous one.
With these definitions, we can write the inverse DFT (iDFT) as follows:
\begin{equation}
\label{eq:iDFT}
f({\vec{n}})=\frac{1}{dx^3}\sum_{\vec{m}}{f}(\vec{\kappa}_{\vec{m}})\text{ }e^{+i\frac{2\pi}{N}\vec{m}\cdot\vec{n}}.
\end{equation}
\subsection{Discrete Laplacian and the effective momenta }

\label{sec:modifieddr}
In this section, we describe the effect of the discretization on the propagation of Fourier modes on the lattice. We will mainly consider the following standard definition of discrete Laplacian operator \cite{press1986numerical}:
\begin{equation}
\label{eq:discretelaplacian}
[\nabla^2f](\vec{n})=\frac{1}{(\Delta x)^2}\sum_{\alpha=\pm1}\biggl(f(\vec{n}+ \alpha\vec{e}_1)+f(\vec{n}+ \alpha\vec{e}_2)+f(\vec{n}+\alpha \vec{e}_3)-3f(\vec{n})\biggr), 
\end{equation}
where $\vec{e}_1=(1,0,0),\text{ }\vec{e}_2=(0,1,0),\text{ }\vec{e}_3=(0,0,1)$. This Laplacian converges to the continuous one for $\Delta x \rightarrow 0$, and for a finite $\Delta x$ it has a second order truncation error $O(\Delta x^2)$ with respect to the continuous one.
\subsubsection*{The effective momenta}
\begin{figure}
	\centering
	
	\begin{tikzpicture}
	\node (img) {\includegraphics[width=8cm]{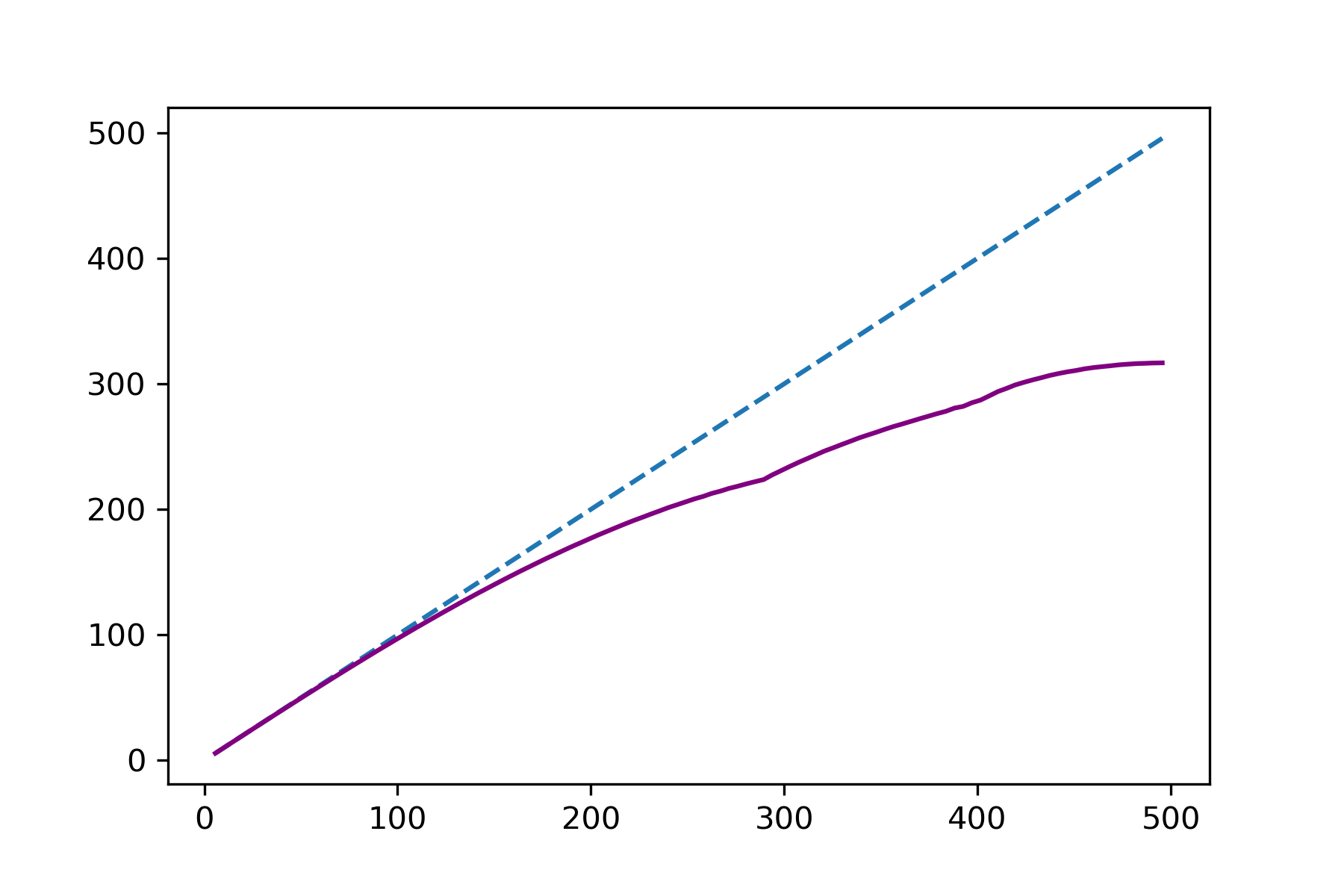}};
	
	\node [rotate=0,text width=0.01cm,align=center] at (-4.5,0){ $k_{\text{eff}}$};
	\node [text width=0.01cm,align=center] at (-0.3,-2.8){ $\kappa$};

	\end{tikzpicture}

	\caption{The dispersion relation of modes on the lattice. On the $y$-axis, we show $k_{\text{eff}}$ obtained from \cref{eq:keff}, while on the $x$-axis we show the lattice modes of \cref{eq:modes}. The departure from the diagonal is a manifestation of the modified dispersion relation induced by lattice spacing.}
	\label{fig:dispersion}
\end{figure}

 In continuous space, the Fourier transform (FT) of the Laplacian operator for differential equations is quite simple and reads:
\begin{equation}
\nabla^2\phi(\vec{x})\quad\xrightarrow{\text{FT}}\quad-k^2\phi(\vec{k}).
\end{equation}
As it is well known, this relation gets modified on the lattice \cite{press1986numerical}, where we transform the field with the DFT. It can be easily derived from  \cref{eq:discretelaplacian} and  \cref{eq:iDFT} that:
\begin{equation}
\label{eq:ftkeff}
\normalfont [\nabla^2\phi](n)\quad \xrightarrow{\text{DFT}}\quad-k_{\text{eff}}^2(\vec{\kappa}_{\vec{m}}){\phi}({\vec{\kappa}_{\vec{m}}}) ,
\end{equation}
where we introduced the effective modes $k_{\text{eff}}$ as:
\begin{equation}
\label{eq:keff}
k_{\text{eff}}(\vec{\kappa}_{\vec{m}}) = \frac{2}{dx}\sqrt{\sin^2\left(\frac{\pi m_1}{N}\right)+\sin^2\left(\frac{\pi m_2}{N}\right)+\sin^2\left(\frac{\pi m_3}{N}\right)}.
\end{equation}
Contrarily to what happens in the continuous case, this relation differs significantly from the value of the modes of the reciprocal lattice defined in \cref{eq:modes} $	k_{\text{eff}}(\vec{\kappa}_{\vec{m}}) \neq \vec{\kappa}_{\vec{m}}$. Indeed, $\kappa$ and $k_{\text{eff}}$ are only equal in the limit $m_1,m_2,m_3\ll N$. In \cref{fig:dispersion} we show the difference between $\kappa$ and $k_{\rm eff}$ for a lattice with $N=128$ and $L=1.4$. The one-dimensional quantities in the plot are obtained from \cref{eq:keff} and \cref{eq:modes} averaging over spherical bins on the lattice.
Note that the expression for $k_{\rm eff}$ of \cref{eq:keff} depends on the definition of the lattice Laplacian of \cref{eq:discretelaplacian}. A different choice of the numerical stencil for the Laplacian would lead to a different expression for $k_{\rm eff}$, as discussed below. In \cref{sec:results} we will introduce other stencils for the Laplacian operator and discuss the consequences on the dynamics of the simulation.
\subsubsection*{Evolution of perturbations during inflation}
We can interpret \cref{eq:ftkeff} as a modified dispersion relation induced by the discrete spacing on the modes propagating on the lattice. Indeed, if we look for example at the equation for a free, massless scalar field on the lattice 
\begin{equation}
\partial^2_\tau{\phi}({\vec{n}})=[\nabla^2\phi](\vec{n}),
\end{equation}
we can operate a DFT to obtain:
\begin{equation}
\partial^2_\tau{\phi}(\vec{\kappa}_{\vec{m}})=-k^2_{\rm eff}(\vec{\kappa}_{\vec{m}}){\phi}(\vec{\kappa}_{\vec{m}}).
\end{equation}
From this equation, we can see that modes will propagate with energy $\omega(\kappa)=k_{\rm eff}\neq \kappa$, which is different from the usual $\omega(k)=k$ of continuous space. In other words, the dynamics of fields on the lattice is different from the one of continuous space.

Let us now analyze the effects of discretization on the evolution of perturbations during inflation. In analogy to the continuous case, let us introduce the following discretized Mukhanov-Sasaki variable $v({\vec{\kappa}_{\vec{m}}})=a{\phi}(\vec{\kappa}_{\vec{m}})$. Let us focus on $\kappa>0$, as $\kappa=0$ describes the evolution of the background. Moreover, let us assume an exact de-Sitter dynamics, similar to what is done in \cref{sec:quantum}. Then, if we operate the DFT on $\cref{eq:fullcont}$, and we expand it to second order in $v({\vec{\kappa}_{\vec{m}}})$, we obtain\footnote{In this equation, $m$ has nothing to do with the lattice index $\vec{m}$. We apologize with the reader for the abuse of notation.}:
\begin{equation}
\label{eq:discMS}
\partial^2_\tau v(\vec{\kappa}_{\vec{m}}) +(k_{\text{eff}}^2(\vec{\kappa}_{\vec{m}})+m^2_{\rm eff}(\tau))v(\vec{\kappa}_{\vec{m}})=0,\quad\quad m^2_{\rm eff}= m^2-\frac{2}{\tau^2},
\end{equation}
where $m=V^{\prime\prime}(\phi)$. 

This equation looks very similar to its continuous counterpart \cref{eq:MS}, except that we have $k^2_{\rm eff}$ instead of $\kappa^2$ inside the parenthesis\footnote{Note that we do not use this equation to evolve $\phi$ on the lattice, as we use the full nonlinear equation \cref{eq:disc}. For the moment, we are just trying to understand how the discretization is expected to influence the dynamics on the lattice.}. This difference reflects our previous intuition: perturbations on the lattice and in continuous space evolve differently, due to discretization. Despite this difference, we notice from \cref{eq:MS} and \cref{eq:discMS} that the dynamics of Fourier modes on the lattice is equivalent to the one of the continuous mode functions $\delta\phi(\vec{k},\tau)$ defined in \cref{eq:quantization}, if we interpret $k_{\rm eff}$ as the physical modes actually probed by the lattice simulation, instead of $\kappa$. This suggests the following equivalence principle: 
$$k_{\rm eff}\leftrightarrow k.$$
In other words, what happens on the lattice at scales $\kappa$ will reflect what happens to the inflationary universe at scales $k=k_{\rm eff}(\kappa)$, and not at scales $k=\kappa$.
This is why we called it \textit{effective} momentum. As we will see in \cref{sec:results}, this equivalence principle turns out to be very useful in interpreting the outputs of the simulation and in computing observables such as the power spectrum from the code.

This modified dispersion relation will also have consequences on the effective spacial resolution of the simulation. Instead of probing modes up to\footnote{More details about $\kappa_{\rm max}$ can be found in \cref{sec:output}.} $k_{\text{lat},\text{max}}=2\pi\sqrt{3} N_{\text{Nyquist}}/L$, where $N_{\text{Nyquist}}=N/2$, it will probe physical modes up to:
\begin{equation}
\label{eq:effmax}
k_{\text{eff},\text{max}} = \frac{2}{dx}\sqrt{3\sin^2\left(\frac{\pi N_{\text{Nyquist}}}{N}\right)}=\frac{2\sqrt{3}}{dx}=\frac{2}{\pi}\kappa_{\rm{max}}.
\end{equation}
This means that the effective range of physical modes evolved by the simulation will be reduced by a factor of $2/\pi\simeq0.64$.

As already mentioned, a different definition of the Laplacian would lead to a different expression for $k_{\rm eff}$ and to a different value of $k_{\rm eff,max}$. In \cref{sec:diff} we show the comparison between the $k_{\rm eff}$ associated with different stencils and we discuss the consequences on the dynamics of the simulation.


\section{Spacetime evolution}
\label{sec:gravity}
\subsection{The role of gravity}
In deriving equation \cref{eq:fullcont} from the action of \cref{eq:action}, we assumed an unperturbed, spatially flat, FLRW metric in conformal time:
\begin{equation}
	ds^2=\bar{g}_{\mu\nu}dx^\mu dx^\nu=a^2(\tau)\left(-d\tau^2+d\vec{x}^2\right).
\end{equation}
This metric describes a perfectly homogeneous universe without spatial perturbations. During inflation, however, the energy content of the universe is perturbed, and we should include metric perturbations of the form $g_{\mu\nu}=\bar{g}_{\mu\nu}+\delta g_{\mu\nu}$. As discussed in \cref{sec:cpt}, we can use the gauge redundancy to set $\delta g_{ij}=0$. However, $\delta g_{i0}$ and $\delta g_{00}$ do play a role in the evolution of perturbations. At linear order in perturbation theory, the role of metric perturbations in the evolution of the field content is slow-roll suppressed. To see this, we can expand the relevant terms in the action \cref{eq:action} to second order in perturbations around the background. The first relevant term is:
\begin{align}
	&\sqrt{-g}g^{\mu\nu}\partial_\mu \phi \partial_\nu \phi=\sqrt{-\bar g}\bar{g}^{\mu\nu}\partial_\mu\phi \partial_\nu \phi \,+\\
	&\quad\quad\quad\quad+{\sqrt{-\bar g}} \partial_0 \bar \phi\left[2 \delta\phi^\prime\delta g^{00}+2\partial_i\delta\phi\delta g^{0i}\right]+{\sqrt{-\bar g}} (\partial_0 \bar \phi)^2\left[\frac{1}{2}\delta g^{00} + O\left((\delta g^{00})^2\right)\right],\nonumber
\end{align}
where the $O(\delta g_{00} ^2)$ term in the second squared brackets does not contain any power of $\delta\phi$. The second relevant term is
\begin{align}
	 & \sqrt{-g}V(\phi)=\sqrt{-\bar g} V(\phi)-\\&-\sqrt{-\bar g}\frac{1}{2a^2}V^\prime(\bar\phi)\delta g_{00}\delta\phi+\sqrt{-\bar g}V(\bar\phi)\left[-\frac{1}{2a^2}\delta g_{00}+\frac{1}{8a^4}(\delta g_{00})^2\right]\nonumber.
\end{align}
From these equations, we can see that the interactions between metric perturbations $\delta g_{0\mu}$ and field perturbation $\delta\phi$ are slow-roll suppressed either by a factor of $\partial_0 \bar\phi$ or by $V^\prime(\bar{\phi})$. This means that, at leading order in slow-	roll, metric perturbations remain decoupled during inflation. 

As we discussed at the end of \cref{sec:scalarpert}, if we take into account these interactions between $\delta\phi$ and $\delta g_{0\mu}$, they will result in slow-roll suppressed corrections to the effective mass of the inflaton in \cref{eq:MS}, which will affect the scale dependence of the power spectrum. As we neglect metric perturbations, our simulation will not be able to capture this effect. 

\subsection{The Friedmann equations}
\label{sec:friedmann}
In the light of the assumption above, the evolution of the metric is solely described by the Friedmann equations:
\begin{align}
\label{eq:friedmann}
&\mathcal{H}^2=\frac{1}{3}\langle\rho\rangle a^2\\
&\frac{d^2a}{d\eta^2}=\frac{1}{6}\left(\langle\rho\rangle-3\langle p \rangle\right)a^3,
\end{align}
where $\langle \rho \rangle $ and $\langle p\rangle $ are the mean energy-density and pressure contained in the lattice, which are computed as an average of the full $\rho$ and $p$ over the $N^3$ points of the cubic lattice. To derive $\rho$ and $p$ we first compute the stress-energy tensor from the action:
\begin{equation}
\label{eq:tmn}
T_{\mu\nu}=\frac{-2}{\sqrt{-g}}\frac{\delta S}{\delta g^{\mu\nu}}.
\end{equation}
The density and pressure are then extracted as follows:

\begin{equation}
\begin{split}
\rho=-T^0_{\hspace{1mm}0}=\frac{	(\partial_\tau\phi)^2}{2a^2}+\frac{(\partial_i\phi)^2}{2a^2}+V(\phi),\\
p=\frac{1}{3}\sum_iT^i_{\hspace{1mm}i}=\frac{(\partial_\tau\phi)^2}{2a^2}-\frac{(\partial_i\phi)^2}{6a^2}-V(\phi).
\end{split}
\label{eq:ED}
\end{equation}
In the code, the gradient term $(\partial_i \phi)$ is evaluated in the following way:
$$[(\partial_i\phi)^2](\vec{n})\underset{\rm code}{=}-\phi(\vec{n})[\nabla^2 \phi](\vec{n}).$$
The right- and left-hand sides of this equation are different up to a total derivative. This corresponds to an integration by parts in the action before taking the derivative of \cref{eq:tmn}. In this way, we can use the definition of the Laplacian used to evolve the equations to evolve the derivatives as well. Moreover, evaluating the Laplacian is computationally less expensive than computing the absolute value of the gradient term. This trick is inherited from LATTICEEASY. 

To evolve the metric of the universe, either of the Friedman \cref{eq:friedmann} can be used. The choice is arbitrary, and we use the second of these equations to determine the evolution of the scale factor in the simulation. This will allow us to use the first one as an energy conservation check, as we will see in \cref{sec:results}. This is similar to what is done in LATTICEEASY.

\section{Initial conditions}
\label{sec:IC}
 We now describe how we set the initial conditions on the lattice. Although we do not split in background and perturbation quantities in the evolution of the system, at the initial time we operate such a splitting. The background values will determine the point on the background inflationary trajectory. The fluctuations around the background values reflect the quantum nature of the inflaton field, as mentioned in \cref{sec:lattice}. 
\subsection{Background quantities}
\label{sec:initialbackground}
The initial background values of the inflaton $\bar{\phi}_{\rm in}$ and its velocity $\bar{\phi}^\prime_{\rm in}$ are set by the background inflationary trajectory. Their explicit values will depend on the inflaton potential, that we will choose in \cref{sec:results}. The scale factor $a$ is simply set to $1$ at the beginning of the simulation, while its derivative in program time $a^\prime$ is computed via the first Friedman equation \eqref{eq:friedmann} using only the background energy density and pressure of the field and neglecting gradient contributions:
$$3a^\prime_{\rm in}=\bar\rho = \frac{	({\bar\phi}^\prime)^2}{2} + V(\bar\phi).$$
Then, after the field fluctuations are generated (as described in the next section), the value of $a_{\rm in}^\prime$ is updated to include the gradient term, computed as a lattice average of $(\partial_i\phi)^2/2$. Note that quantum vacuum sub-horizon fluctuations should not contribute to the Friedmann equations. However, we still include them in generating the initial value of $H$, and this is a consequence of our semi-classical approximation. This is important in order to evolve the discrete system in a consistent way. Indeed, neglecting gradient contributions will result in an effective residual curvature in the second Friedmann equation, that we use to evolve the scale factor during the simulation. Moreover, including the gradient term in a consistent way allows us to check that energy is conserved in the discrete system, i.e. to ensure that there are no numerical errors propagating on the lattice during the simulation. More about the energy conservation check can be found in \cref{app:energy}.

\subsection{Quantum fluctuations}
\label{sec:initialconditions}
We now explain how we generate the initial field perturbations on the lattice. 
 The first step is defining the discrete version of \cref{eq:quantization}:
\begin{align}
\label{eq:discquantization}
\hat\phi({\vec{n}})=\sum_{\vec{m}}\Bigl[\hat{a}_{\vec{m}}\text{ }u(\vec\kappa_{\vec{m}})\text{ }e^{i\frac{2\pi}{N}\vec{n}\cdot\vec{m}}+\hat{a}_{\vec{m}}^\dagger\text{ }u^{\dagger}(\vec\kappa_{\vec{m}})\text{ }e^{-i\frac{2\pi}{N}\vec{n}\cdot\vec{m}}\Bigr]=\frac{1}{dx^3}\sum_{\vec{m}}{\hat\phi}(\vec\kappa_{\vec{m}})\text{ }e^{+i\frac{2\pi}{N}\vec{n}\cdot\vec{m}}.
\end{align}
In the second equality, we show the comparison with the lattice definition of Fourier modes $\hat{\phi}(\vec\kappa_{\vec{m}})$, which are momentarily promoted to quantum operators. Here, we introduced the discrete quantum creation and annihilation operators:
\begin{equation}
[\hat a_{\vec{m}},\hat a^\dagger_{\vec{m}^\prime}]=\delta_L(\vec{m},\vec{m}^\prime)=\frac{1}{L^3}\delta(\vec{m},\vec{m}^\prime).
\end{equation}
$u(\vec\kappa_{\vec{m}})$ are the discrete mode functions, which are the lattice counterparts of $\delta\phi(\vec{k})$ defined in \cref{eq:quantization}. We start the simulation when the comoving size of the box is smaller than the Hubble horizon, $L<1/(aH)$. In this case, the inflaton is in its Bunch-Davies vacuum:
\begin{equation}
\label{eq:discBD}
u(\vec\kappa)=\frac{L^{3/2}}{a\sqrt{2\omega_{\vec{\kappa}}}}e^{- i \omega_{\vec{\kappa}}\tau},\quad\quad\quad \omega_{\vec{\kappa}}^2=k_{\rm eff}^2(\vec{\kappa})+m^2,
\end{equation}
where $m$ is the mass of the inflaton. There are two differences between this expression and the Bunch-Davies vacuum in continuous space. The first is a normalization factor of $L^{3/2}$, which is commonly introduced to correct for the finite volume of space. To understand this, we take the two-point function of the field:
\begin{equation}
\langle\hat \phi_{\vec{i}}^2\rangle=	\langle 0|\hat \phi_{\vec{i}}^2|0\rangle=\sum_{\vec{m},\vec{m}^\prime}\delta_L(\vec{m},\vec{m}^\prime)u(\vec\kappa_{\vec{m}})u(\vec\kappa_{\vec{m}^\prime})=\frac{1}{L^3}\sum_{\vec{l}}|u(\vec\kappa_{\vec{m}})|^2.
\end{equation}
We can clearly see that this scales as $L^{-3}$ due to the presence of the finite-volume delta function $\delta_L$. If we want the quantity $\langle \hat \phi^2 \rangle$, and two-point functions in general, to be independent of the physical size of the lattice, we have to normalize the mode functions by a factor of $L^{3/2}$.

The second difference is the presence of $k_{\rm eff}$ instead of $\kappa$ inside the mode frequency $\omega_{\kappa}$. This is done to make the initial fluctuations compatible with the discrete Mukhanov-Sasaki \cref{eq:discMS}, and takes into account the modified dispersion relation caused by the discretization. Note that in LATTICEEASY, and all other lattice simulations in the context of reheating, the initial fluctuations are usually generated using $\kappa$ instead of $k_{\rm eff}$ in \cref{eq:discBD}. 

In practice, the lattice will be only approximately sub-horizon at the beginning of the simulation, i.e. $L\lesssim aH$. For this reason, it is better to use the following expression for the discrete mode functions in order to correct for the finite size of the lattice:
\begin{equation}
\label{eq:discmodes}
u({\vec{\kappa}_{\vec m}},\tau)=L^{3/2} \frac{\sqrt{-\pi \tau}}{2a}H_\nu^{(1)}(-k_{\rm{eff}}(\vec\kappa_{\vec m})\text{ }\tau).
\end{equation}
This expression reduces to \cref{eq:discBD} for most of the modes. 

Now that the discrete mode functions are defined, we need to generate the field fluctuations on the lattice. As already mentioned, in our classical simulation we do not solve for the full quantum operator. Instead, we take a statistical point of view, interpreting the quantum creation and annihilation operators as stochastic variables that take different values at each realization. In this picture the creation and annihilation operators of \cref{eq:discquantization} are initiated as:
\begin{equation}
a_{\vec{m}}=e^{i 2\pi Y_{\vec{m}}}\sqrt{-\ln(X_{\vec{m}})/2},
\end{equation}
where $X_{\vec{m}}$ and $Y_{\vec{m}}$ are random variables uniformly distributed between 0 and 1 for each $\vec{m}$. From \cref{eq:discquantization}, we can see that this is equivalent to generating the Fourier modes of the field as Gaussian random numbers with variance $|u(\vec\kappa_{\vec m})|^2$ as follows:
\begin{equation}
\label{eq:randommodes}
\phi(\vec\kappa_{\vec m})=e^{i 2\pi Y_{\vec{m}}} \sqrt{-\ln(X_{\vec{m}})} \text{ }u(\vec\kappa_{\vec m}),
\end{equation}
where $u(\vec\kappa_{\vec m})$ is given by \cref{eq:discmodes} with $a=1$ and $\tau=0$. Note that we dropped the hat symbol $\hat \cdot$ from $\hat a$ and $\hat \phi$, as we now think of these quantities as classical realizations of a stochastic process. From here, we first apply the iDFT \cref{eq:iDFT} and then add the background value of the inflaton to obtain the initial field configuration on the lattice. The fluctuations of the time derivative of the scalar field $\partial_\tau{\phi}(\vec\kappa_{\vec m})$ are generated in the same way using the time derivative of the mode functions $\partial_\tau{u}(\vec\kappa_{\vec m})$ and using the same realizations of $X_{\vec{l}}$ and $Y_{\vec{l}}$. Note that we do not adopt the same procedure of LATTICEEASY for generating the initial field configuration, which is known to have a bug, as first noticed in Ref. \cite{Frolov_2008}. 
\section{Numerical integrator}
\label{sec:numint}
We now introduce the time integrator for the equations of motion. Before proceeding, we first operate the following rescaling of space and time coordinates:
 \begin{equation}
 	\label{eq:rescaling}
 	d\tau\rightarrow d\tilde{\tau}=Ba^sd\tau,\quad x\rightarrow \tilde{x}=Bx.
 \end{equation}
This is similar to the rescaling adopted in LATTICEEASY, and it is done to make the equations numerically stable. After the rescaling, the equations of motion are the following\footnote{We avoid writing the lattice point $\vec{n}$ explicitly. Prime derivatives within this section are with respect to rescaled conformal time $f^\prime\equiv df/d\tilde\tau$. This is different from the notation in the rest of the manuscript, where primes denote derivative with respect to conformal time \textit{without} the rescaling.}:
\begin{align}
\label{eq:eoms}
\begin{split}
&		\phi^{\prime\prime}+(2+s)\frac{a^\prime}{a} {\phi}^\prime-a^{-2s}[\tilde{\nabla}^2\phi]+a^{2-2s}\frac{\partial \tilde{V}}{\partial \phi}=0,\\
&a^{\prime\prime}=-s\frac{{a^\prime}^2}{a}+\frac{1}{3}a^{-2s+3}(\langle \tilde \rho \rangle -3 \langle \tilde p \rangle),
\end{split}
\end{align}
where the tilde $\tilde \cdot$ represents a variable normalized by the constant $B$, e.g. $\tilde{V}=V/B$.

To evolve these equations, we use a Runge-Kutta 4th order integrator (RK4) with an adaptive time step \cite{press1986numerical}. To do so, we first transform the second-order system of equations to a first-order one, introducing the variables $\phi^v\equiv\phi^{\prime}$ and $a^v\equiv a^{\prime}$. Then, our system of $N^3+1$ second-order equations becomes a set of first-order $2N^3+2$ equations, that we can write in the following form:
\begin{align*}
&{\phi^v}^{\prime}=F_{\phi^v}(\phi^v,\phi,a^v,a),\\
&\phi^{\prime}\,\,=F_\phi(\phi^v,\phi,a^v,a),\\
&{a^v}^{ \prime}=F_{a^v}(\phi^v,\phi,a^v,a),\\
&a^{\prime}\,\,=F_a(\phi^v,\phi,a^v,a),
\end{align*}
where, for example:
\begin{align*}
&F_{\phi^v}(\phi^v,\phi,a^v,a) = -(2+s)\frac{a^v}{a}, {\phi}^v+a^{-2s}[\tilde{\nabla}^2\phi]-a^{2-2s}\frac{\partial \tilde{V}}{\partial \phi}\\
&F_\phi(\phi^v,\phi,a^v,a) = \phi^v.
\end{align*}
To evolve the system with RK4 for a finite step $\Delta \tilde\tau$, we first define the following quantities:
\begin{align*}
k^{(1)}_f&={\Delta \tilde\tau} F_f(\phi^v,\phi,a^v,a),\\
k^{(2)}_f&=\Delta \tilde\tau F_f\left(\phi^v+\frac{1}{2}k^{(1)}_{\phi^v},\phi+\frac{1}{2}k^{(1)}_{\phi},a^v+\frac{1}{2}k^{(1)}_{a^v},a+\frac{1}{2}k^{(1)}_{a}\right),\\
k^{(3)}_f&=\Delta \tilde\tau F_f\left(\phi^v+\frac{1}{2}k^{(2)}_{\phi^v},\phi+\frac{1}{2}k^{(2)}_{\phi},a^v+\frac{1}{2}k^{(2)}_{a^v},a+\frac{1}{2}k^{(2)}_{a}\right),\\
k^{(4)}_f&=\Delta \tilde\tau F_f\left(\phi^v+k^{(3)}_{\phi^v},\phi+k^{(3)}_{\phi},a^v+k^{(3)}_{a^v},a+k^{(3)}_{a}\right),
\end{align*}
where $f\in\{\phi^v,\phi,a^v,a\}$. Then, the field configuration at a time $\tilde\tau_0+\Delta \tilde\tau$ is obtained from the field configuration at a time $\tilde\tau_0$ in the following way:
\begin{equation}
f(\tilde\tau_0+\Delta\tilde\tau)=f(\tilde\tau_0)+\frac{1}{6}k^{(1)}_f+\frac{1}{3}k^{(2)}_f+\frac{1}{3}k^{(3)}_f++\frac{1}{6}k^{(4)}_f.
\end{equation}
At the initial time, we fix an initial $\Delta \tilde\tau_0$. This will be typically of order $10^{-4}$, corresponding to a physical comoving time of $\Delta\tau_0=B\Delta\tilde\tau_0$. In all the cases discussed in this thesis, we set the rescaling factor to be the mass of the inflaton $B=m=V^{\prime\prime}(\phi)$. Then, during the simulation, we adapt the time step in the following way $\Delta \tilde\tau=a^{{s-1}}\Delta\tilde\tau_0$. This makes the time step constant in cosmic time, which is defined by $s=1$.

\section{Lattice outputs}
\label{sec:output}

We now describe how outputs are computed from the code. We mainly discuss quantities related to scalar perturbations, such as power spectrum and bispectrum of $\phi$. The computation of these quantities in our code differs from all other examples in the literature (such as LATTICEEASY). The main difference is that our procedure takes into account the discretization, so that the final spectra are independent of the lattice implementation and can be compared directly with analytical computations. Moreover, we discuss how energy conservation is checked in our code.
\subsubsection*{Background quantities}
As usual in lattice simulations, background quantities are simply computed as averages over the $N^3$ points of the lattice. We output quantities such as the average of the field $\bar{\phi}$, its derivative $\bar{\phi}^\prime$, and the energy density and pressure of the field $\bar \rho$ and $\bar p$.

\subsubsection*{Power spectrum}
To compute the power spectrum from the simulation, we first take the DFT to obtain $|{\phi}(\vec\kappa_{\vec m})|^2$. Then, after normalizing by a factor $L^{-3}$ to get the physical power spectrum of the mode functions (see the discussion in \cref{sec:initialconditions}), we average over spherical bins to obtain the one-dimensional isotropic power spectrum $P_\ell$. This is done by averaging $|{\phi}(\vec\kappa_{\vec m})|^2$ over all lattice points such that $\text{int}(|\vec{m}|)=\ell$, where $\ell$ is the bin number. Then, a comoving momentum $\kappa_{\ell}$ is associated to each bin by averaging the absolute value of \cref{eq:modes} over the bin. Note that the procedure for associating the momentum to each bin is different from the one of LATTICEEASY, where the momenta associated to the bins are simply $\kappa_\ell=2\pi\ell/L$. This leads to a distortion in the output momenta of LATTICEEASY, which is independent of $N$ and can lead to a difference of up to $20\%$ in the IR\footnote{Note that this distortion can also be relevant in generating the initial conditions.}. We output the power spectrum for modes only up to the Nyquist frequency $\kappa_{\rm Nyquist}=2\pi \sqrt{3} N_{\rm Nyquist}/L$, where $N_{\rm Nyquist}=N/2$,  because they contain all the physical information.

The dimensionless power spectrum $\mathcal{P}_{\phi}$ is obtained as $P_\ell\,\kappa_{\ell}^2/(2\pi^2)$, and it is plotted against the bin momentum $\kappa_\ell$ (left plots of \cref{fig:finalPS,fig:finalPS_step}). This power spectrum is expected to be different from the analytical expectation due to discretization effects. However, as we discuss in \cref{sec:resultsslow}, we can successfully reproduce the results of the continuous theory at all scales if we interpret $k_{\rm eff}$ of \cref{eq:keff} as the physical momentum of the lattice simulation. For this reason, we multiply the dimension-full power spectrum $P_\ell$ by $k_{\rm eff,\ell}^2/(2\pi^2)$ instead of $\kappa_{\ell}^2/(2\pi^2)$, where $k_{\rm eff,\ell}$ is obtained averaging \cref{eq:keff} over the same spherical bins. Plotting this quantity against the effective momenta of each bin $k_{\rm eff,\ell}$ gives the same result of the linear theory in continuous space  (right plots of \cref{fig:finalPS,fig:finalPS_step}).
\subsubsection*{Bispectrum}
In the single-field model studied in this chapter, scalar perturbations are nearly Gaussian, as we discussed in \cref{sec:scalarpert}. This will not be true in the more complicated axion-U(1) model discussed in the second part of the thesis. For this reason, it is useful to compute the inflationary three-point function $\langle \phi(\vec k_1)\phi(\vec k_2)\phi(\vec k_3)\rangle$ from the code, often called bispectrum. This is the first computation of a bispectrum from a lattice code in the context of primordial cosmology.


Due to statistical isotropy, the three-point function is different from zero only when the three momenta $\vec k_i$ form a closed triangle $\vec{k}_1+\vec{k}_2+\vec{k}_3=0$. This makes it a function of two three-momenta $\vec k_1$ and $\vec{k_2}$, hence the name bispectrum. In this thesis, we will only compute the bispectrum on equilateral configurations $|\vec k_1|=|\vec k_2|=|\vec k_3|$. The equilateral bispectrum is a function of one parameter $k$:
 \begin{equation}
 \label{eq:equib}
 	\mathcal{B}_\phi(k)\equiv\langle \phi(\vec{k_1})\phi(\vec{k_1})\phi(\vec{k_3})\rangle,
 \end{equation}
 where $k=|\vec k_1|=|\vec k_2|=|\vec k_3|$ and $\vec k_3=-\vec{k}_1-\vec{k}_2$. 
 
To compute this quantity from the code, we first take a discrete number of bins defined by the bin number $\ell\in \{1,...,N_{\rm Nyquist}\}$. To each bin, we associate a lattice momentum $\kappa_\ell$ and an effective momentum $k_{\rm eff,\ell}$ through a spherical binning of \cref{eq:modes} and \cref{eq:keff}. This is similar to what is done above for the power spectrum. For each of these bins, we need to count all lattice triangles $\vec\kappa_{\vec{m_1}}$, $\vec\kappa_{{\vec{m_2}}}$ and $\vec\kappa_{{\vec{m_3}}}$ such that  $\ell=\text{int}(|\vec{m}_1|)=\text{int}(|\vec{m}_2|)=\text{int}(|\vec{m}_3|)$ {and} $\vec{m}_3=-\vec{m}_1-\vec{m}_2$. For each bin $\ell$, the quantity $\mathcal{B}_{\phi,\ell}$ is obtained as an average of the product $\phi(\vec \kappa_{{\vec{m_1}}})\phi(\vec \kappa_{{\vec{m_2}}})\phi(\vec\kappa_{{\vec{m_3}}})$ over all these triangles. The final result is plotted as a function of the effective momentum of each bin $\mathcal{B}_\phi(k_{\rm eff})$, in order to obtain a lattice estimate for \cref{eq:equib}. This is similar to what is done above for the power spectrum, and takes into account the modified dispersion relation induced by the discretization. Note that the numerical implementation of the triangle counting presented here might be nontrivial due to the reality of $\phi$, which requires $\phi(\vec{\kappa})=\phi^*(-\vec{\kappa})$.
\subsubsection*{Energy conservation}
\label{app:energy}
In the simulation, the scale factor is evolved using the second Friedmann equation. This allows using the first Friedmann equation to check energy conservation in the code, which serves as a test for the numerical accuracy of the time integrator. To do so, we define the following quantity:
\begin{equation}
\label{eq:cons}
E = \frac{3\mathcal{H}^2}{\rho a^2}.
\end{equation}
We check that this quantity is close to 1 throughout the numerical integration. This ensures that the numerical errors are under control. This energy conservation check is inherited from LATTICEEASY. In \cref{sec:energy}, we show the energy conservation checks for all the cases considered in this chapter and discuss other methods to assess the accuracy of the time integrator. 

\section{Results of the simulation}
\label{sec:results}
We now proceed showing the results of the simulation. We study the single-field model of inflation in two cases. The first is a standard slow-roll potential for the inflaton. The second is a similar potential, but with a small step added on top of it. 
All numerical values in this section are given in Planck units $M_{\rm Pl}=1$. Details about the procedure to compute power spectra in our code can be found in \cref{sec:output}.
\subsection{Slow-roll potential}
\label{sec:resultsslow}
For simplicity, we take the following harmonic potential for the inflaton:
\begin{equation}
V(\phi)=\frac{1}{2}m^2\phi^2,
\end{equation}
where $m=0.51\cdot10^{-5}$. This value is chosen to roughly match the observed power spectrum of curvature perturbation $\mathcal{P}_{\mathcal{R}}\simeq25\times 10^{-10}$. As mentioned in \cref{sec:scalarpert}, this particular inflationary potential is disfavored by CMB observations. What we discuss, however, does not depend on the particular choice of inflationary potential, which can be freely set in the lattice simulation.
\subsubsection*{Background evolution}
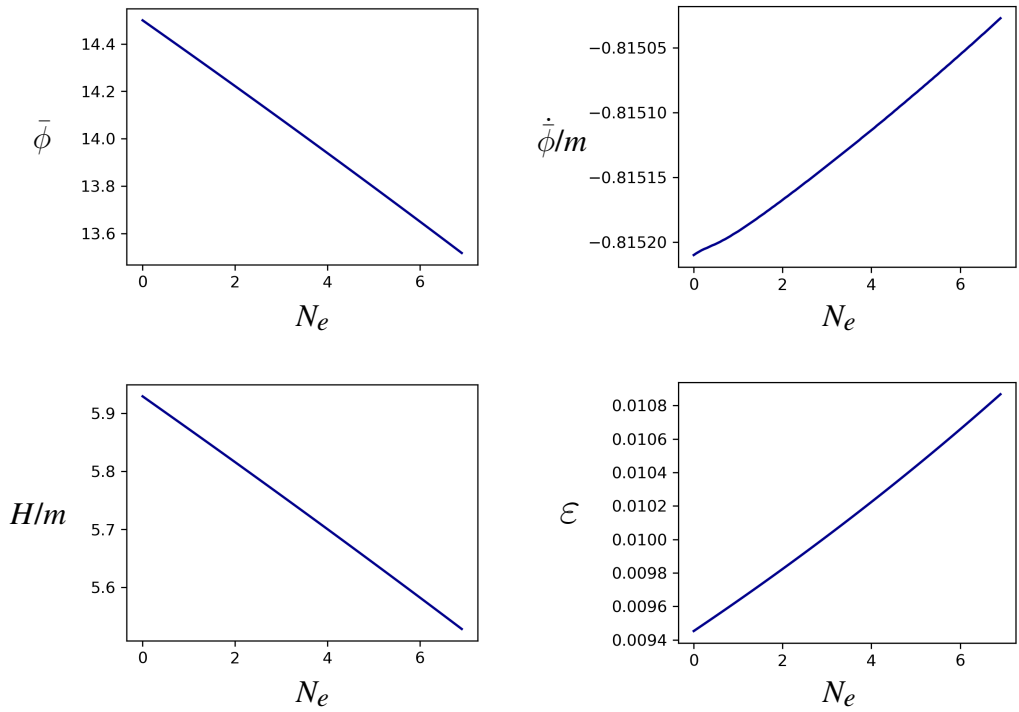
\begin{figure}
	\centering
	
	\begin{tikzpicture}
	\node (img) {\includegraphics[width=6cm]{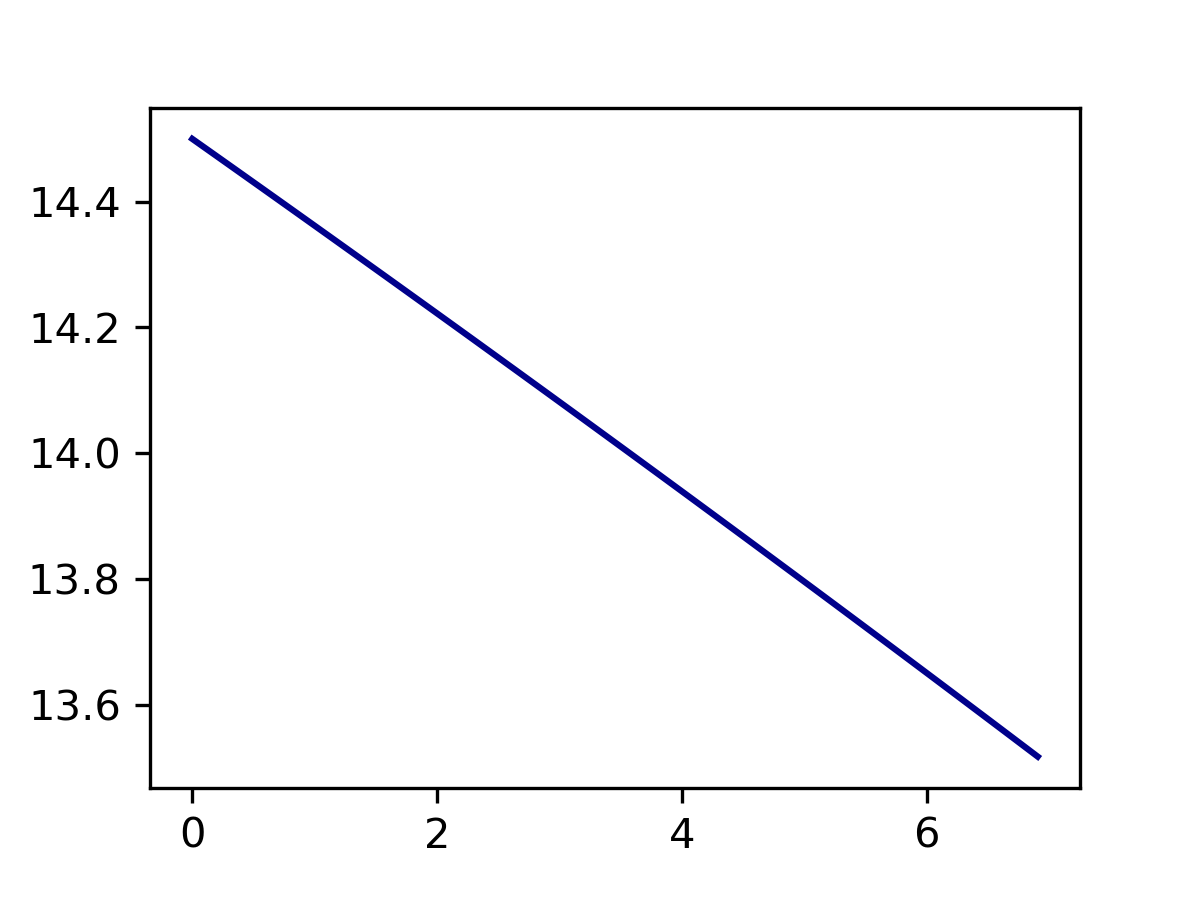}};
	
	\node [rotate=0,text width=0.01cm,align=center] at (-3.5,0){ $\bar{\phi}$};
	\node [text width=0.01cm,align=center] at (0,-2.4){$N_e$};

	\node (img2) at (7,0) {\includegraphics[width=6.4cm]{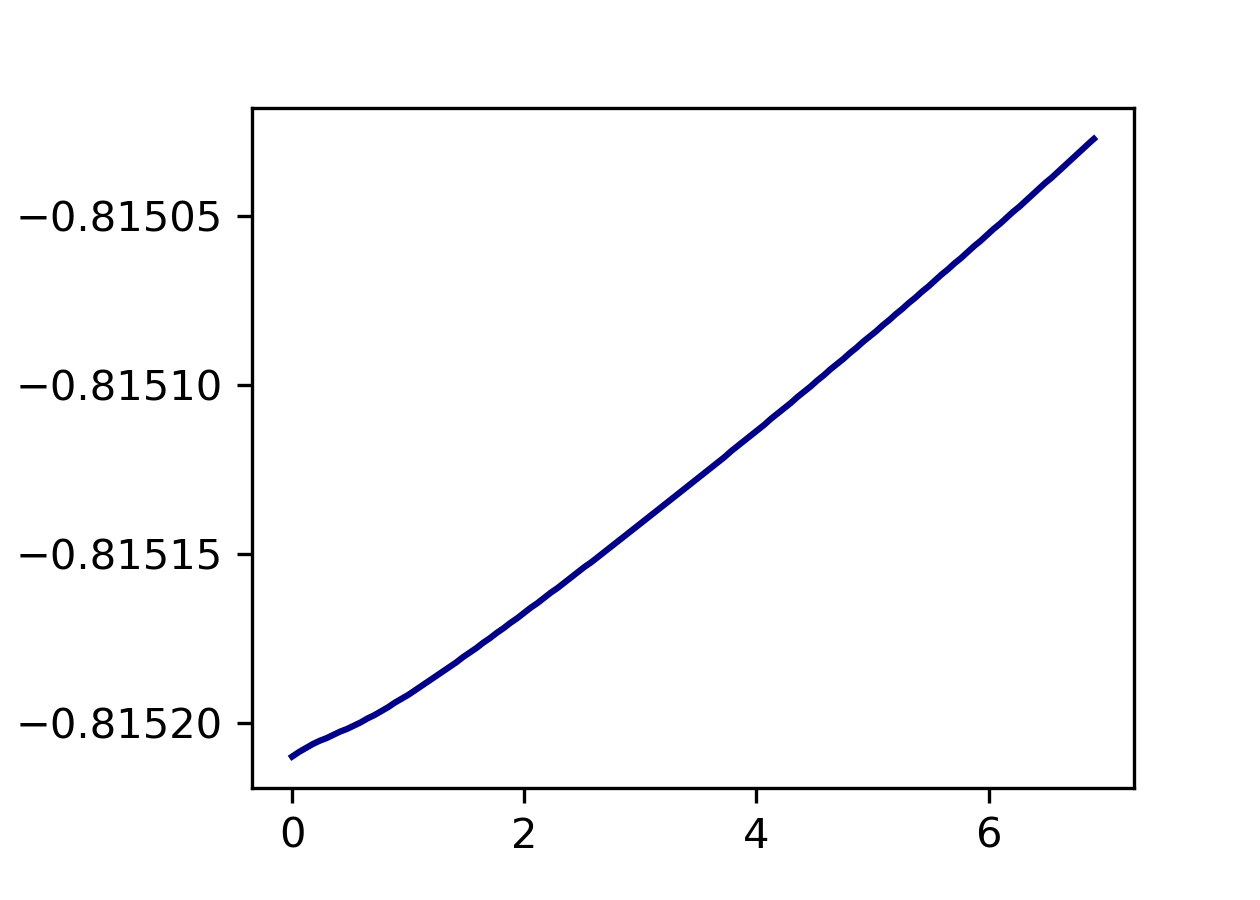}} ;
	
	\node [rotate=0,text width=0.01cm,align=center] at (-3.8+7,0){ ${\dot{\bar{\phi}}}/{m}$};
	\node [text width=0.01cm,align=center] at (0+7,-2.4){$N_e$};

	\node (img3) at (0,-5) {\includegraphics[width=6cm]{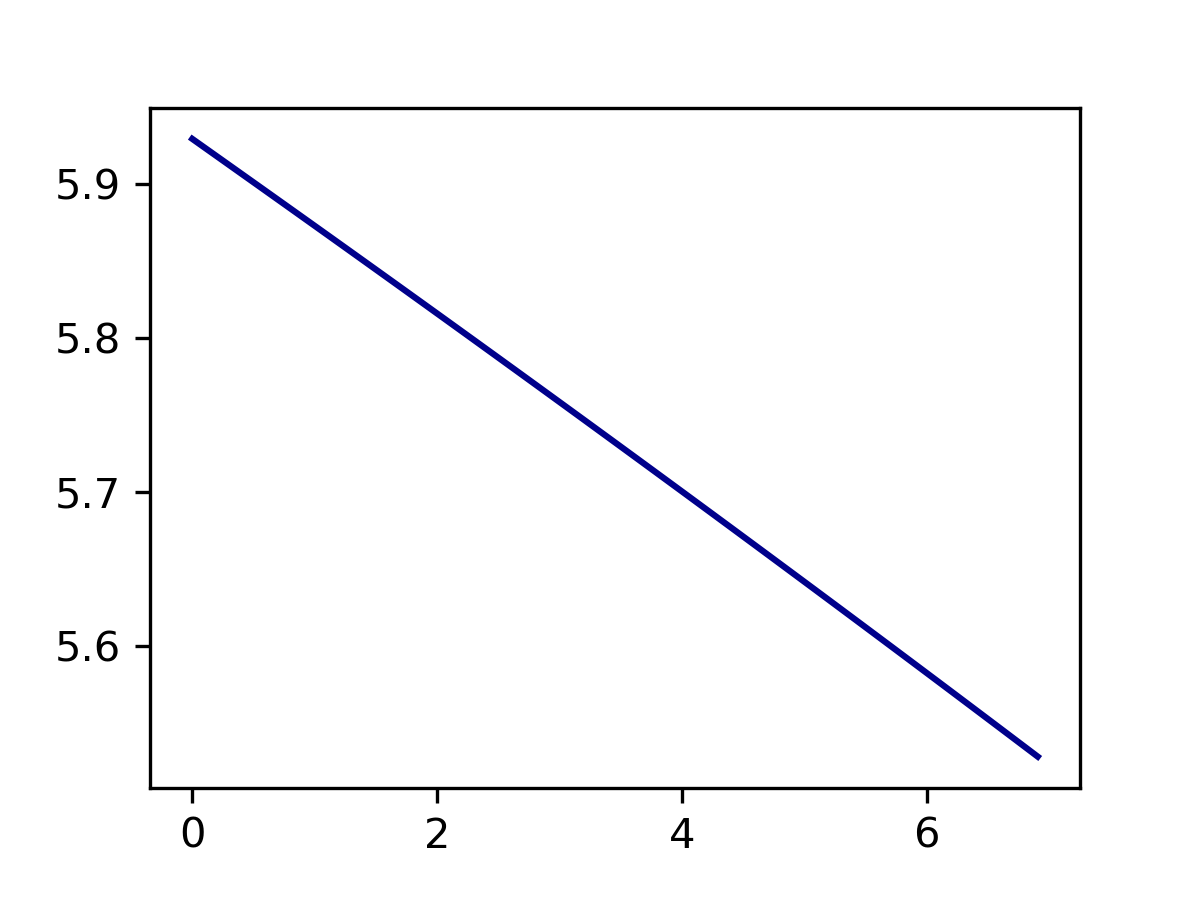}};
	
	\node [rotate=0,text width=0.01cm,align=center] at (-3.8,0-5){ $H/m$};
	\node [text width=0.01cm,align=center] at (0,-2.4-5){$N_e$};

	\node (img4) at (7,-5) {\includegraphics[width=6.4cm]{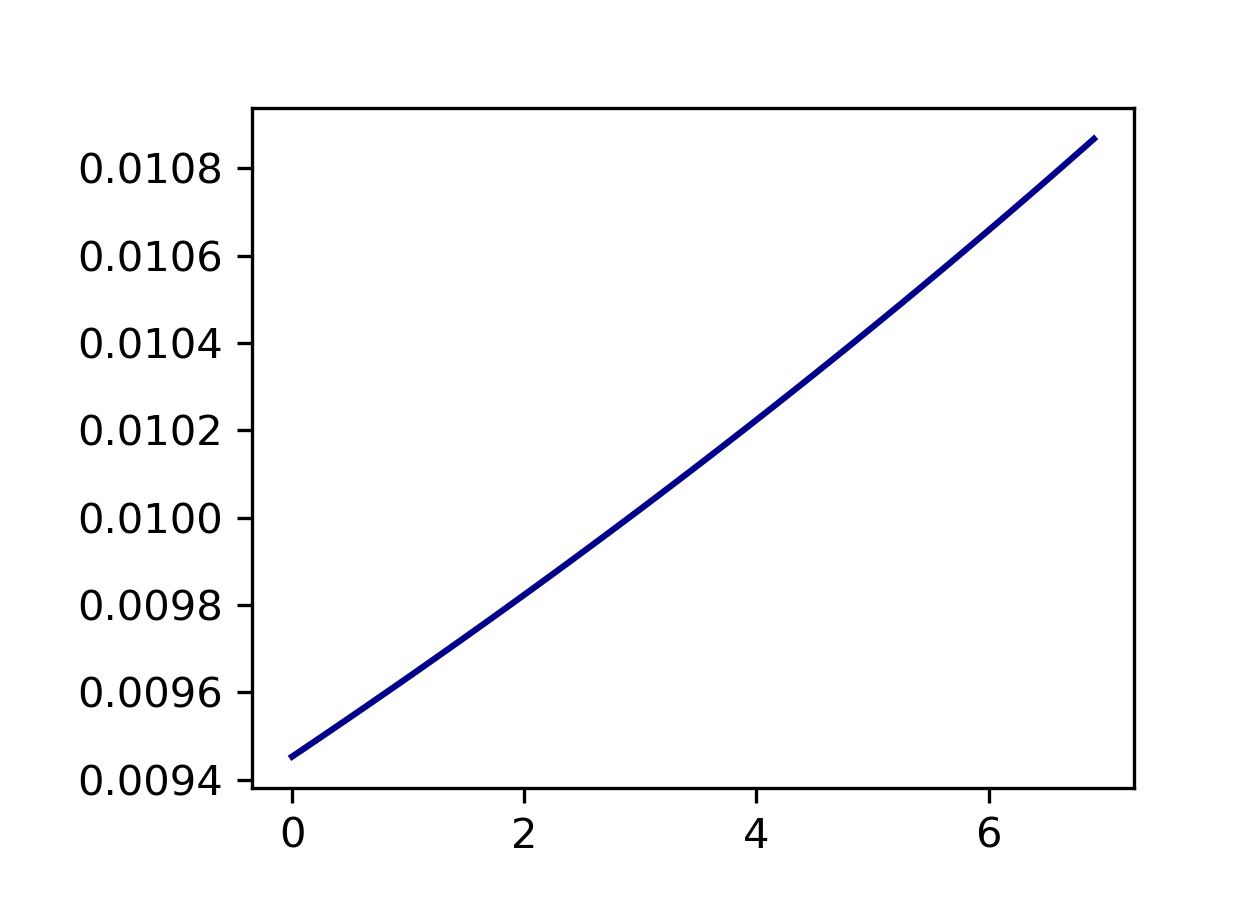}} ;
	
	\node [rotate=0,text width=0.01cm,align=center] at (-3.5+7,0-5){ \Large $\varepsilon$};
	\node [text width=0.01cm,align=center] at (0+7,-2.4-5){$N_e$};

	\end{tikzpicture}

	\caption{Evolution of background quantities during the simulation. From top left: the background value of the inflaton $\bar\phi$, its derivative in cosmic time $\dot{\bar{\phi}}$, the Hubble parameter $H$ and the slow-roll parameter $\epsilon$.}
	\label{fig:backgroundvalues}
\end{figure}

The initial average value of the inflaton is chosen to be $\bar{\phi}_{\rm in}=14.5$. Its velocity is determined by solving the background Klein-Gordon equation \eqref{eq:KG}, and it is given by $\bar{\phi}^\prime_{\rm in}=-0.8152m$. With these values, the Universe is in the middle of the inflationary phase, and there are $N_e\simeq53$ e-folds\footnote{We set as a convention $N_e=0$ at the beginning of the simulation.} left before the end of inflation. The system is evolved until $a=10^3$ ($N_e\simeq6.9$) which means that at the end of the simulation we will still be in the inflationary phase.
In \cref{fig:backgroundvalues} we show the evolution of the background value of the inflaton $\bar{\phi}$ and its velocity $\dot{\bar{\phi}}$ as functions of the number of e-folds $N_e$. As explained in \cref{sec:output}, these quantities are computed from the simulation as averages over the $N^3$ points of the lattice. In the same plot, we also show the evolution of $H$ and $\varepsilon$ . 
From these plots we clearly see that we are in the middle of the inflationary phase, being $\varepsilon \ll1$ and $\dot{{\bar\phi}}\simeq \text{constant} \ll V(\phi)$. 
\subsubsection*{Perturbations}
We now come to the dynamics of field fluctuations and to the importance of the modified dispersion relation discussed in \cref{sec:modifieddr}. We show results from a run of the code with $L=1.4/m$ and $N^3=128^3$. This translates to:
\begin{equation}
\kappa_{\rm min}=\frac{2\pi}{L}\simeq 4.49 m \simeq 0.76H_{\rm in}, \quad\quad \kappa_{\rm max}=k_{\rm in} \frac{\sqrt{3}}{2}N\simeq 84.5 H_{\rm in},
\end{equation}
where $H_{\rm in}=\mathcal{H}_{\rm in}$ is the initial value of the Hubble rate. The modes are almost all sub-horizon at the beginning of the simulation. We evolve the system until $a=10^3$, which means that the modes are all super-horizon at the end of the simulation.
\begin{figure}
	\centering
	
	\begin{tikzpicture}
	\node (img) {\includegraphics[width=8.cm]{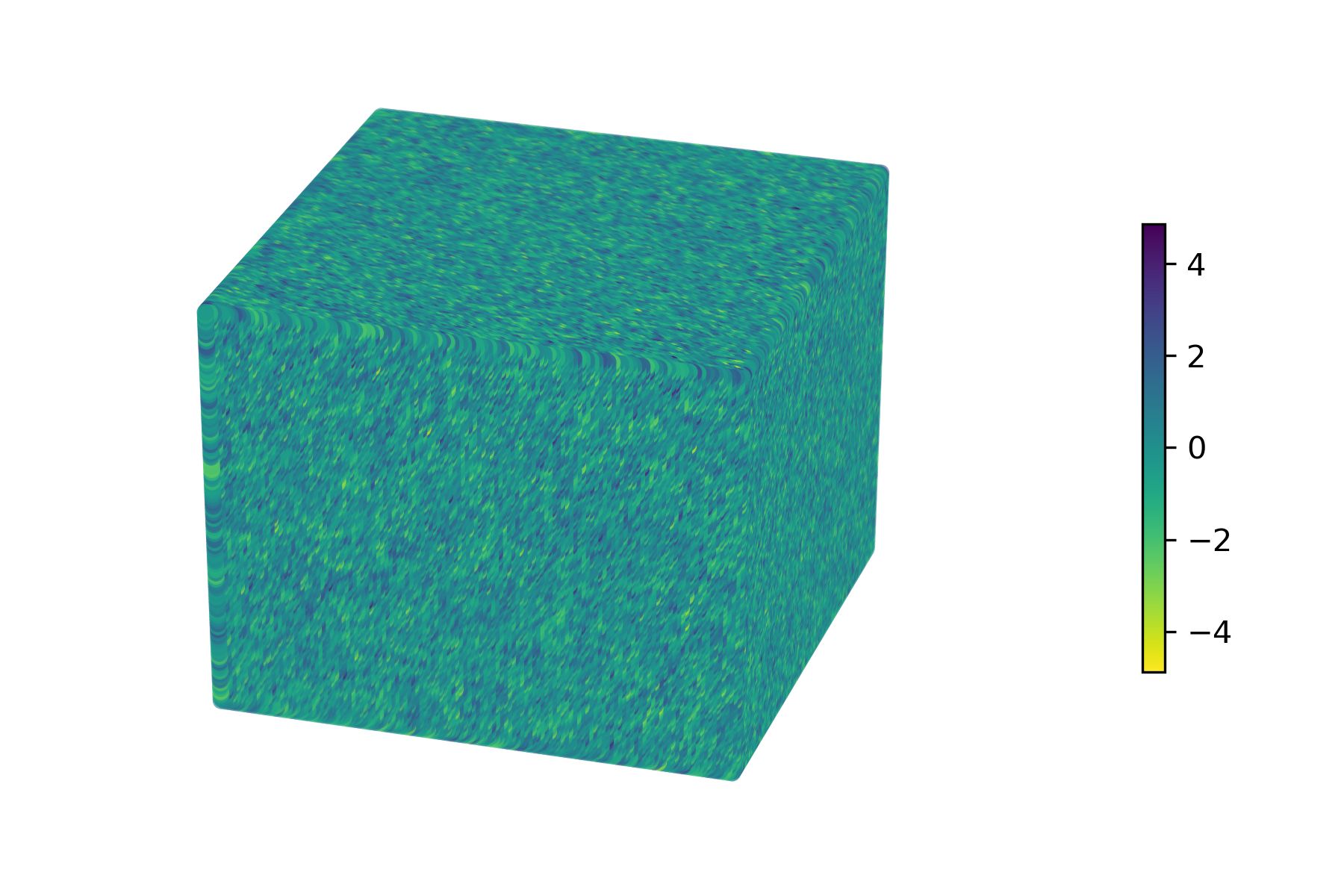}};

	\node [text width=0.01cm,align=center] at (2.2,0.){$\frac{\delta\phi}{\sigma}$};

	\node (img2) at (7.4,0) {\includegraphics[width=8.cm]{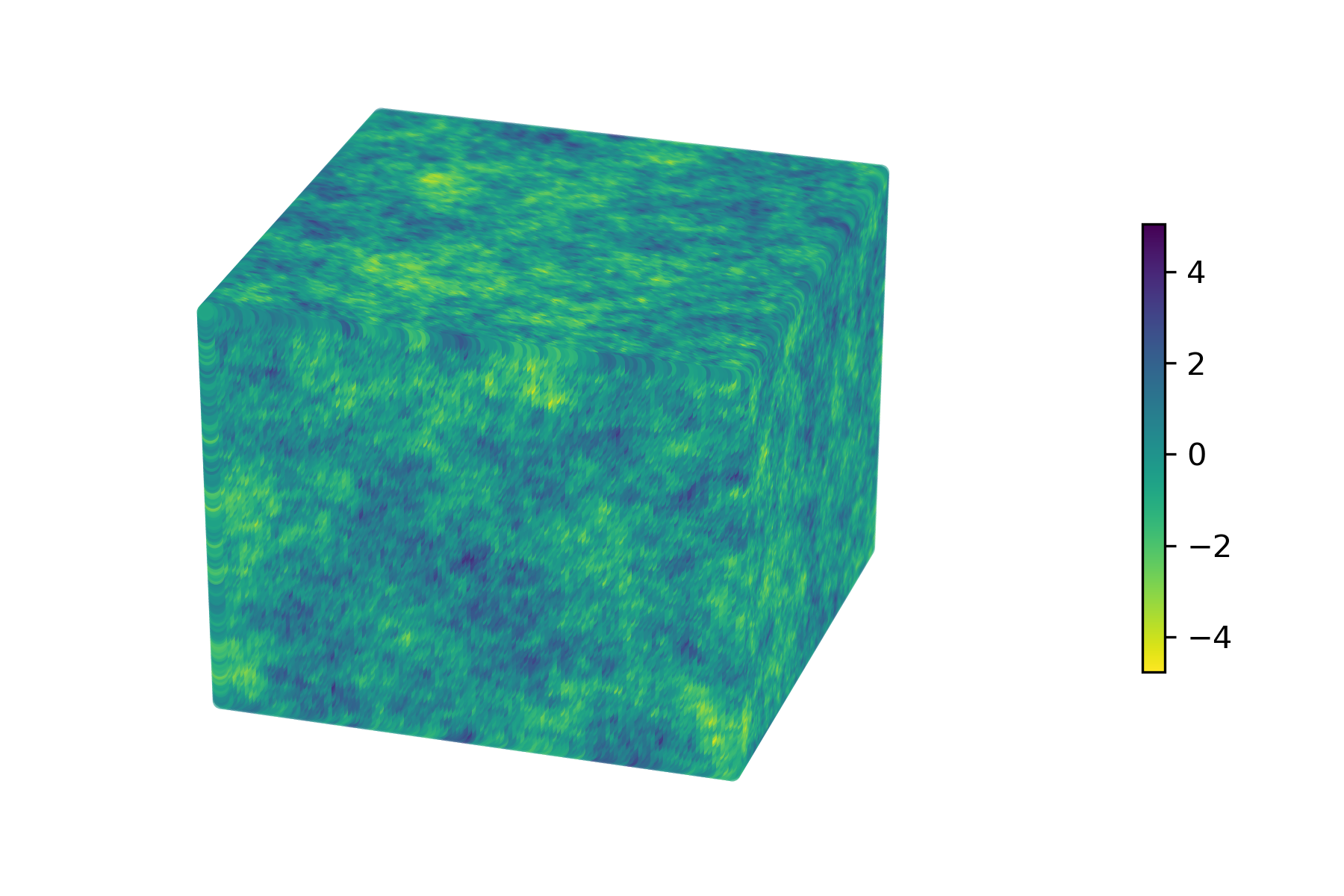}} ;
	
	\node [text width=0.01cm,align=center] at (0+7.4+2.2,0){$\frac{\delta\phi}{\sigma}$};

	\end{tikzpicture}

	\caption{Scalar field fluctuation $\delta\phi$ in real space, normalized by the standard deviation $\sigma=\sqrt{\langle \delta\phi^2 \rangle}$. The left plot shows the fluctuations at the initial time. The right plot shows the nearly scale-invariant fluctuations at the final time.}
	\label{fig:box}
\end{figure}
For illustrative purposes, in \cref{fig:box} we show the map of the scalar field fluctuation $\delta\phi=\phi-\bar\phi$ in real space. The left panel shows the Bunch-Davies UV-peaked fluctuations at the beginning of the simulation. The right panel shows the nearly scale-invariant fluctuations at the final time.

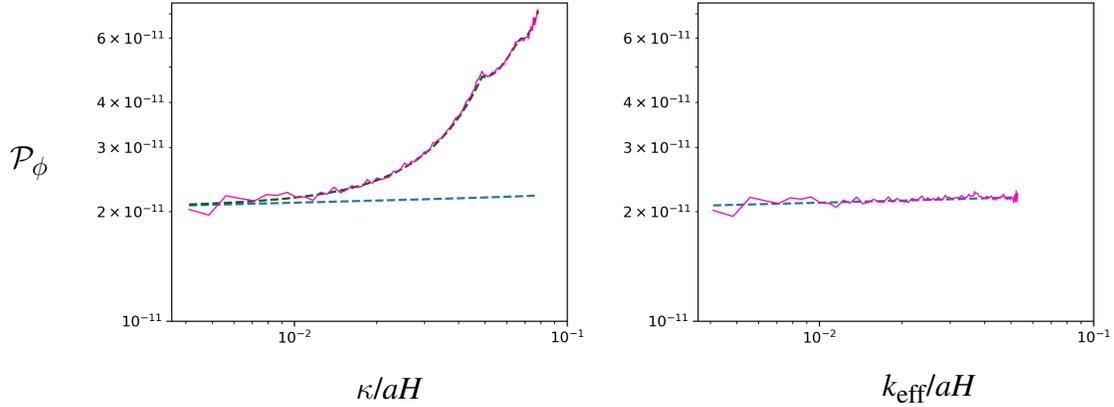
\begin{figure}
	\centering
	
	\begin{tikzpicture}
	\node (img) {\includegraphics[width=7.cm]{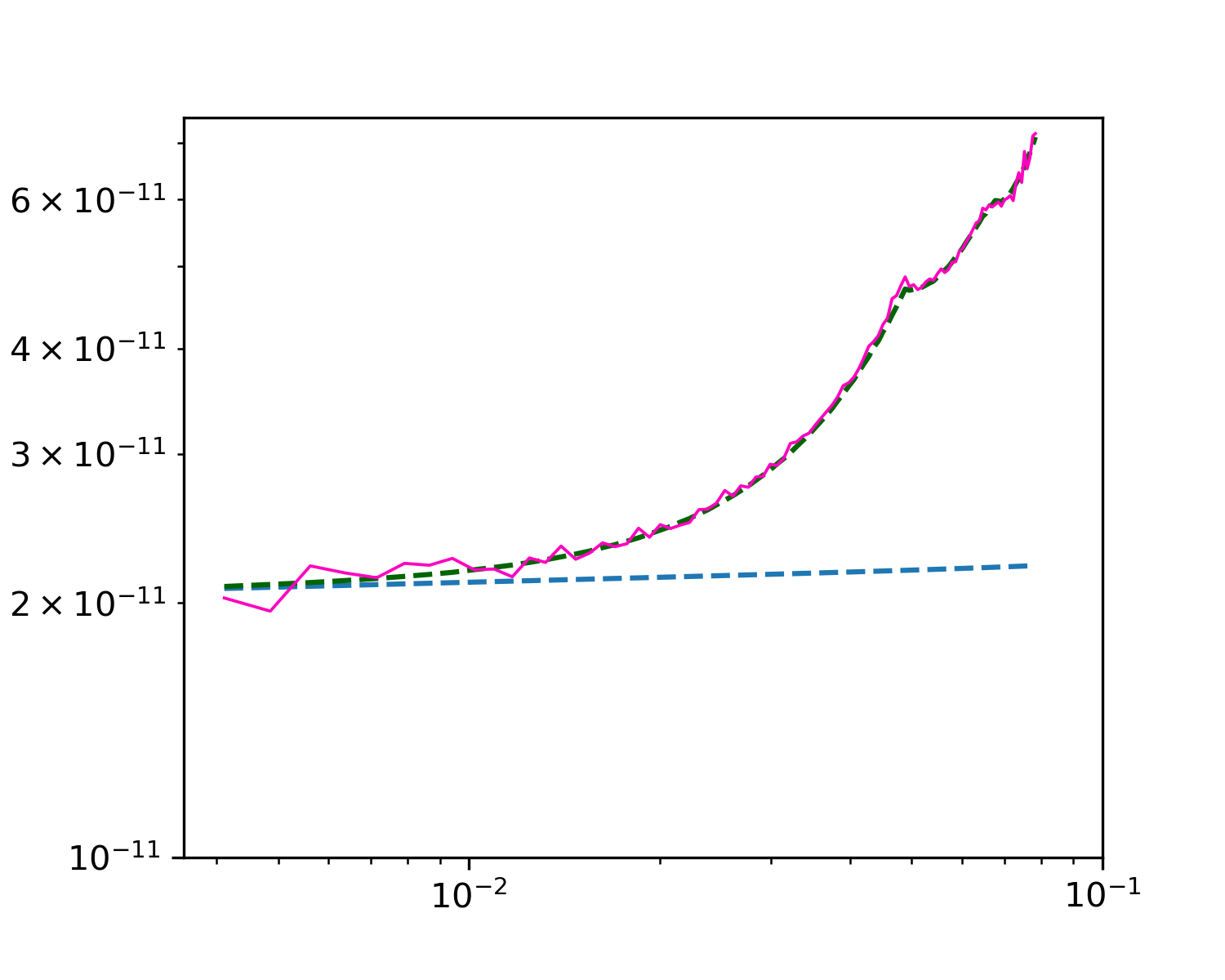}};
	
	\node [rotate=0,text width=0.01cm,align=center] at (-4.6,0){ $\mathcal{P}_{\phi}$};
	\node [text width=0.01cm,align=center] at (0,-3){$\kappa/aH$};

	\node (img2) at (7,0) {\includegraphics[width=7.cm]{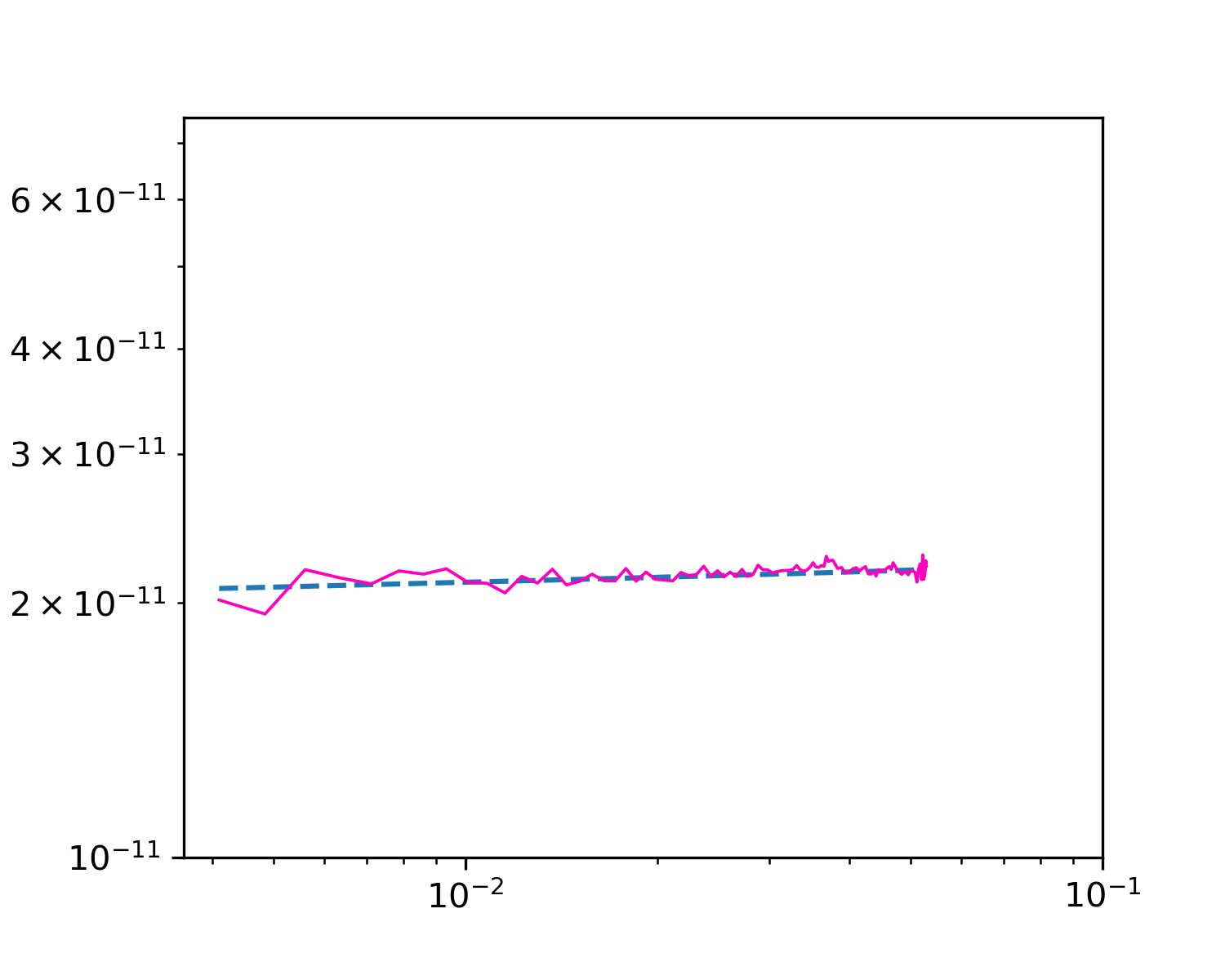}} ;
	
	\node [text width=0.01cm,align=center] at (0+7,-3){$k_{\text{eff}}/aH$};

	\end{tikzpicture}

	\caption{The final power spectrum computed from the lattice simulation (magenta line) compared to the theoretical prediction of \cref{eq:thprediction} (blue dotted line). In the left panel, we show the naive lattice result, while in the right panel we show the same result after taking into account the modified dispersion relation discussed in \cref{sec:modifieddr}.
		The green line in the left panel is the theoretical prediction for the discrete dynamics, as computed from \cref{eq:discMS}.}
	\label{fig:finalPS}
\end{figure}

In the left panel of \cref{fig:finalPS} we show the dimensionless power spectrum of the inflaton $\mathcal{P}_\phi$ at the end of the simulation, plotted against lattice modes $\kappa$ of \cref{eq:modes}. We compare the power spectrum computed from the simulation with the prediction for discrete dynamics, which is obtained by solving the discrete version of the Mukhanov-Sasaki \cref{eq:discMS} and is shown as a green line. From this plot, we can see that the discrete power spectrum is quite different from the almost scale-invariant power spectrum of the continuous theory, given by \cref{eq:thprediction} and depicted as a blue line in the plot. This is a manifestation of the different dynamics between discrete and continuous space. However, as we discussed in \cref{sec:modifieddr}, continuous and discrete dynamics are equivalent if we interpret $k_{\rm eff}$ of \cref{eq:keff} instead of $\kappa$ as the physical modes probed by the lattice simulation. 


In the right panel of \cref{fig:finalPS} we show the power spectrum from the simulation computed interpreting $k_{\rm eff}$ as physical modes and we compare it to the theoretical prediction of \cref{eq:thprediction}.
This equivalence principle allows us to reproduce with precision the nearly scale-invariant spectrum of single-field inflation. 
Note that interpreting $k_{\rm eff}$ as the physical modes will also reduce the resolution in Fourier space, which is computed from \cref{eq:effmax} and it is given by $k_{\text{eff,max}}\simeq 53.50H_{\rm in}$. 
In \cref{sec:diff} we also show the results of simulations with different stencils for the discrete Laplacian and compare the corresponding effective momenta.

\begin{figure}
	\centering
	
	\begin{tikzpicture}
	\node (img) {\includegraphics[width=14cm]{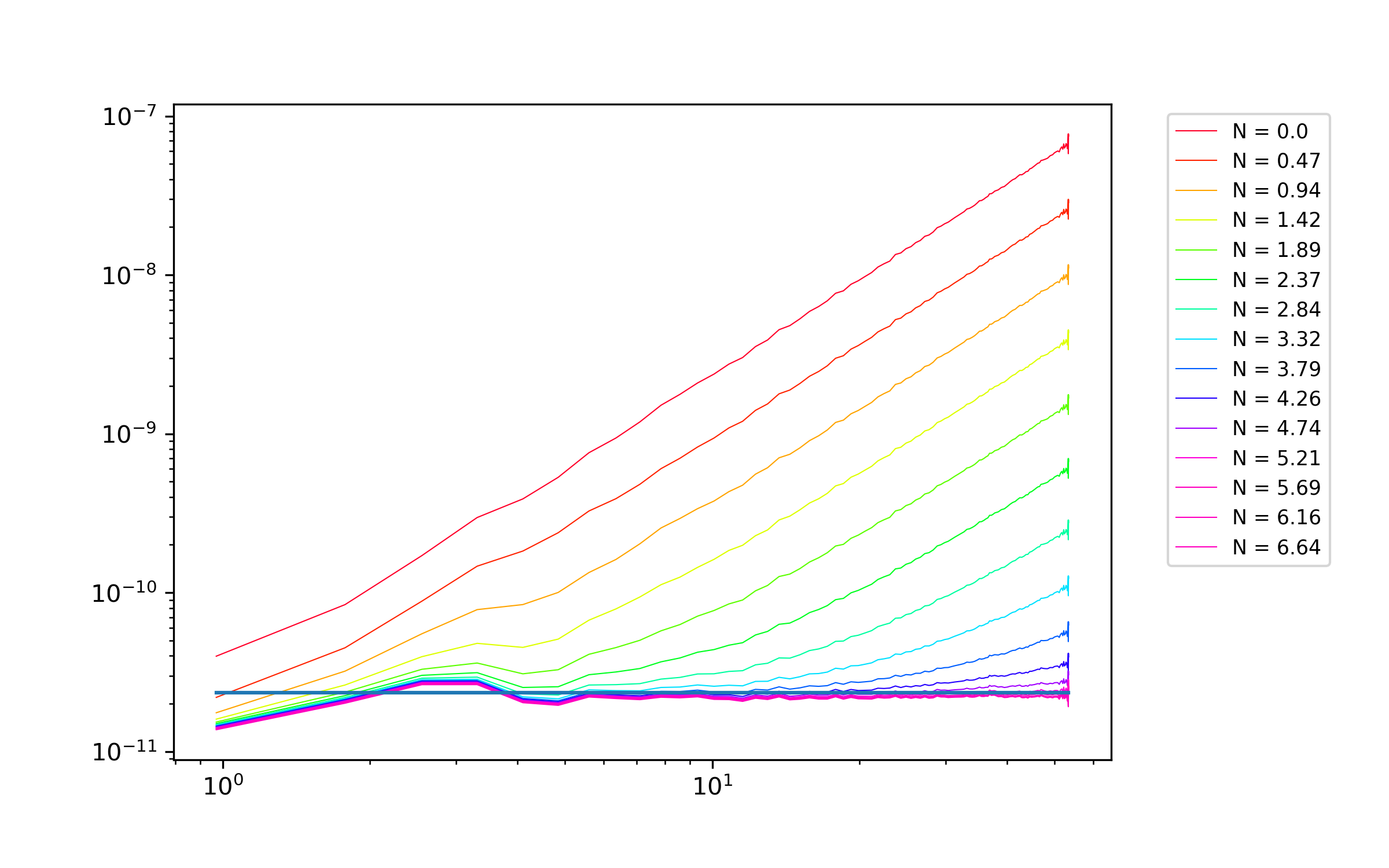}};
	\node [rotate=0,text width=0.01cm,align=center] at (-7,0){ $\mathcal{P}_{\phi}$};
	\node [text width=0.01cm,align=center] at (-.5,-4.2){$k_{\text{eff}}/H_i$};
	
	\end{tikzpicture}

	\caption{The evolution of the power spectrum of inflaton perturbation during the simulation. The colors go from early times (red) to late times (purple). The blue line represents the theoretical prediction for the final power spectrum as computed from \cref{eq:thprediction}.}
	\label{fig:ps}
	
\end{figure}

In \cref{fig:ps} we show the evolution of the power spectrum during the simulation, plotted at different times as a function of physical modes and going from the early-time Bunch-Davies state to the final scale-invariant state.

\subsection{Step potential}
\label{sec:resultsstep}
As a further example, in this section we show the results of the code for a model with potential:
\begin{equation}
V(\phi)=\frac{1}{2}m^2\phi^2\left[1+ s\tanh\left(\frac{\phi-\phi_{\rm step}}{d}\right) \right].
\end{equation}
This potential is analogous to the harmonic potential of the last section, but with a step localized at $\phi_{\rm step}$. This model has been studied in Ref. \cite{Adams_2001}, where the authors show that the presence of the step causes oscillations in the power spectrum of scalar perturbations. Here we show results for the same parameters of the last section. The only difference here is that we use $L=0.6/m$ as comoving size of the box, which corresponds to $k_{\text{eff,max}}\simeq124.84 H_{\rm in}$. Moreover, we have three extra parameters $s$, $d$, and $\phi_{\rm step}$. We choose $\phi_{\rm step} = 14.35$, and we run the simulation with different values of $s$ and $d$.
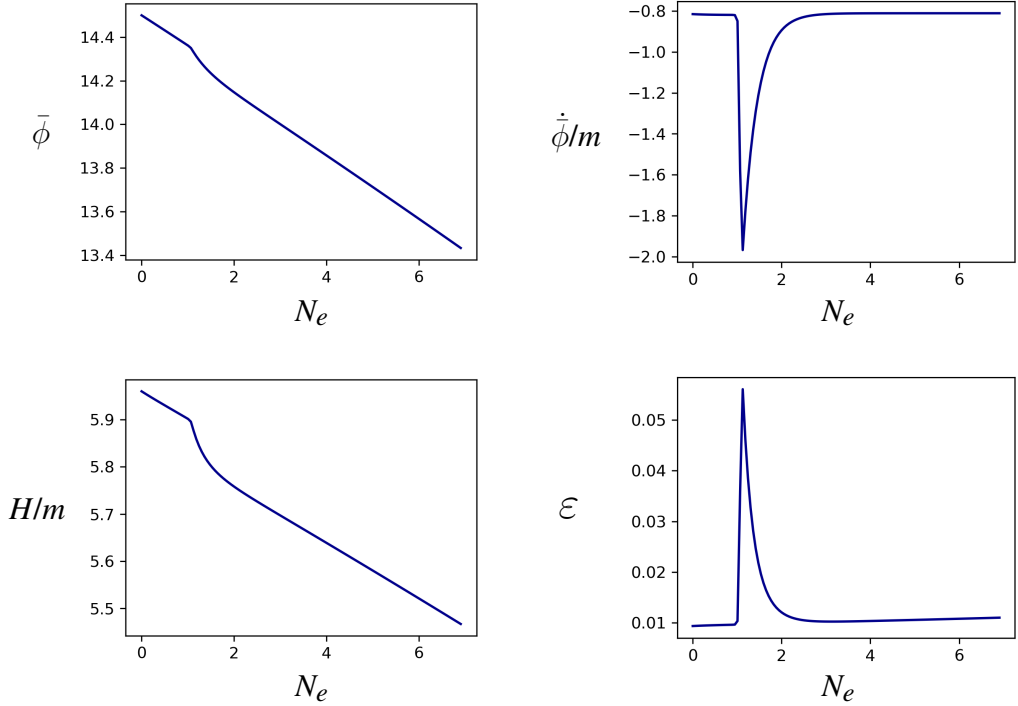
\begin{figure}
	\centering
	
	\begin{tikzpicture}
	\node (img) {\includegraphics[width=6cm]{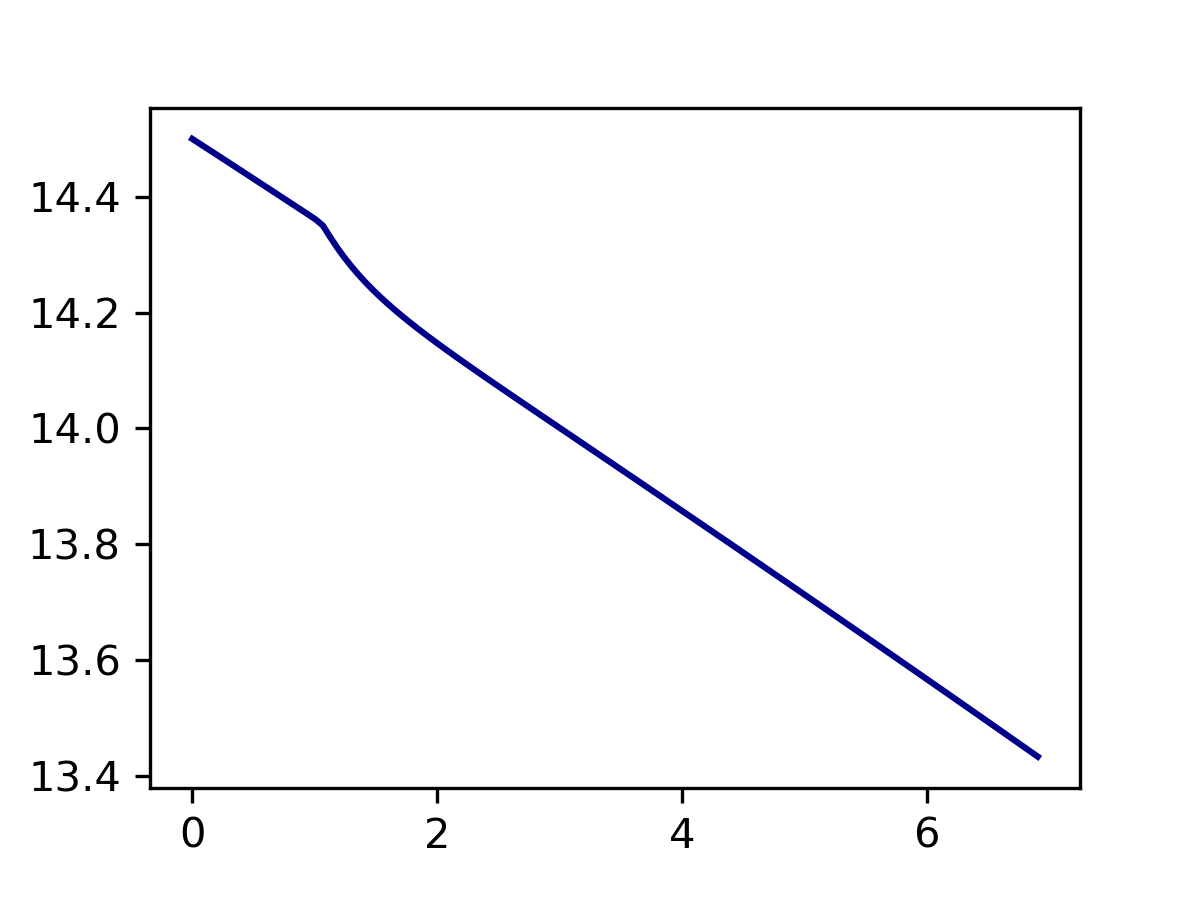}};
	
	\node [rotate=0,text width=0.01cm,align=center] at (-3.5,0){ $\bar{\phi}$};
	\node [text width=0.01cm,align=center] at (0,-2.4){$N_e$};

	\node (img2) at (7,0) {\includegraphics[width=6.4cm]{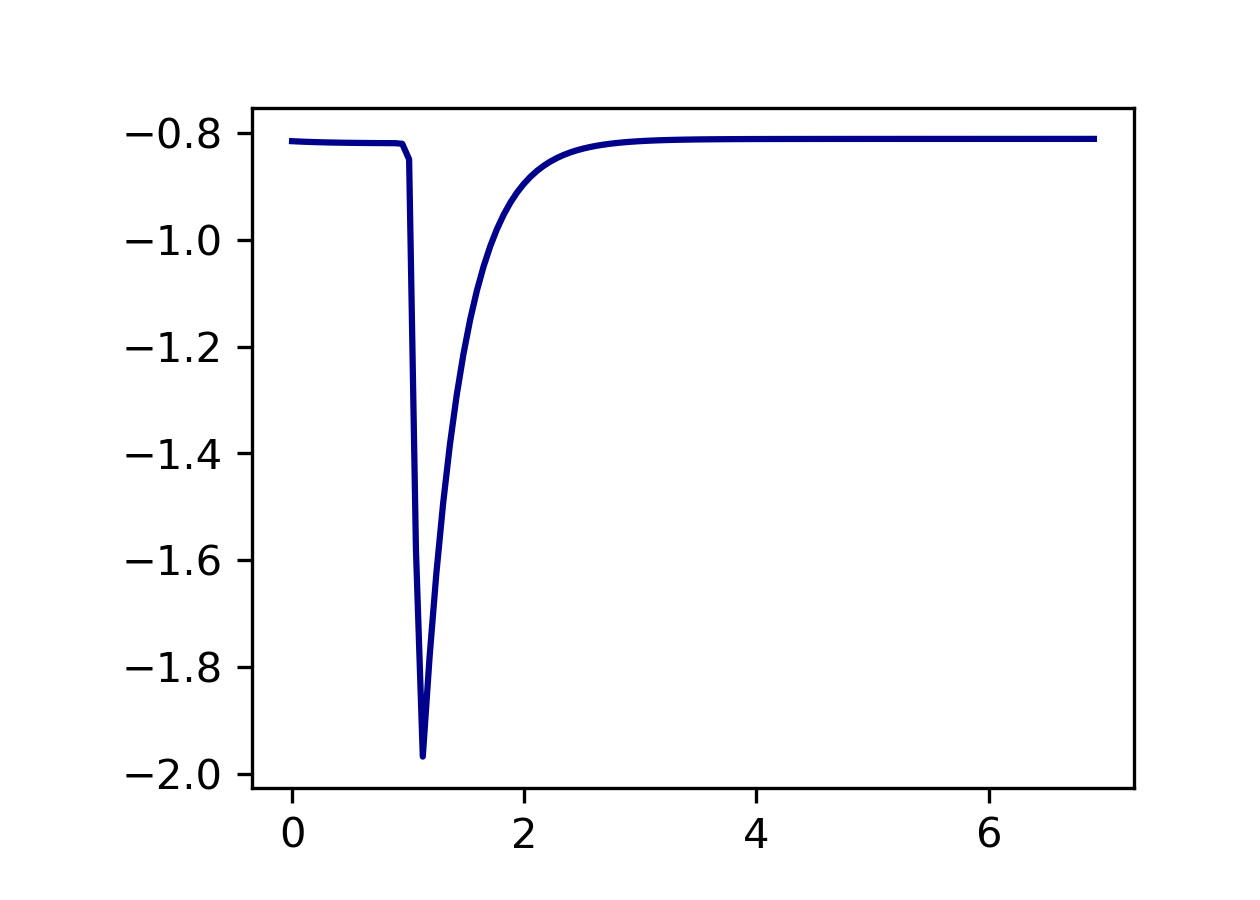}} ;
	
	\node [rotate=0,text width=0.01cm,align=center] at (-3.6+7,0){ ${\dot{\bar{\phi}}}/{m}$};
	\node [text width=0.01cm,align=center] at (0+7,-2.4){$N_e$};

	\node (img3) at (0,-5) {\includegraphics[width=6cm]{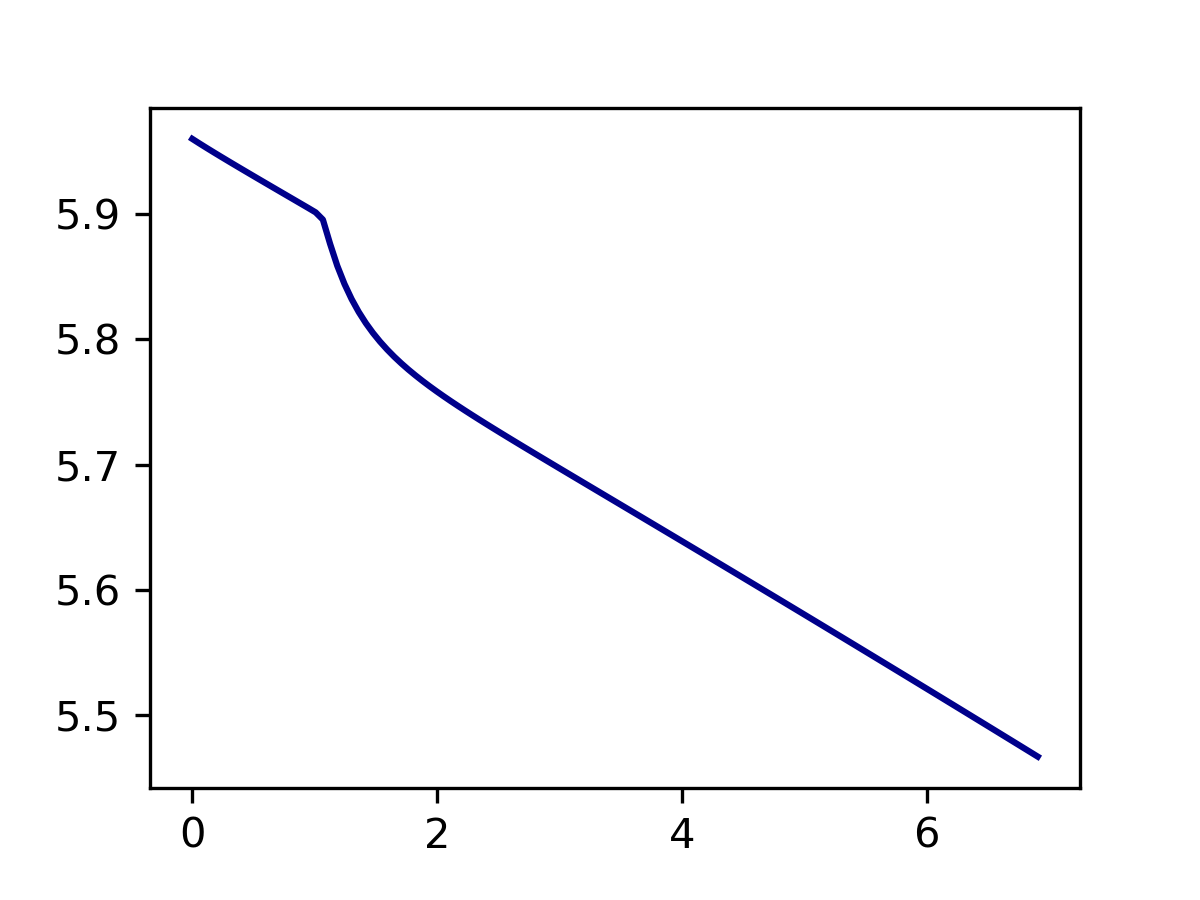}};
	
	\node [rotate=0,text width=0.01cm,align=center] at (-3.8,0-5){ $H/m$};
	\node [text width=0.01cm,align=center] at (0,-2.4-5){$N_e$};

	\node (img4) at (7,-5) {\includegraphics[width=6.4cm]{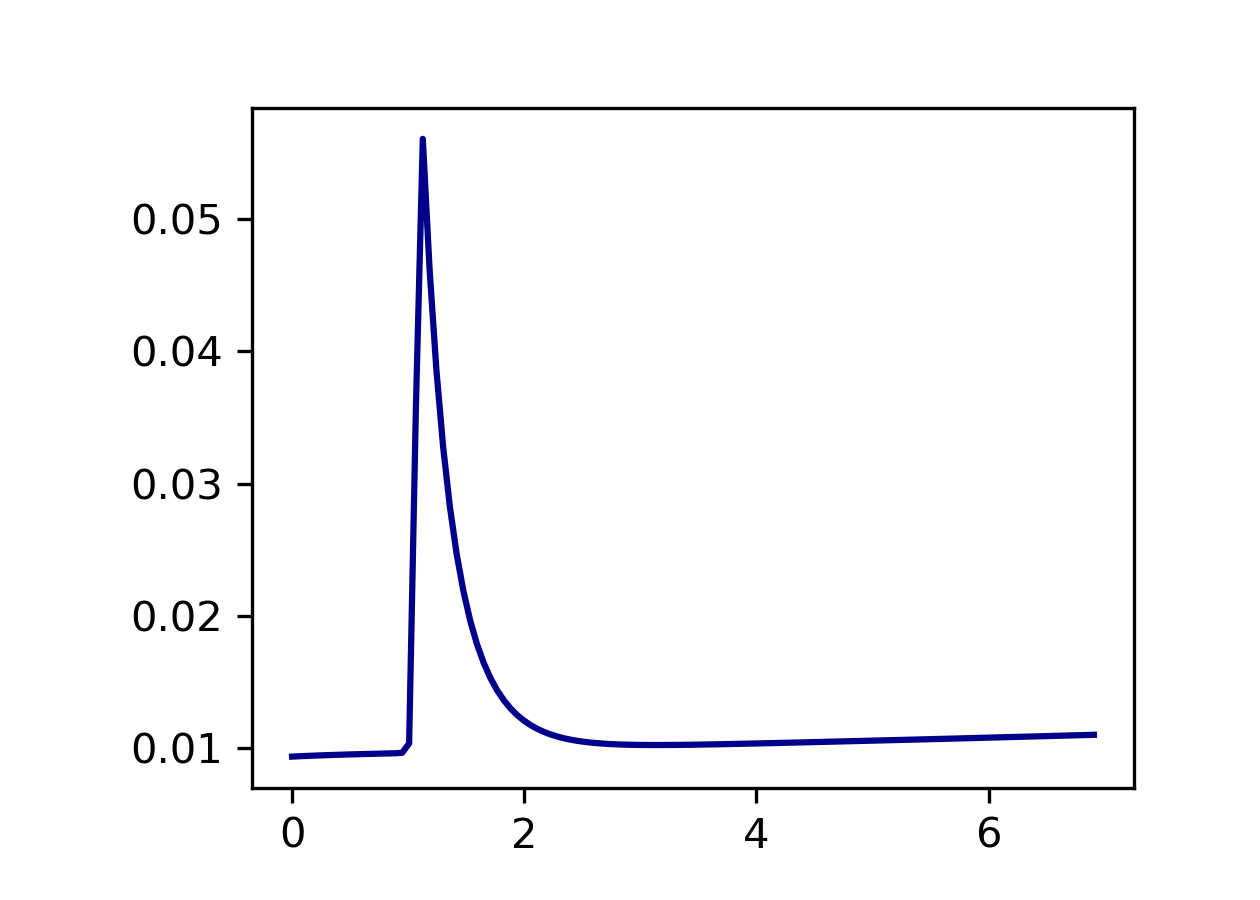}} ;
	
	\node [rotate=0,text width=0.01cm,align=center] at (-3.5+7,0-5){ \Large $\varepsilon$};
	\node [text width=0.01cm,align=center] at (0+7,-2.4-5){$N_e$};

	\end{tikzpicture}

	\caption{Background quantities during the simulation for a potential with a step with $s=0.01$, $d=0.005$ and $\phi_{\rm step}=14.35$. From top left: the background value of the inflaton $\bar\phi$, its derivative in cosmic time $\dot{\bar{\phi}}$, the Hubble parameter $H$ and the slow-roll parameter $\epsilon$.}
	\label{fig:backgroundvalues_step}
\end{figure}

In \cref{fig:backgroundvalues_step} we show background quantities in the case $s=0.01$, $d=0.005$. We can see here that the step in the potential causes a bump in all the background quantities, but without changing significantly the slow-roll dynamics of the inflaton. Indeed, $\varepsilon$ is still much smaller than 1 during the simulation and the departure of the inflaton from the slow-roll trajectory is relatively small. 

\begin{figure}
	\centering
	
	\begin{tikzpicture}
	\node (img) {\includegraphics[width=12cm]{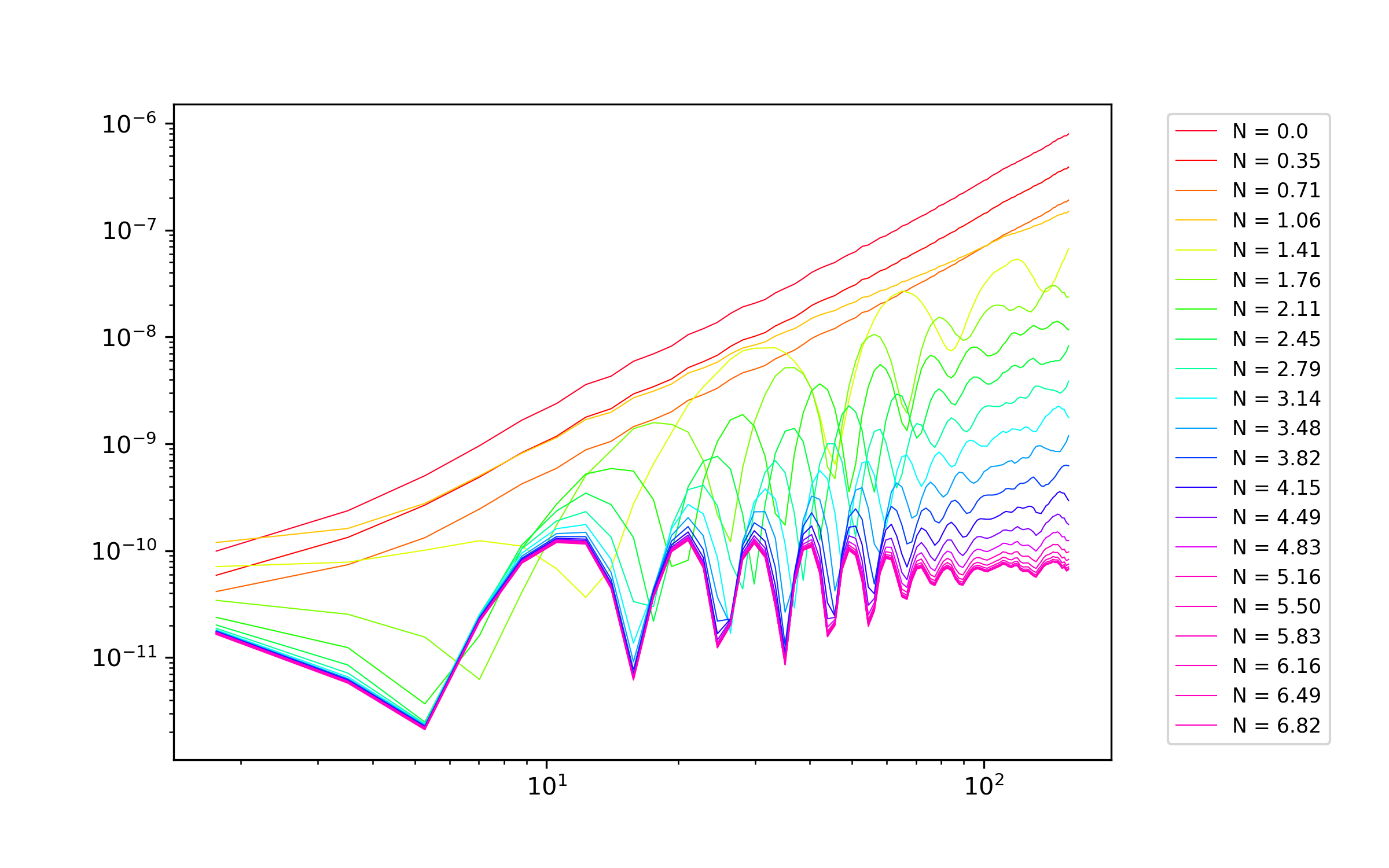}};
	\node [rotate=0,text width=0.01cm,align=center] at (-6.3,0){ $\mathcal{P}_{\phi}$};
	\node [text width=0.01cm,align=center] at (-.5,-3.5){$k_{\text{eff}}/H_i$};

	\end{tikzpicture}

	\caption{The evolution of the power spectrum of inflaton perturbation during the simulation for the step potential. The colors go from early times (red) to late times (purple). We show the result for $s=0.01$ and $d=0.005$.}
	\label{fig:ps_step}
	
\end{figure}

In \cref{fig:ps_step} we show the evolution of the power spectrum during the simulation for $s=0.01$ and $d=0.005$. Here we can clearly see that the presence of the step introduces oscillations in the power spectrum.

\begin{figure}
	\centering
	
	\begin{tikzpicture}
	\node (img) {\includegraphics[width=7.cm]{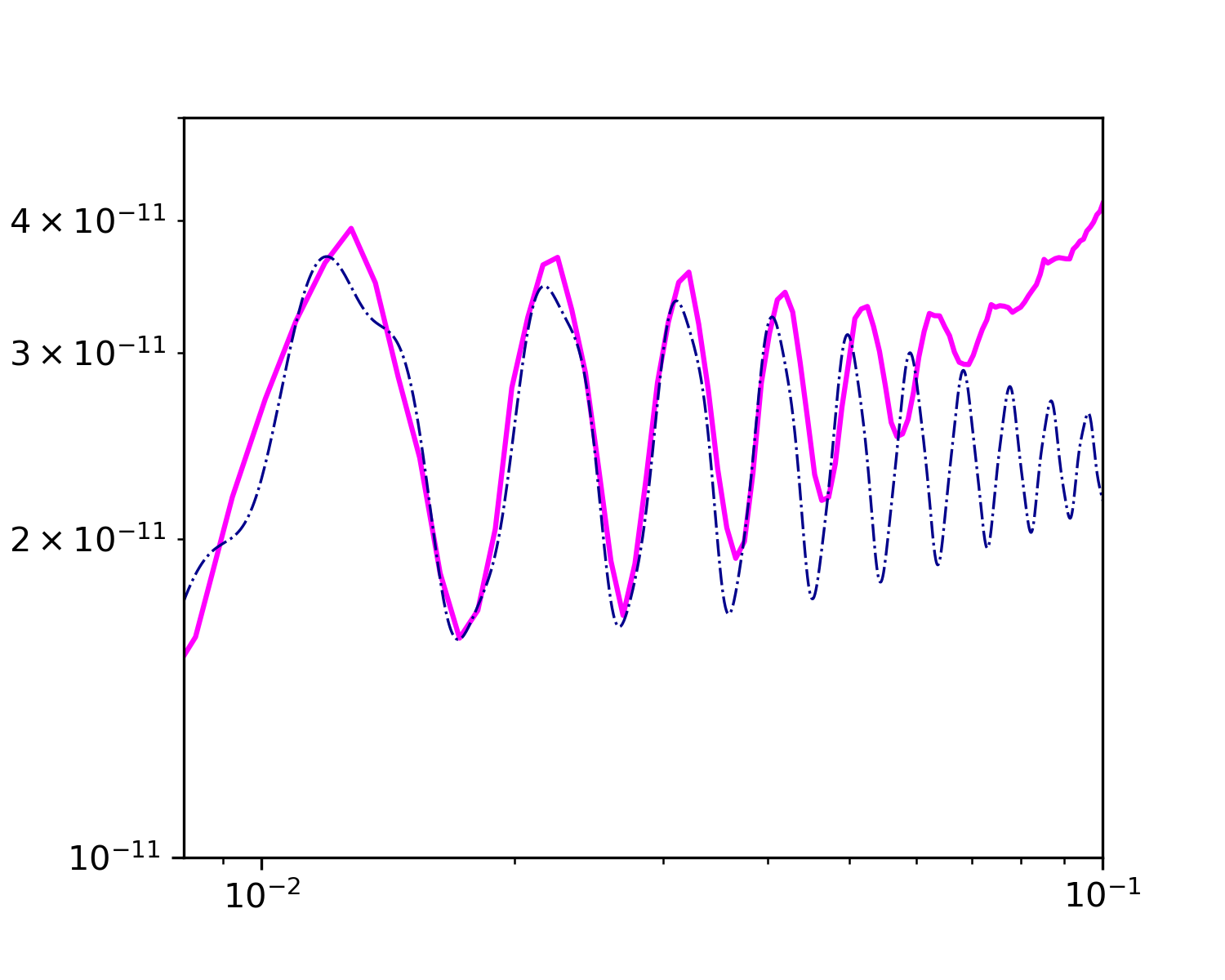}};
	
	\node [rotate=0,text width=0.01cm,align=center] at (-4.5,0){ $\mathcal{P}_{\phi}$};
	\node [text width=0.01cm,align=center] at (0,-2.7){$\kappa/aH$};

	\node (img2) at (7,0) {\includegraphics[width=7.cm]{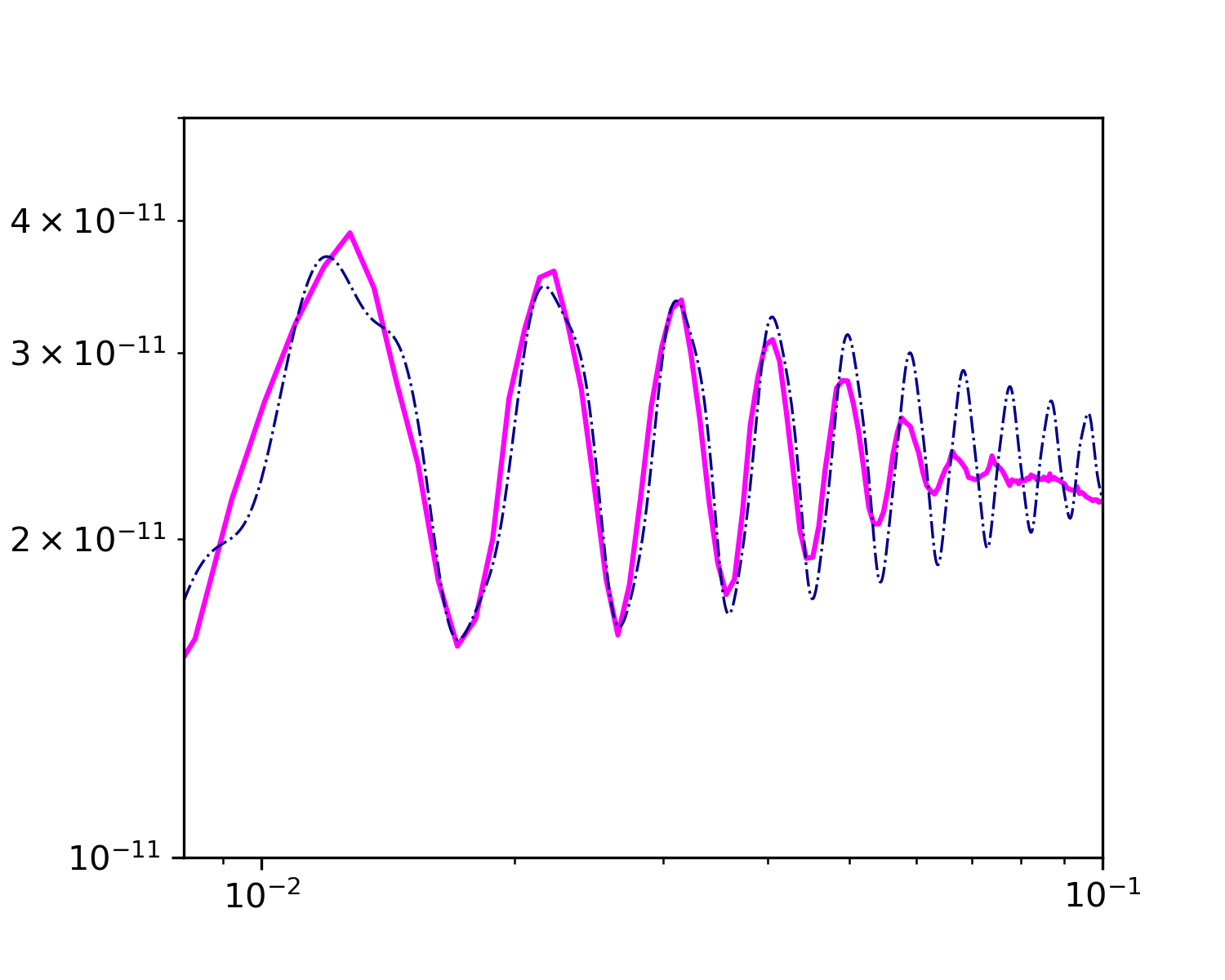}} ;
	
	\node [text width=0.01cm,align=center] at (0+7,-2.7){$k_{\text{eff}}/aH$};

	\end{tikzpicture}

	\caption{The final power spectrum computed from the lattice simulation (magenta line) compared with the theoretical prediction computed with a linear code that solves numerically the Mukhanov-Sasaki \cref{eq:MS} (blue dotted line). The step parameters are $s=0.001$ and $d=0.005$ for this plot. In the right panel, we show the result obtained taking into account the modified dispersion relation discussed in \cref{sec:modifieddr}, while on the left we show the naive result before taking into account the equivalence $k\leftrightarrow k_{\rm eff}$.}
	\label{fig:finalPS_step}
\end{figure}
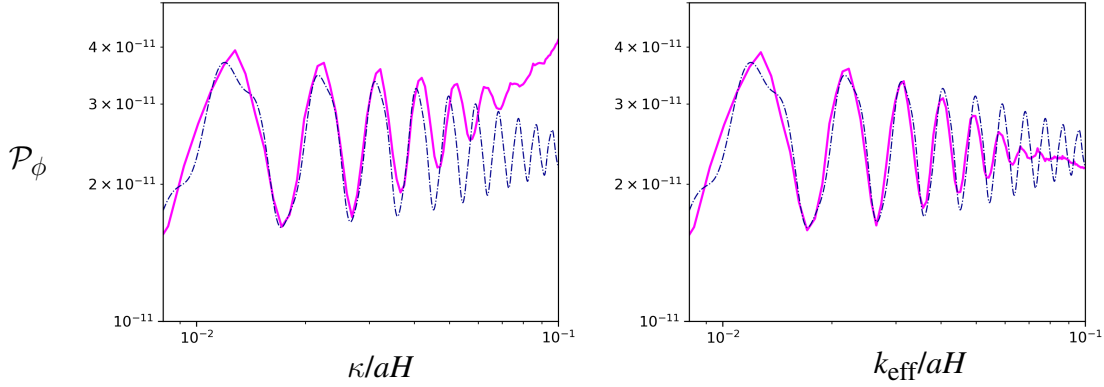

In \cref{fig:finalPS_step} we show the final power spectrum of a simulation run with $s=0.001$, $d=0.005$ and we compare it with the result obtained by solving the Mukhanov-Sasaki equation \eqref{eq:MS} with a numerical integrator, which serves as a theoretical prediction.  For this simulation, we increased the number of lattice points to $N^3=256^3$ and the box size to $L=1.2/m$ in order to improve the spatial resolution. In the right panel of this figure, we show the result obtained by interpreting $k_{\rm eff}$ the physical modes, while in the left panel we show the power spectrum without this identification. From the right plot, we can see that the matching between the theoretical prediction and the lattice simulation is not perfect, in particular for the largest modes of the simulation. However, the lattice code is able to correctly reproduce the oscillations, that have the same frequency and a similar amplitude compared to the theoretical prediction. 
In this example, we can again see that interpreting $k_{\rm eff}$ as the physical modes allows us to obtain a more precise result.







\subsection{Different stencils for the Laplacian operator}
\label{sec:diff}
We now consider different stencils for the Laplacian and discuss their effects on the evolution of perturbations. We refer to the Laplacian of \cref{eq:discretelaplacian} and its corresponding effective momentum as $L^{(2)}[\phi]_{i_1,i_2,i_3}$ and $k^{(2)}_{\rm eff}$, where the $2$ refers to the second order of the stencil.
The first one we consider is the following 4th order stencil, which has a similar structure of $L^{(2)}[\phi]$ but involves more points:
\begin{equation}
L^{(4)}[\phi]_{i_1,i_2,i_3}=\frac{1}{dx^2}\sum_{a_1,a_2,a_3}c_{a_1,a_2,a_3}\phi_{i_1+a_1,i_2+a_2,i_3+a_3},
\end{equation}
where the only non-zero coefficients are $c_{\pm1,0,0}=c_{0,\pm1,0}=c_{0,0,\pm1}=4/3$, $c_{\pm2,0,0}=c_{0,\pm2,0}=c_{0,0,\pm2}=-1/12$ and $c_{0,0,0}=-15/2$.

\begin{figure}
	\centering
	
	\begin{tikzpicture}
	\node (img) {\includegraphics[width=12cm]{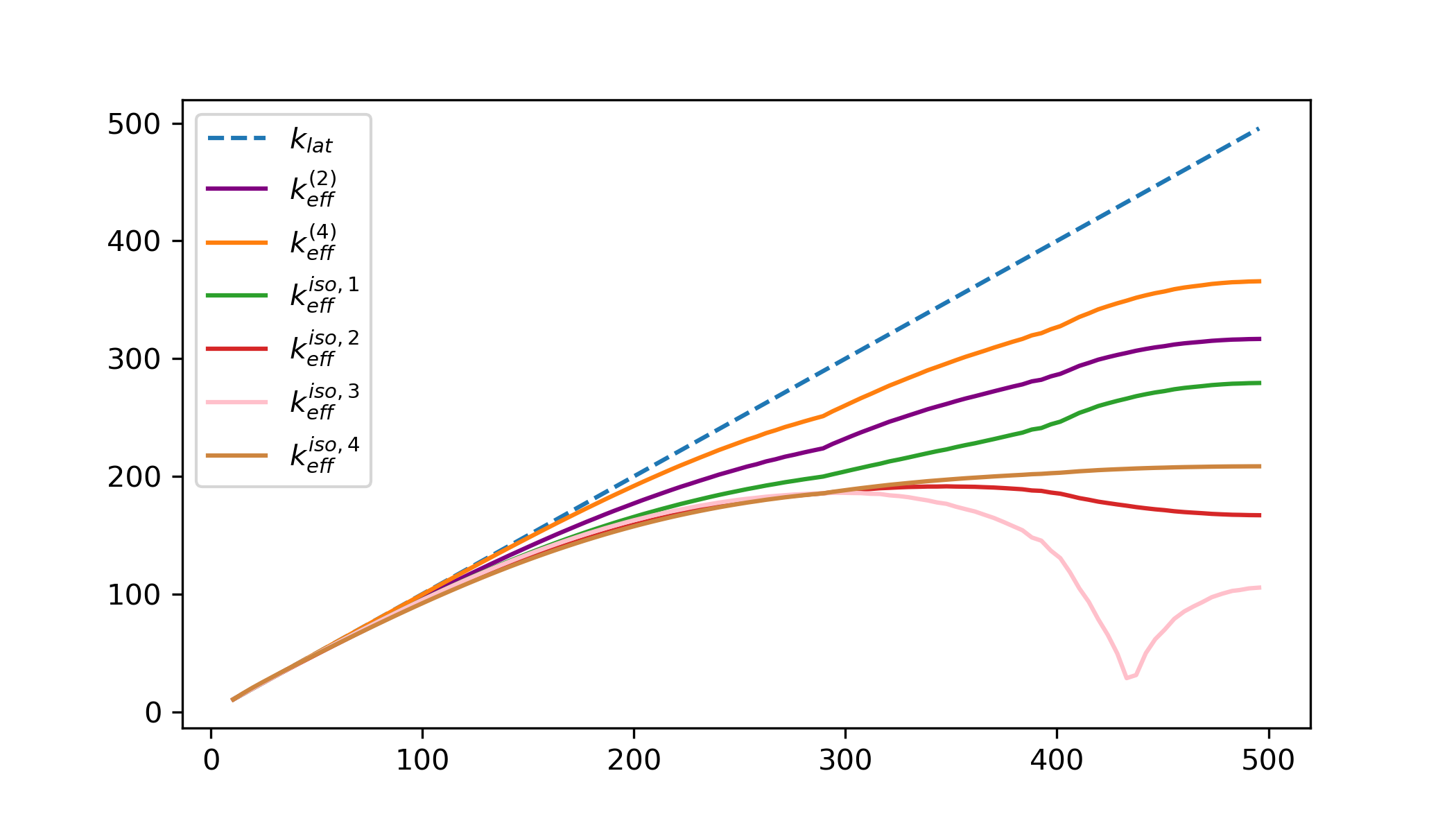}};
	\node [text width=0.01cm,align=center] at (-6.5,0){$k_{\rm eff}/m$};
	\node [text width=0.01cm,align=center] at (-.5,-3.5){$\kappa/m$};

	\end{tikzpicture}

	\caption{Comparison of the effective momenta coming from different Laplacian operators defined in \cref{sec:diff} for a lattice with $N=128$ and $L=1.4/m$.}
	\label{fig:compare}
	
\end{figure}

Next, we consider the isotropic second-order stencils defined in Ref. \cite{stencils}. We display the coefficients associated to these stencils as:
\begin{equation}
\begin{bmatrix}
c_{1,1,1}      & c_{0,1,1} & c_{-1,1,1}  \\
c_{1,0,1}      & c_{0,0,1} & c_{-1,0,1}   \\
c_{1,-1,1}       & c_{0,-1,1}   & c_{-1,-1,1}  
\end{bmatrix}
\begin{bmatrix}
c_{1,1,0}      & c_{0,1,0} & c_{-1,1,0}  \\
c_{1,0,0}      & c_{0,0,0} & c_{-1,0,0}   \\
c_{1,-1,0}       & c_{1,-1,0}   & c_{-1,-1,0}  
\end{bmatrix}
\begin{bmatrix}
c_{1,1,-1}      & c_{0,1,-1} & c_{-1,1,-1}  \\
c_{1,0,-1}      & c_{0,0,-1} & c_{-1,0,-1}   \\
c_{1,-1,-1}       & c_{1,-1,-1}   & c_{-1,-1,-1}  
\end{bmatrix}.
\end{equation}
With this convention, we can display the 4 isotropic stencils of Ref. \cite{stencils} as:
\begin{equation}
\quad \quad \quad\,\, L^{\rm iso,1}[\phi]: \begin{bmatrix}
1/12     & 0& 1/12   \\
0      & 2/3 & 0   \\
1/12     & 0  & 1/12 
\end{bmatrix}
\begin{bmatrix}
0     &  \frac{2}{3} & 0 \\
2/3   & -14/3 & 2/3  \\
0      &  \frac{2}{3}   &0
\end{bmatrix}
\begin{bmatrix}
1/12    & 0& 1/12    \\
0      & 2/3 & 0   \\
1/12     & 0  &1/12   
\end{bmatrix}
\quad\quad\quad\quad\quad\quad\quad
\end{equation}
\begin{equation}
\quad L^{\rm iso,2}[\phi]: \begin{bmatrix}
0   & 1/6& 0  \\
1/6     & 1/3 &  1/6   \\
0      &  1/6  & 0
\end{bmatrix}
\begin{bmatrix}
1/6     &  1/3 &  1/6 \\
1/3    & -4 &  1/3  \\
1/6     &  1/3   & 1/6
\end{bmatrix}
\begin{bmatrix}
0   &  1/6 &0   \\
1/6    & 1/3 &  1/6  \\
0      &  1/6  & 0  
\end{bmatrix}
\quad\quad\quad\quad\quad
\end{equation}
\begin{equation}
\quad L^{\rm iso,3}[\phi]: \begin{bmatrix}
-1/12   & 1/3& -1/12  \\
1/3     & 0 &  1/3   \\
-1/12      &  1/3  & -1/12
\end{bmatrix}
\begin{bmatrix}
1/3     &  0 &  1/3 \\
0   & -10/3 &  0  \\
1/3     &  0   & 1/3
\end{bmatrix}
\begin{bmatrix}
-1/12   &  1/3 &-1/12   \\
1/3    & 0 &  1/3  \\
-1/12     &  1/3  & -1/12  
\end{bmatrix}
\end{equation}
\begin{equation}
L^{\rm iso,4}[\phi]: \begin{bmatrix}
1/30   & 1/10& 1/30  \\
1/10    & 7/15 &  1/10   \\
1/30      &  1/10  & 1/30
\end{bmatrix}
\begin{bmatrix}
1/10     &  7/15 &  1/10\\
7/15   & -64/15 &  7/15  \\
1/10     &  7/15   & 1/10
\end{bmatrix}
\begin{bmatrix}
1/30   &  1/10 &1/30   \\
1/3    & 7/15 &  1/3  \\
1/30     &  1/10  & 1/30  
\end{bmatrix}.
\end{equation}
For each stencil $L^{i}$ we refer to its corresponding effective momentum as $k_{\rm eff}^{i}$. We avoid writing the lengthy expressions for all the effective momenta, but we plot them in \cref{fig:compare} for a lattice with $N=128$ and $L=1.4/m$.
All the $k^{\rm iso,i}_{\rm eff}$ are real, with the exception of $k^{\rm iso, 3}_{\rm eff}$ which becomes purely imaginary around $\kappa\simeq435m$ (we show the absolute value of $k^{\rm iso, 3}_{\rm eff}$ in the plot). From this plot we can see that only $L^{(4)}$ performs better than $L^{(2)}$ in terms of $k_{\rm eff,max}$ and in terms of overall deviation from $\kappa$, while the other isotropic stencils are significantly worse in this sense. The isotropic stencils, however, might perform better from other points of view. For example, these stencils do not have directional dependence in the second order truncation term in real space \cite{stencils}, contrarily to $L^{(2)}$ and $L^{(4)}$.

\begin{figure}
	\centering
	
	\begin{tikzpicture}
	\node (img) {\includegraphics[width=8.cm]{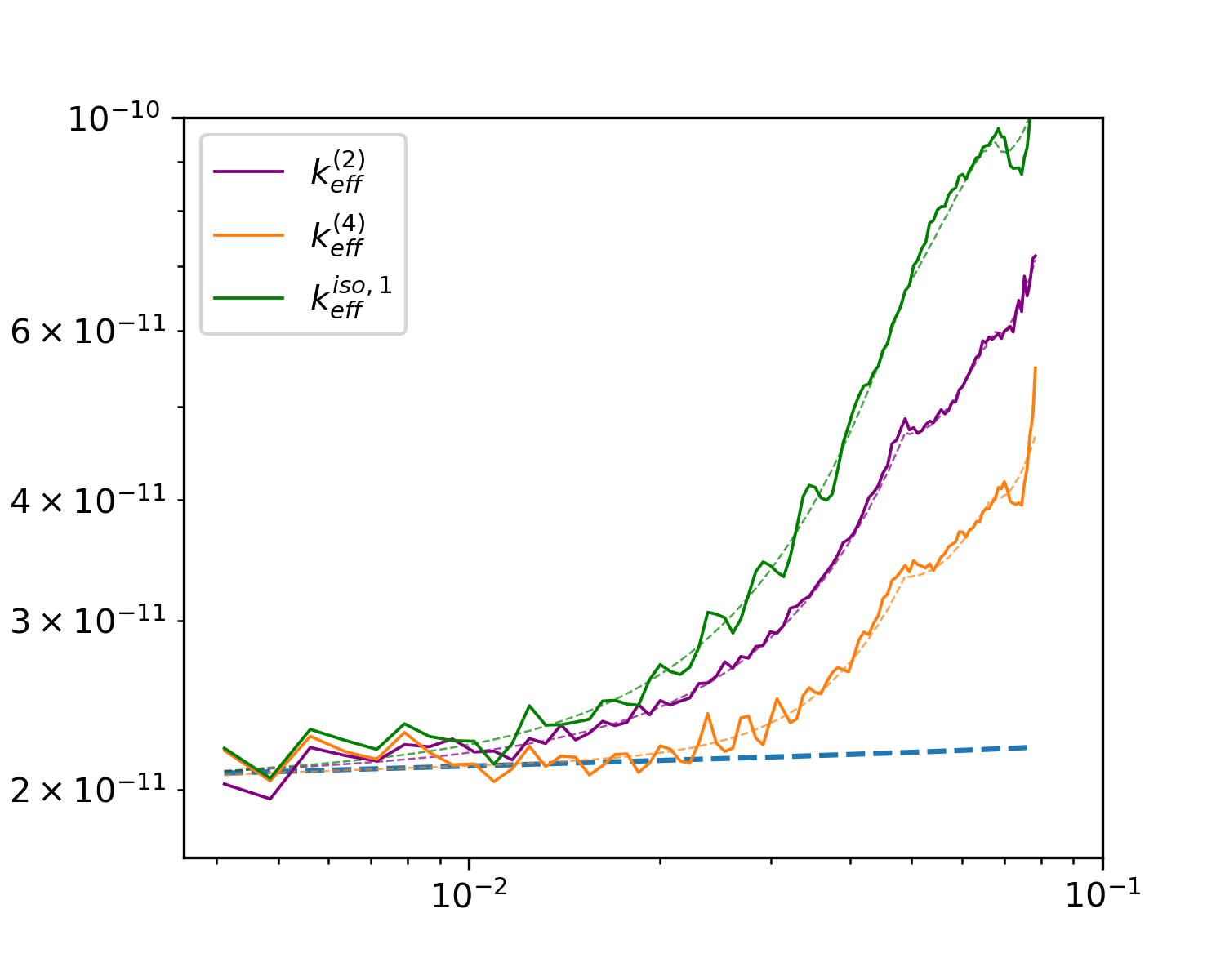}};
	
	\node [rotate=0,text width=0.01cm,align=center] at (-4.6,0){ $\mathcal{P}_{\phi}$};
	\node [text width=0.01cm,align=center] at (0,-3){$\kappa/aH$};

	\node (img2) at (7.5,0) {\includegraphics[width=8.cm]{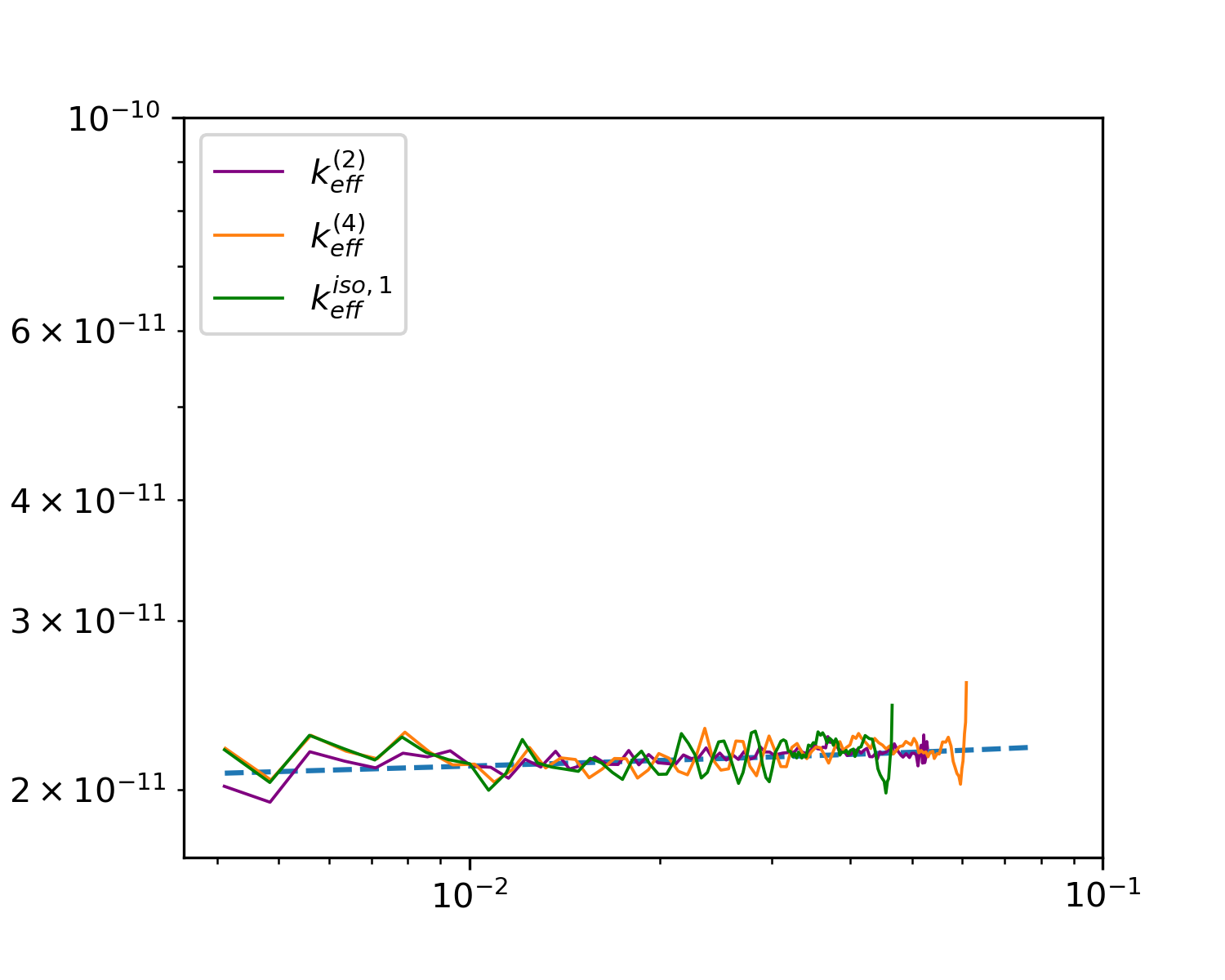}} ;
	
	\node [text width=0.01cm,align=center] at (0+7,-3){$k_{\text{eff}}/aH$};

	\end{tikzpicture}

	\caption{The final power spectrum computed from the lattice simulation for different stencils for the Laplacian operator. In the left panel, we show the lattice results without taking into account the modified dispersion relation and compare it to the continuous result (blue dashed line). The dashed lines in the left panel are the predictions for discrete dynamics computed from \cref{eq:discMS}. In the right panel, we show the results after the identification $k_{\rm eff}\leftrightarrow k$.}
	\label{fig:finalPS_stencils}
\end{figure}
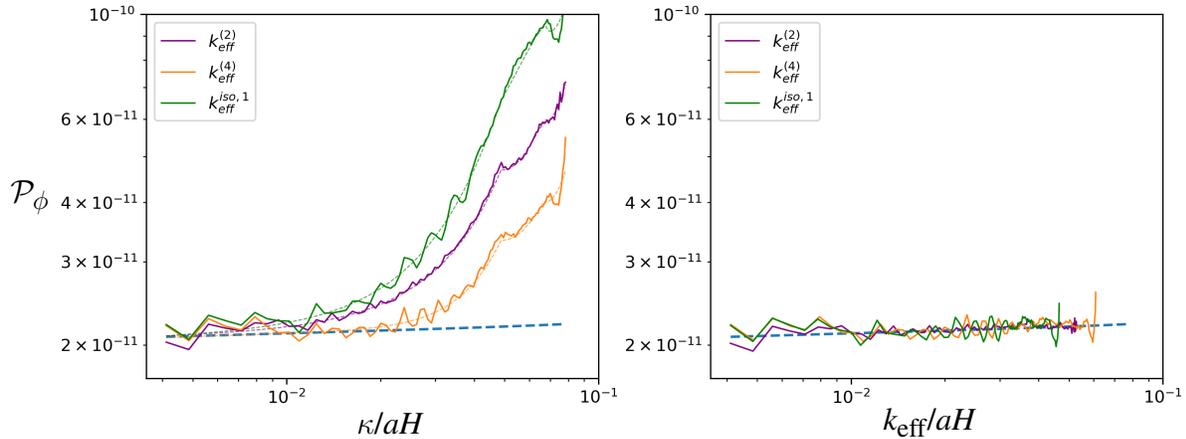

In \cref{fig:finalPS_stencils} we show the final power spectrum computed from simulations with different stencils for the Laplacian. We run these simulations with the $\frac{1}{2}m^2\phi^2$ potential and with the same parameters of \cref{sec:resultsslow}. We compare results from $L^{(2)}$, $L^{(4)}$ and, the isotropic stencil $L^{\rm iso,1}$. This figure is analogous to \cref{fig:finalPS}, and the dashed lines in the left plot are the analytical predictions for discrete dynamics computed from \cref{eq:discMS}. In all these cases, we can see that the identification $k_{\rm eff}\leftrightarrow k$ allows us to recover the continuous result with good precision (right panel of \cref{fig:finalPS_stencils}). The same result can be obtained with the other isotropic stencils $L^{iso,i}$, that we do not show in order to make the plots more readable. For $L^{\rm iso, 2}$ and $L^{\rm iso, 3}$, however, this is true only up to a certain momentum cutoff after which $k_{\rm eff}(\kappa)$ starts decreasing, making the modes unphysical (see \cref{fig:compare}).

\subsection{Energy conservation and numerical accuracy}
\label{sec:energy}
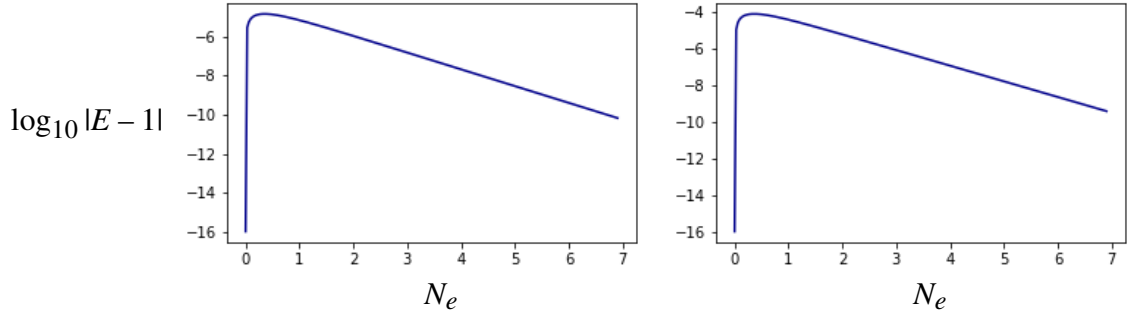
\begin{figure}
	\centering
	
	\begin{tikzpicture}
	\node (img) {\includegraphics[width=7cm]{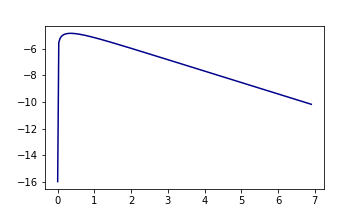}};
	
	\node [rotate=0,text width=0.01cm,align=center] at (-5.5,0){ $\log_{10}{|E-1|}$};
	\node [text width=0.01cm,align=center] at (0,-2.3){$N_e$};

	\node (img2) at (6.5,0) {\includegraphics[width=7cm]{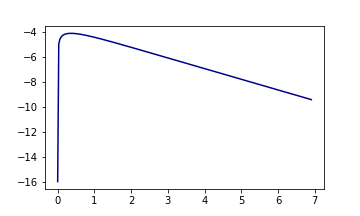}} ;
	
	\node [text width=0.01cm,align=center] at (+6.5,-2.3){$N_e$};

	\end{tikzpicture}

	\caption{Energy violation during the simulation in the case of the RK4 integrator introduced in \cref{sec:numint}, which is the numerical integrator employed in most of this thesis. On the left, we show the result for the standard slow roll potential of \cref{sec:resultsslow}, while on the right we show the result for the step potential of \cref{sec:resultsstep} with $s=0.01$ and $d=0.005$.}
	\label{fig:energy}
\end{figure}

We now discuss energy conservation during the numerical integration. In this chapter, we employed a Runge-Kutta explicit fourth-order integrator (RK4). In \cref{fig:energy} we show a plot of the quantity $E$ defined in \cref{eq:cons} during the simulation. The left panel shows the energy conservation plot in the case of the slow-roll potential of \cref{sec:resultsslow}. The right panel shows the same plot for the step potential of \cref{sec:resultsstep} with $s=0.01$ and $d=0.005$. 

From this figure, we can see that energy violation shows a peak at early times, and then progressively decreases. This small initial energy violation is a consequence of the fact that we used an explicit Runge-Kutta integrator, which is not symplectic and is not expected to preserve energy with good precision. Energy violation is larger during the first e-folds of evolution due to the nature of the initial Bunch-Davies state. Indeed, Bunch-Davies fluctuations are UV-peaked and rapidly oscillating, making it harder to numerically solve their evolution with good precision. Moreover, the typical size of the fluctuations is much larger at early times. This can be seen in \cref{fig:finalPS,fig:finalPS_step}, where the power spectrum at early times is UV-peaked and much larger. This initial energy violation is larger in the case of the step potential, and this is due to the higher UV-cutoff of this simulation. Usually, a good criterion for the energy conservation check is that the violation $|E-1|$ should be smaller than the typical size of fluctuations $\delta\phi$, which is given by the $\sigma=\sqrt{\langle \delta\phi^2\rangle}$ used in \cref{fig:box}. This quantity is roughly $\sim 10^{-2}$ at the beginning of the simulation and $\sim 10^{-4}$ at the end. This means that energy conservation satisfies this simple criterion in our code.

\begin{figure}
	\centering
	
	\begin{tikzpicture}
\node (img) {\includegraphics[width=7cm]{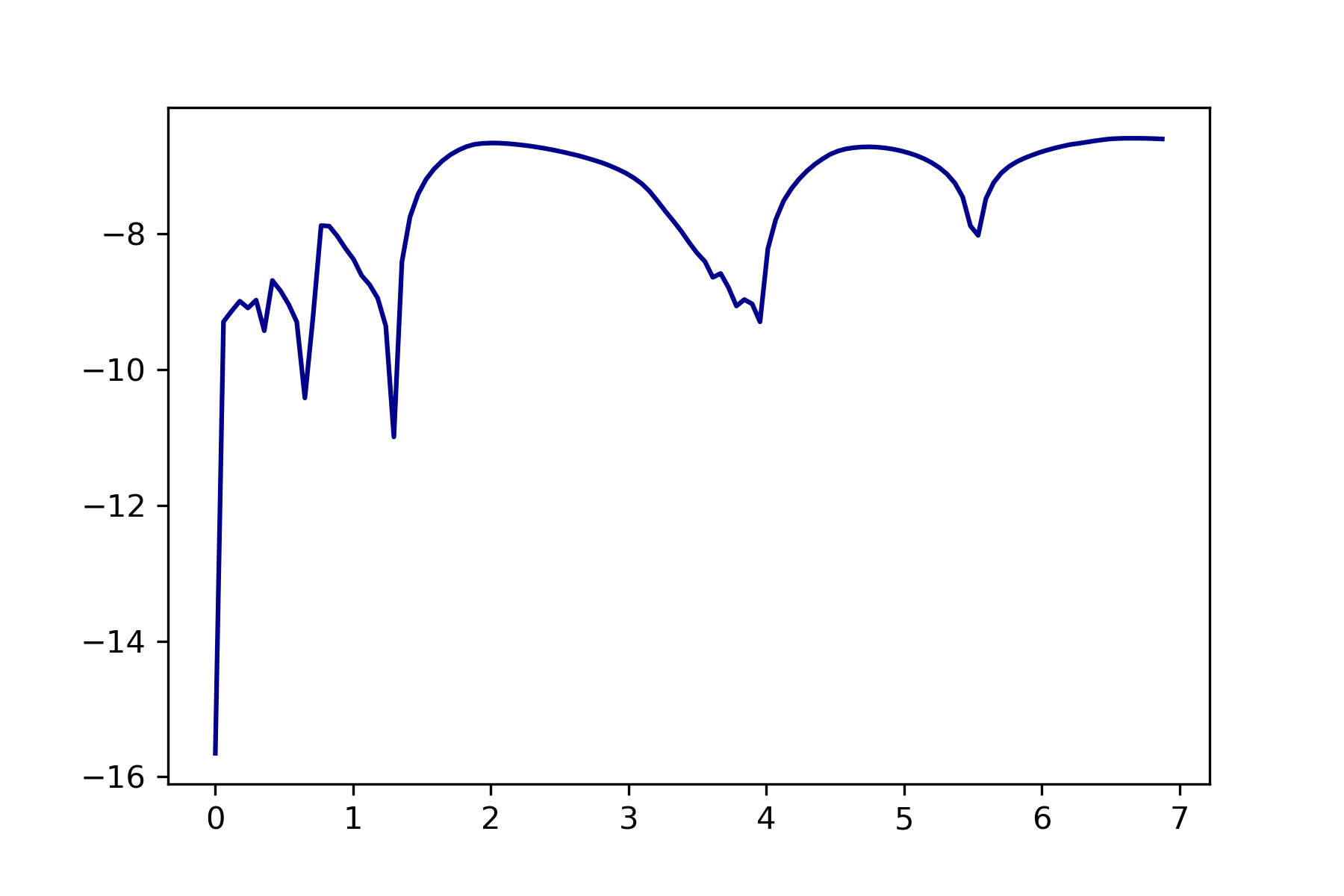}};

\node [rotate=0,text width=0.01cm,align=center] at (-5.5,0){ $\log_{10}{|E-1|}$};
\node [text width=0.01cm,align=center] at (0,-2.5){$N_e$};

\node (img2) at (6.5,0) {\includegraphics[width=7cm]{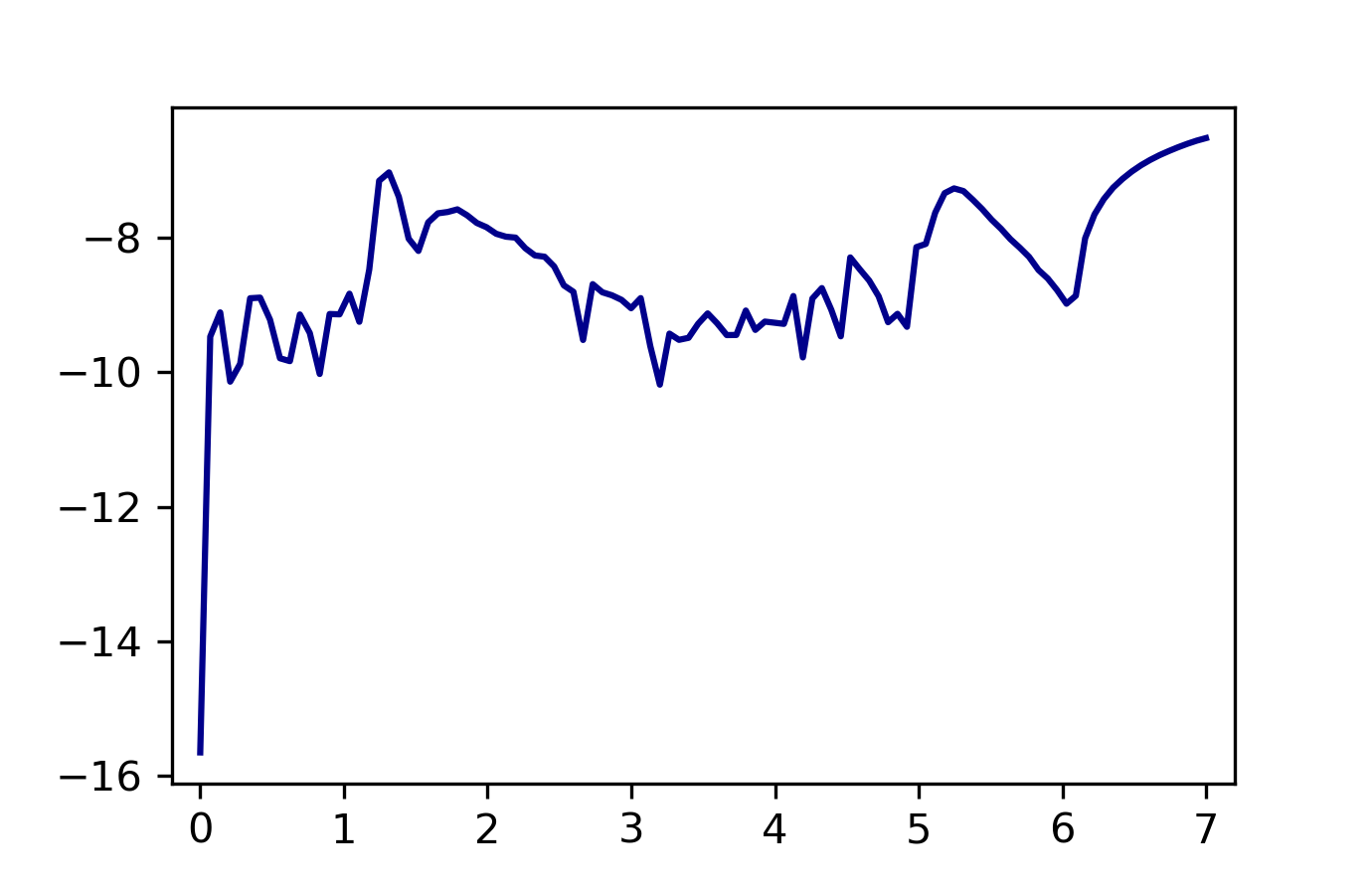}} ;

\node [text width=0.01cm,align=center] at (+6.5,-2.5){$N_e$};

\end{tikzpicture}

	\caption{Energy violation during the simulation, in the case of the leapfrog integrator inherited from LATTICEEASY \cite{latticeeasy}. On the left, we show the result for the standard slow-roll potential of \cref{sec:resultsslow}, while on the right we show the result for the step potential of \cref{sec:resultsstep} with $s=0.01$ and $d=0.005$.}
	\label{fig:energy_leap}
\end{figure}
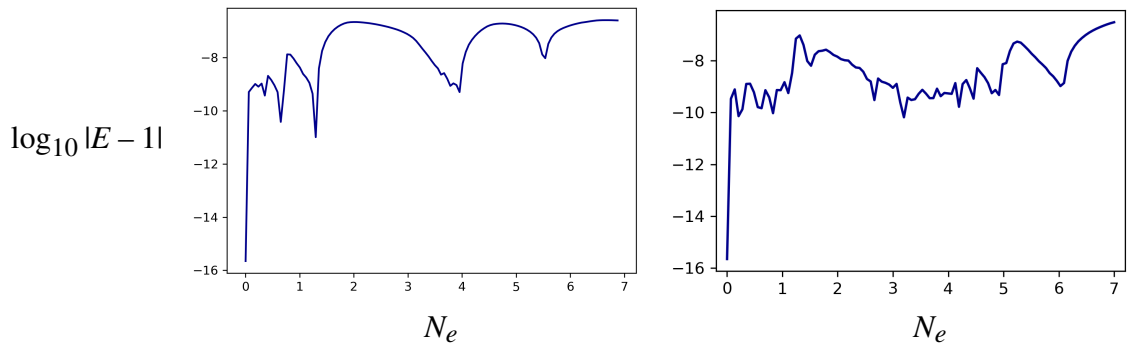
In order to check that this small energy violation at the beginning of the simulation is not related to the way we generate initial conditions on the lattice, we run again the simulations discussed in this chapter using a symplectic leap-frog time integrator, which is the one originally used in LATTICEEASY\footnote{See the documentation of LATTICEEASY for details about this numerical integrator.}. In this case, the simulation leads to the same results regarding the power spectrum and all other quantities shown in this chapter. In \cref{fig:energy_leap} we show the same energy conservation plot in the case of the symplectic leap-frog integrator. From this plot, we see that energy violation is much smaller in this case due to the symplectic nature of the integrator. Moreover, there is no initial peak in the energy violation, contrary to the RK4 integrator considered above. This ensures that the initial small energy violation of \cref{sec:energy} is not related to the spatial discretization or to the way initial conditions are generated, but it is simply related to the time integrator being not symplectic. We employed a Runge-Kutta non-symplectic integrator because we want to use a similar methodology to study the more complicated axion-U(1) inflationary model. This will be the subject of the second part of the thesis. For this model, writing a symplectic integrator is nontrivial because it means that one has to deal with more complicated \textit{implicit} time integrators. This is why we introduced the RK4 integrator in this chapter and discussed energy conservation in this case.

Another way to check the numerical accuracy of the Runge-Kutta integrator is to study the time-step convergence. This consists in varying the simulation (initial) step $\Delta\tau_0$ and checking that the final result does not depend on this quantity. If we can find a wide range of $\Delta\tau_0$ values that give the same final result, then we are confident that the result is physical and that numerical errors are under control. We use this check, together with the energy conservation discussed above, to test the accuracy of the numerical integration. In all cases discussed in this work (also the ones in \cref{sec:axionsim}), we checked time-step convergence by changing the time step of a factor of 100. This is done by increasing \textit{and} decreasing the initial $\Delta\tau_0$ by a factor of 10, and checking that the final result does not change.

\part{Axion-U(1) inflation}
\cleardoublepage
\chapter{Axion-U(1) inflation: a short review}
\label{sec:axioninf}
In the first part of the thesis, we focused on the most minimal realization of inflation, where only one scalar field is involved. 
 Although observations are consistent with this simple model, in this second part we consider a more complicated scenario. We assume that, together with the inflaton $\phi$, a U(1) gauge field $A_\mu$ is present during inflation. This field is coupled to the inflaton through a Chern-Simons interaction $\phi F\tilde F$. In this setup, the inflaton is a pseudo-scalar axion field, hence the name axion-U(1) inflation.

As we will see, the coupling between the inflaton and the gauge field substantially alters the phenomenology of inflation. In some situations, it leads to strong backreaction effects, invalidating the perturbation theory approach that is typically used to study inflationary observables. In the next chapter, we are going to study this system with a lattice simulation. This will allow us to reveal unknown aspects of this model that are beyond the reach of standard perturbation theory computations. In this chapter, we introduce this more complicated inflationary model and summarize some well-known results related to its phenomenology. 


\section{The axion-U(1) model of inflation}
In this section, we introduce the axion-U(1) model of inflation. We first briefly discuss some theoretical motivations for introducing this non-minimal scenario of inflation.

\subsection{Motivation}
\label{sec:motivation}
The idea that the inflaton might be an axion field is related to one of the main theoretical problems of the minimal slow-roll scenario introduced in \cref{sec:inflation}. From the observations of the power spectrum of scalar perturbations $\mathcal{P}_{\zeta}$, we know that the inflaton self-coupling must be very small $m=V^{\prime\prime}(\phi)\sim 10^{-5} M_{\rm Pl}$. 
This makes it very hard to achieve a slow-roll potential for the $N_e\simeq 60$ e-folds needed for inflation. Indeed, radiative corrections introduce new contributions to the potential that are expected to spoil its flatness. These radiative corrections are a consequence of the UV sensitivity of inflation, and expected to be of the order of the cutoff of the theory, i.e. the Planck mass $M_{\rm Pl}$.

This problem is a hint of new physics beyond the minimal model of inflation presented in \cref{sec:inflation}. A possible solution is assuming that the inflaton is a pseudo-Nambu-Goldstone boson (PNGB) \cite{PhysRevLett.65.3233}, analogous to the axion field introduced to solve the strong CP problem of quantum chromo dynamics (QCD)\footnote{We will call this field \textit{axion}, even if it is not necessarily the QCD axion. Sometimes this is called an axion-like field in the literature.}. The PNGB appears in the action with derivative term only $\partial_\mu \phi\partial^\mu \phi$. Therefore, it enjoys a global shift symmetry $\phi\rightarrow \phi+ \text{constant}$. Moreover, it is naturally coupled to gauge fields via Chern-Simons interaction $\phi F\tilde F$, where $F$ is the strength tensor of the gauge field and $\tilde F$ its Hodge dual (see later for an explicit definition of $F$ and $\tilde F$).
The PNGB axion field acquires a potential $V(\phi)$ through nonperturbative instanton configurations of the gauge field. This potential breaks the shift symmetry of the axion. Therefore, it is protected against radiative corrections, making the axion field an appealing candidate for the inflaton field responsible for the expansion of the early universe. 

The shape of the axion potential $V(\phi)$ depends on the particular model. In the case of the QCD axion, the potential is of the form:
\begin{equation}
	V(\phi)=\Lambda^4 \left[1-\cos\left(\frac{\phi}{f}\right)\right].
\end{equation}
This potential breaks the global shift symmetry of the axion down to a discrete symmetry $\phi\rightarrow\phi+2\pi f$. Unfortunately, this potential can be a successful inflaton candidate only if $f> M_{\rm Pl}$ \cite{Arkani-Hamed:2003xts}, which means that the symmetry breaking scale is above the Planck scale, where the effective field theory description is expected to break. This problem can be avoided by invoking extra dimensions \cite{Arkani-Hamed:2003xts} or several axion fields \cite{Kim:2004rp,Dimopoulos:2005ac,Easther:2005zr}. Moreover, various power-law potentials of the form
\begin{equation}
V(\phi)=\mu^{4-n}\phi^n
\end{equation}
can be obtained in string theory realizations of inflation known as axion monodromy \cite{McAllister:2014mpa}. In any case, the aspects discussed in this thesis will not depend on the particular shape of the axion potential. Therefore, we keep the symmetry-breaking potential $V(\phi)$ general, except assuming that it is of the right shape to give rise to slow-roll inflation. 







\subsection{The model}

In the light of the discussion above, we assume that the inflaton is a PNGB axion field coupled to some gauge field. For simplicity, we consider the case where $\phi$ is coupled to an Abelian U(1) field $A_\mu$. This model of inflation is described by the following Lagrangian:
\begin{align}
S=\int d^4x \sqrt{-g}\Biggl[&\frac{M^2_{Pl}}{2}R-\frac{1}{2}(\partial_\mu\phi)^2
-V(\phi)-\frac{1}{4}F_{\mu\nu}F^{\mu\nu}-\frac{\alpha}{4f}\phi F_{\mu\nu}\tilde{F}^{\mu\nu}\Biggr],
\label{eq:actionA}
\end{align}
where $\phi$ is the inflaton and $V(\phi)$ its slow-roll potential, which we keep general for the rest of this chapter. $F$ is the strength tensor of the gauge field $F_{\mu\nu}=\partial_\mu A_\nu-\partial_\nu A_\mu$, with dual $\tilde{F}^{\mu\nu}=\epsilon^{\mu\nu\rho\sigma}/2F_{\rho\sigma}$. The Levi-Civita tensor $\epsilon_{\mu\nu\rho\sigma}$ is defined such that $\epsilon^{0123}=1/\sqrt{-g}$.  From now on everything will be expressed in reduced Planck mass units $M_{Pl}\equiv1$.

This model has been extensively studied in the literature due to its interesting phenomenology. In the rest of this chapter, we summarize some well-known results and explain why a lattice simulation is necessary to understand the observational implications of this model.
\section{Gauge field particle production}
\label{sec:gfparticles}
The main consequence of the Chern-Simons interaction $\phi F\tilde F$ during inflation is an exponential production of gauge field particles, that influence several inflationary observables. We will start by describing this exponential production. To study the dynamics of this system with perturbation theory, we assume an unperturbed FLRW metric:
\begin{equation}
\label{eq:spacetime}
ds^2=a^2(\tau)(-d\tau^2+d\vec{x}^2).
\end{equation}
and set all perturbations of the metric to zero $\delta g_{\mu\nu}=0$. This is the usual approximation when dealing with the axion-U(1) model of inflation. We will motivate it in \cref{sec:gravityA}. Similarly to \cref{sec:inflation}, we split the inflaton field in a background plus perturbation $\phi(\vec{x},\tau)=\bar\phi(\tau)+\delta\phi(\vec{x},\tau)$. On the contrary, we assume that the gauge field has no background value $\langle A_\mu \rangle = 0$, so that $A_\mu=\delta A_\mu$.

To study the production of the gauge field, we first quantize it as follows:
\begin{align}
\begin{split}
\label{eq:quantizationA}
{\vec{A}}(\tau,\vec{x})=\sum_{\lambda=\pm}\int\frac{d^3k}{(2\pi)^{3/2}}\Biggl[\vec{\epsilon}_\lambda(\vec{k})  A_{\lambda}(\tau,\vec{k})\hat a_{\vec{k}}\,e^{-i\vec{k}\cdot\vec{x}}\Biggr]  +h.c.,
\end{split}
\end{align}
where $h.c.$ means Hermitian conjugate and the creation and annihilation operators satisfy the commutation relation:
\begin{equation}
		[\hat a_{\vec{k}},\hat a^\dagger_{\vec{k}^\prime}]=\delta(\vec{k}-\vec{k}^\prime).
\end{equation}
 $\vec\epsilon_\pm$ are the polarization vectors, defined by the relation:
\begin{align}
\begin{split}
\label{eq:epsilon}
&{\vec\epsilon_\lambda}^{\,*}(\vec{k})\cdot 	\vec\epsilon_{\lambda^\prime}(\vec{k})=\delta_{\lambda,\lambda^\prime},\quad \vec{k}\cdot\vec\epsilon_\pm(\vec k)=0,\\ &\vec{k}\times\vec\epsilon_\pm(\vec k)=\mp i k \vec{\epsilon}_\pm(\vec k).
\end{split}
\end{align}
Plugging this decomposition in the action  \eqref{eq:actionA} and neglecting metric perturbations, one can derive the following equation at linear order for the gauge polarizations $A_\pm$ \cite{Anber_2006, Anber_2010}:
\begin{equation}
\label{eq:cont_modes}
A_{\pm}^{\prime\prime}+\left(k^2\pm k \bar\phi^\prime\frac{\alpha}{f}\right)A_{\pm}=0,
\end{equation}
where $\bar\phi$ is the background value of the inflaton\footnote{We omit the explicit derivation, that we perform in the case of a discrete spacetime in \cref{sec:discA}.}. If the gauge coupling $\alpha/f$ is equal to zero, the gauge field simply oscillates and is never excited. If it is different from zero, the background velocity of the inflaton causes a tachyonic growth of one of the two polarizations of the gauge field for $k<\bar\phi^\prime\alpha/f$. Without loss of generality, we assume that $ \bar\phi^\prime>0$, so that the growing polarization is $A_-$. It is important to introduce the following parameter to describe the growth of gauge field modes:
\begin{equation}
\label{eq:xi}
	\xi = \frac{\alpha \dot{\bar\phi}}{2fH}.
\end{equation}
This quantity is slowly varying during inflation, as it is defined in terms of $\dot{\bar\phi}$ and $H$. Gauge field modes $A_-(k)$ such that $k<2\xi aH$ are excited by the inflationary background. As we will see, in all relevant applications $\xi\gtrsim O(1)$, so that the gauge field growth occurs close to horizon crossing. The fact that only one of the gauge field polarizations is excited is a symptom of the parity-violating nature of the term $F\tilde F$ in the action.

We can use a simple numerical integrator to solve \cref{eq:cont_modes} and determine the evolution of gauge polarizations starting from their Bunch Davies vacuum, that in the case of the gauge field reads:
\begin{equation}
\label{eq:BDA}
A_{\pm}(\vec{k},\tau)=\frac{1}{\sqrt{2k}}e^{- i \omega_{k}\tau},\quad\quad\quad -k\tau\gg 1.
\end{equation}
In order to get an analytical solution, one can assume a de Sitter background $\tau=-1/(aH)$ with $H=\dot{a}/a=\text{constant}$. In this case, the term in the bracket can be rewritten as $k\bar\phi^\prime\alpha/f=-2\xi/\tau$, and one can write an analytical solution for the growing mode \cite{Anber_2006, Anber_2010}:
\begin{equation}
\label{eq:ex_sol}
A_-(k,\tau)=\frac{1}{\sqrt{2k}}\left[G_0(\xi,-k\tau)+iF_0(\xi,-k\tau)\right],
\end{equation}
where $F_\ell$ and $G_\ell$ are the Coulomb wave functions. This solution is well approximated in the range $(8\xi)^{-1}<k/aH<2\xi$ by the following expression  \cite{Barnaby_2011_Large}:
\begin{equation}
\label{eq:ex_sol_approx}
A_-(k,\tau)=\frac{1}{\sqrt{2k}}\left(-\frac{k\tau}{2\xi}\right)e^{\pi\xi-2\sqrt{-2\xi k\tau}},
\end{equation}
that makes it evident that $A_-$ is exponentially enhanced. This exponential production will eventually influence inflationary observables, such as the curvature perturbation $\zeta$. This is the subject of the next section.

\section{Scalar perturbations}
The exponential production of gauge field particles affects the evolution of inflationary scalar perturbations. This is described by the following equation \cite{Barnaby_2011}:
\begin{equation}
\label{eq:gfparticle}
\left(\frac{\partial^2}{\partial\tau^2}+2\mathcal{H}\frac{\partial}{\partial\tau}-\nabla^2+a^2m^2\right)\delta\phi(\vec{x},\tau)=a^2\frac{\alpha}{f}\left(F_{\mu\nu}\tilde F^{\mu\nu}-\langle F_{\mu\nu}\tilde F^{\mu\nu}\rangle\right),
\end{equation}
which can be easily derived from the action of \cref{eq:actionA} as we are neglecting metric perturbations. This equation shows that the gauge field acts as a source for $\delta\phi$. In order to determine how the gauge field affects the spectrum of scalar perturbations, one has to solve this equation. This is a nontrivial task, that has been performed in the literature in the approximation $\xi=\text{constant}$. In the following, we summarize these well-known results without showing the detailed computation. 

As it is customary in the literature, we show results for the curvature perturbation $\zeta$ defined in \cref{eq:zeta}, computed in the spatially flat gauge as:
\begin{equation}
	\zeta \simeq H\frac{\delta\phi}{\dot \phi}.
\end{equation}
As discussed in \cref{sec:cpt}, this expression is valid on super-horizon scales and at leading order in slow-roll expansion. In the case of the axion-U(1) model, it is valid as long the energy density contained in the gauge field is small compared to the total energy density, which remains true in all the cases discussed in this thesis.

\subsection{Power spectrum}
Plugging the approximate solution of \cref{eq:ex_sol_approx} into \cref{eq:gfparticle}, and assuming a constant $\xi$, an involved computation leads to the following solution for the power spectrum of the curvature perturbation $\zeta$ on super-horizon scales \cite{Anber_2010,Barnaby_2011_Large, Barnaby_2011, Anber_2012}:
\begin{equation}
\label{eq:ps_th}
\mathcal{P}_{\zeta}(k)\simeq\mathcal{P}_{\rm vac}+\mathcal{P}_{\rm vac}^2f_2(\xi)e^{4\pi\xi},\quad\quad k\ll aH, \quad\quad \xi=\text{constant},
\end{equation}
where $\mathcal{P}_{\rm vac}=H^4/(2\pi{\dot\phi})^2$ is the vacuum contribution of \cref{eq:thprediction_simple}, and $f_2$ is a function computed in Ref. \cite{Barnaby_2011}, that can be approximated for large $\xi$ as:
\begin{equation}
	f_2(\xi)\simeq\frac{7.5\cdot 10^{-5}}{\xi^6},\quad\quad\xi\gg1.
\end{equation}
This result shows that the power spectrum of the curvature perturbation is exponentially sensitive to the parameter $\xi$. If we do not want to spoil the observed power spectrum of scalar perturbations, $\xi$ has to remain $O(1)$ at large scales corresponding to CMB experiments. In \cref{sec:obsA} we will discuss some quantitative bounds on $\xi$ and $\alpha/f$.
\subsection{Bispectrum}
An important feature of the scalar perturbations sourced by the gauge field is their non-Gaussian statistics. This is expected, as the source term $F \tilde F$ in \cref{eq:gfparticle} is bilinear in the field $A_{\mu}$. Using a similar computation in the $\xi=\text{constant}$ approximation, the bispectrum of this model has been also estimated on super-horizon scales $k\ll aH$ \cite{Barnaby_2011}:
\begin{equation}
\label{eq:bis_th}
	\langle \zeta(\vec{k})\zeta(\vec{k}_2)\zeta(\vec{k}_3)\rangle=\frac{3}{10}(2\pi)^{5/2}\mathcal{P}_{\rm vac}^3 e^{6\pi\xi}\frac{\delta^D(\vec{k}+\vec{k}_2+\vec{k}_3)}{k^6}\frac{1+x_2^3+x_3^3}{x_2^3 x_3^3}f_3(\xi,x_2,x_3),
\end{equation}
where $x_i=k_i/k$, $\delta^{D}$ is the Dirac delta function and $f_3$ is a function that can be found in Ref. \cite{Barnaby_2011}. As usual, the bispectrum is different from zero only on triangular shapes $\vec{k}_1+\vec{k}_2+\vec{k}_3=0$. The bispectrum of \cref{eq:bis_th} peaks on equilateral shapes $x_1=x_2=0$. In this case $f_3$ can be approximated for large $\xi$ as \cite{Barnaby_2011}:
\begin{equation}
\label{eq:f3}
f_3(\xi,1,1)\simeq\frac{2.8\cdot 10^{-7}}{\xi^9},\quad\quad\xi\gg1.
\end{equation}
Similar to the power spectrum, also the bispectrum grows exponentially with $\xi$. As we discussed in \cref{sec:scalarpert}, observations are consistent with Gaussian scalar perturbations. This puts a stringent constraint on $\xi$ at CMB scales, which we will discuss later in \cref{sec:obsA_bisp} .
\section{Gravitational waves}
Although we are mostly interested in scalar perturbations in this thesis, let us also briefly mention the consequences of the gauge field production on the tensor sector. The gauge field acts as a source for tensor perturbations, affecting the power spectrum of gravitational waves $\mathcal{P}_{\rm GW}$. This power spectrum can be decomposed in right-handed and left-handed components, corresponding to the two helicities of the graviton $\mathcal{P}_{\rm GW}=\mathcal{P}_{\rm GW,L}+\mathcal{P}_{\rm GW,R}$. As a result of the parity-violating source $F\tilde F$, one has $\mathcal{P}_{\rm GW,L}\neq\mathcal{P}_{\rm GW,R}$. The two components can be computed as \cite{Barnaby_2011}:
\begin{equation}
\label{eq:GW}
	\mathcal{P}_{\rm GW,\lambda}=\frac{H^2}{\pi^2}\left(\frac{k}{k_0}\right)^{n_T}\left[1+H^2f_{\lambda}(\xi)e^{4\pi\xi}\right],
\end{equation}
where $n_T=-2\varepsilon$, $\lambda=\rm L,R$ and $f_{L,R}$ are two functions computed in \cite{Barnaby_2011}. In analogy to the scalar power spectrum, the final result is the sum of a vacuum contribution and a term sourced by the gauge field, which grows exponentially with $\xi$. For large $\xi$, the two functions $f_\lambda$ can be approximated as:
\begin{equation}
	f_L(\xi)\simeq\frac{4.3\cdot 10^{-7}}{\xi^6},\quad	f_R(\xi)\simeq\frac{9.2\cdot 10^{-10}}{\xi^6},\quad\quad \xi\gg1,
\end{equation} 
where one can see that, if the sourced part dominates, gravitational waves are chiral $P_{\rm GW}\simeq P_{\rm GW,L}$. This is a consequence of the parity-violating nature of the source $F\tilde F$.
\section{Backreaction}
\label{sec:backreaction}
The well-known results presented in this section are only valid if the exponential production of the gauge field does not affect the overall background trajectory. This is ensured as long as two conditions are satisfied. First, the gauge field should not influence the background slow-roll trajectory of the inflaton. This can be quantified using the mean field equation for the inflaton background $\bar\phi$, which can be easily derived from the action \eqref{eq:actionA} as:
\begin{equation}
	\label{eq:KGA}
	\partial_\tau^2\bar \phi+2\mathcal{H}\partial_\tau{\bar\phi}+a^2 V^\prime(\bar\phi)=a^2\frac{a}{f}\langle F_{\mu\nu}\tilde F^{\mu\nu}\rangle.
\end{equation}
This equation is very similar to the Klein-Gordon \cref{eq:KG}, but with an extra source term $\langle F\tilde F\rangle$. The gauge field does not affect the inflationary background as long as the potential term dominates $\alpha/f \langle F_{\mu\nu}\tilde F^{\mu\nu}\rangle \ll V^\prime(\phi)$. This translates into the following bound \cite{Barnaby_2011}:
\begin{equation}
\label{eq:bounds}
\frac{H^2}{26\pi|\dot{\phi}|}\xi^{-3/2}e^{\pi\xi}\ll 1.
\end{equation} 
Another requirement for the validity of the results presented so far is that the energy density contained in the gauge field is negligible with respect to the total energy density $\rho_{\rm GF}\ll \rho_{\rm tot}$, so that the gauge field has no effect on the quasi-de Sitter expansion. This requires \cite{Barnaby_2011}:
\begin{equation}
\label{eq:bounds_other }
H\ll 146\xi^{3/2}e^{-\pi\xi}.
\end{equation}
If these bounds are violated, background and perturbation quantities can not be treated separately and one needs to take into account the backreaction of the produced gauge field particles on the overall trajectory. This is not a requirement on the consistency of the theory, but simply a computational constraint due to the breakdown of the perturbation theory approach. In the next chapter, we are going to develop a lattice simulation for the axion-U(1) system. This will allow us to study the axion-U(1) dynamics when these conditions are violated. The first of these bounds, \cref{eq:bounds}, constitutes the most stringent constraint in all cases considered in this work. Therefore, we will use this equation to determine whether perturbation theory is reliable or not.

\section{Observational constraints}
\label{sec:obsA}  
We now discuss the current observational constraint on the axion-U(1) model. We summarize the well-known results, that we will revisit in the next chapter in light of the results of our lattice simulation.
\subsection{CMB scales}
\label{sec:obsA_bisp}  
At CMB scales, the most stringent bound comes from the non-Gaussian statistics of the curvature perturbation. The bound is derived assuming that the curvature perturbation for this model can be expanded as a perturbation around a Gaussian field as follows:
\begin{equation}
\label{eq:fnl}
	\zeta(\vec{x})=\zeta_g(\vec{x})+\frac{3}{5}f_{\rm NL}\left[\zeta^2_g(\vec{x})-\langle\zeta_g(\vec{x})\rangle^2\right],
\end{equation}
where $\zeta_g$ is a Gaussian field, and non-Gaussianity is entirely described by the parameter $f_{\rm NL}$. This parametrization results in the following three-point function:
\begin{equation}
	\label{eq:bis_th_fnl}
	\langle \zeta(\vec{k})\zeta(\vec{k}_2)\zeta(\vec{k}_3)\rangle=\frac{3}{10}(2\pi)^{5/2}\mathcal{P}_{\zeta}^2 f_{\rm NL} {\delta^D(\vec{k}+\vec{k}_2+\vec{k}_3)}\frac{k^2+k_2^3+k_3^3}{k^2k_2^2k_3^2}.
\end{equation}
Comparing this expression with \cref{eq:bis_th}, we can define an effective momentum-dependent $f_{\rm NL}$ parameter for the axion-U(1) model \cite{Barnaby_2011}:
\begin{equation}
	f^{\rm(eff)}_{\rm NL}(\xi,x_1,x_2)=\frac{f_3(\xi,x_1,x_2)\mathcal{P}^3_{\rm vac}e^{6\pi\xi}}{\mathcal P^2_{\zeta}}.
\end{equation}
Assuming that equilateral configurations $x_2=x_3=1$ dominate the signal, one can evaluate this parameter at CMB scales using the observed $\mathcal{P}_{\zeta}\simeq 22\cdot10^{-10}$ and the expression \eqref{eq:f3} for $f_3$:
\begin{equation}
f_{\rm NL}^{\rm (equil.)}|_{\rm CMB}\simeq 5.7\cdot 10^{10}\mathcal{P}^3_{\rm vac}\frac{e^{6\pi\xi}}{\xi^9}.
\end{equation}
CMB constraints on non-Gaussianity set $f_{NL}\lesssim 100$, implying $\xi_{\rm CMB}\lesssim2.55$. Assuming a quadratic potential $V(\phi)=\frac{1}{2}m^2\phi^2$, this gives an upper bound for the gauge coupling\footnote{$M_{\rm Pl}=1$ in our units} $\alpha/f<32$.
\subsection{Smaller scales}
\label{sec:obsA_PBH}
The non-Gaussian statistics discussed above give the most stringent bound on the axion-U(1) model at CMB scales. We now discuss how this model can leave observational imprints on smaller scales, corresponding to modes exiting the horizon later during the inflationary epoch. Although physics at these scales is much less experimentally constrained, the effects of the Chern-Simons interaction are typically much stronger. This is because $\dot\phi$ usually increases during inflation, as the inflaton slowly approaches the minimum of its potential. This makes $\xi$ slowly grow during the inflationary epoch. Due to the exponential sensitivity to $\xi$, even an $O(1)$ change in this parameter can drastically change inflationary observables at late times. We now discuss separately the main observational consequences.

\subsubsection*{Primordial Black Holes}
Even if $\xi_{CMB}$ is small enough to be consistent with CMB data, later during inflation $\xi$ can increase enough to make $\mathcal{P}_{\zeta}\sim 10^{-2}$, exceeding the threshold for primordial black holes (PBH) production \cite{PhysRevD.87.103506, Garcia-Bellido:2016dkw}. PBH are formed after inflation through the gravitational collapse of high-density Hubble-sized patches at horizon re-entry. 

Estimating the amount of PBH produced from the axion-U(1) model is a challenging task. The main problem is that such high values of the power spectrum are typically associated with the breakdown of the perturbative assumption of \cref{eq:bounds}. In other words, if $\xi$ is large enough to efficiently produce PBH, perturbation theory breaks and one needs to take into account the complicated interplay between background and perturbations to compute $P_{\zeta}$. Another problem is that the statistics of the curvature perturbation for this model is expected to be non-Gaussian, making the computation of PBH particularly difficult. Indeed, PBH production highly depends on the shape of the probability distribution of $\zeta(\vec{x})$ in real space, which is needed to count how many regions of spacetime collapse into black holes at horizon re-entry. Unfortunately, the exact distribution of $\zeta$ in real space is unknown, both in the weak and strong backreaction regimes of the theory. In the next chapter, we will see how our lattice approach can solve both of these problems.

 Despite these issues, stringent bounds on the gauge coupling $\alpha/f$ have been obtained in the literature from PBH production \cite{PhysRevD.87.103506, Garcia-Bellido:2016dkw}. To obtain such bounds, one has to make the following assumptions:
 \begin{enumerate}
 	\item First, one has to assume that the analytical estimate for $\mathcal{P}_\zeta$ of $\cref{eq:ps_th}$ is approximately valid even when the perturbative description breaks down, i.e. when the bound of \cref{eq:bounds} is violated. 
 	
 	\item Second, the curvature perturbation is assumed to be approximated by a non-Gaussian $\chi^2$ distribution in the strong backreaction phase:
 	\begin{equation}
 		\zeta(\vec{x})\simeq\zeta^2_g(\vec{x})-\langle\zeta_g(\vec{x})\rangle^2,
 	\end{equation}
 	where $\zeta_g$ is a Gaussian field. This assumption is motivated by the fact that the source term $F\tilde F$ is bilinear in the field $A_\mu$. 
 \end{enumerate}
 Given these assumptions, Refs.  \cite{PhysRevD.87.103506, Garcia-Bellido:2016dkw} obtained a bound of $\xi_{\rm CMB}\lesssim 1.66$, corresponding to $\alpha/f\lesssim 23$ for the quadratic potential. If this bound is violated, the production of PBH is so abundant to spoil the observed fraction of energy density contained in matter, leading to an overclosure of the Universe. This is because PBH energy density redshifts like matter $\rho_{PBH}\propto a^{-3}$. 
 
 This PBH bound is much more stringent than the one obtained from the statistics of the CMB mentioned above. This is an indication that the axion-U(1) model can not be a successful candidate for small non-Gaussianity at large scales, as this would lead to an overproduction of PBH. In the next chapter, we are going to study the axion-U(1) system with a lattice simulation, allowing us to consistently take into account the backreaction of perturbations on the background dynamics. This will allow testing the validity of the assumptions above. As we will see, our results invalidate the second of these assumptions, with major observational implications.

\subsubsection*{Gravitational waves background}
Another consequence of the fact that $\xi$ grows during inflation is the production of tensor perturbations, which is also exponentially sensitive to this parameter. If the tensor power spectrum becomes large enough, one can hope to observe it in the form of a stochastic gravitational waves (GW) background using upcoming experiments. Unfortunately, if one takes values satisfying the PBH bound $\xi_{\rm CMB}\lesssim 1.66$, the signal is too weak to be detected by future missions such as LISA \cite{LISA1,Bartolo:2016ami}, advanced LIGO \cite{LIGOScientific:2016fpe} or PTA-SKA \cite{1990pta,5136190,Kramer:2004hd}. For this model to generate a GW signal observable by these experiments, and at the same time compatible with the PBH bound, one needs to modify the minimal axion-U(1) model \cite{Garcia-Bellido:2016dkw}. For example, a sizable GW signal can be obtained assuming some \textit{ad-hoc} modified axion potential, many different gauge fields or a gauge field that couples to a spectator axion field instead of the inflaton \cite{Garcia-Bellido:2016dkw,Campeti:2022acx,Campeti:2022vom}. At the end of the next chapter, we are going to revisit this conclusion in the light of the results of the lattice simulation. 

The axion-U(1) model of inflation also predicts a sizable amount of gravitational waves from the preheating phase just after the end of inflation. This signal has been used to constrain the axion-gauge coupling down to $\alpha/f\lesssim 15$ \cite{2020axiin,2020axiin2}. In this thesis, we choose to focus only on bounds from inflationary physics. The main motivation is that, as explained for example in Refs. \cite{domcke2020resonant,2020axiin2}, preheating bounds strongly depend on the dynamics of the final e-folds of inflation, which is still unknown for higher values of $\alpha/f$.

\cleardoublepage
\newpage\null
\chapter{Lattice simulations of axion-U(1) inflation}
\label{sec:axionsim}
In this chapter, we present a lattice simulation for the axion-U(1) model of inflation introduced in \cref{sec:axioninf}. Although gauge field theories have been already simulated in the context of reheating \cite{Figueroa_2018,Cuissa_2019,Figueroa_2019,figueroa2021cosmolattice,Deskins_2013,Adshead_2015,Adshead_2016,Adshead_2018,Braden_2010,2020axiin,2020axiin2,2021axiin}, this is the first simulation of an axion-gauge system during the inflationary epoch. The content of this chapter represents an extension of the single-field simulation presented in \cref{sec:inflationsim}, which contains many of the technical details regarding the lattice methodology.

The results from the simulations are mostly contained in \cref{sec:results_weak,sec:results_strong}. We show results for two different cases. In the first case, we study the axion-U(1) model in the regime of negligible backreaction, where the exponential production of gauge field particles is not strong enough to influence the background inflationary dynamics. In the second case, we study the axion-U(1) dynamics in the strong backreaction regime, characterized by the breakdown of perturbation theory. 

This chapter is based on Refs. \cite{caravano2021lattice,Caravano:2022epk}, and it is organized in the following way: in \cref{sec:aoma} we derive the nonlinear equations of motion that are used to evolve the system; in \cref{sec:discA} we discuss the choice of the discretization scheme, and its effects on the dynamics of the gauge field on the lattice; in \cref{sec:results_weak,sec:results_strong} we present the results from the simulation in the two cases mentioned above; finally, in \cref{sec:observations} we discuss the implications of our results and their physical interpretation.

\section{Nonlinear equations of motion}
\label{sec:aoma}
When simulating gauge field theories, there are different ways of dealing with the discretization procedure. A first approach is to write a discretized action that enjoys a discretized version of the gauge symmetry through the use of \textit{link variables} \cite{Figueroa_2018,Cuissa_2019,Figueroa_2019,figueroa2021cosmolattice}. This is similar to what is done in the context of lattice quantum chromodynamics (QCD) simulations. A second approach is to discretize the system directly at the level of the equations of motion \cite{Deskins_2013,Adshead_2015,Adshead_2016,Adshead_2018,Braden_2010,2020axiin,2020axiin2,2021axiin}. In this work, we follow the second approach. Therefore, our starting point is deriving the classical nonlinear equations of motion from the action of \cref{eq:actionA}. In analogy to \cref{sec:inflationsim}, we assume an unperturbed FLRW metric in conformal time to derive the equations. We will justify later this assumption. 

The first equation is the one for the inflaton, which is obtained similarly to the single-field case of \cref{sec:lattice}:
\begin{equation}
\label{eq:eom1}
	\phi^{\prime\prime}+2\mathcal{H}{\phi^\prime}-\nabla^2\phi+a^2\frac{\partial V}{\partial \phi}=-a^2\frac{\alpha}{4f}F_{\mu\nu}\tilde F^{\mu\nu},
\end{equation}
where, as usual, $\nabla^2\equiv \partial_j\partial_j$.
The equations of motion for the gauge field in are the following \cite{Deskins_2013}:
\begin{equation*}
\partial_\rho(\sqrt{-g}F^{\rho\sigma})+\frac{\alpha}{f}	\partial_\rho(\sqrt{-g}\phi\tilde{F}^{\rho\sigma})=0.
\end{equation*}
Let us compute separately the cases $\sigma=0$ and $\sigma=i$:
\begin{itemize}
	\item[$\mathbf{\sigma=0}$: ] 
	
	The first term of this equation can be written as: $$	\partial_\rho(\sqrt{-g}F^{\rho 0})=\partial_j(\sqrt{-g}g^{jj}g^{00}F_{j 0})=-(\partial_j\partial_jA_0-\partial_j\partial_0A_j),$$
	while the second one reads:
	\begin{align*}
	\partial_\rho(\sqrt{-g}\phi\tilde{F}^{\rho0})&=\frac{1}{2}\partial_k(\sqrt{-g}\epsilon^{k0ij}F_{ij})=\frac{1}{2}\sqrt{-g}\epsilon^{k0ij}\partial_k\phi F_{ij}\\&=\sqrt{-g}\epsilon^{k0ij}\partial_k\phi \partial_iA_j=-\epsilon_{ijk}\partial_k\phi \partial_iA_j.
	\end{align*}
	Here, we defined $\epsilon_{ijk}=\epsilon^{0ijk}\sqrt{-g}$ so that $\epsilon_{123}=+1$. Therefore, the equation for $\sigma=0$ reads:
	\begin{equation}
	\partial_j\partial_jA_0-\partial_j\partial_0A_j+\frac{\alpha}{f}\epsilon_{ijk}\partial_k\phi \partial_iA_j=0.
	\label{eq:eom2}
	\end{equation}
	
	\item[$\mathbf{\sigma=i}$: ] 
	
	The first term is:
	\begin{align*}
	\partial_\rho(\sqrt{-g}F^{\rho i})&=\partial_\rho(\sqrt{-g}g^{\rho\rho}g^{ii}F_{\rho_i})=-\partial_0F_{0i}+\partial_mF_{mi}\\&=\partial_0(\partial_iA_0-\partial_0A_i)+\partial_m(\partial_mA_i-\partial_iA_m).
	\end{align*}
	The second one reads:
	\begin{align*}
	\partial_\rho(\sqrt{-g}\phi\tilde{F}^{\rho i})=&\frac{1}{2}\sqrt{-g}\epsilon^{\rho i \alpha \beta}\partial_\rho\phi F_{\alpha\beta}=\frac{1}{2}\sqrt{-g}\epsilon^{0 i jk}\partial_0\phi F_{jk}+\\&+\frac{1}{2}\sqrt{-g}\epsilon^{m i 0k}\partial_m\phi F_{0k}+\frac{1}{2}\sqrt{-g}\epsilon^{m i k0}\partial_m\phi F_{k0}=\\&\sqrt{-g}\epsilon^{0 i jk}\partial_0\phi \partial_jA_k-\sqrt{-g}\epsilon^{0 i jk}\partial_j\phi (\partial_0A_k-\partial_kA_0),
	\end{align*}
	where we used $\partial_0(\sqrt{-g}\epsilon^{\mu\nu\rho\sigma})=0$. Therefore, the $\sigma=i$ equation reads:
	\begin{align}
	\begin{split}
	\partial_0(\partial_iA_0-\partial_0A_i)+\partial_m(\partial_mA_i-\partial_iA_m)+\\+\frac{\alpha}{f}\epsilon_{i jk}\partial_0\phi \partial_jA_k-\frac{\alpha}{f}\epsilon_{i jk}\partial_j\phi (\partial_0A_k-\partial_kA_0)=0
	\end{split}
	\label{eq:eom3}
	\end{align}
	
\end{itemize}
In the end, we have \cref{eq:eom1,eq:eom2,eq:eom3}, one for the scalar field and one for each component of the gauge field. In order to solve these equations, we first need to fix the gauge. This is the topic of the next section.
\subsection{Gauge fixing}
\label{sec:gauge}
We choose to work in the Lorenz gauge, defined by: 
\begin{equation}
\label{eq:gauge}
	\partial^\mu A_\mu=0\quad\Longrightarrow\quad\partial_iA_i=a^2\partial_0A_0.
\end{equation}
In this gauge, the equations read:.
\begin{align}
\label{eq:eomsa}
&		\partial_0^2\phi+2\mathcal{H}\partial_0{\phi}-\nabla^2\phi+a^2\frac{\partial V}{\partial \phi}+\frac{\alpha}{a^2f}\epsilon_{ijk}\partial_jA_k(\partial_0A_i-\partial_iA_0) =0\\
&\partial_0^2{A}_0-	\nabla^2A_0-\frac{\alpha}{f}\epsilon_{ijk}\partial_k\phi \partial_iA_j=0\\
&\partial_0^2{A}_i-\nabla^2A_i+\frac{\alpha}{f}\epsilon_{i jk}\partial_j\phi (\partial_0A_k-\partial_kA_0)-\frac{\alpha}{f}\epsilon_{i jk}\partial_0\phi \partial_jA_k=0,
\end{align}
where we also expanded the right-hand side of the first equation to express it in terms of the gauge field $A_\mu$. To solve this system of equations, we associate $N^3$ values to the inflaton $\phi(\vec{n})$ and to each of the 4 components of the gauge field $A_0(\vec{n})$ and $A_{i}({\vec{n}})$. We then evolve these $5N^3$ values (plus their time derivatives) using a RK4 integrator, as explained in the next section. This is analogous to what is done in the single-field case.

If we use these equations to evolve the system, however, we are not enforcing the gauge constraint of \cref{eq:gauge} to be preserved during the evolution. In other words, we evolve the 4 components of $A_{\mu}$ as independent components, while they should not be independent. For this reason, we need to check by hand that the gauge symmetry is preserved during the evolution. This is done in \cref{sec:energy+gauge} for all the cases considered in this work.

\subsection{Equations of motion for the simulation}
As already mentioned, we follow a very similar approach to the single-field case of \cref{sec:inflationsim}. We associate field values to the $N^3$ points of the cubic lattice. We adopt the same rescaling of the single-field case, defined in \cref{eq:rescaling}. After this rescaling, the equations of motion for the simulation read\footnote{In these equations, contrarily to the rest of this manuscript, primes denote derivatives with respect to rescaled conformal time $\tilde{\tau}$.}:
\begin{align}
\label{eq:final1}
&		\phi^{\prime\prime}(\vec{n})+(2+s)\frac{a^\prime}{a} {\phi}^\prime(\vec{n})-a^{-2s}[\tilde{\nabla}^2\phi](\vec{n})+a^{2-2s}\frac{\partial \tilde{V}}{\partial\nonumber \phi}(\vec{n})+\\&\quad\quad\quad\quad\quad\quad\quad\quad+\frac{\alpha}{f}a^{-2s-2}\epsilon_{ijk}[{\tilde\partial}_jA_k](\vec{n})\left(a^sA_i^\prime(\vec{n})-[\tilde{\partial}_iA_0](\vec{n})\right) =0\\
\label{eq:final2}
&{A}^{\prime\prime}_0(\vec{n})+s\frac{a^\prime}{a} {A_0}^\prime(\vec{n})-a^{-2s}[\tilde\nabla^2A_0](\vec{n})\,-\frac{\alpha}{f}a^{-2s}\epsilon_{ijk}[\tilde{\partial}_k\phi(\vec{n}) [\tilde{\partial}_iA_j](\vec{n})=0\\
\label{eq:final3}
&{A}^{\prime\prime}_i(\vec{n})+s\frac{a^\prime}{a} {A_i}^\prime(\vec{n})-a^{-2s}[\tilde\nabla^2A_i](\vec{n})-\frac{\alpha}{f}a^{-s}\epsilon_{i jk}\phi^\prime(\vec{n}) [\tilde{\partial}_jA_k](\vec{n})+\nonumber\\&\quad\quad\quad\quad\quad\quad\quad\quad\quad+\frac{\alpha}{f}a^{-2s}\epsilon_{i jk}[\tilde{\partial}_j\phi](\vec{n}) \left(a^sA_k^\prime(\vec{n})-[\tilde{\partial}_kA_0](\vec{n})\right)=0.
\end{align}
We solve these equations, together with the Friedmann equation for the scale factor (see the next section), using an RK4 integrator with an adaptive time-step. This is conceptually analogous to what is done in \cref{sec:numint}. For this reason, we avoid writing the explicit expressions for the numerical integrator and refer to \cref{sec:numint} for details.
\subsection{Spacetime evolution}
\label{sec:gravityA}
In analogy to the single-field case, the Universe is evolved using the Friedmann equations and neglecting metric perturbations. In the single-field case, this was justified by the fact that metric perturbations remain decoupled at leading order in slow-roll. We now see how a similar argument applies to the axion-U(1) model. 
\subsubsection*{The role of metric perturbations}
Neglecting metric perturbations is the usual approximation employed in the literature when dealing with the axion-U(1) model. In this section, we summarize the well-known arguments behind this assumption and refer to the literature for a more detailed computation.

The Chern-Simons term $\sqrt{-g}\phi F_{\mu\nu}\tilde F^{\mu\nu}$ does not contain any coupling between metric perturbations and field perturbations. This is because the Levi-Civita totally antisymmetric tensor $\epsilon^{\mu\nu\rho\sigma}$ contains a factor of $1/\sqrt{-g}$. Therefore, the leading coupling between metric perturbations and field fluctuations comes from the Maxwell term $\sqrt{-g}F_{\mu\nu}F^{\mu\nu}$ in the action:
\begin{equation}
\mathcal L \supset	\frac{1}{2a^2}\delta g_{00}F_{\mu\nu}F^{\mu\nu}.
\end{equation} 
This term is cubic in perturbations, and it is expected to contribute to the bispectrum of this model. This effect, however, is slow-roll suppressed. To prove this, one has to write the $00-$component of the Einstein equations at second order in perturbation theory and use it to solve for $\delta g_{00}$ as a function of $\phi$ and $A_{\mu}$. We avoid showing the detailed computation, which can be found for example in Ref. \cite{Barnaby_2011}. The result is:
\begin{equation}
\sqrt{-g}F_{\mu\nu}F^{\mu\nu} = \frac{{\bar \phi}^\prime}{2\mathcal{H}}\delta\phi\left(-\frac{1}{2}A_i^\prime A_i^\prime-\frac{1}{4}F_{ij}F_{ij}+\partial^{-2}\partial_0\partial_i(F_{ij}A^{\prime}_j)\right) ,
\end{equation}
where $\partial^{-2}$ is the inverse Laplacian operator. From this equation, we can see that all the interactions in the cubic action induced by the metric are slow-roll suppressed with respect to the Chern-Simons interaction $\alpha/f\phi F_{\mu\nu}\tilde F^{\mu\nu}$, as long as the coupling $\alpha/f$ is large enough. In practice, this translates into the bound\footnote{We remind the reader that we are working in reduced Planck mass units, in which $M_{\rm Pl}\equiv 1$.}:
\begin{equation}
	\alpha/f\gg \frac{\bar \phi^\prime}{\mathcal{H}}\simeq\sqrt \epsilon,
\end{equation}
which remains true for all the observationally relevant cases considered in this manuscript.
Therefore, gravitational effects on the bispectrum of this model are negligible. 

Later in this chapter, we will use the simulation to compute higher order correlators of the curvature perturbation such as $\langle \zeta^4 \rangle$ and $\langle \zeta^5\rangle$. We assume that gravitational effects are unimportant in characterizing these correlators, although we do not have a formal proof of this statement. The decoupling of gravitational effects from the matter content at all orders in perturbation theory is a common assumption when working with models of inflation involving large non-Gaussianity. This fact has been conjectured in Ref. \cite{Leblond_2011} based on a simple physical argument: gravitational interactions typically lead to a very small amount of non-Gaussianity. Therefore, the field dynamics of models with large non-Gaussianity is expected to be decoupled from the gravitational sector. Proving this conjecture is beyond the scope of this thesis. We consider this as our working assumption, that will need to be proved and/or verified in the future.


\subsubsection*{The Friedmann equation and the stress-energy tensor}
To evolve the scale factor for the Universe, we need to compute the contribution of the gauge field to the stress-energy tensor. A simple computation leads to:
\begin{equation}
\label{eq:stress_GF}
T^{(GF)}_{\mu\nu}=F_{\mu\alpha}F_{\nu\beta}g^{\alpha\beta}-\frac{g_{\mu\nu}}{4}F_{\alpha\beta}F^{\alpha\beta},
\end{equation}
where $GF$ stands for gauge field, as the contribution from the scalar sector was already computed in \cref{sec:friedmann}. From the stress-energy tensor, we can read off the energy density and pressure as follows:
\begin{align}
\begin{split}
&\rho_{\rm GF}=-T^0_{\hspace{1mm}0}=\frac{1}{2a^4}(\partial_0A_i-\partial_iA_0)^2+\frac{1}{4a^4}(\partial_iA_j-\partial_jA_i)^2,\\
&p_{\rm GF}=\frac{1}{3}\sum_iT^i_{\hspace{1mm}i}=\frac{1}{6a^4}(\partial_0A_i-\partial_iA_0)^2+\frac{1}{12a^4}(\partial_iA_j-\partial_jA_i)^2.
\end{split}
\label{eq:EDA}
\end{align}
We can easily notice that $\rho_{\rm GF}=3p_{\rm GF}$. This means that, as expected, the gauge field behaves as radiation and does not contribute to the second Friedmann equation that we use to evolve the scale factor. Nevertheless, we need to evaluate $\rho_{\rm GF}$ in order to check energy conservation in our code. This is done using the first Friedmann equation, as explained in \cref{sec:output}. 
\section{Discretization scheme}
\label{sec:discA}
The purpose of this section is to define the discretization scheme employed to solve \cref{eq:final1,eq:final2,eq:final3}. This corresponds to defining the lattice Laplacian $[\nabla^2 f](\vec{n})$ and the 1-dimensional spatial derivative $[\partial_j f](\vec{n})$ for all fields $f$ living on the lattice. As we will in \cref{sec:consequences}, the choice of the discretization scheme has a great impact on the dynamics of the axion-U(1) system on the lattice. Nevertheless, we will be able to find a consistent discretization scheme that allows us to reproduce the continuous dynamics of this model with good precision.

Before proceeding, let us define the gauge polarizations on the lattice $A_{\pm}$. These are obtained from the discrete Fourier transform of the gauge field as follows:

\begin{equation}
\label{eq:pol}
\vec{A}(\vec{n})=\sum_{\vec{m}}\sum_{\lambda=\pm}\vec{\epsilon}_\lambda(\vec\kappa_{\vec{m}})A_\lambda(\vec\kappa_{\vec{m}})\text{ }e^{-i\frac{2\pi}{N}\vec{m}\cdot\vec{n}},
\end{equation}
where $\vec\epsilon_\lambda(\vec\kappa)$ are the polarization vectors defined in \cref{eq:epsilon}. The explicit definition of $\vec \epsilon_\pm$ used in the lattice code can be found in \cref{sec:ICA}.
\subsection{Consequences of the discretization}
To find a suitable discretization scheme, we first study how the evolution of perturbations during inflation is affected by the discretization. This is similar to what is done in \cref{sec:modifieddr} for the single-field case and will be the subject of this section.






\label{sec:consequences}
We start with the following standard definitions of lattice Laplacian and one-dimensional derivative with second order truncation errors $O(\Delta x^2)$ \cite{press1986numerical}:
\begin{align}
\label{eq:lapl_old}
&[\nabla^2f](\vec{n})=\frac{1}{(\Delta x)^2}\sum_{\alpha=\pm1}\biggl(f(\vec{n}+ \alpha\vec{e}_1)+f(\vec{n}+ \alpha\vec{e}_2)+f(\vec{n}+\alpha \vec{e}_3)-3f(\vec{n})\biggr), \\
&[\partial_j f](\vec{n})=\frac{ f(\vec{n}+\vec{e}_j)- f(\vec{n}-\vec{e}_j)}{2 \Delta x} \label{eq:1d}.
\end{align}
where $\vec{e}_1=(1,0,0),\text{ }\vec{e}_2=(0,1,0),\text{ }\vec{e}_3=(0,0,1)$. The Laplacian is the same employed for the single-field case, and it was defined in \cref{eq:discretelaplacian}. In Fourier space, these discrete derivative operators result in the following effective momenta:
\begin{align}
\label{eq:klapl}
[\nabla^2 f](\vec{n})\,&\longrightarrow\,{k}^2_{\rm lapl},\quad\vec{k}_{\rm lapl}=\frac{2}{\Delta x}\sin\left({\vec{\kappa}_{\vec{m}}}\frac{\Delta x}{2}\right),\\
\label{eq:ksd}
[\partial_j f](\vec{n})\,&\longrightarrow\,\vec{k}_{\rm sd},\quad\,\,\,\vec{k}_{\rm sd}\text{ }\,=\frac{1}{\Delta x}\sin\left({\vec{\kappa}_{\vec{m}}}\Delta x\right).
\end{align}
We want to understand how the discretization influences the growth of the gauge field on the lattice induced by the background inflaton velocity. This growth is described by \cref{eq:final3}, that we rewrite with\footnote{The rescaling variables are defined in \cref{eq:rescaling}. This choice makes $\tilde\nabla=\nabla$.} $s=0$, $B=0$, and neglecting the term involving spatial derivatives of the inflaton $\partial_j \phi$:
\begin{align}
{A}^{\prime\prime}_i(\vec{n})-[\nabla^2A_i](\vec{n})-\frac{\alpha}{f}\epsilon_{i jk}\phi^\prime(\vec{n}) [\tilde{\partial}_jA_k](\vec{n})=0.
\end{align}
Applying a discrete Fourier transform, we can rewrite this equation in terms of the gauge field polarizations $A_{\pm}$ making use of \cref{eq:klapl,eq:ksd}:
\begin{align}
\label{eq:eomtogh}
\vec{\epsilon}_+A_{+}^{\prime\prime}+\vec{\epsilon}_-A_{-}^{\prime\prime}+k^2_{\rm lapl}(\vec{\epsilon}_+A_++\vec{\epsilon}_-A_-)+i\frac{\alpha}{f}\phi^{\prime}\vec{k}_{\rm sd}\times(\vec{\epsilon}_+A_++\vec{\epsilon}_-A_-)=0.
\end{align}
In analogy to what is done in continuous space, we want to obtain two separate equations for $A_+$ and $A_-$. To achieve this goal, let us use the following properties of mixed cross and scalar products:
\begin{align}
\label{eq:vec1}
&\vec V_1\times \vec V_2\times\vec V_3=(\vec V_1\cdot \vec V_3) \vec V_2-( \vec V_1\cdot  \vec V_2) \vec V_3,	\\
\label{eq:vec2}
&(\vec V_1\times \vec V_2) \cdot \vec V_3=(\vec V_3\times \vec V_1) \cdot \vec V_2=(\vec V_2\times \vec V_3) \cdot \vec V_1,
\end{align}
valid $\forall$ vectors $V_i\in \mathbb{C}^3$. Using the last relation of \cref{eq:epsilon}, we can write
\begin{equation}
\vec	\epsilon^{\, *}_{\pm}(\vec\kappa)=\mp i \frac{\vec{\kappa}}{\kappa}\times\vec\epsilon^{\, *}_{\pm}(\vec\kappa).
\end{equation}
We can use this equation, together with  \cref{eq:vec2}, to write:
\begin{equation}
\vec{\epsilon}^{\, *}_\pm(\vec{\kappa}) \,\cdot\,\left(\vec{V}\times\vec{\epsilon}_\ell(\vec{\kappa})\right)=\mp i\frac{\vec\kappa}{\kappa}\cdot \left(\vec\epsilon^{\, *}_\pm(\kappa) \times \vec V \times \vec\epsilon_\ell(\vec\kappa)\right),
\end{equation}
where $\vec{V}$ is a generic $\mathbb{C}$-vector. 
Next, we use \cref{eq:vec1} and the first and second of \cref{eq:epsilon} to rewrite this expression as follows:
\begin{equation}
\label{eq:id}
\vec{\epsilon}^{\, *}_\pm(\vec{\kappa}) \,\cdot\,\left(\vec{V}\times\vec{\epsilon}_\ell(\vec{\kappa})\right)=\mp i\left( \vec{V}\,\cdot\,\frac{\vec{\kappa}}{|\vec{\kappa}|}\right)\delta_{\ell,\pm}.
\end{equation}
 We can use this relation with $\vec{V}=\vec{k}_{\rm sd}$ to obtain two separate equations for the polarization modes of the gauge field $A_{\pm}$. Indeed, multiplying \cref{eq:eomtogh} by $\vec{\epsilon}_\pm(\vec{\kappa})^\ast$ we obtain:
\begin{equation}
\label{eq:latt_modes}
A^{\prime\prime}_\pm+\left(k^2_{\rm lapl}\pm\frac{\alpha}{f}\phi^{\prime}\vec{k}_{\rm sd}\,\cdot\,\frac{\vec{\kappa}}{|\vec{\kappa}|}\right)A_\pm=0,
\end{equation}
which describes the evolution of $A_{\pm}$ on the lattice.
This equation is very different from its continuous version of \cref{eq:cont_modes}. This is due to the different effective momenta emerging from the definition of the Laplacian and of the one-dimensional derivative. 

To illustrate this effect, let us take a lattice of $N=512$ and $L=4$. In \cref{fig:dispersion} we show a comparison between $\kappa$, $k_{\rm lapl}$ and $k_{\rm sd}$, which are computed as one-dimensional quantities through a spherical binning on the lattice of \cref{eq:modes,eq:klapl,eq:ksd}. From this plot, we can see that $k_{\rm sd}$ (red line) is strongly suppressed with respect to $k_{\rm lapl}$ (yellow line) for most of the scales, and in particular for the largest modes of the simulation.  Indeed, $k_{\rm sd}$ approaches zero for large $\kappa$. From \cref{eq:latt_modes} we can see that this results in an unphysical behavior of the gauge field, which will not be growing on the smaller scales of the lattice\footnote{The reader should keep in mind that we do not use \cref{eq:latt_modes} to evolve fields on the lattice. Indeed, we use the Euler-Lagrange  equations in real space defined in section \cref{sec:aoma}, and in this section we are only studying with analytical tools what is expected to happen on the lattice.}. In fact, it is easy to see that the departure of lattice modes $A_\pm(\vec\kappa)$ from the continuous solution is exponentially sensitive to the difference between $k_{\rm lapl}$ and $k_{\rm sd}$. In Section \ref{sec:results_weak} we will discuss in detail the consequences of this effect showing the results of a lattice simulation with spatial derivatives defined as in \cref{eq:lapl_old,eq:1d}.
\begin{figure}
	\centering
	
	\begin{tikzpicture}
	\node (img) {\includegraphics[width=9cm]{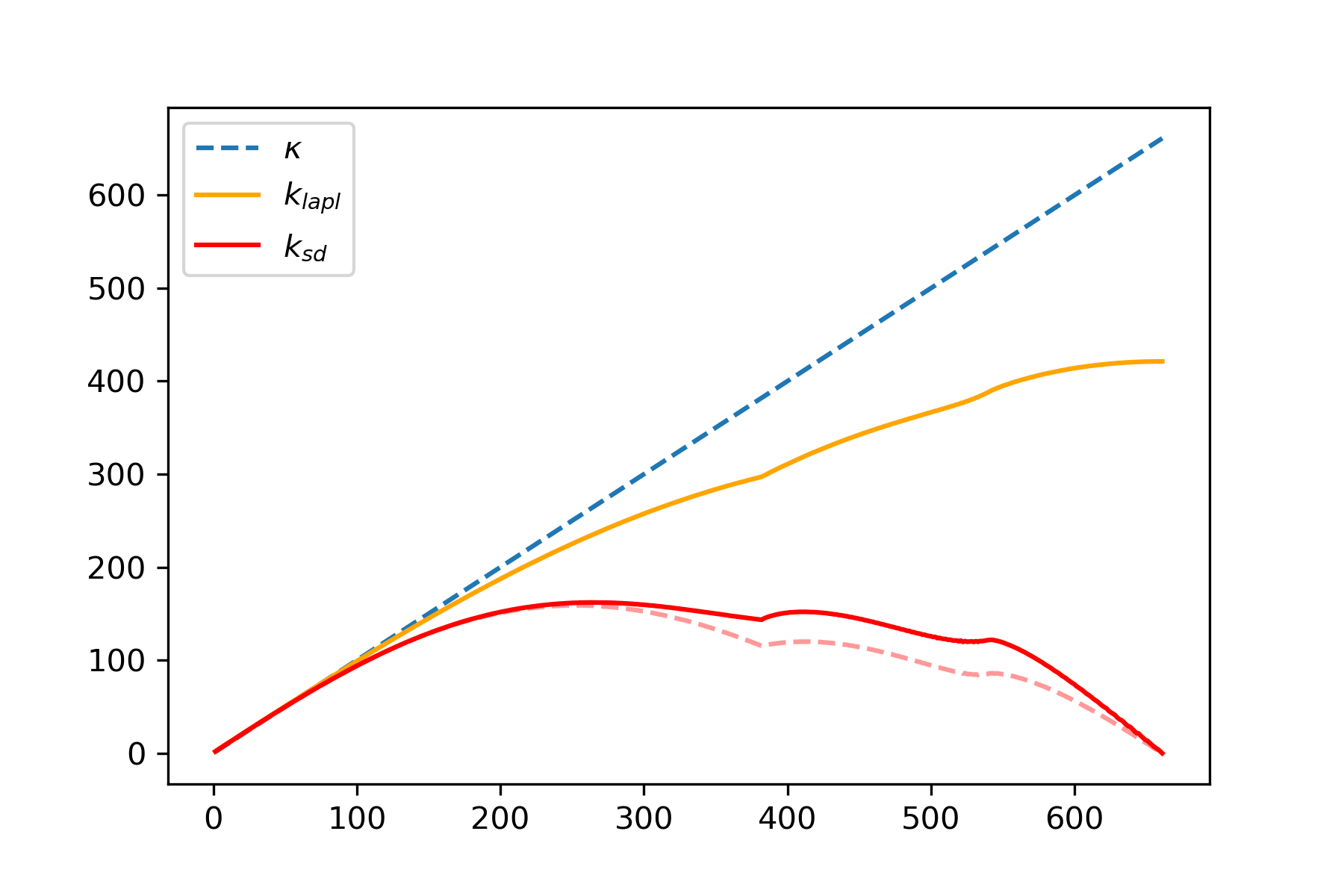}};
	
	\node [text width=0.01cm,align=center] at (0,-3){ $\kappa$};

	\end{tikzpicture}

	\caption{Plot of the different effective momenta emerging from the definitions of the second order centered Laplacian of \cref{eq:lapl_old} (yellow) and the second order centered spatial derivative of \cref{eq:1d} (red). These one-dimensional quantities are obtained from \cref{eq:modes,eq:klapl,eq:ksd} through a spherical binning. The red dashed line shows the quantity $\vec{k}_{\rm sd}\cdot \vec{\kappa}/|\kappa|$.}
	\label{fig:dispersionA}
\end{figure}
\subsection{Choice of the discretization scheme}
\label{sec:consistent}

In order to correctly evolve the gauge field on the lattice, we need to find a discretization scheme for the Laplacian and for the one-dimensional derivative such that $k_{\rm sd}=k_{\rm lapl}$. Indeed, in this case, \cref{eq:latt_modes} will be very similar to its continuous version of \cref{eq:cont_modes}, with the only exception of having $\vec{k}_{\rm sd}\cdot\vec{\kappa}/|\vec{\kappa}|$ inside the bracket instead of $|k_{\rm sd}|$. Later in this section, we will discuss this difference, which turns out to be negligible in the evolution of the gauge field. To achieve our goal, we define the Laplacian in the following way:
\begin{align}
\label{eq:lapl}
[\nabla^2 f](\vec{n})=\frac{1}{(2\Delta x)^2}\sum_{\alpha={\pm2}}\biggl(f(\vec{n}+ \alpha\vec{e}_1)+f(\vec{n}+ \alpha\vec{e}_2)+f(\vec{n}+\alpha \vec{e}_3)-3f(\vec{n})\biggr).
\end{align}
This choice corresponds to defining the Laplacian in a consistent way with respect to the one-dimensional spatial derivative. Indeed, once we fix \cref{eq:1d} as the one-dimensional derivative, this expression for the Laplacian follows by requiring that $\nabla^2=\partial_j \partial_j$ on the lattice:
\begin{equation}
[\nabla^2 f](\vec{n})=\frac{[\partial_j f](\vec{n}+\vec{e}_j)-[\partial_j f](\vec{n}-\vec{e}_j)}{2\Delta x}.
\end{equation}
As the Chern-Simons interaction $\phi F\tilde F$ involves spatial derivatives, this will turn out to be a good feature in simulating this model. Indeed, it allows a consistent comparison between the Laplacian terms $\nabla ^2 $ and the one-derivative terms $\partial_j$ in \cref{eq:final1,eq:final2,eq:final3}, to which the gauge field growth is exponentially sensitive.


Using the Laplacian of \cref{eq:lapl}, together with the definition of the one-dimensional derivative of \cref{eq:1d}, will result in the same effective momenta:
\begin{equation}
\label{eq:samemodes}
\vec{k}_{\rm eff}\equiv \vec{k}_{\rm lapl}=\vec{k}_{\rm sd}=\frac{1}{\Delta x}\sin\left({\vec{\kappa}_{\vec{m}}}\Delta x\right).
\end{equation}
Note that, contrarily to \cref{eq:lapl_old}, the Laplacian of \cref{eq:lapl} only employs the next to neighboring points of $\vec{n}$, instead of the neighboring points directly. For this reason, we can interpret this choice as thinking of the $N^3$ cubic lattice as an effective lattice with $N^3_{\rm eff}=(N/2)^3$ points, and where the extra intermediate points are only needed to compute spatial derivatives in a way that is consistent with the problem at hand. As a consequence, only roughly the lower half of the Fourier modes will be physical and we will put a hard cutoff on the lattice in order to exclude the upper part of the spectrum. We choose this cutoff to be the value at which $k_{\rm sd}(\kappa)$ starts decreasing (in \cref{fig:dispersion} this happens roughly around $\kappa\simeq 250$). 

Thanks to this equivalence between $k_{\rm sd}$ and $k_{\rm lapl}$, \cref{eq:latt_modes} will be much closer to its continuous counterpart of \cref{eq:cont_modes}. As mentioned, the only difference is the term inside the brackets where, instead of $|\vec{k}_{\rm sd}|$, we have $\vec{k}_{\rm sd}\cdot \vec \kappa/|\vec \kappa|$. 
We checked by solving numerically the linear equation \eqref{eq:latt_modes} that the presence of this scalar product causes a negligible difference in the evolution of the gauge field. Therefore we neglect this effect and assume $|\vec{k}_{\rm sd}|\simeq \vec{k}_{\rm sd}\,\cdot\, \vec{\kappa}/|\kappa|$.
In \cref{fig:dispersion} we show the difference between these two quantities, which are depicted respectively as a red and a red dashed line in the plot. From this plot, we can see that the difference between the two is quite small, and it is negligible in the relevant part of the spectrum (below $\kappa=250$). The validity of this approximation will be confirmed by the results of the simulation. However, as we will see in Section \ref{sec:pol}, this approximation can be avoided by using a different definition of the $\epsilon_\pm$ vectors.

In the end, we can use the results of this section to write a solution for the gauge field growth on the lattice in analogy to the continuous case:
\begin{equation}
\label{eq:ex_sol_lat}
A^{\rm(lat)}_-(\kappa,\tau)\simeq\frac{1}{\sqrt{2k_{\rm eff}}}\left[G_0(\xi,- k_{\rm eff}\tau)+iF_0(\xi,- k_{\rm eff}\tau)\right]
\end{equation}
This solution is similar to \cref{eq:ex_sol} but with $k_{\rm eff}$ instead of $k$. In an analogy with the single-field case of \cref{sec:inflationsim}, the lattice solution is equivalent to the continuous one using the equivalence principle ${k}_{\rm eff}\leftrightarrow k$. As we will see, this identification turns out to be very useful when interpreting the outputs of the simulation and comparing them to the analytical results.

The strategy that we adopted to achieve the same effective momenta is not necessarily unique. Another way, for example, would be to keep the same definition for the the Laplacian operator and use the following $O(\Delta x^4)$ stencil for the one-dimensional derivative:
\begin{align}
\label{eq:1d_b}
[\partial_j f]^{(4)}(\vec{n})=\frac{1}{12 \Delta x} \Biggl[-\frac{1}{6}f(\vec{n}+\vec{e}_j)+8f(\vec{n}+\vec{e}_j)- 8f(\vec{n}-\vec{e}_j)+ \frac{1}{6}f(\vec{n}-\vec{e}_j)\Biggr].
\end{align}
This leads to the following effective momentum:
\begin{equation}
\vec{k}^{(4)}_{\rm sd}=\frac{1}{\Delta x}\left[\frac{4}{3}\sin\left({\vec{\kappa}_{\vec{m}}}\Delta x\right)-\frac{1}{6}\sin\left(2{\vec{\kappa}_{\vec{m}}}\Delta x\right)\right].
\end{equation}
With this choice, we still have $\vec k_{\rm lapl}\neq \vec k^{(4)}_{\rm sd}$, but this time we can find a larger range of modes for which $\vec k_{\rm lapl}\simeq \vec k^{(4)}_{\rm sd}$. This range constitutes roughly the lower half of the spectrum, in a similar way to the strategy above. However, this is achieved only approximately, and the $O(\Delta x^4)$ derivative of \cref{eq:1d_b} is computationally more expensive. For these reasons, we prefer to stick to the first strategy. 

\subsection{Initial conditions for the gauge field}
\label{sec:ICA}
Fluctuations of the inflaton field at the initial time are generated as described in \cref{sec:inflationsim}. The gauge field is also initiated in a similar way.
We start from the definition of the discretized version of \cref{eq:quantizationA}:
\begin{align}
\begin{split}
\label{eq:disc_quantization}
&\hat{\vec{A}}(\vec{n})=\sum_{\lambda=\pm}\sum_{\vec{m}}\biggl[\vec{\epsilon}_\lambda(\vec{\kappa}_{\vec{m}})u_{\lambda}(\vec{\kappa}_{\vec{m}})a_{\vec{m}}e^{-i\frac{2\pi}{N}\vec{n}\cdot\vec{m}}+h.c.\biggr], \\ &[a_{\vec{m}},a^\dagger_{\vec{m}^\prime}]
=\frac{1}{L^3}\delta(\vec{m},\vec{m}^\prime).
\end{split}
\end{align}
We start the simulation when the simulation box is sub-horizon $L\lesssim 1/(aH)$, so that $u_{\pm}$ are approximately in their Bunch-Davies vacuum at the beginning of the simulation:
\begin{equation}
\label{eq:discmodesA}
u_{\pm}(\vec{\kappa}_{\vec{m}})=\frac{L^{3/2}}{\Delta x^3}\frac{1}{\sqrt{2k_{\rm eff}}}e^{- ik_{\rm eff}\tau }.
\end{equation}
In this expression, we used $k_{\rm eff}$ instead of $\kappa$ so that the initial conditions of the simulation are compatible with the lattice solution of \cref{eq:ex_sol_lat}.  The normalization factors $L^{3/2}$ and $1/\Delta x^3$ are explained in \cref{sec:initialconditions}.
Since the growth of the gauge field occurs only approximately at horizon crossing, a few modes will already be tachyonic at the beginning of the simulation. For this reason, we will initiate $u_-$ taking into account some of the tachyonic growth, i.e.
\begin{align}
\label{eq:inmodes}
u_{-}(\vec{\kappa}_{\vec{m}})=\frac{L^{3/2}}{\Delta x^3}\frac{1}{\sqrt{2k_{\rm eff}}}\biggl[G_0(\xi,- k_{\rm eff}\tau)+iF_0(\xi,- k_{\rm eff}\tau)\biggr],
\end{align}
where we made use of the solution of \cref{eq:ex_sol_lat} for the discrete dynamics. This expression reduces to \eqref{eq:discmodesA} for $-k\tau\gg2\xi$, which will be true for most of the modes. Once the mode functions $u_\pm$ are specified, the field configuration on the lattice is generated in Fourier space as statistical realization of a random process, using the same procedure explained in \cref{sec:initialconditions}.
After this, the gauge field $\vec{A}(\vec{n})$ in real space is obtained from \cref{eq:pol} making use of the polarization vectors $\vec\epsilon_\pm(\kappa)$ defined in \cref{eq:epsilon}. We use the following explicit definition of the polarization vectors \cite{kalo}:
\begin{equation}
	\vec\epsilon_{\pm}(\vec\kappa)=\frac{\left(-\kappa_1\kappa_3\pm i\kappa_2 \kappa\,,\,-\kappa_2\kappa_3\mp i\kappa_1 \kappa\,,\,\kappa^2-\kappa_3^2\right)^T}{\sqrt{2\kappa^2(\kappa^2-\kappa^2_3)}},
\end{equation}
where $\vec\kappa=(\kappa_1,\kappa_2,\kappa_3)^T$ and $T$ means that we are showing the transpose of the vector. Note that this expression is divergent in a one-dimensional line of the Fourier transformed lattice defined by $\kappa=\kappa_3$. For these points, we adopt an analogous definition of $\epsilon_\pm$ that is divergent on points such that $\kappa_1=\kappa$.

The $A_0$ component of the gauge field is set to zero at the beginning of the simulation. Note that this is not compatible with the gauge choice $\partial^\mu A_\mu=0$. However, as we will see in \cref{sec:gauge}, this leads to a negligible violation of the gauge condition.

\section{Results of the simulation: negligible backreaction}
\label{sec:results_weak}
In this section, we show the results of the code in the case of negligible backreaction and compare some of them with the results of perturbation theory. Similarly to \cref{sec:inflationsim}, we 
assume a simple quadratic potential for the inflaton: $$V(\phi)=\frac{1}{2}m^2\phi^2,$$ with $m=0.51\cdot10^{-5}$. As mentioned in \cref{sec:motivation}, this kind of potential can be generated in string theory realizations of axion monodromy inflation \cite{McAllister:2014mpa}. Although the quadratic potential is disfavored by the latest Planck-BICEP/Keck results \cite{Planck:2018jri,BICEP:2021xfz}, the aspects discussed in this chapter do not depend on the particular shape of $V(\phi)$, as long as it provides the slow-roll dynamics needed for inflation. We leave a detailed study on the dependence of inflationary potential for future work.

The system is initiated far from the end of inflation. This is determined by the background values of the inflaton, that we set to\footnote{Note that this is different to what is done in \cref{sec:results}, where we assumed $\phi_{in}>0$ and $\phi^{\prime}_{in}<0$.} ${\phi}_{\rm in}=-14.5$ and ${\phi}^\prime_{\rm in}=0.8152m$, where $m=0.51\cdot10^{-5}$. With this choice, the system is initiated $53$ e-folds before the end of inflation. We run a simulations with a lattice of $N^3=256^3$ number of points and comoving length $L=2/m$. With these values, lattice modes $\kappa$ will range from $\kappa_{\rm min}\simeq 0.6 H_i$ to $\kappa_{\rm max}\simeq 118 H_i$, where $H_i$ is the Hubble parameter at the beginning of the simulation. In this section, we mainly consider $\alpha/f=42$ as the value of the gauge field coupling. This value is excluded by CMB observations but allows us to better compare the results of the simulation with the existing analytical estimates. In \cref{sec:results_strong} we consider a more realistic value of the coupling.

\begin{figure}
	\centering

	\begin{tikzpicture}
	\node (img) {\includegraphics[width=6cm]{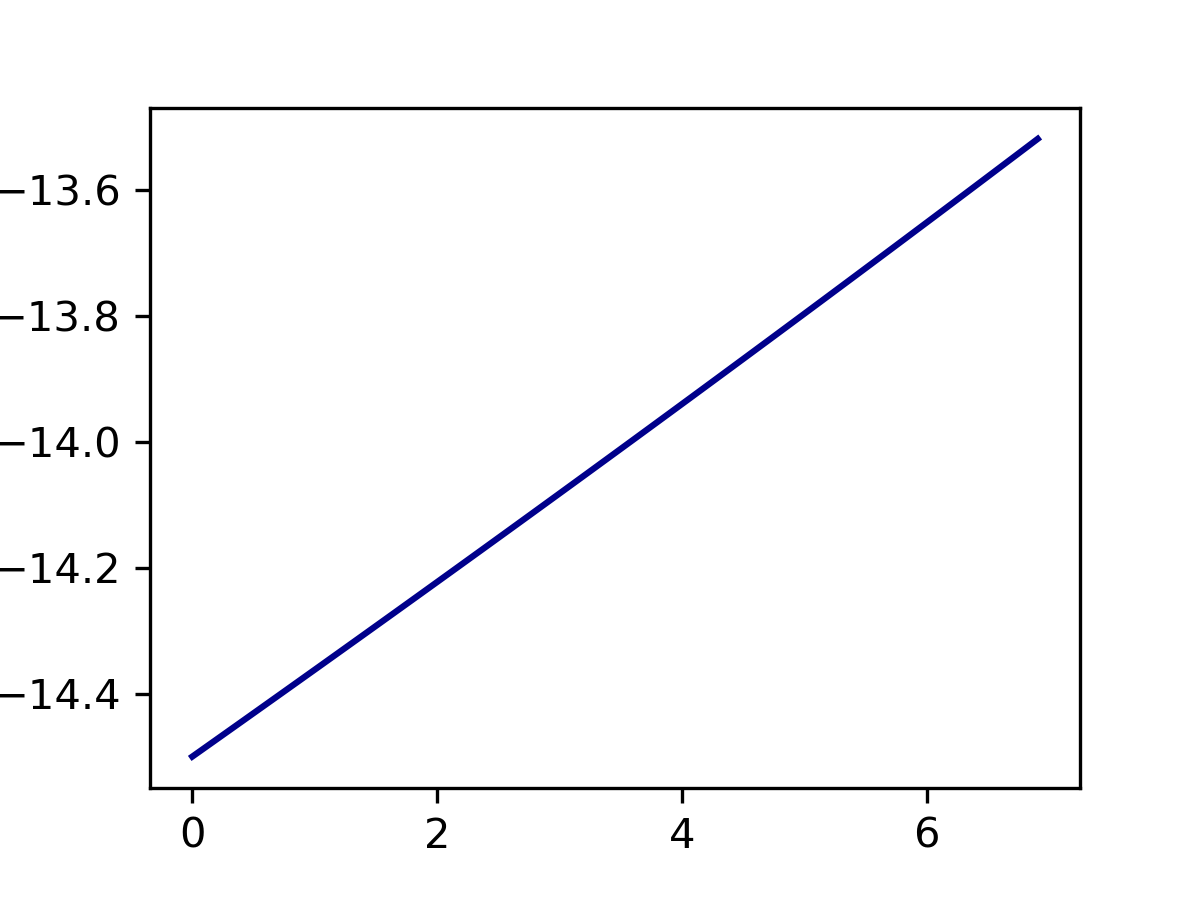}};
	
	\node [rotate=0,text width=0.01cm,align=center] at (-3.5,0){ $\bar{\phi}$};
	\node [text width=0.01cm,align=center] at (0,-2.4){$N_e$};

	\node (img2) at (7,0) {\includegraphics[width=6.4cm]{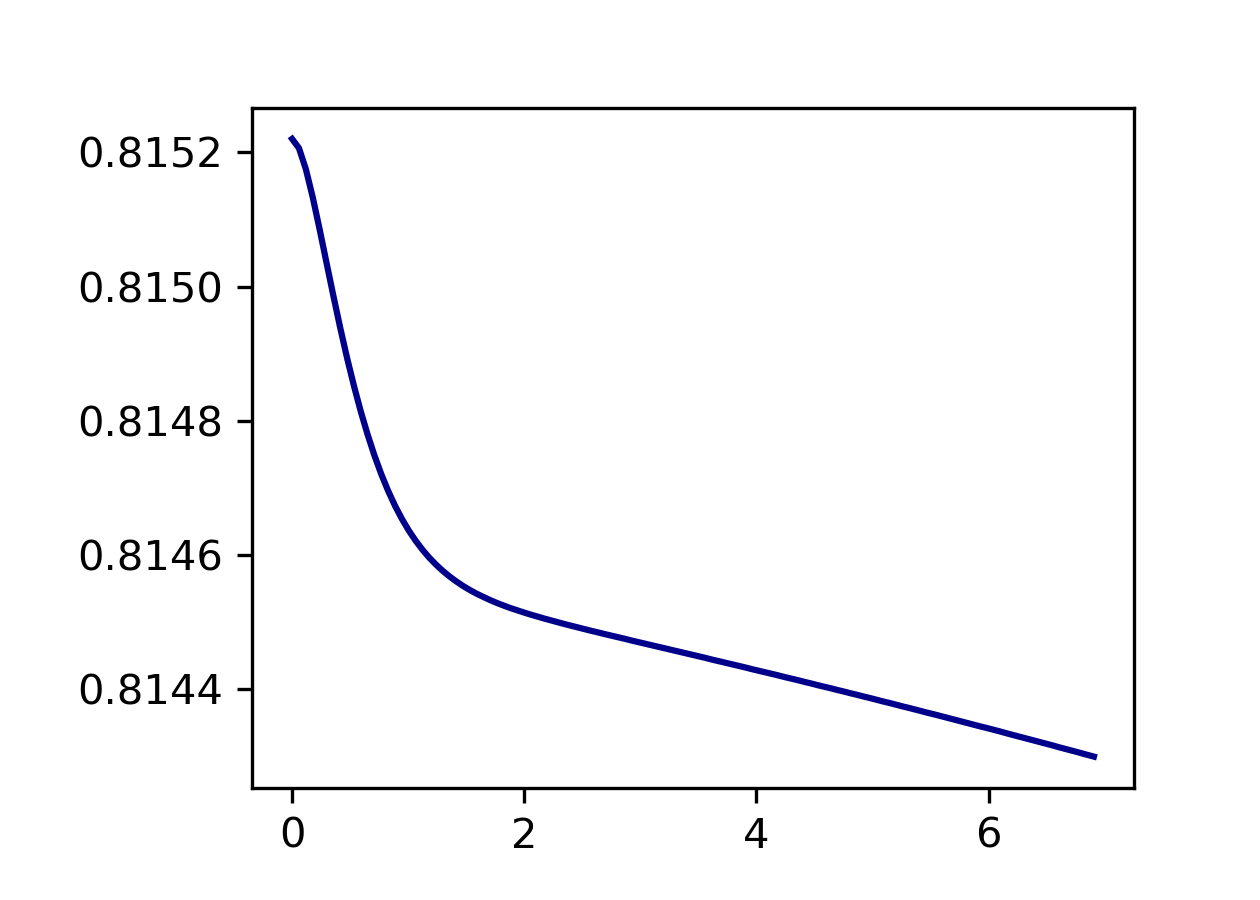}} ;
	
	\node [rotate=0,text width=0.01cm,align=center] at (-3.8+7,0){ ${\dot{\bar{\phi}}}/{m}$};
	\node [text width=0.01cm,align=center] at (0+7,-2.4){$N_e$};

	\node (img3) at (0,-5) {\includegraphics[width=6cm]{H.png}};
	
	\node [rotate=0,text width=0.01cm,align=center] at (-3.8,0-5){ $H/m$};
	\node [text width=0.01cm,align=center] at (0,-2.4-5){$N_e$};

	\node (img4) at (7,-5) {\includegraphics[width=6.4cm]{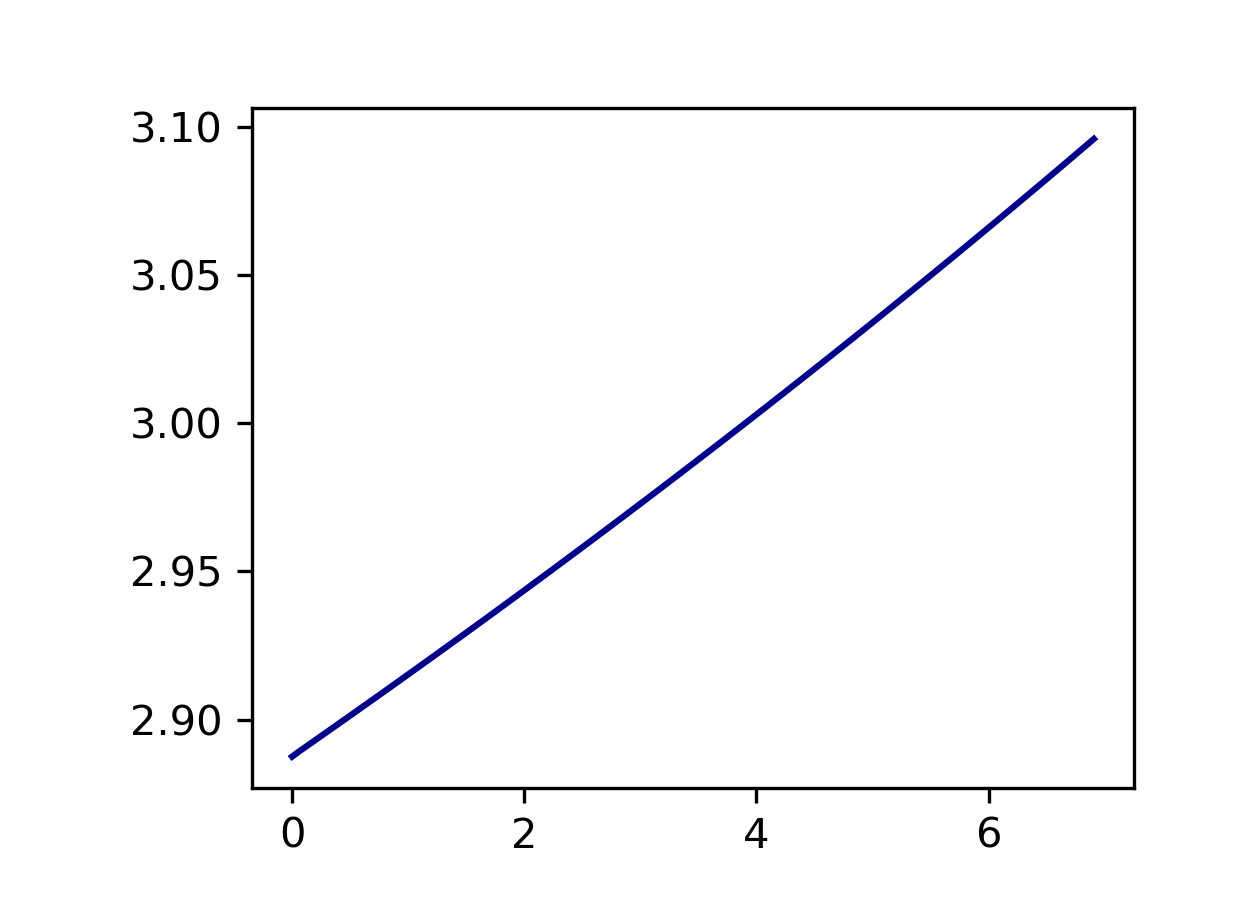}} ;
	
	\node [rotate=0,text width=0.01cm,align=center] at (-3.5+7.3,0-5){  $\xi$};
	\node [text width=0.01cm,align=center] at (0+7,-2.4-5){$N_e$};

	\end{tikzpicture}

	\caption{Plot of the background value of the inflaton (top left), its velocity (top right), the Hubble parameter $H$ (bottom left) and $\xi$ (bottom right) during the simulation in the case of negligible backreaction.}
	\label{fig:backgroundvaluesA}
\end{figure}

We evolve the system for $N_e\simeq6$ e-folds, which makes the simulation box super-horizon $L\gg 1/(aH)$ at the end of the simulation. In \cref{fig:backgroundvaluesA} we show the background value of the inflaton and its velocity in cosmic time, together with the Hubble parameter $H$ and $\xi$ during the evolution. From these plots, we see that the background slow-roll trajectory is not affected by the excitation of the gauge field.

The main result of this section is providing a full characterization the statistics of the curvature perturbation $\zeta$ for this model, and it is mainly contained in \cref{sec:PSA,sec:highA}. To obtain these results, we employ the discretization scheme developed in \cref{sec:discA}, which is defined by \cref{eq:lapl} and \cref{eq:1d}. Before proceeding, we first demonstrate the importance of the choice of the discretization scheme, and how it affects the growth of the gauge field on the lattice. To do so, we study the differences between our discretization scheme and the conventional scheme given by the Laplacian of \cref{eq:lapl_old}. 

As we discussed in \cref{sec:gauge}, we do not enforce the gauge constraint to be satisfied exactly. Therefore, we need to check that it is approximately satisfied during the evolution. We perform this check in \cref{sec:energy+gauge}, where we also discuss energy conservation for all the simulations shown in this chapter. 

\subsection{Discretization scheme and gauge field growth}
In this (somewhat technical) section we show the effects of the discretization on the growth of the gauge field on the lattice.
 The main quantity that we want to reproduce is the power spectrum of the growing mode of the gauge field. We show results for $\alpha/f=42$, which sets $\xi\simeq 2.9$ at the initial time. What we discuss in this section, however, does not depend on the particular value of the axion-gauge coupling, and we tested the simulation in the range $1<\alpha/f<80$ leading to the same results.

\begin{figure}
	\centering
	
	\begin{tikzpicture}
	\node (img) {\includegraphics[width=15cm]{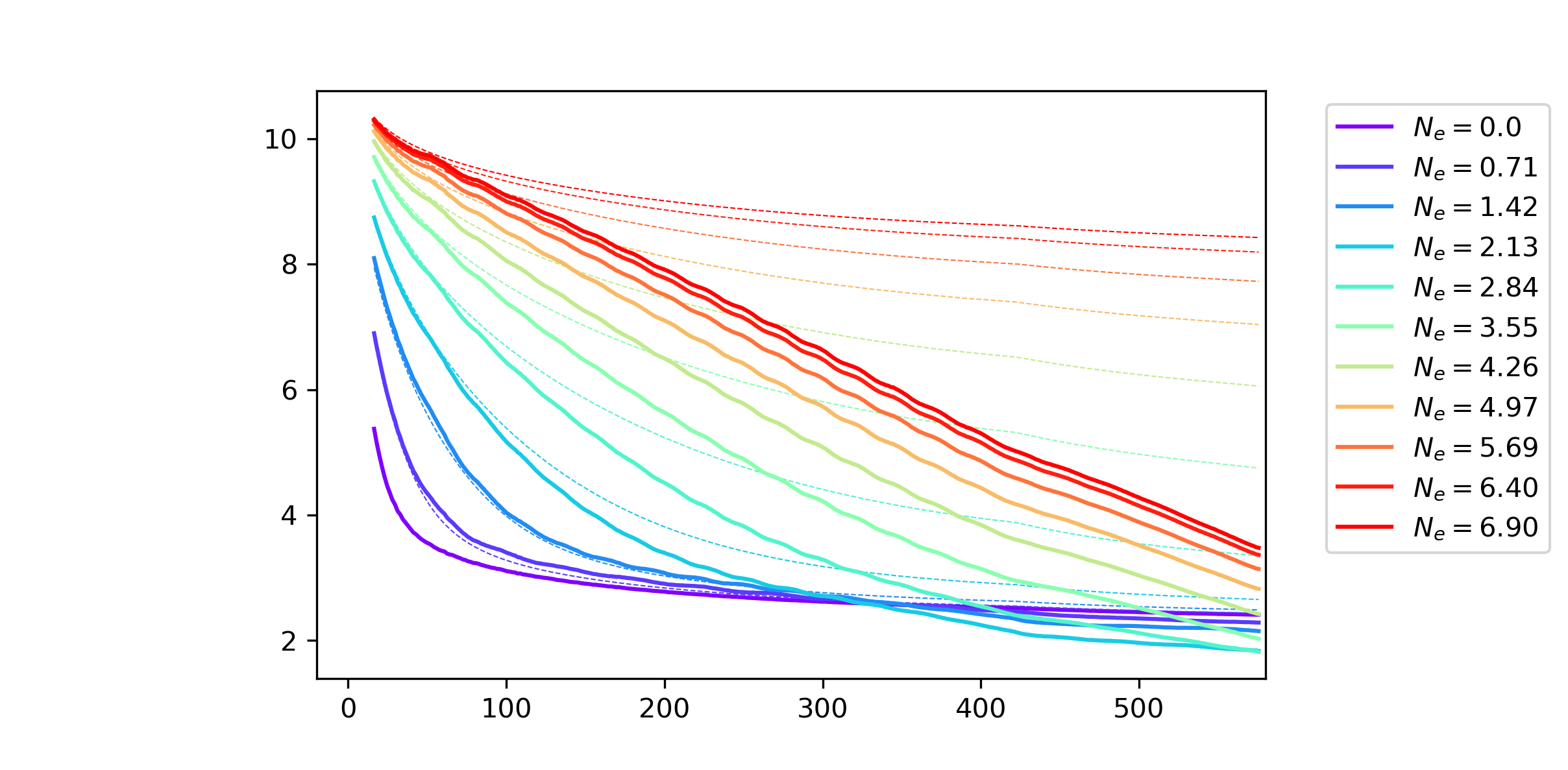}};
	
	\node [text width=0.01cm,align=center] at (0,-3.5){ $\kappa$};
	\node [text width=0.01cm,align=center] at (-7.5,0){ $\log_{10}{|A_-|^2}$};

	\end{tikzpicture}

	\caption{Plot of the power spectrum of the growing mode of the gauge field, computed from a lattice simulation with a $O(\Delta x^2)$ spatial discretization scheme defined by \cref{eq:lapl_old,eq:1d}. The solid lines represent the power spectrum computed from the simulation at different times, while the dashed lines represent the analytical expectation from the linear theory.}
	\label{fig:gauge_growth_bad}
\end{figure}

In \cref{fig:gauge_growth_bad} we show the power spectrum\footnote{Power spectra are computed using the procedure explained in \cref{sec:output}.} of $A_-$ computed from a simulation with the scheme defined by \cref{eq:lapl_old,eq:1d}, for which $k_{\rm lalp}\neq k_{\rm sd}$. As we want to compare different discretization schemes, we plot the power spectra as function of lattice momenta $\kappa$ within this section.
 In this plot, the solid lines are the results from the simulation, while the dashed lines represent the expected theoretical power spectrum computed from the linear theory. The theoretical spectra are obtained by solving numerically the linear \cref{eq:cont_modes}, in order to take into account slow-roll corrections to \cref{eq:ex_sol}. From this plot we can see that the simulation is not able to correctly reproduce the growth of the gauge field. Indeed, the gauge field growth is exponentially suppressed with respect to the analytical expectation for most of the modes, affecting both the amplitude and shape of the power spectrum. This expected behavior is a consequence of the exponential sensitivity of the gauge field dynamics to the choice of the spatial discretization scheme, as explained in Section \ref{sec:consequences}. Moreover, the lattice solution shows small oscillations in the form of wiggles at intermediate scales, which are particularly evident at late times (red curves). These oscillations are caused by a misalignment of phase between $A_-$ and its time derivative $A^\prime_-$ at the initial time. This is a consequence of the fact that initial conditions are generated using \cref{eq:inmodes}, which is not compatible with the lattice dynamics for this discretization scheme, as described by \cref{eq:latt_modes}.

\begin{figure}
	\centering
	
	\begin{tikzpicture}
	\node (img) {\includegraphics[width=15cm]{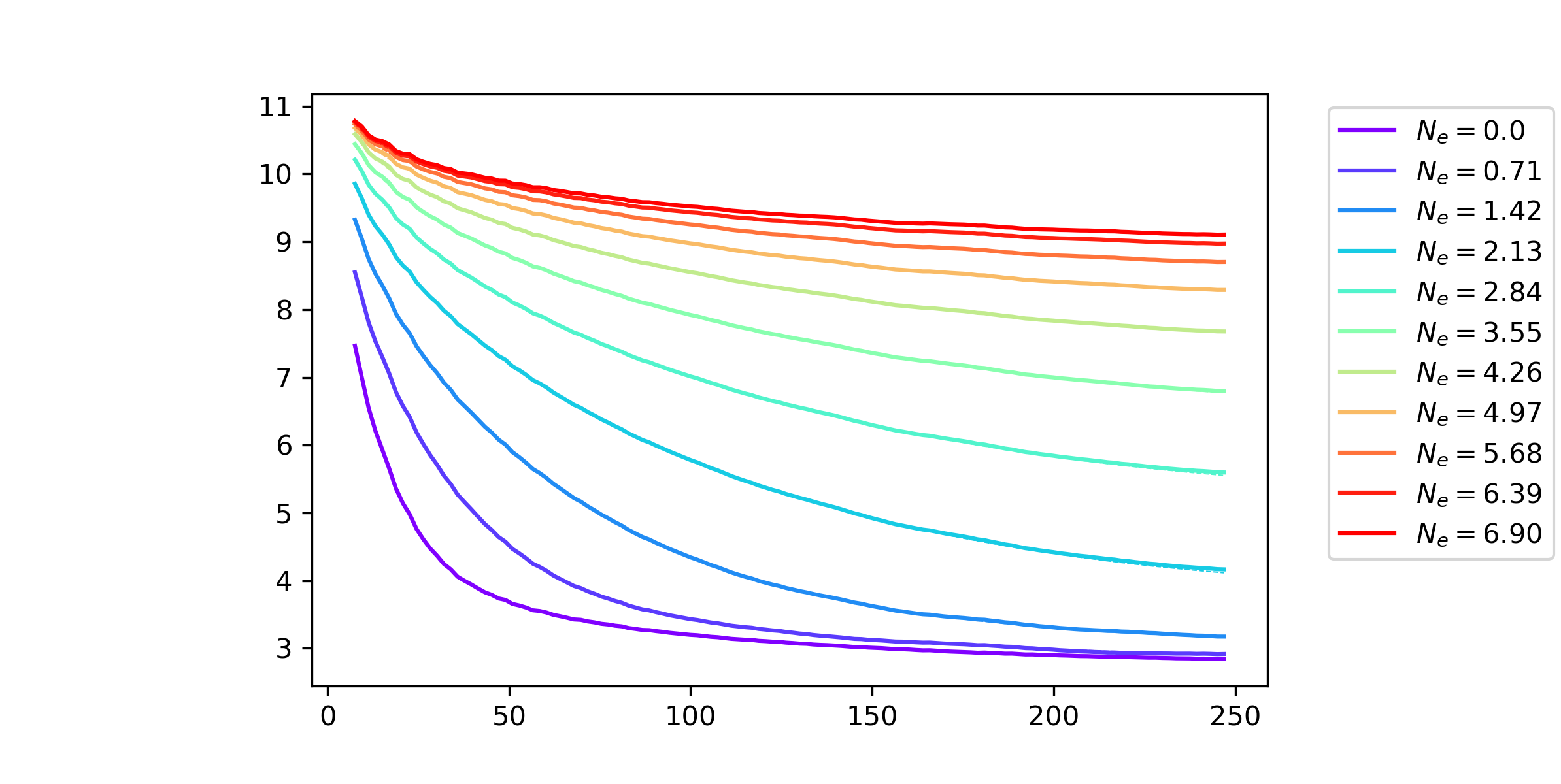}};
	
	\node [text width=0.01cm,align=center] at (0,-3.5){ $\kappa$};
	
	\node [text width=0.01cm,align=center] at (-7.5,0){ $\log_{10}{|A_-|^2}$};
	\end{tikzpicture}

	\caption{Plot of the power spectrum of $A_-$, computed from a lattice simulation with the improved scheme defined by \cref{eq:lapl,eq:1d}. The solid lines are the lattice power spectra, which almost overlap with the expectation from the linear theory depicted as dashed lines (barely visible in this plot).}
	\label{fig:gauge_growth}
\end{figure}

In \cref{fig:gauge_growth} we show the result from a simulation with the improved scheme where the Laplacian is defined as in \cref{eq:lapl}, for which $k_{\rm eff}=k_{\rm lapl}=k_{\rm sd}$.  We can see that the simulation is now able to reproduce the growth of the gauge field with much better precision. Indeed, the lattice result basically overlaps with the spectra from the linear theory (dashed lines), making them barely visible in the plot.
Within this scheme, $k_{\rm eff}$ plays the role of the effective momentum of the simulation, and for this reason the analytical result in this plot is computed from the linear theory thinking of $k_{\rm eff}$ as the physical momentum (i.e. $k_{\rm eff}$ plays the role of the $k$ of the continuous theory).
As we already discussed in \cref{sec:consistent}, the price that we pay when using this discretization scheme is that the upper part of the spectrum is unphysical. For this reason, we set the value of the UV cutoff in this case to be $\kappa_{\rm max}\simeq250/m$, contrarily to $\kappa_{\rm max}\simeq600/m$ of the first scheme above. 

\subsubsection*{Energy density}
\begin{figure}
	\centering
	
	\begin{tikzpicture}
	\node (img) {\includegraphics[width=11cm]{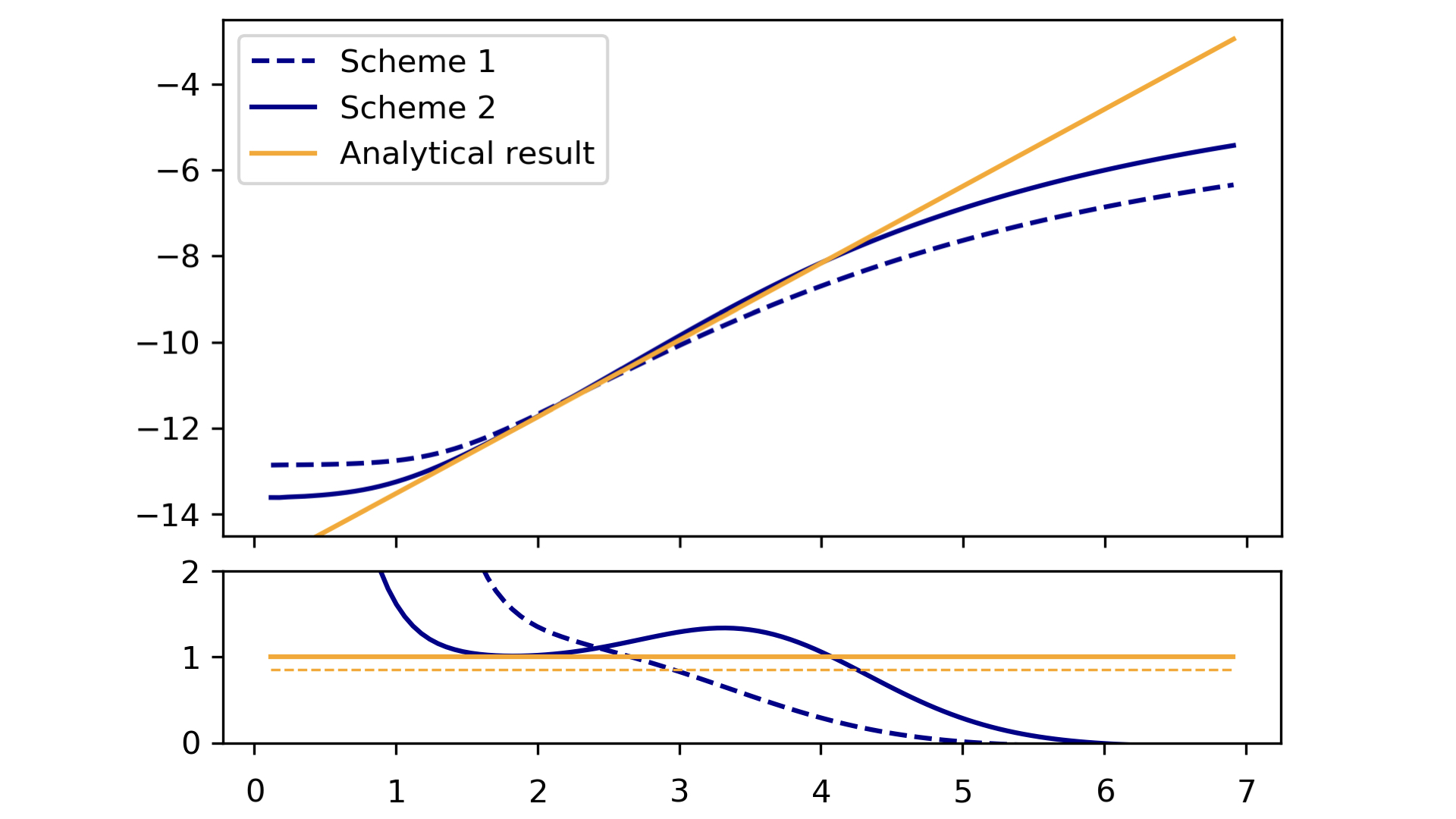}};
	
	\node [rotate=0,text width=0.01cm,align=center] at (-6.6,1.){ $\log_{10}\left(a^4\rho_{\rm GF}\right)$};
	\node [rotate=0,text width=1cm,align=center] at (-5.5,-1.85){ $\rho_{\rm GF}/\rho^{\rm(a)}_{\rm GF}$};
	\node [text width=0.01cm,align=center] at (0,-3.4){$N_e$};

	\end{tikzpicture}

	\caption{Plot of the average value of $\rho_{\rm GF}$ on the lattice. In the top panel we show the value of $\rho_{\rm GF}$, and in the bottom panel we show the same quantity normalized by the analytical result. We call {Scheme 1} the one defined by \cref{eq:lapl_old,eq:1d}, and {Scheme 2} the improved one defined by \cref{eq:lapl,eq:1d}.}
	\label{fig:rhoGF}
\end{figure}
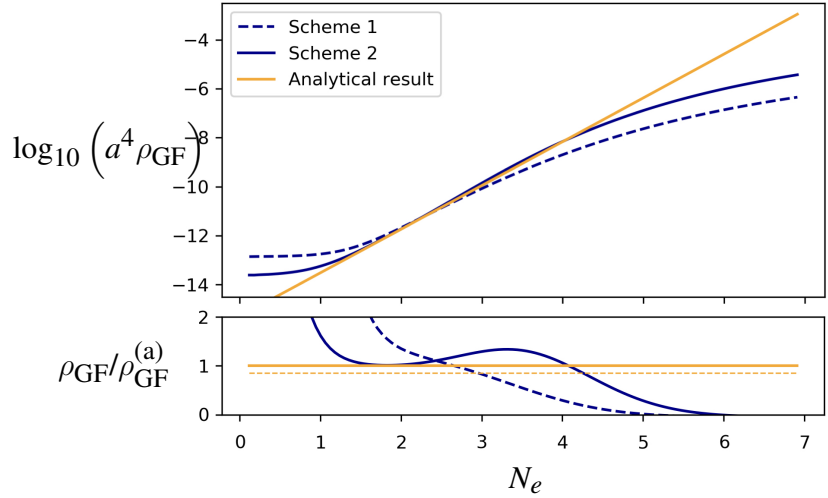

\begin{figure}
	\centering
	
	\begin{tikzpicture}
	\node (img) at (1,0) {\includegraphics[width=11cm]{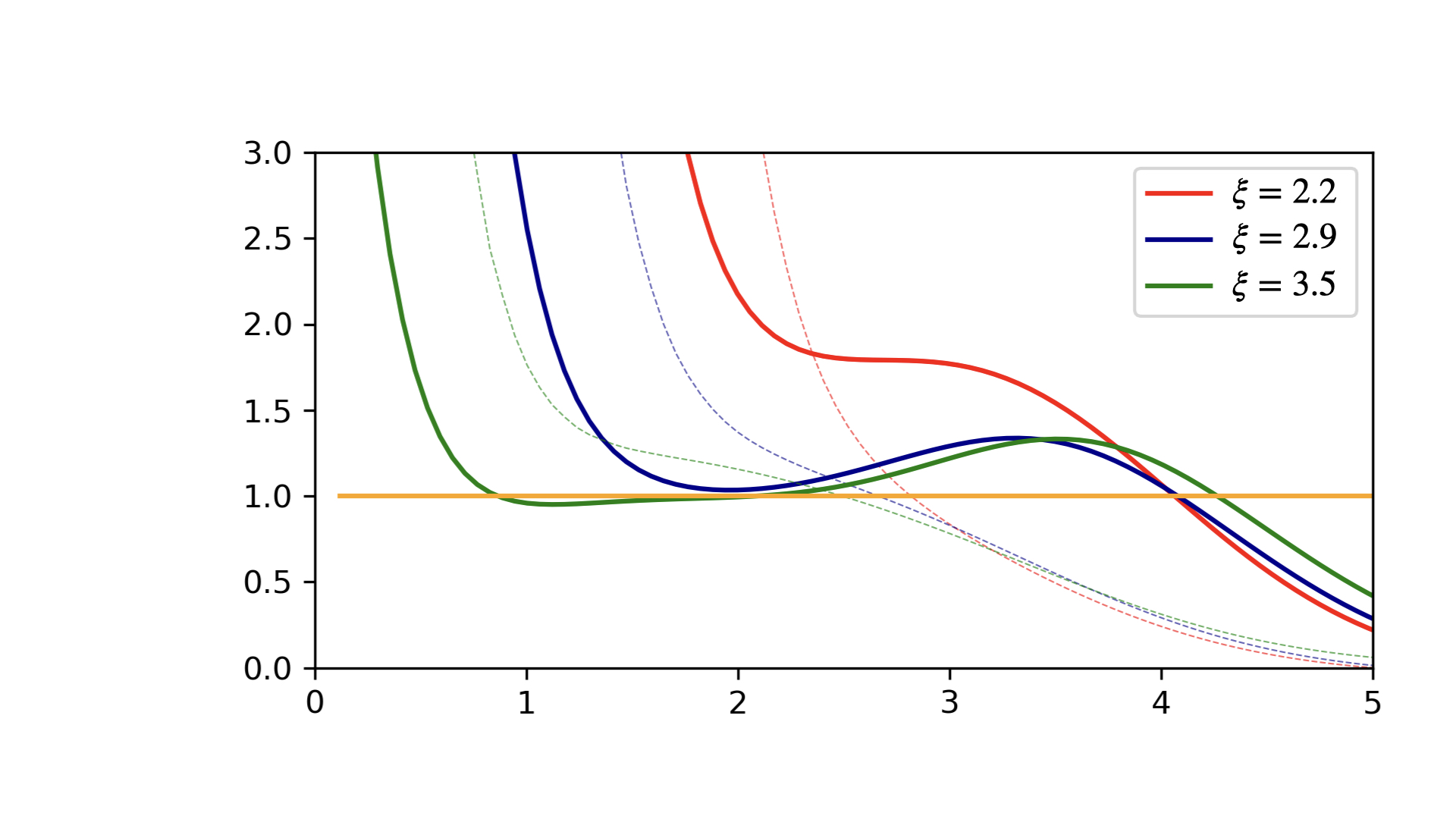}};
	
	
	\node [rotate=0,text width=1cm,align=center] at (-4.3,0){ $\rho_{\rm GF}/\rho^{\rm(a)}_{\rm GF}$};
	\node [text width=0.01cm,align=center] at (1.5,-2.7){$N_e$};

	\end{tikzpicture}

	\caption{Plot of the average value of $\rho_{\rm GF}$ on the lattice divided by the analytical prediction, shown for different values of $\xi$. The $\xi$ in the legend are the values at the beginning of the simulation. The full lines are obtained with the improved scheme of \cref{eq:lapl,eq:1d}, while the dashed ones are obtained with the first scheme of \cref{eq:lapl_old,eq:1d}.}
	\label{fig:rhoGF_xi}
\end{figure}

We now show the effects of the discretization on the real space energy density of the gauge field $\rho_{\rm GF}$. We first show results from the same simulation with $\xi\simeq2.9$. In the upper panel of \cref{fig:rhoGF} we show the evolution of $a^4\rho_{\rm GF}$ computed as an average energy density over the $N^3$ points of the lattice. We show both the results for the first scheme defined by \cref{eq:lapl_old,eq:1d} (dashed blue line) and for the improved scheme of \cref{eq:lapl,eq:1d} (blue line). To check the accuracy of the lattice result, comparing it to an analytical prediction would be valuable. Unfortunately, an exact analytical result does not exist for a time-dependent $\xi$. However, we can compare the lattice results to the ones obtained in the literature for a constant $\xi$, as it is slowly varying during slow-roll. For a constant $\xi\gg1$, the energy density can be approximated as \cite{Anber_2010}:
\begin{equation}
\label{eq:rhogf}
\rho^{\rm(a)}_{GF}\simeq\frac{6!}{2^{19}\pi^2}\frac{H^4}{\xi^3}e^{2\pi\xi}.
\end{equation}
For a finite $\xi$ this value needs to be corrected to account for a rigorous renormalization procedure \cite{Jimenez:2017cdr,Ballardini:2019rqh,Animali:2022lig}. In the upper panel of  \cref{fig:rhoGF} we show the corrected value of Ref. \cite{Ballardini:2019rqh} for $ \rho^{\rm(a)}_{GF}$ (orange line), that we avoid writing explicitly\footnote{	
	More precisely, the $\rho^{\rm(a)}_{GF}(\tau)$ in \cref{fig:rhoGF,fig:rhoGF_xi} shows, for each $\tau$, the $\xi$-constant result of \cite{Ballardini:2019rqh} for the corresponding $\xi(\tau)$.
	
	
	
	
}. 
In the bottom panel of the same figure we show $\rho_{\rm GF}$ from the simulation normalized by the same analytical prediction. The dashed orange line in the bottom panel of \cref{fig:rhoGF} shows the ratio between the expression \eqref{eq:rhogf} and the corrected value.

We can see that the evolution of $\rho_{\rm GF}$ from the simulation can be divided into three phases. During the first phase, from the beginning until $N_e\sim1$, the lattice energy density is much higher than the theoretical expectation due to the classical nature of the simulation. This is because the lattice calculation includes sub-horizon UV-divergent contributions that are subtracted in the analytical computation due to the renormalization procedure. These UV-divergent contributions are not subtracted in the lattice computation, as this would lead to an unphysical spatial curvature in the first Friedmann equations, as explained in \cref{sec:initialbackground}. The initial value of $\rho_{\rm GF}$ is bigger in the case of the first scheme, and this is a consequence of the higher UV-cutoff (see the end of the previous section). After this, there is a second phase until $N_e \sim 4$ in which the lattice result is of the same order of the theoretical value (see more below). This is only true for the second scheme, as the first one significantly underestimates the energy density of the gauge field in real space. During the last e-folds of evolution (after $N_e\sim 5$), there is a last phase in which the energy density is much lower than the analytical value, and this is due to the finite size of the lattice. Indeed, at the end of the simulation all the modes become super-horizon and the gauge field production is suppressed due to the finite spatial resolution of the lattice.

In \cref{fig:rhoGF_xi} we show the ratio between the energy density from the simulation and the analytical prediction for different values of the coupling $\alpha/f$. We mainly focus on the results from the improved scheme, but we also show results from the first scheme as dashed lines in the plot.
Since $\xi$ changes during the evolution due to slow-roll, we show the initial value of $\xi$ in the legend of \cref{fig:rhoGF_xi}.
From this plot we can see that for $\xi\gtrsim 3$ 
we can always find an intermediate time range (roughly $1<N_e<4$) in which the result of the simulation is roughly constant and it is close to the analytical expectation of Ref. \cite{Ballardini:2019rqh}, up to small corrections due to the time-varying $\xi$. This means that, during this intermediate phase, the finite simulation box contains all the relevant excited modes of the gauge field.
The situation is different for smaller values of $\xi$, and this is because the UV-divergent contributions, subtracted in the analytical computation, become more important in this case. This is a consequence of the classical nature of the lattice simulation. Note that the result of Ref. \cite{Ballardini:2019rqh} predicts a negative energy density for $\xi<1.5$, which cannot be achieved using the lattice approach. Although the lattice result for the energy density is only an approximation of analytical result, in particular for small $\xi$, we can see that the second scheme performs significantly better than the first one. Indeed, the second one is never able to achieve a roughly constant $\rho_{\rm GF}$, and the energy density of the gauge field is clearly suppressed in this case.

\subsubsection*{Polarization vectors}
\label{sec:pol}

Let us comment now about the definition of the $\epsilon_{\pm}$ vectors used to project $\vec A$ into the polarizations $A_{\pm}$. As previously mentioned, $\epsilon_{\pm}(\vec\kappa)$ are defined as in continuous space from \cref{eq:epsilon}.
However, having identified $k_{\rm eff}$ as the physical modes of the lattice theory, it might be better to define the polarization states according to $k_{\rm eff} $ instead of $\kappa$. This corresponds to defining the following lattice polarization vectors $\epsilon_{L,\pm}(\vec \kappa)$:
\begin{equation}
\label{eq:epsilon_lat}
{\vec\epsilon_{L,\lambda}}^{\,*}(\vec\kappa)\cdot	\vec\epsilon_{L,\lambda^\prime}(\vec\kappa)=\delta_{\lambda,\lambda^\prime},\quad \vec{k}_{\rm eff}(\vec\kappa)\cdot\vec\epsilon_{L,\pm}(\vec \kappa)=0,\quad \vec{k}_{\rm eff}(\vec\kappa)\times\vec\epsilon_{L,\pm}(\vec \kappa)=\mp i k_{\rm eff}(\vec{\kappa}) \,\vec{\epsilon}_{L,\pm}(\vec \kappa).
\end{equation}
Note that, if we do so, in the derivation of \cref{eq:latt_modes} one ends up with $|\vec k_{\rm sd}|$ inside the bracket instead of $ \vec{k}_{\rm sd}\,\cdot\, \vec{\kappa}/|\kappa|$, which makes the approximation made in \cref{sec:consistent} unnecessary. In other words, in this case we have an exact equality $=$ instead of $\simeq$ in \cref{eq:ex_sol_lat} (exact in the de Sitter approximation, in the same way of \cref{eq:ex_sol}). However, as we see from the results of this section, this approximation does not spoil the accuracy of the lattice simulation, and for this reason we kept the continuous definition of the $\epsilon_{\rm \pm}$ vectors throughout this chapter. 

\subsection{Power spectrum and bispectrum}
\label{sec:PSA}
We know show the results of the simulation regarding power spectrum and bispectrum of the scalar perturbation. Again, we show results for $\alpha/f=42$, which is excluded by CMB observations but allows us to better compare our results with the previous estimates of \cref{eq:ps_th} and \cref{eq:bis_th}.
\begin{figure}
	\centering
	
	\begin{tikzpicture}
	\node (img) {\includegraphics[width=15cm]{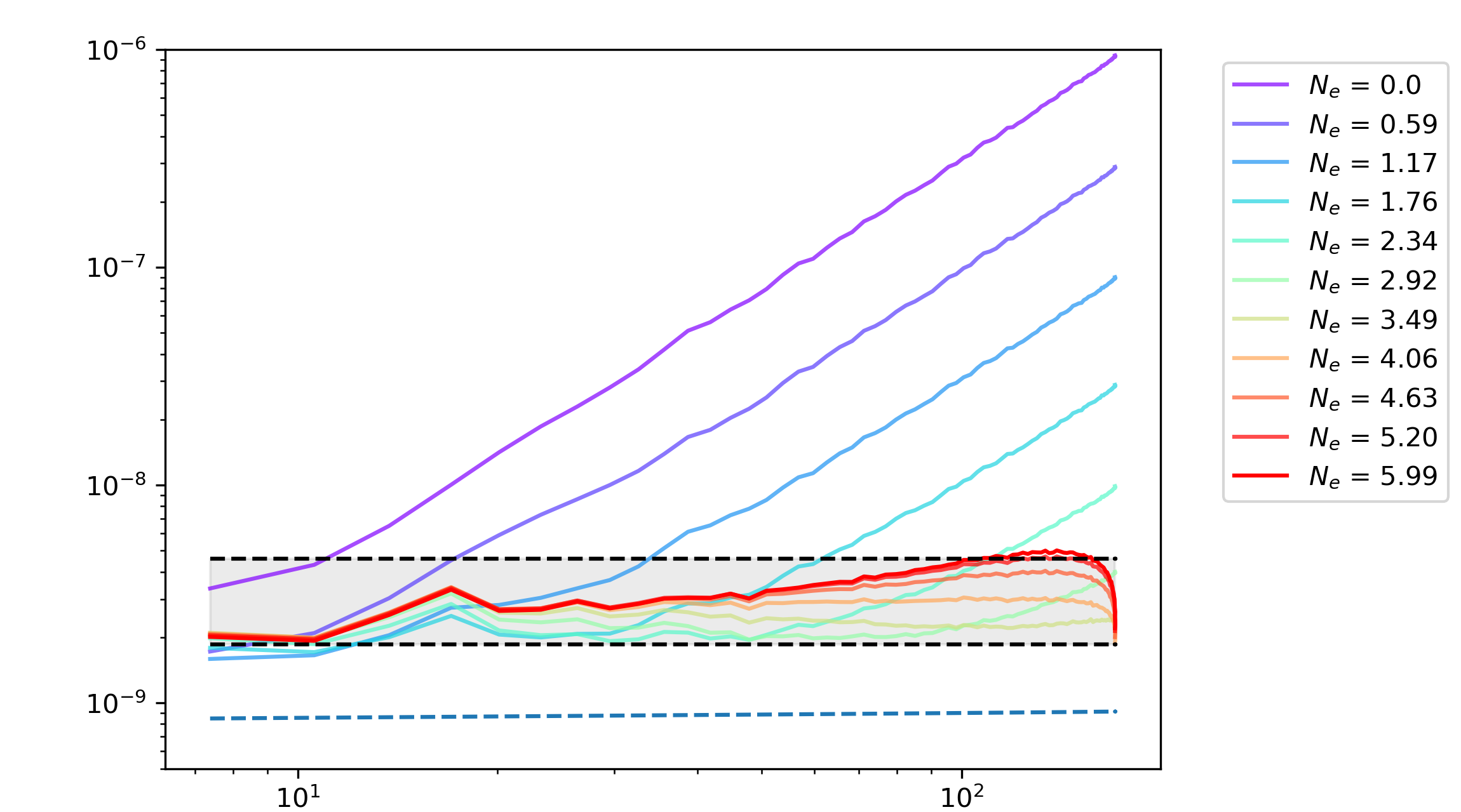}};
	
	\node [text width=0.01cm,align=center] at (-1,-4.6){ $k_{\rm eff}/m$};
	\node [text width=0.01cm,align=center] at (-7.5,0){ $P_{\zeta}$};

	\end{tikzpicture}

	\caption{Power spectrum of $\zeta$ in the case of weak backreaction. The shaded region, delimited by black dashed lines, shows the analytical result of \cref{eq:ps_th}. The blue dashed line shows the vacuum contribution $\mathcal{P}_{\rm vac}$.}
	\label{fig:PS_weak}
\end{figure}
\begin{figure}
	\centering
	
	\begin{tikzpicture}
	\node (img) {\includegraphics[width=13cm]{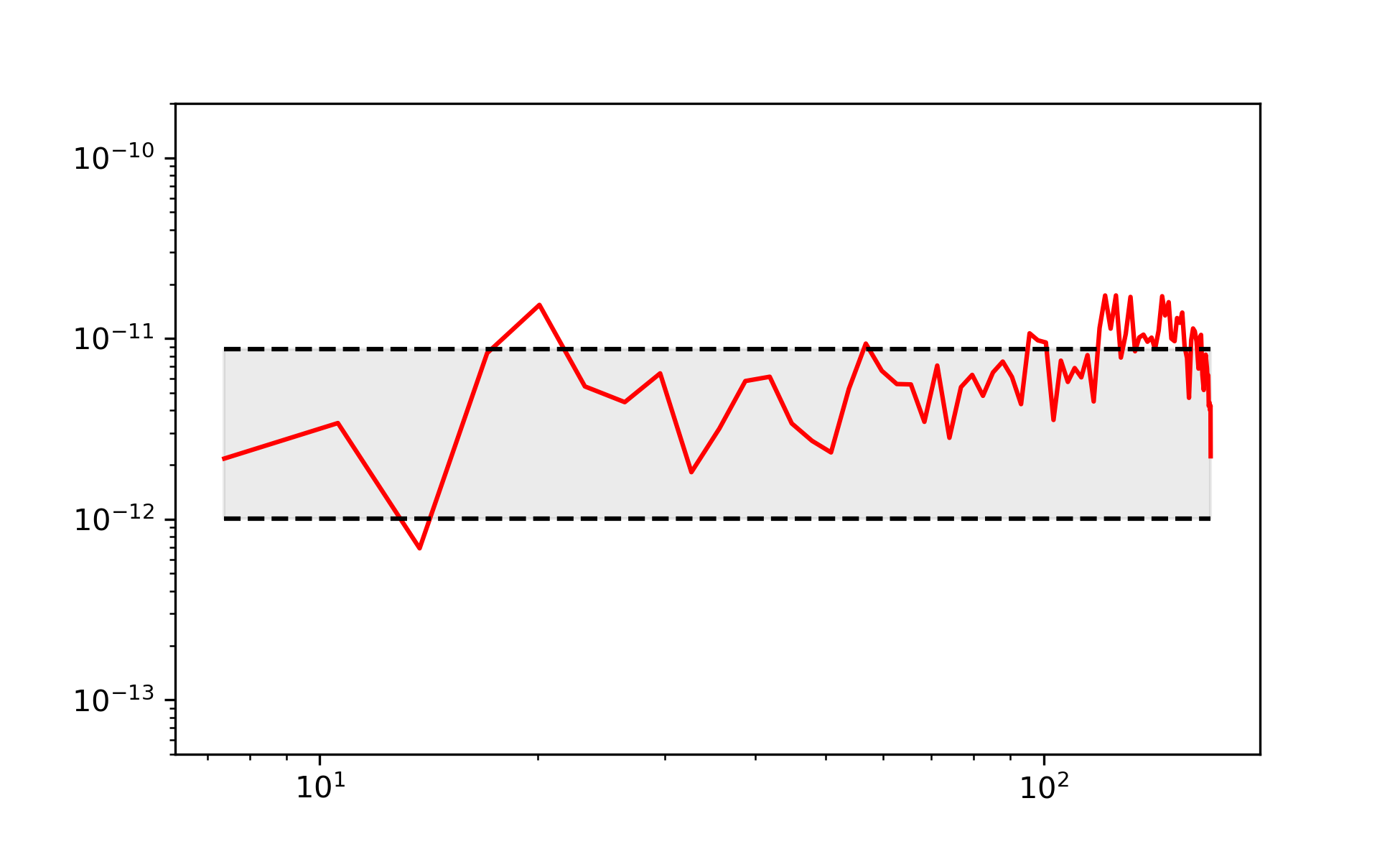}};
	
	\node [text width=0.01cm,align=center] at (0,-3.7){ $k_{\rm eff}/m$};
	\node [text width=0.01cm,align=center] at (-6.7,0){ $k_{\rm eff}^6\,\mathcal{B}_{\zeta}$};

	\end{tikzpicture}

	\caption{Equilateral-shape bispectrum of $\zeta$ in the case of weak backreaction. The shaded region, delimited by black dashed lines, shows the analytical prediction of \cref{eq:bis_th}. }
	\label{fig:bis_weak}
\end{figure}
We show results for the curvature perturbation, which is computed as $\zeta=\frac{H}{\dot {\bar\phi}}\delta \phi$, where $\delta \phi = \phi-\bar\phi$.
\subsubsection*{Power spectrum}
 In \cref{sec:output} we explain how power spectrum and bispectrum of $\phi$ are computed in our code. In \cref{fig:PS_weak} we show the power spectrum at different times during the simulation, going from initial time $N_e=0$ to $N_e\simeq 6$. Colors go from purple (early times) to red (late times). The spectra are plotted against $k_{\rm eff}$, defined in \cref{eq:samemodes} and identified as the physical momentum of the simulation. The dashed blue line shows the prediction for single-field dynamics $\mathcal{P}_{\rm vac}=H^4/(2\pi\dot{{\bar\phi}})^2$. At the final time, all the modes are super-horizon and the power spectrum of $\zeta$ is frozen in time. Therefore, we compare it to the analytical prediction of \cref{eq:ps_th}, which is shown as a shaded region delimited by black dashed lines. This is because the analytical prediction is computed assuming a constant $\xi$. The two dashed black lines show the analytical estimates computed with the initial and final values of $\xi$, which are respectively $\xi\simeq 2.9$ and $\xi\simeq 3.05$. 
 
 We can see that the lattice simulation is able to reproduce with precision the analytical estimates. Indeed, the lattice result clearly interpolates between these two values. Note that the lattice result is less precise for low $k_{\rm eff}$ due to cosmic variance. In order to obtain a more precise result for these modes, we would have to obtain the power spectrum as an average over different realizations of the lattice simulation. Moreover, for large  $k_{\rm eff}$ there is a drop in the power spectrum from the lattice. This is unphysical, and it is caused by the fact that $\zeta$ at these small scales is sourced by gauge field modes beyond the lattice UV-cutoff.$  $
\subsubsection*{Bispectrum}
We now show results for the bispectrum. This is the first time a bispectrum is evaluated using a lattice code, and it is one of the original results of this thesis regarding the lattice methodology. As explained in \cref{sec:output}, we evaluate the bispectrum $B(\vec k_1,\vec k_2,\vec k_3)$ on equilateral configurations $k=|\vec k_1|=|\vec k_2|=|\vec k_3|$. Therefore, we can plot the bispectrum as a function of one momentum $\mathcal{B}_\zeta(k)$, defined in \cref{eq:equib}. In principle, we could also evaluate the bispectrum on different shapes, but we focus on the equilateral one in this work. The main reason for doing so is that the bispectrum is expected to peak on equilateral configurations. In the next section, we use the simulation to compute the real space three-point function $\langle \zeta^3(\vec{x}) \rangle$, which contains the sum of all bispectrum shapes. 

In \cref{fig:bis_weak} we show the bispectrum from the simulation computed at the final time\footnote{Note that, in order to compare the bispectrum from the code with analytical computations, one has to keep track of $(2\pi)^{3/2}$ factors in the definition of continuous Fourier transform, which is arbitrary. In our case, this required to check step by step the computation made in Ref. \cite{Barnaby_2011}, in order to understand which factors of $2\pi$ in the final result of \cref{eq:bis_th} are an artifact of the convention adopted for the Fourier transform.  Moreover, one has to keep track of factors of $(\Delta x)^3$ and $N^3$ appearing in the definition of the discrete Fourier transform in \cref{eq:DFT}. This is much more tricky than in the case of the power spectrum. In order to ensure that we correctly kept track of all these factors, we checked the results of the simulation for different values of $\alpha/f$.}. This time, we avoid showing the bispectrum at all times because it is computationally expensive, and because the bispectrum is negligible at the beginning of the simulation. Similarly to the power spectrum, we compare the bispectrum from the simulation with the analytical result of $\cref{eq:bis_th}$, which is shown as a shaded region delimited by black dashed lines representing the prediction for the initial and final values of $\xi$. Also in this case, the simulation is able to recover the analytical result with good precision. Note that cosmic variance effects are larger than in the case of the power spectrum, and they affect also intermediate scales of the simulation. This is expected, as three-point functions are typically more noisy than two-point functions.
 In order to obtain a more precise prediction for the bispectrum, we would have to compute it as an average over different bispectra coming from independent random realizations of the simulation. This would be relevant when comparing results from the simulation with real data, which is beyond the scope of the thesis.
\subsection{Higher order statistics and PDF of $\zeta$}
\label{sec:highA}
\begin{figure}
	\centering
	
	\begin{tikzpicture}
	\node (img) {\includegraphics[width=8.cm]{boxin.png}};

	\node [text width=0.01cm,align=center] at (2.2,0.){$\frac{\zeta}{\sigma}$};

	\node (img2) at (7.4,0) {\includegraphics[width=8.cm]{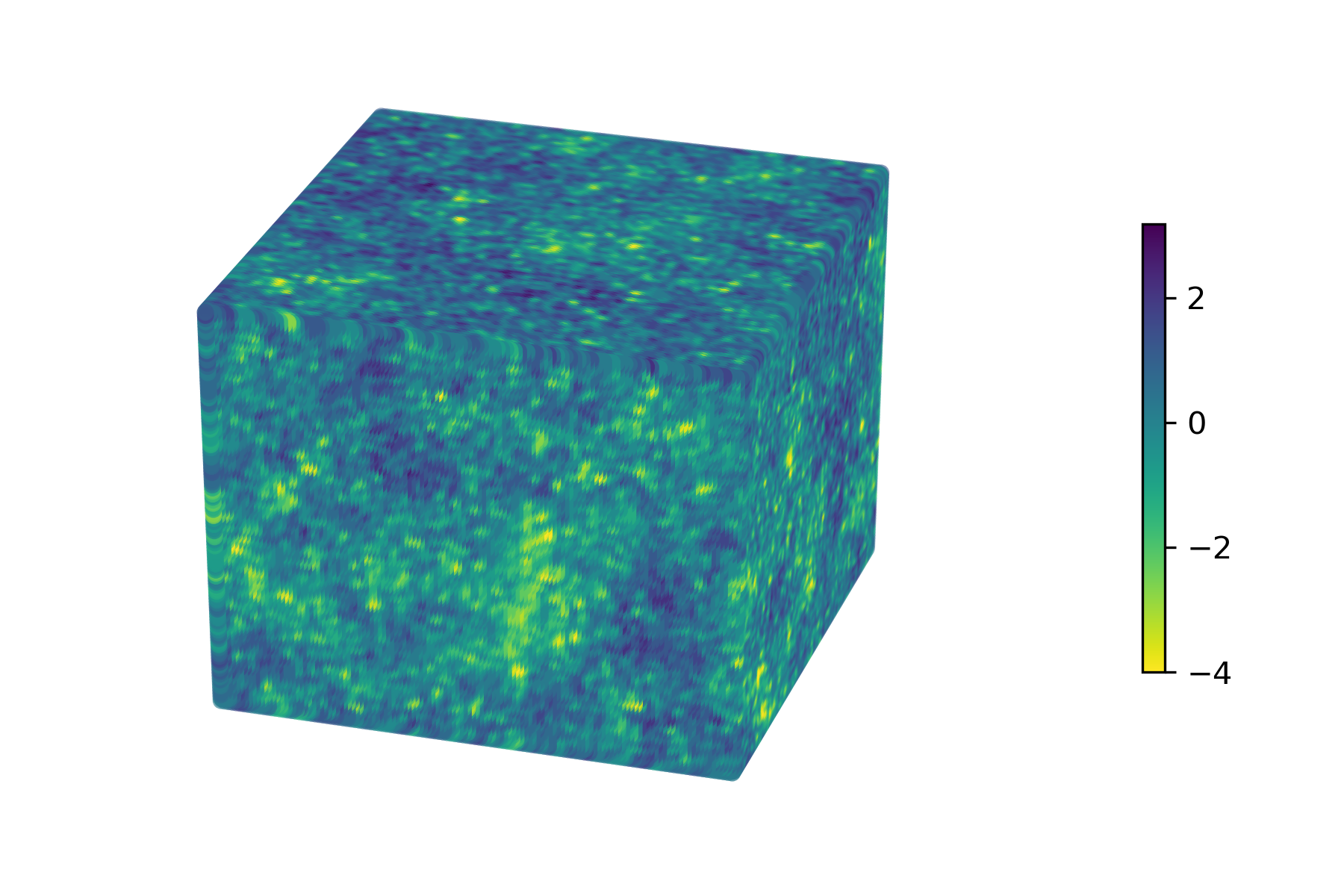}} ;
	
	\node [text width=0.01cm,align=center] at (0+7.4+2.2,0){$\frac{\zeta}{\sigma}$};

	\end{tikzpicture}

	\caption{Plot of the curvature perturbation $\zeta$ in real space, normalized by the standard deviation $\sigma=\sqrt{\langle \zeta^2 \rangle}$. The left panel shows the fluctuations at the initial time, and the right panel the ones at the final time. In the right panel, lattice points with values $\zeta<-4\sigma$ are saturated to better show the comparison between this case and the single-field case of \cref{fig:box}. Note that it is equivalent to show this plot for $\delta\phi$ and for $\zeta$, as we normalize by the standard deviation.}
	\label{fig:boxA}
\end{figure}

\begin{figure}
	\centering
	
	\begin{tikzpicture}
	\node (img) {\includegraphics[width=12cm]{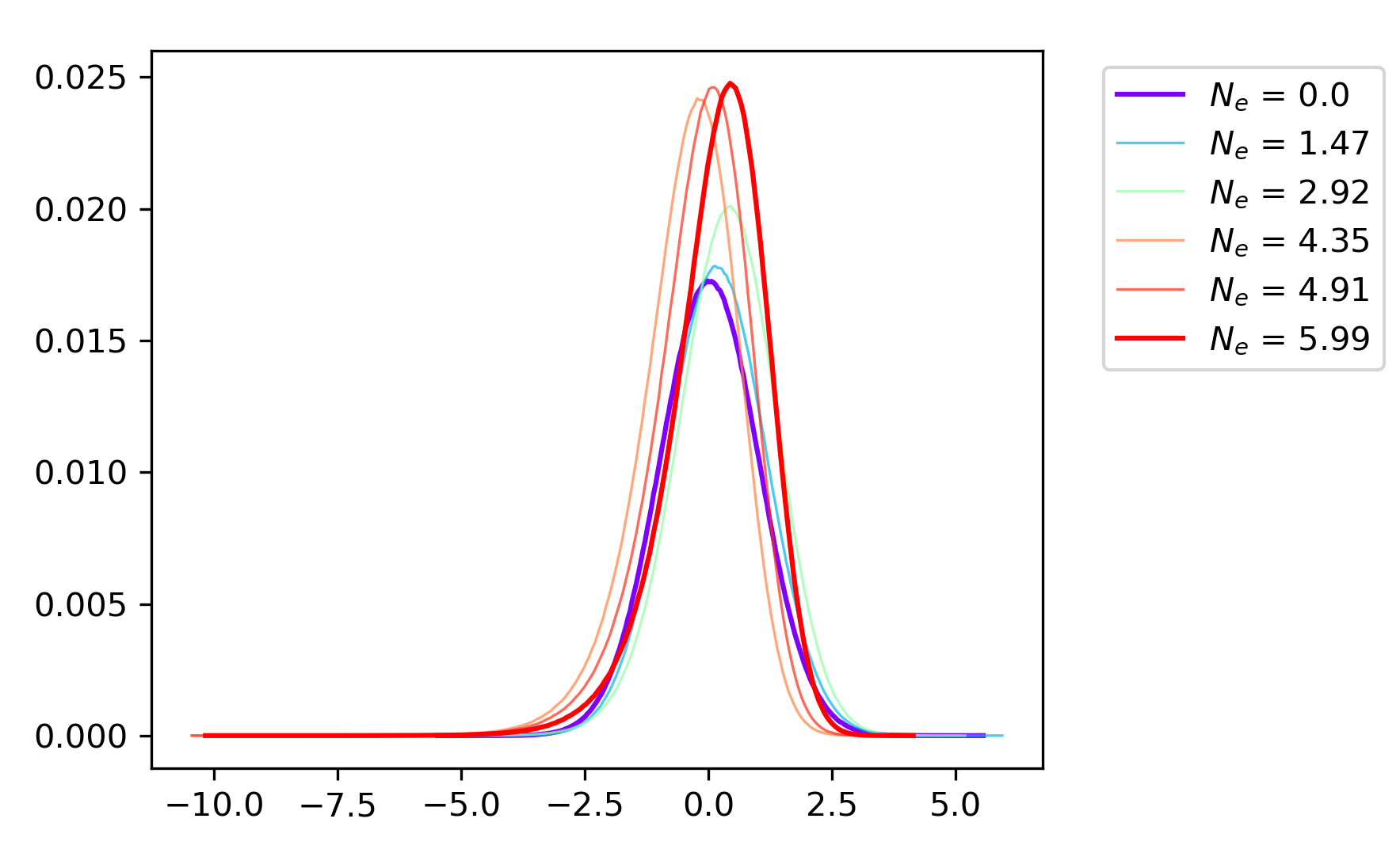}};
	
	\node [text width=0.01cm,align=center] at (-1,-4.3){ $\zeta/\sigma$};

	\end{tikzpicture}

	\caption{Normalized histograms of values $\zeta$ across the $N^3$ points of the lattice, in the case of negligible backreaction. The $x$-axis is normalized by the standard deviation of each histogram. Different colors represent different times, as shown in the legend. }
	\label{fig:hist_weak}
\end{figure}

\begin{figure}
	\centering
	
	\begin{tikzpicture}
	\node (img) {\includegraphics[width=12cm]{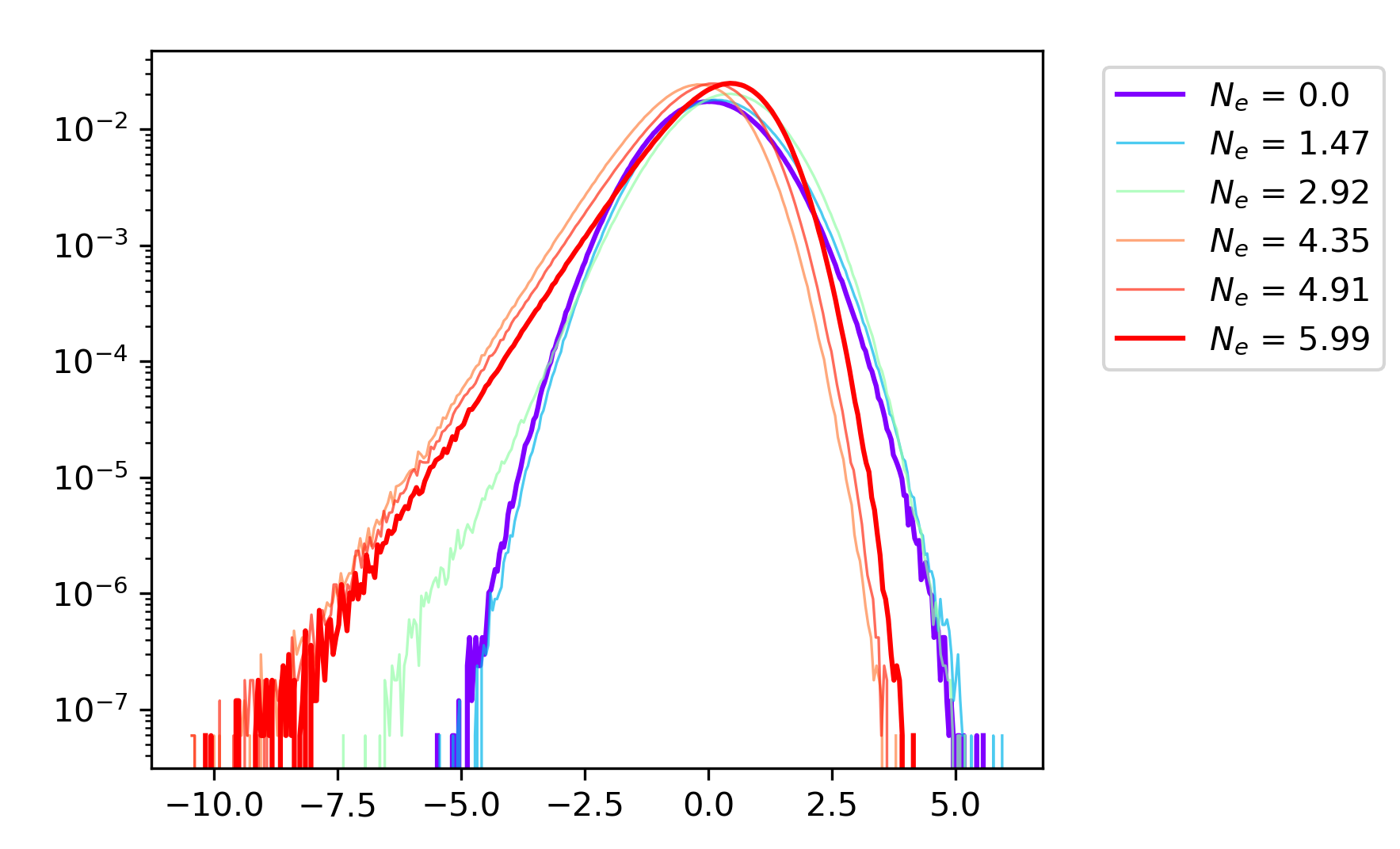}};
	
	\node [text width=0.01cm,align=center] at (-1,-4.3){ $\zeta/\sigma$};
	
	\end{tikzpicture}

	\caption{Same plot of \cref{fig:hist_weak} but with a logarithmic scale on the $y$-axis, in order to focus on the shape of the tail. }
	\label{fig:hist_weak_log}
\end{figure}

Thanks to the lattice approach, we do not only have access to the power spectrum and bispectrum of $\zeta$. The simulation allows to obtain a real space picture of the inflationary Universe, that we show in \cref{fig:boxA}. We can use this information to get the one-point probability density function (PDF) of $\zeta$ in real space. In \cref{fig:hist_weak} we show the PDF of $\zeta$ at different times, computed as the normalized histogram of the values of $\zeta$ across the $N^3$ points of the lattice. At the beginning, the PDF is well approximated by a Gaussian distribution. Non-Gaussianity is sourced by the gauge field, and it appears at late times as a pronounced tail in the distribution of inflaton values. The tail is on the left because we choose the inflaton to have a positive background velocity $\dot{\bar\phi}>0$. If $\dot{\bar\phi}<0$, we would get the same histogram with the $x$-axis inverted. In order to better show the shape of the tail, in \cref{fig:hist_weak_log} we show the same figure but with a logarithmic scale on the $y$-axis. The extreme part of the tail shows some noise, due to the few number of points per bin in the histogram.


 To quantify non-Gaussianity, we compute the cumulants of the PDF. These are defined as the connected part of the correlators of $\zeta$ in real space \cite{Bernardeau:2001qr}:
\begin{equation}
\label{eq:momenta}
\kappa_3 = \frac{\langle \zeta^3 \rangle}{\sigma^{3}},\quad \kappa_4=\frac{\langle \zeta^4 \rangle-3\sigma^4}{\sigma ^{4}},\quad \kappa_5=\frac{\langle \zeta^5 \rangle-10 \langle \zeta^3 \rangle\sigma^2}{\sigma ^{5}},
\end{equation}
which we normalized by powers of $\sigma^2=\langle \zeta^2 \rangle$ to make them dimensionless. In principle, we can compute higher order cumulants as well, but we just evaluate these three as a proof of concept.

\begin{figure}
	\centering
	
	\begin{tikzpicture}
	\node (img) {\includegraphics[width=9cm]{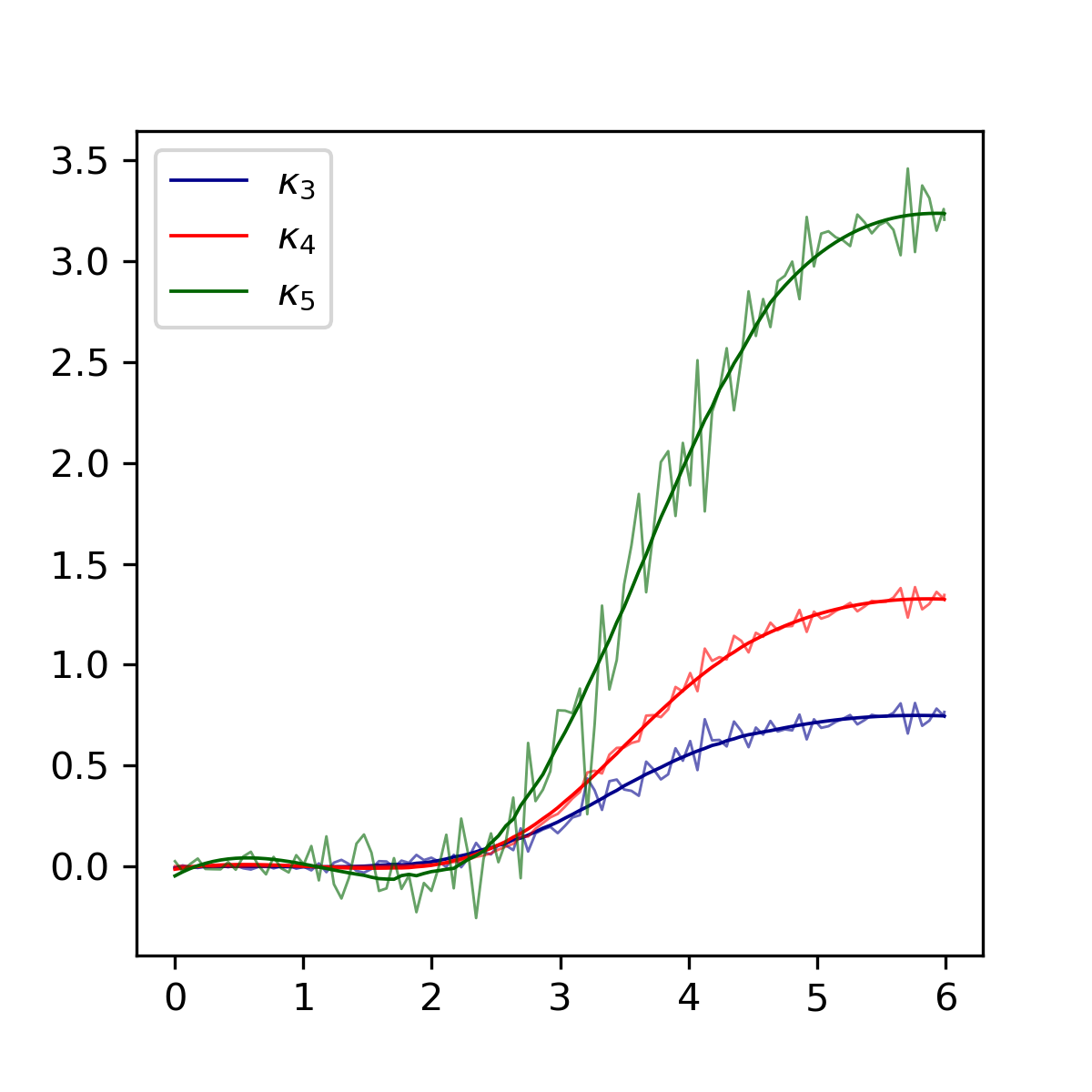}};
	
	\node [text width=0.01cm,align=center] at (0,-4.3){ $N_e$};
	
	\end{tikzpicture}

	\caption{Evolution of the cumulants $\kappa_i$ defined in \cref{eq:momenta} during the simulation with negligible backreaction. The noisy curves show the result from the simulation. Instead of averaging over different realizations of the simulation, we eliminate this noise by applying a smoothing kernel.}
	\label{fig:cum_weak}
\end{figure}

In \cref{fig:cum_weak} we show the evolution of the cumulants during the simulation. These cumulants show some noise, which is related to the statistical uncertainty and is larger for higher order correlators. Instead of reducing this noise taking an average over many realizations of the simulation, we apply a smoothing kernel to the curves, obtaining the noiseless curves in the plot. We can see that higher order correlators are very important in describing the PDF of $\zeta$. In fact, one has $$\kappa_5>\kappa_4>\kappa_3>1$$ at the final time. The fact that higher order cumulants get larger and larger is a symptom that the curvature perturbation sourced by the gauge field can not be seen as an expansion around a Gaussian distribution. We will discuss about the observational implications of this fact in \cref{sec:observations}. Note that the full characterization of the statistics of $\zeta$ provided in this section is beyond the reach of standard perturbative computations. This shows that lattice simulations can be a powerful tool to study the inflationary Universe, even when perturbation theory is still reliable.

\section{Results of the simulation: strong backreaction}
\label{sec:results_strong}
In this section, we focus on the large $\xi$ regime, where the production of the gauge field is expected to influence the background inflationary dynamics. We focus on a more realistic value of the gauge field coupling $\alpha/f=25$, 
so that the imprints of the Chern-Simons coupling on $\zeta$ are unobservable at CMB scales \cite{Anber_2010,Barnaby_2011_Large, Barnaby_2011, Anber_2012}. Due to the slow-roll motion, $\xi$ monotonically increases during inflation and at some point the Universe enters a nonlinear phase.

We start the simulation when $\phi=-5.5$. 
With this choice, the Universe is still in the weak backreaction phase at the beginning of the simulation. Then, after roughly 2 e-folds, the system enters a strong backreaction phase where the bound of \cref{eq:bounds} is violated and \cref{eq:ps_th} gives $\mathcal{P}_\zeta\sim 0.1$, which indicates a breakdown of perturbativity. 

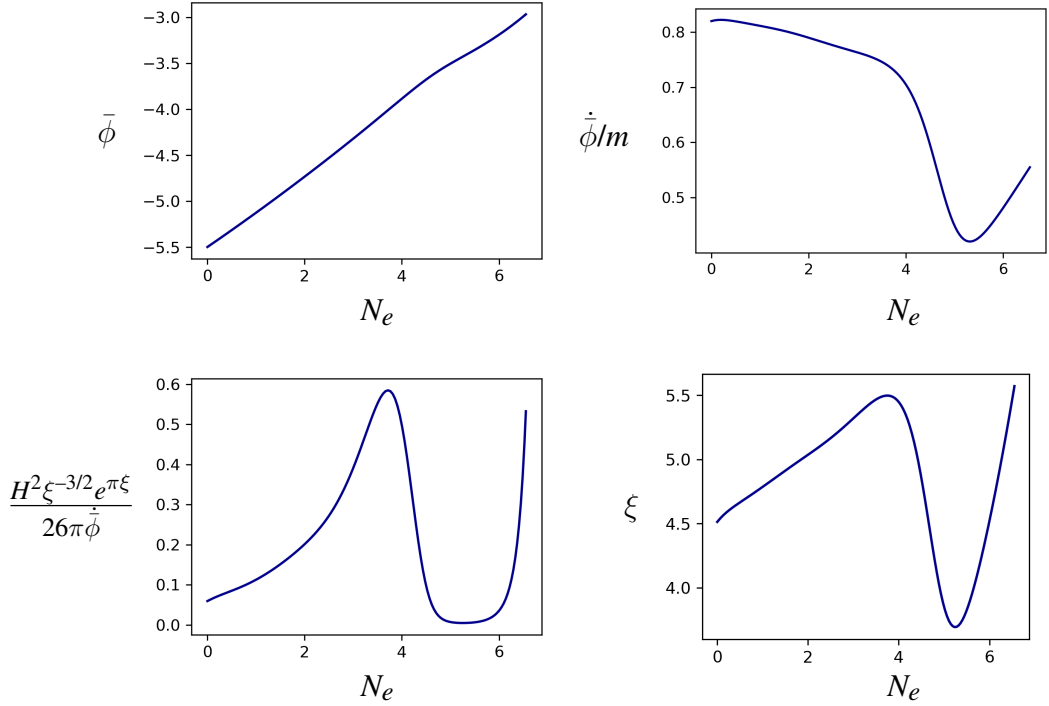
\begin{figure}
	\centering

	\begin{tikzpicture}
	\node (img) {\includegraphics[width=6cm]{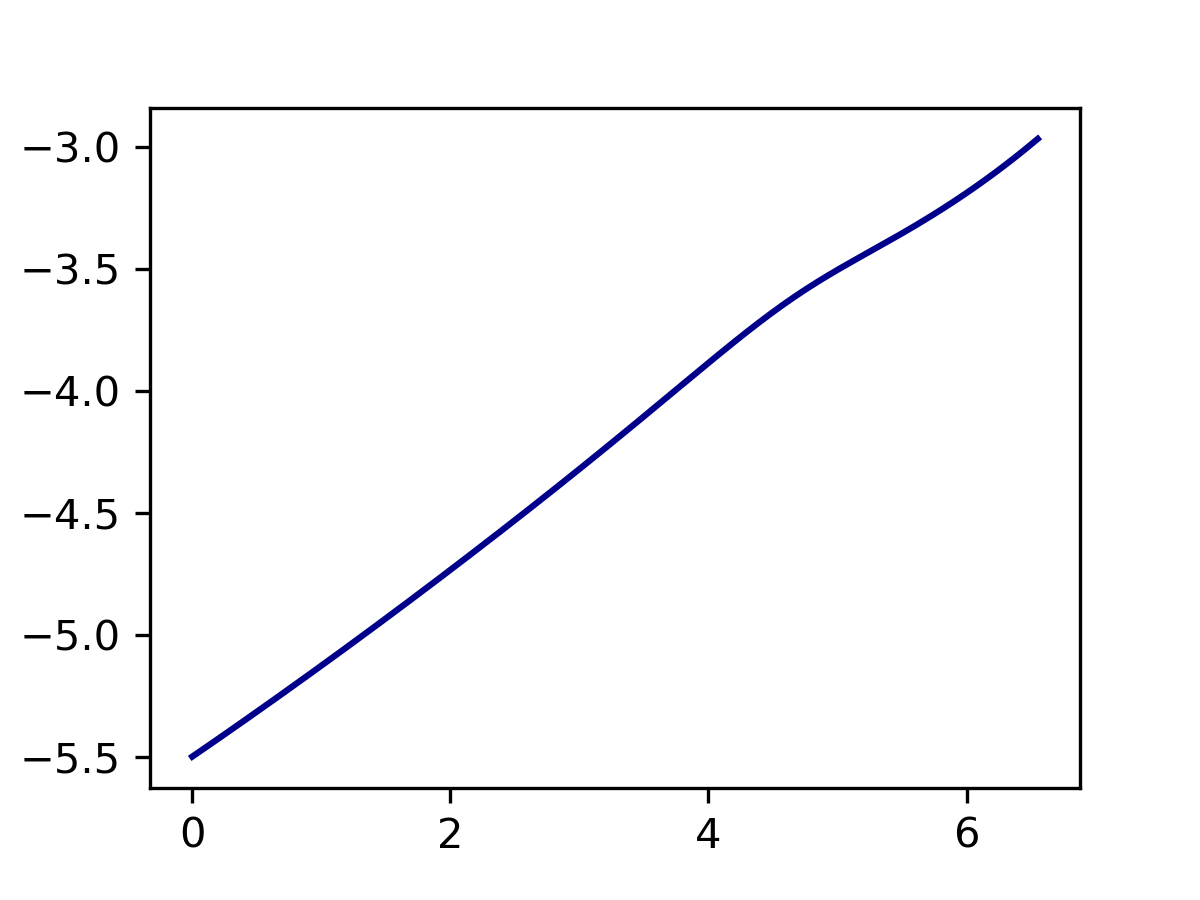}};
	
	\node [rotate=0,text width=0.01cm,align=center] at (-3.5,0){ $\bar{\phi}$};
	\node [text width=0.01cm,align=center] at (0,-2.4){$N_e$};

	\node (img2) at (6.7,0) {\includegraphics[width=6.cm]{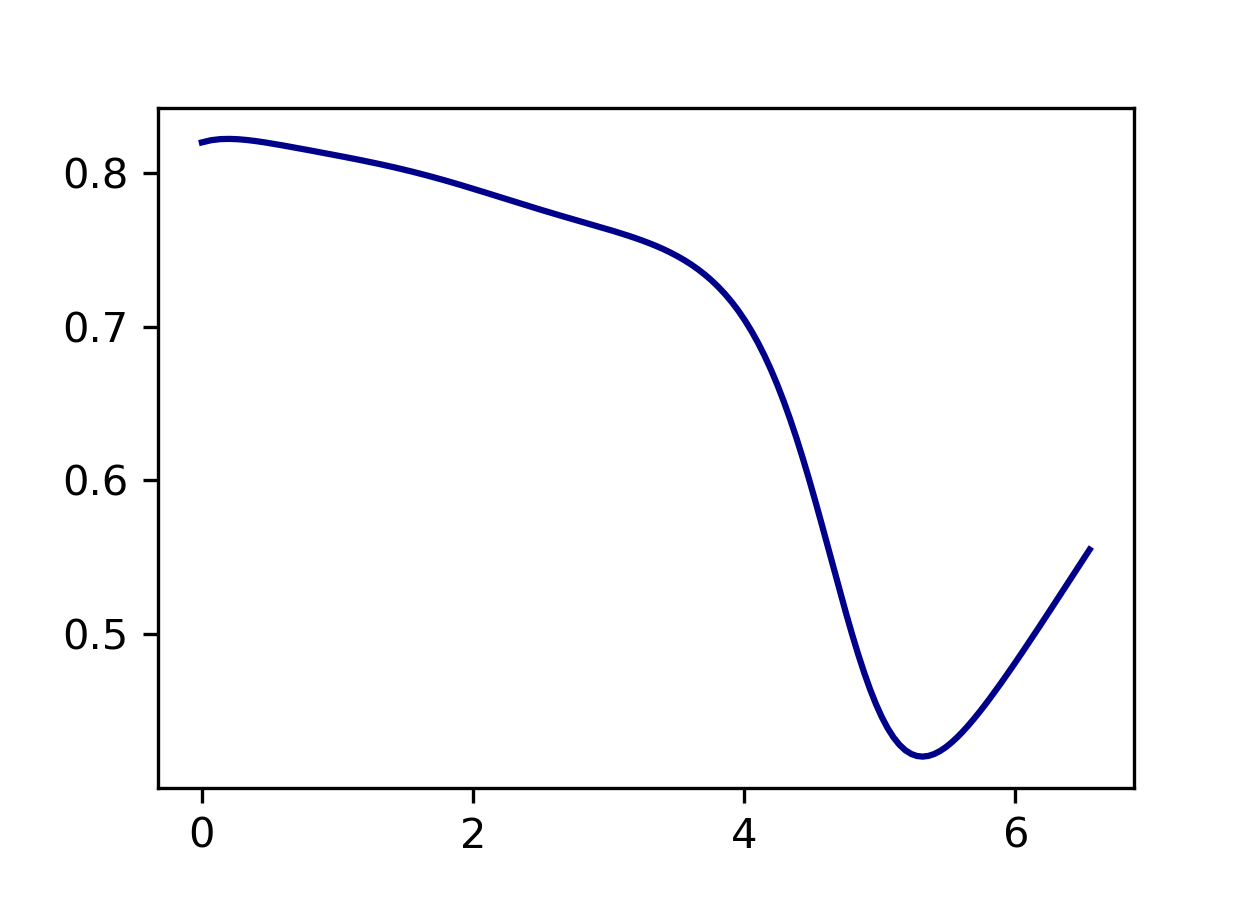}} ;
	
	\node [rotate=0,text width=0.01cm,align=center] at (-3.8+6.7,0){ ${\dot{\bar{\phi}}}/{m}$};
	\node [text width=0.01cm,align=center] at (0+7,-2.4){$N_e$};

	\node (img3) at (0,-5) {\includegraphics[width=6cm]{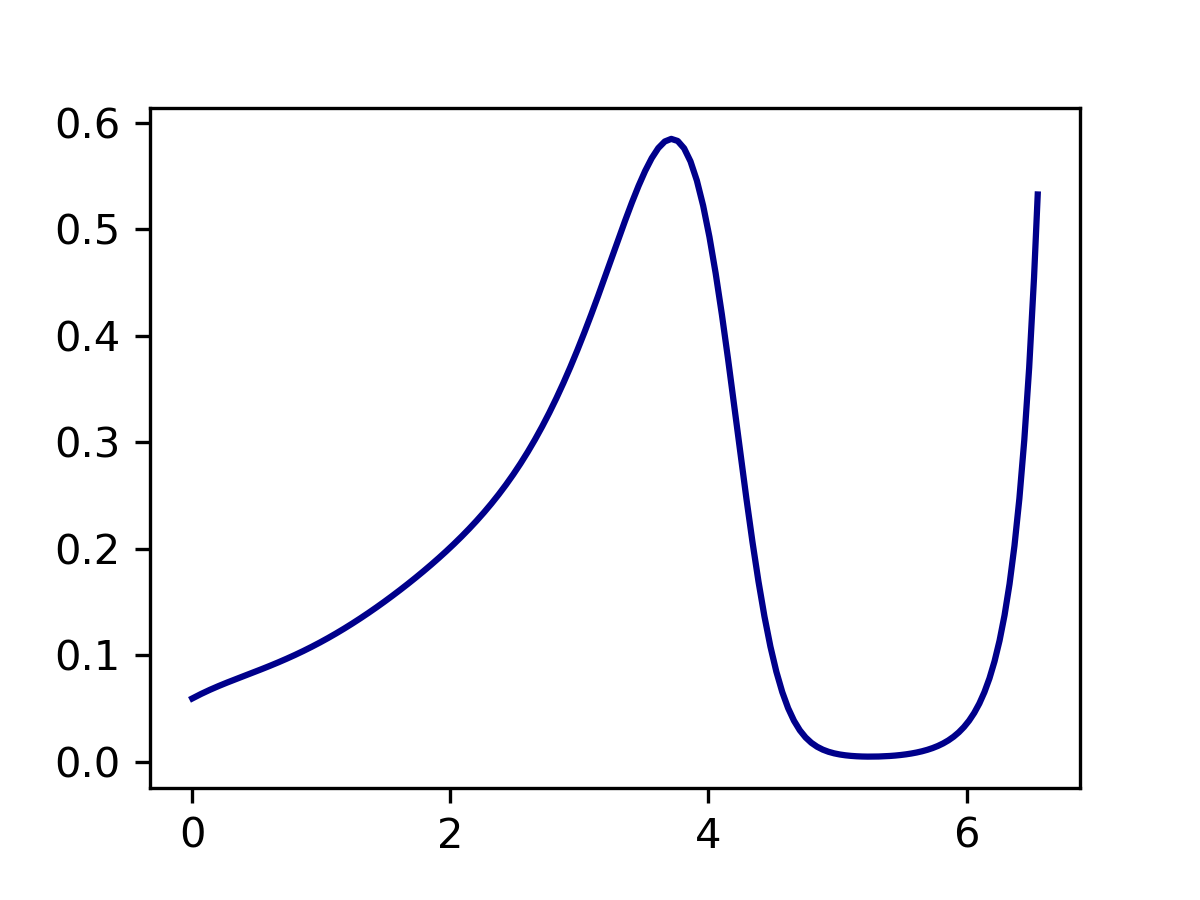}};
	
	\node [rotate=0,text width=0.01cm,align=center] at (-4.7,0-5){ $\frac{H^2\xi^{-3/2}e^{\pi\xi}}{26\pi \dot{{\bar\phi}}}$};
	\node [text width=0.01cm,align=center] at (0,-2.4-5){$N_e$};

	\node (img4) at (6.4,-5) {\includegraphics[width=6.2cm]{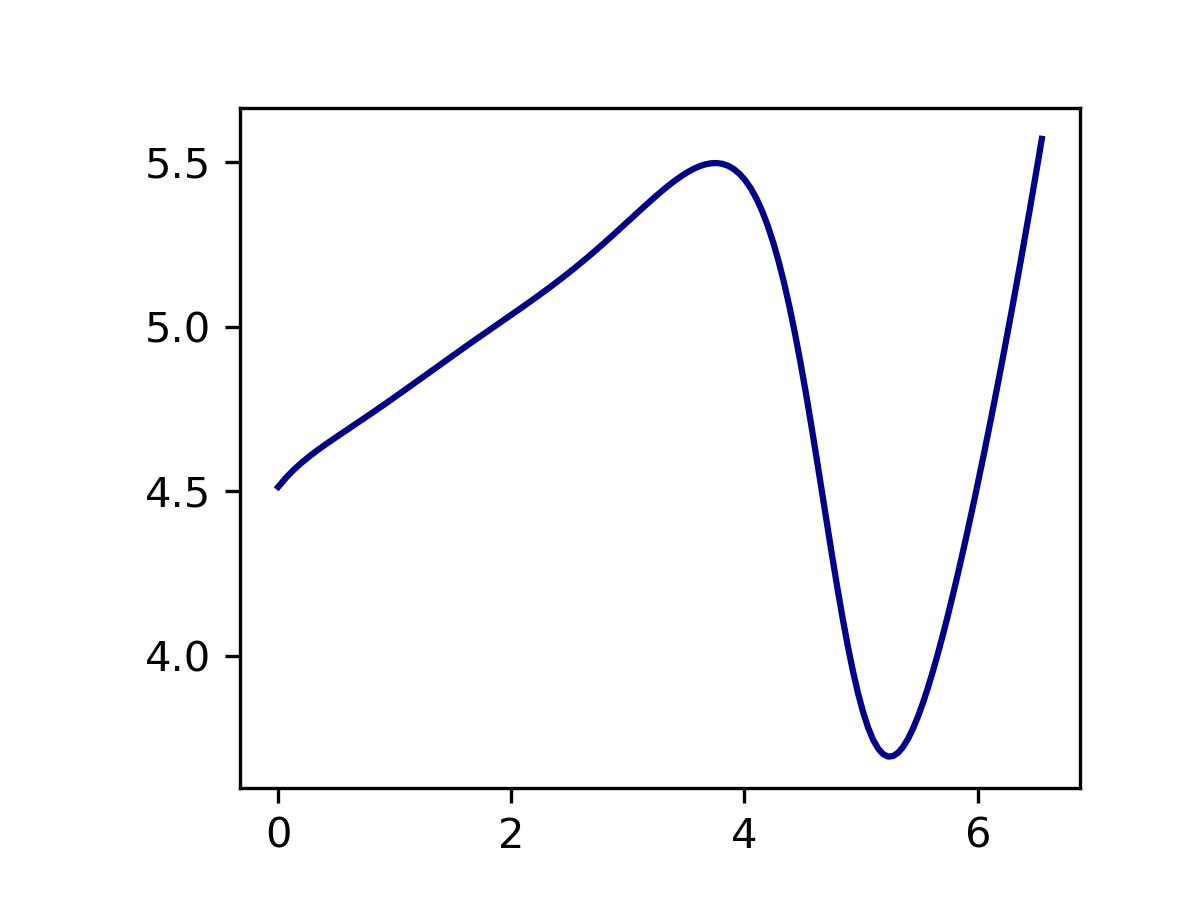}} ;
	
	\node [rotate=0,text width=0.01cm,align=center] at (-3.5+6.7+0.3,0-5){  $\xi$};
	\node [text width=0.01cm,align=center] at (0+7,-2.4-5){$N_e$};

	\end{tikzpicture}

	\caption{Plot of background value of the inflaton (top left), its velocity (top right), the bound of \cref{eq:bounds} (bottom left) and $\xi$ (bottom right) during the simulation in the case of strong backreaction.}
	\label{fig:backgroundvaluesAstrong}
\end{figure}

We mainly focus on results from a simulation run with $(N,L)=(256,1.5/m)$, but we tested our simulation also with other values of $(N,L)$ to ensure that our results (like the background trajectories of \cref{fig:backgroundvaluesAstrong}) are physical and do not depend on the IR and UV cutoff of the simulation. Moreover, we check the accuracy of the time integrator by studying energy conservation and time-step convergence. This is done in \cref{sec:energy+gauge}, where we also check that the gauge condition $\partial^\mu A_\mu=0$ is approximately satisfied during the evolution.

\subsection{Background quantities}
In \cref{fig:backgroundvaluesA} we can clearly see that the system shows a departure from the slow-roll trajectory. In the bottom left plot of \cref{fig:backgroundvaluesA}, we show the value of the bound of \cref{eq:bounds}. We can see that at the beginning of the simulation the bound is still satisfied, but later during the simulation it becomes of order one. This causes a departure from the slow-roll trajectory of all other quantities.
In particular, we find the departure as an oscillatory behavior in the parameter $\xi$. This behavior is intuitive. In order to see why, we can take \cref{eq:eom1} and neglect the gradient term and ${\phi}^{\prime\prime}$.
\begin{equation}
\phi^\prime\simeq -\frac{a^2}{2H}\frac{\partial V}{\partial \phi} - \frac{a^2}{2H}\frac{\alpha}{4f}F_{\mu\nu}\tilde F^{\mu\nu}.
\end{equation}
When $\xi$ becomes large enough (i.e. when the bound of \cref{eq:bounds} is broken), the $F\tilde F$ term gets comparable to the potential term. From this relation, we can see that this has the effect of decreasing the value $\phi^{\prime}$. This lowers the value of $\xi$, making $F\tilde F$ small again and brings the system momentarily closer to the slow roll trajectory, where $F\tilde F$ grows again. Oscillations of similar period and size were already predicted by previous studies \cite{Cheng:2015oqa,Notari:2016npn,DallAgata:2019yrr,domcke2020resonant}, which explored backreaction effects using semi-analytical tools.
 In particular, the oscillation we observe has a half-period of $\Delta N_e/2 \sim 1.6\,$e-folds, which is compatible with the one predicted in Ref. \cite{domcke2020resonant}. This behavior has been recently confirmed by an analytical computation in Ref. \cite{Peloso:2022ovc}.
  
Another consequence of the backreaction is that, after $6.5$ e-folds of evolution, the background inflaton value is $\phi=-3.02$. This value would be reached after $5.4$ e-folds of evolution if the backreaction were negligible, which means that the backreaction significantly delays the background dynamics by roughly $1.1$ e-folds during the simulation time.
\subsection{Statistics of $\zeta$}
\begin{figure}
	\centering
	
	\begin{tikzpicture}
	\node (img) {\includegraphics[width=9cm]{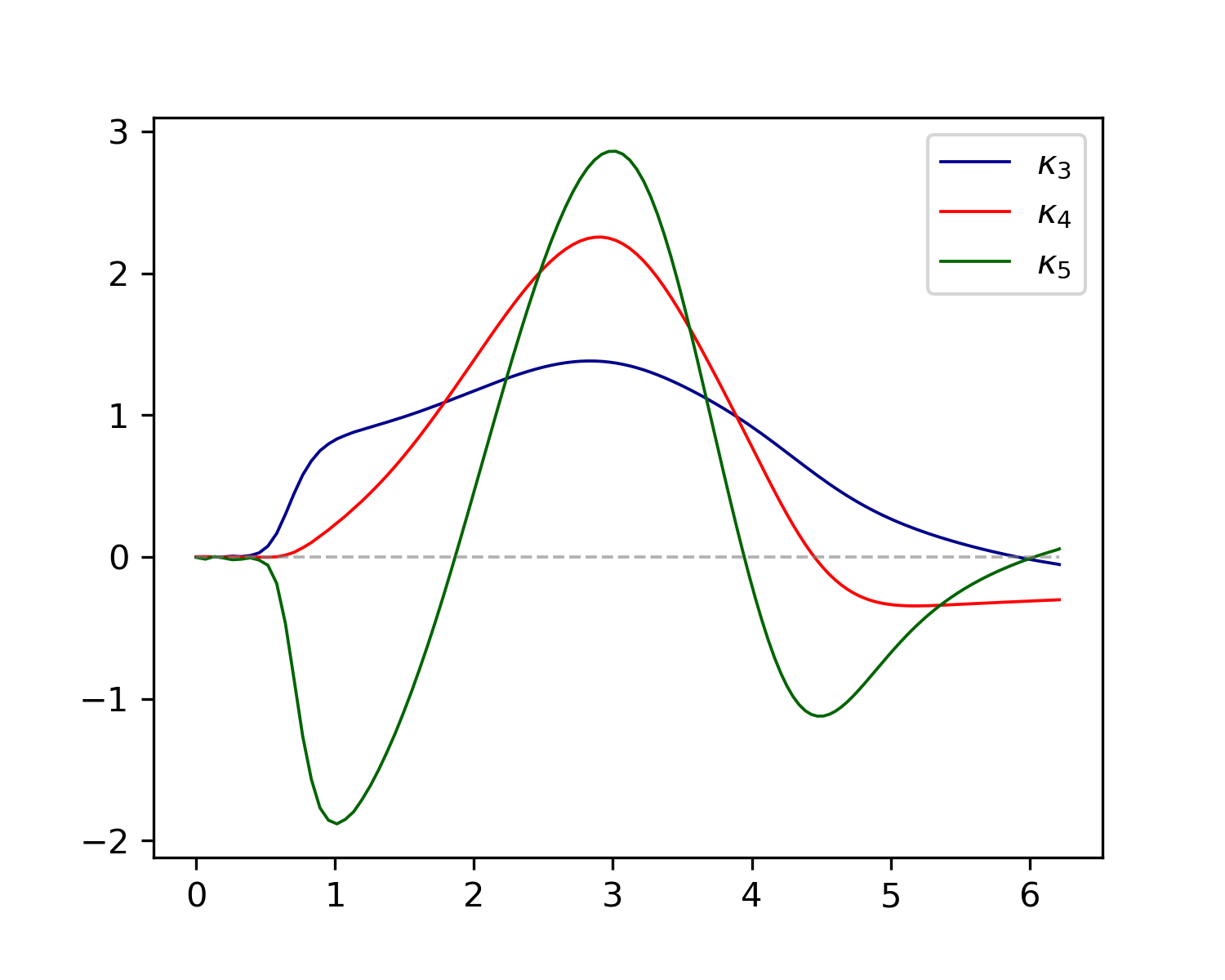}};
	
	\node [text width=0.01cm,align=center] at (0,-4.){ $N_e$};
	
	\end{tikzpicture}

	\caption{Evolution of the cumulants $\kappa_i$ defined in \cref{eq:momenta} during the simulation with strong backreaction. }
	\label{fig:cum_strong}
\end{figure}

We now focus on the statistics of the comoving curvature perturbation in this large backreaction regime.
  \begin{figure}
	\centering
	
	\begin{tikzpicture}
	\node (img) {\includegraphics[width=12cm]{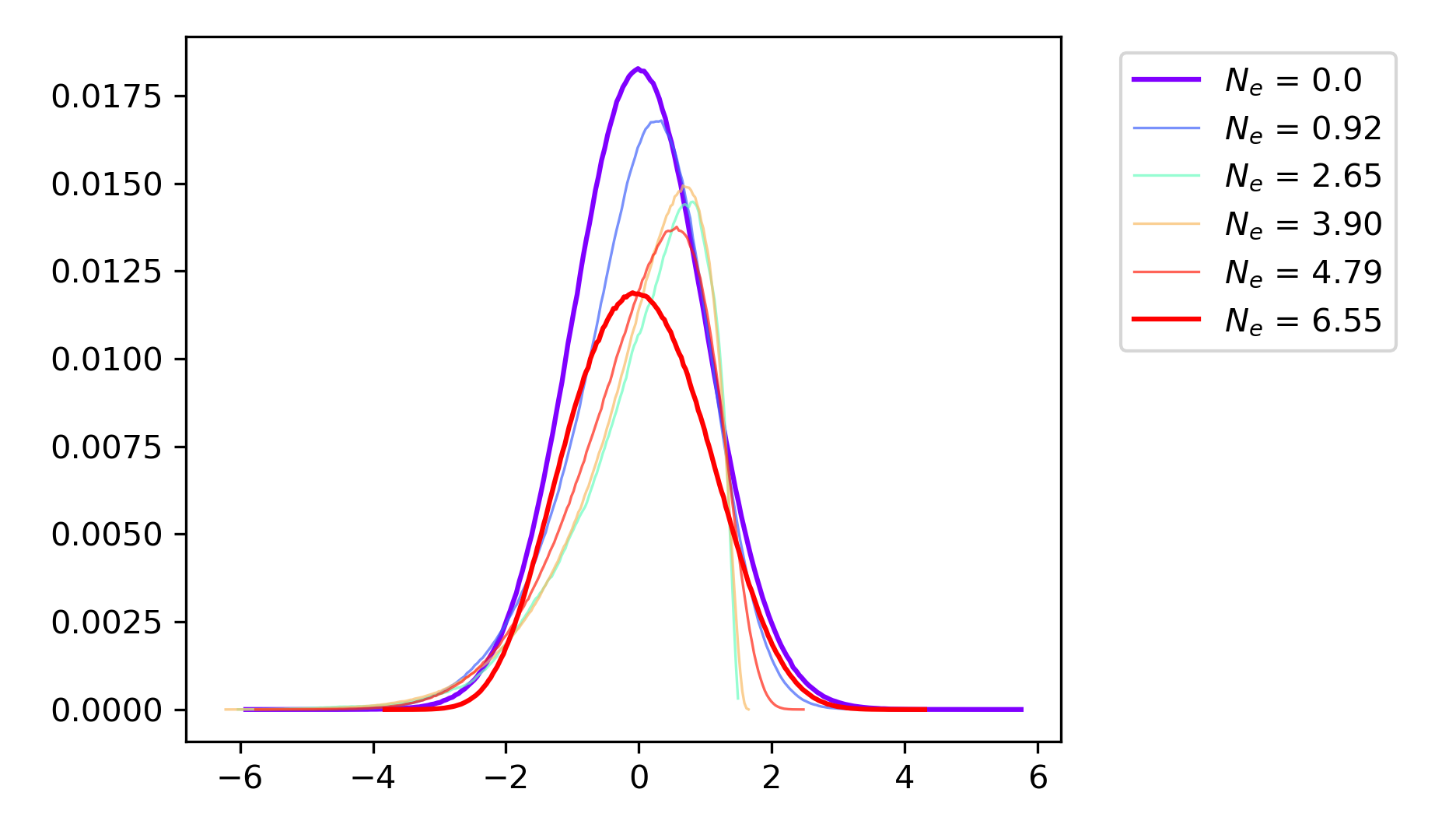}};
	
	\node [text width=0.01cm,align=center] at (-1,-4.3){ $\zeta/\sigma$};
	
	\end{tikzpicture}

	\caption{Normalized histograms of values $\zeta$ across the $N^3$ points of the lattice, in the case of strong backreaction. The $x$-axis is normalized by the standard deviation of each histogram. Different colors represent different times. }
	\label{fig:hist_strong}
\end{figure}
\begin{figure}
	\centering
	
	\begin{tikzpicture}
	\node (img) {\includegraphics[width=12cm]{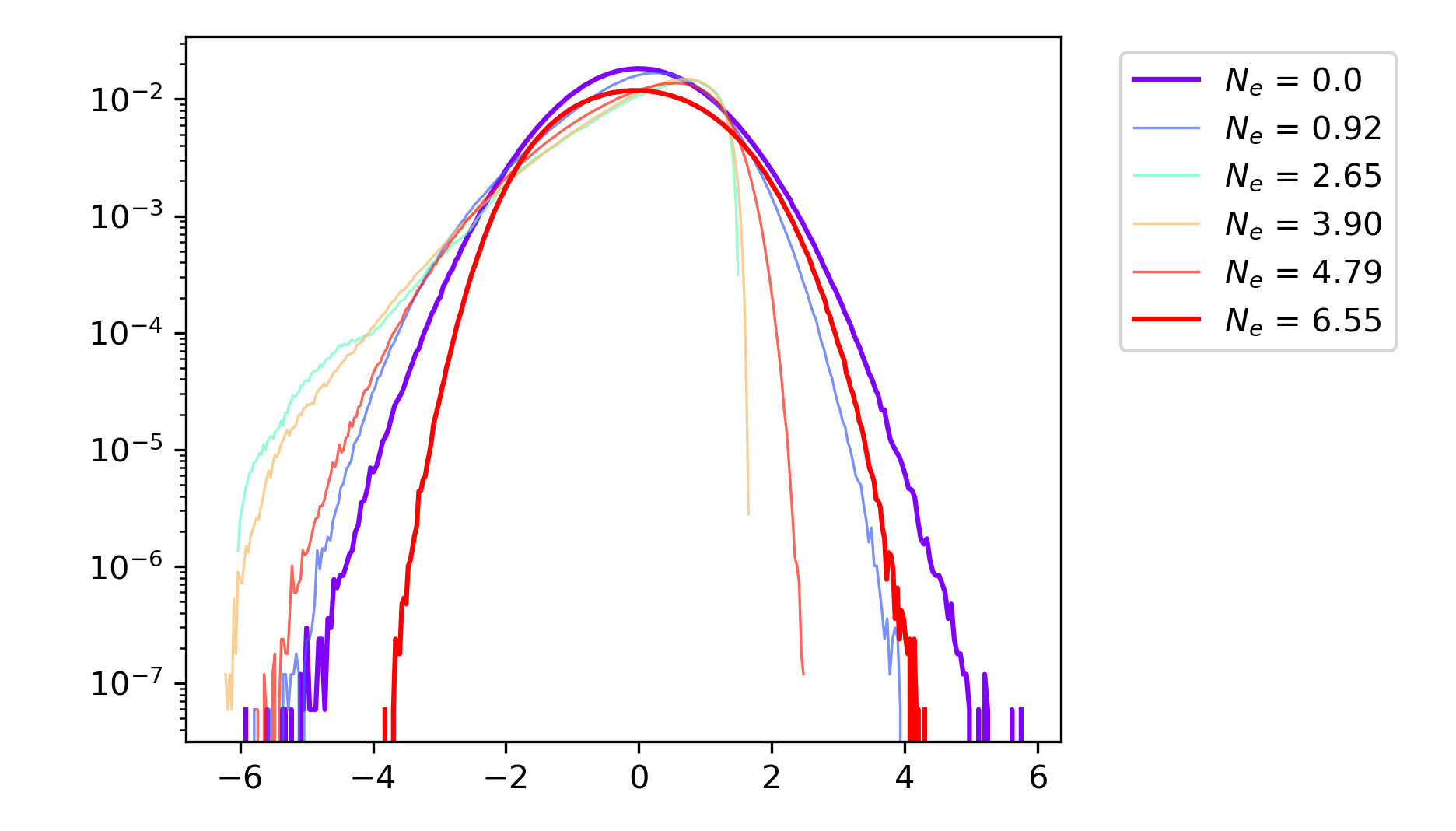}};
	
	\node [text width=0.01cm,align=center] at (-1,-4.3){ $\zeta/\sigma$};
	
	\end{tikzpicture}

	\caption{Same plot of \cref{fig:hist_strong} but with a logarithmic scale on the $y$-axis, in order to focus on the shape of the tail characterizing the distribution at intermediate times. }
	\label{fig:hist_strong_log}
\end{figure}
In \cref{fig:hist_strong} we plot the one-point probability distribution function (PDF) of $\zeta$ at different times during the simulation, computed as the normalized histogram of the values of $\zeta$ across the $N^3$ points of the cubic lattice. In \cref{fig:hist_strong_log}, we show the same plot but with a logarithmic scale on the $y$-axis. To quantify non-Gaussianity, in \cref{fig:cum_strong} we plot the cumulants defined in \cref{eq:momenta} as a function of time during the simulation\footnote{This time, there is no need to apply a smoothing kernel (like we did in \cref{fig:cum_weak}) to eliminate the noise, which is much smaller due to the fact that perturbations are much larger in this case.}. 

From these plots we can see that, although non-Gaussianity grows in the first e-folds of the simulation, it is strongly suppressed when the system enters the strong backreaction phase. The cumulants $\kappa_i$ grow until $N_e\simeq 3$, together with the departure of the PDF from a Gaussian shape. After this moment, the cumulants are strongly suppressed, and the distribution converges again into a Gaussian. From \cref{fig:cum_strong}, we can see that $\kappa_4$ reaches a small non-zero value at the end of the simulation, instead of dropping to (approximately) zero like the other cumulants. The fact that higher order cumulants have non-vanishing (small) values at the end of the simulation is expected, and it is a consequence of the fact that we are simulating the transition between linear and nonlinear dynamics. Indeed, non-Gaussianity is still very large during most of the simulation, and all the modes who exit the horizon during (roughly) the first half of the simulation are frozen in time and remain correlated. Therefore, the simulation box is expected to be non-Gaussian on the large scales of the lattice. The fact that $\kappa_4$ has a larger final values with respect to other cumulants is nontrivial, and we interpret it as a consequence of the complicated background dynamics. Moreover, we can see that $\kappa_5$ shows a nontrivial oscillatory behavior during the simulation. 

The suppression of non-Gaussianity in the strong backreaction regime is unexpected, and has remarkable observational consequences. Indeed, it is commonly believed in the literature that a large value of $\xi$, characterizing the nonlinear regime, would naturally lead to large non-Gaussianity. This is discussed in \cref{sec:observations}, where we also interpret the suppression of non-Gaussianity based on analytical arguments. 

\begin{figure}
	\centering
	
	\begin{tikzpicture}
	\node (img) {\includegraphics[width=14cm]{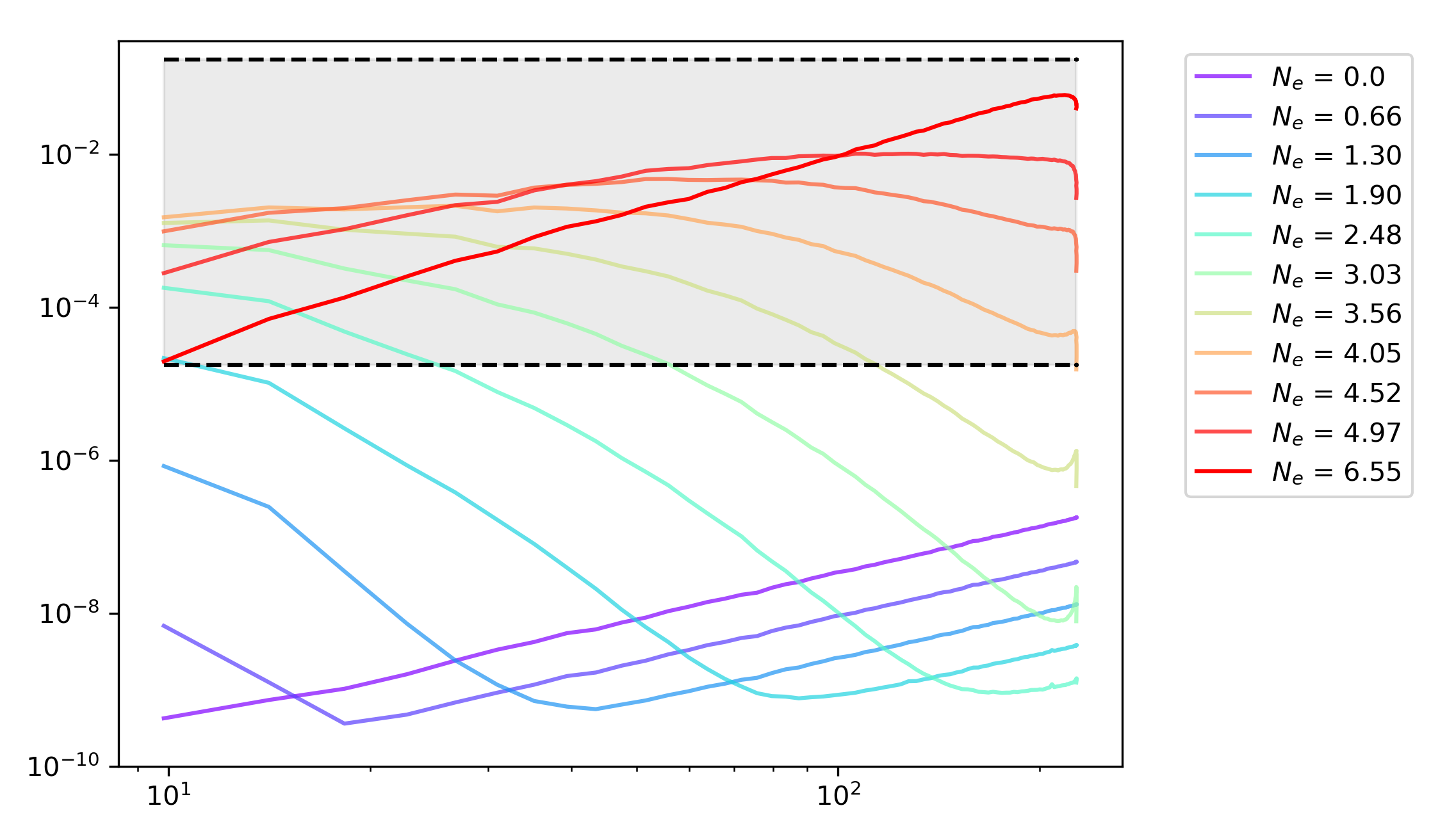}};
	
	\node [text width=0.01cm,align=center] at (-1,-4.4){ $k_{\rm eff}/m$};
	\node [text width=0.01cm,align=center] at (-7.5,0){ $P_{\zeta}$};

	\end{tikzpicture}

	\caption{Power spectrum of $\zeta$ in the case of strong backreaction. The shaded region, delimited by black dashed lines, shows the analytical estimate of \cref{eq:ps_th} computed using the initial and final values of $\xi$ from the lattice simulation.}
	\label{fig:ps_strong}
\end{figure}

The nearly Gaussian nature of the scalar perturbations in this regime means that the statistics is mainly described by the power spectrum, which we show in \cref{fig:ps_strong} at different times during the simulation. Although the analytical estimates are not reliable in this regime, we still compare the power spectrum with \cref{eq:ps_th} using the initial and final values of $\xi$ from the simulation. From this figure, we can see that the analytical result of \cref{eq:ps_th} still offers a reliable estimate of the power spectrum. Of course, one would not be able to compute this power spectrum just by using perturbation theory, as the final value of $\xi$ is the result of the complicated background dynamics captured by the lattice simulation.

 From \cref{fig:ps_strong} we see that the oscillatory behavior of $\xi$ does not leave any trace on the power spectrum at the final time.
This is caused by the fact that the period of oscillation of $\xi$ is smaller than the times it takes for the relevant modes to leave the horizon. Indeed, although the power spectrum develops a peak around $N_e\simeq 5$, this feature is quickly washed away by the fact that $\xi$ grows again at the end of the simulation. Note that the behavior of $\xi$ during the final $\sim 1$ e-fold of evolution might be influenced by the finite UV resolution of the lattice. The backreaction from UV modes beyond the lattice cutoff could change the final value of $\xi$, affecting the value of the power spectrum for the most UV modes of the lattice. Therefore, this behavior has to be confirmed by higher-resolution simulations, that are able of capturing more than 1 oscillation in $\xi$. This is possible but requires a more efficient parallelization of the lattice computation, which is beyond the scope of this thesis and will be subject of future work.




\subsection{Gauge constraint and energy conservation}
\label{sec:energy+gauge}
In this section we discus energy conservation and the gauge constraint for all the simulations presented in this chapter. 
\subsubsection*{Gauge constraint}

\begin{figure}
	\centering
	
	\begin{tikzpicture}
	\node (img) {\includegraphics[width=9cm]{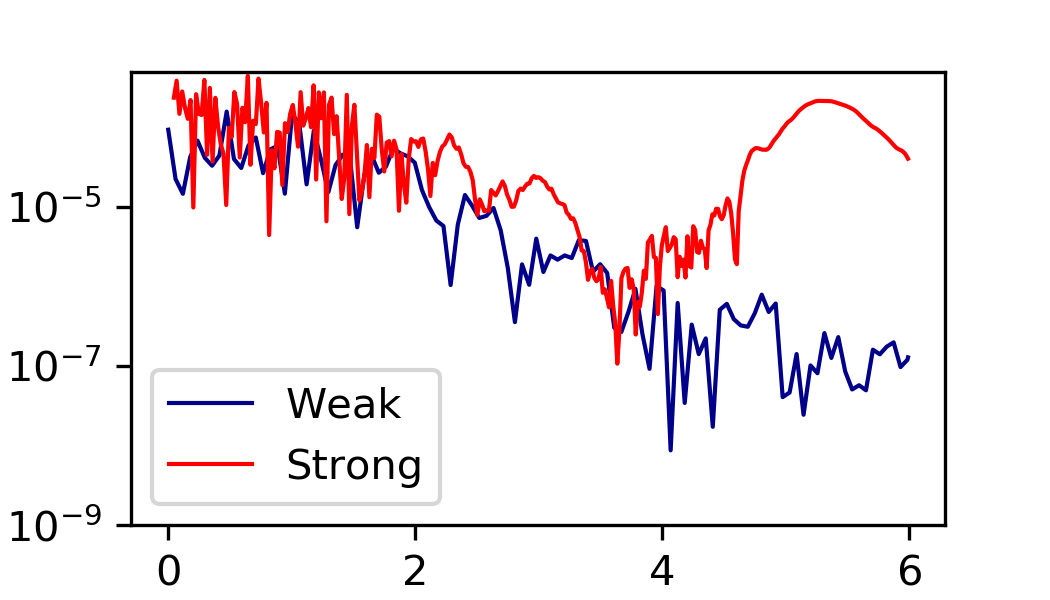}};
	
	\node [rotate=0,text width=0.01cm,align=center] at (-5.4,0){ $G$};
	\node [text width=0.01cm,align=center] at (0,-3.){$N_e$};

	\end{tikzpicture}

	\caption{Plot of the quantity $G(\tau)$ defined in \cref{eq:gauge_check}, quantifying the departure from the Lorenz gauge $\partial^\mu A_\mu=0.$. The blue curve shows the evution of this quantity in the case of negligible backreaction studied in \cref{sec:results_weak}, while the red one shows the result in the case of strong backreaction of \cref{sec:results_strong}. }
	\label{fig:gauge}
\end{figure}

As discussed in \cref{sec:gauge}, in our simulation we do not require the Lorenz gauge 	$\partial^\mu A_\mu=0$ to be exactly preserved on the lattice. Indeed, we evolve the four components of $A_\mu$ as independent degrees of freedom. In order to ensure that there are no unphysical degrees of freedom propagating on the lattice, we need to check by hand that the gauge condition is approximately preserved during the numerical integration. In order to do so, we define the following dimensionless quantity:
\begin{equation}
\label{eq:gauge_check}
G(\tau)=\frac{\partial^\mu A_\mu}{\sqrt{\sum_\rho |\partial^\rho A_\rho|^2}}.
\end{equation}
This is analogous to what is done in lattice simulations in the context of preheating \cite{Deskins_2013,Adshead_2015}.
This quantity has to remain small at all times in order to ensure the conservation of the gauge constraint. In \cref{fig:gauge} we plot the evolution of this quantity in the case of negligible backreaction of \cref{sec:results_weak} (blue line) and in the case of strong backreaction considered in this section (red line).

 In both cases, we can see that the gauge constraint is well preserved throughout the evolution. In the case of strong backreaction, the violation of the gauge constraint $G(\tau)$ slightly grows during the nonlinear phase. Nevertheless, the value of $G(\tau)$ is always small compared to the beginning of the simulation. The fact that the gauge condition is weakly violated at the initial time is a consequence of the fact that we set $A_0=0$ as initial condition.

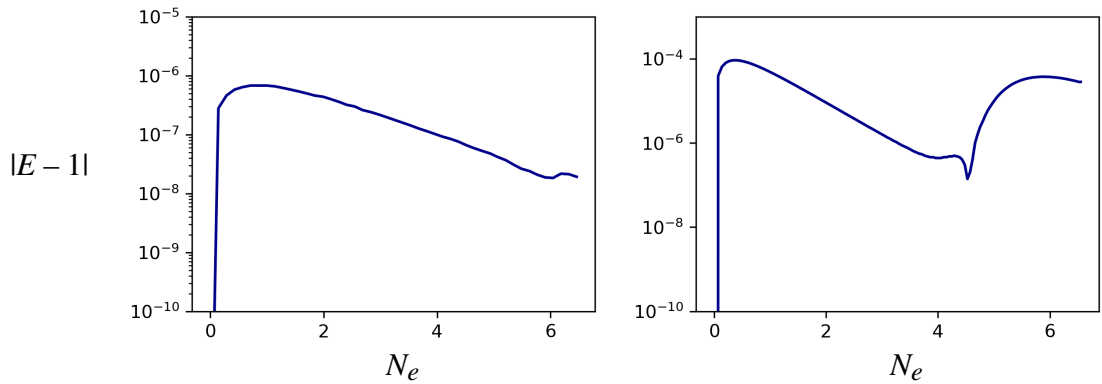
\begin{figure}
	\centering
	
	\begin{tikzpicture}
	\node (img) {\includegraphics[width=6.9cm]{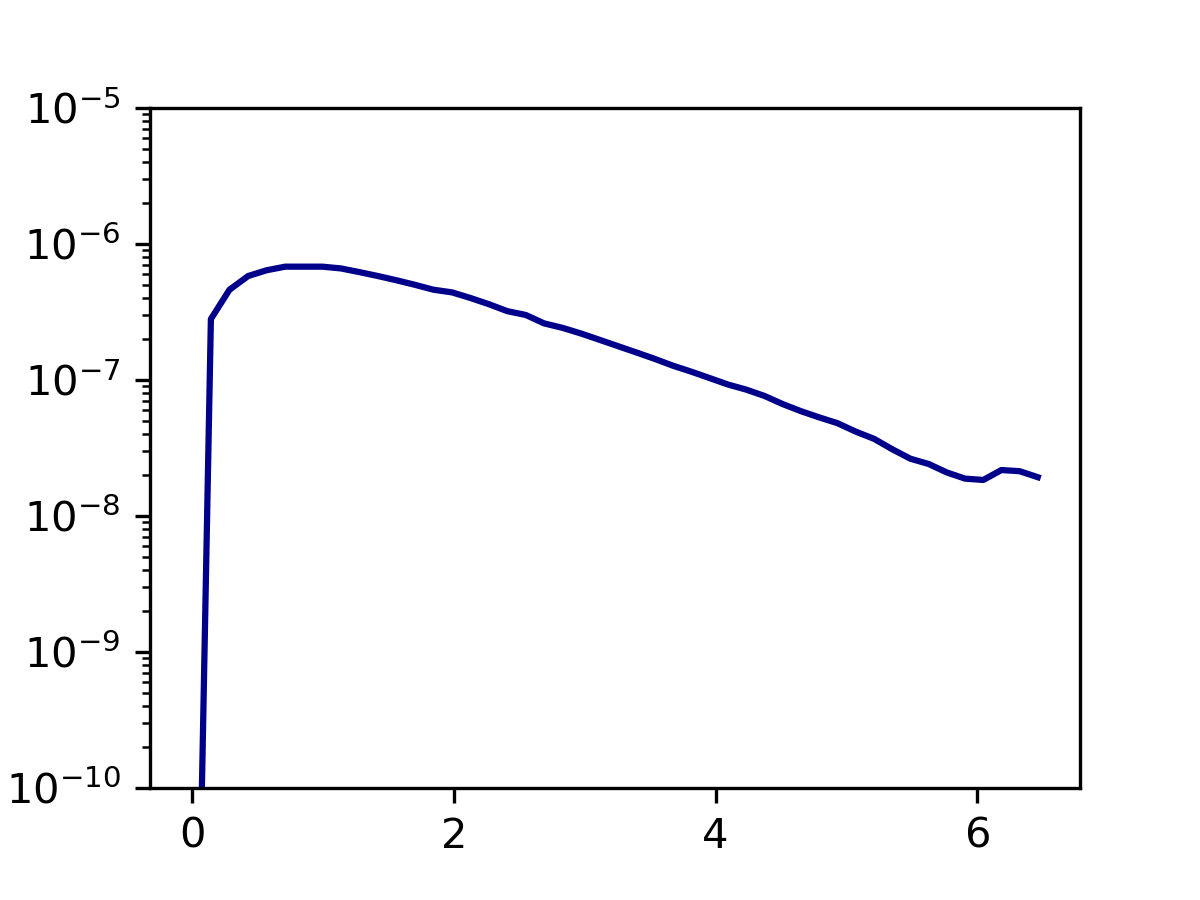}};
	
	\node [rotate=0,text width=0.01cm,align=center] at (-5.,0){ ${|E-1|}$};
	\node [text width=0.01cm,align=center] at (0,-2.7){$N_e$};

	\node (img2) at (6.7,0) {\includegraphics[width=6.9cm]{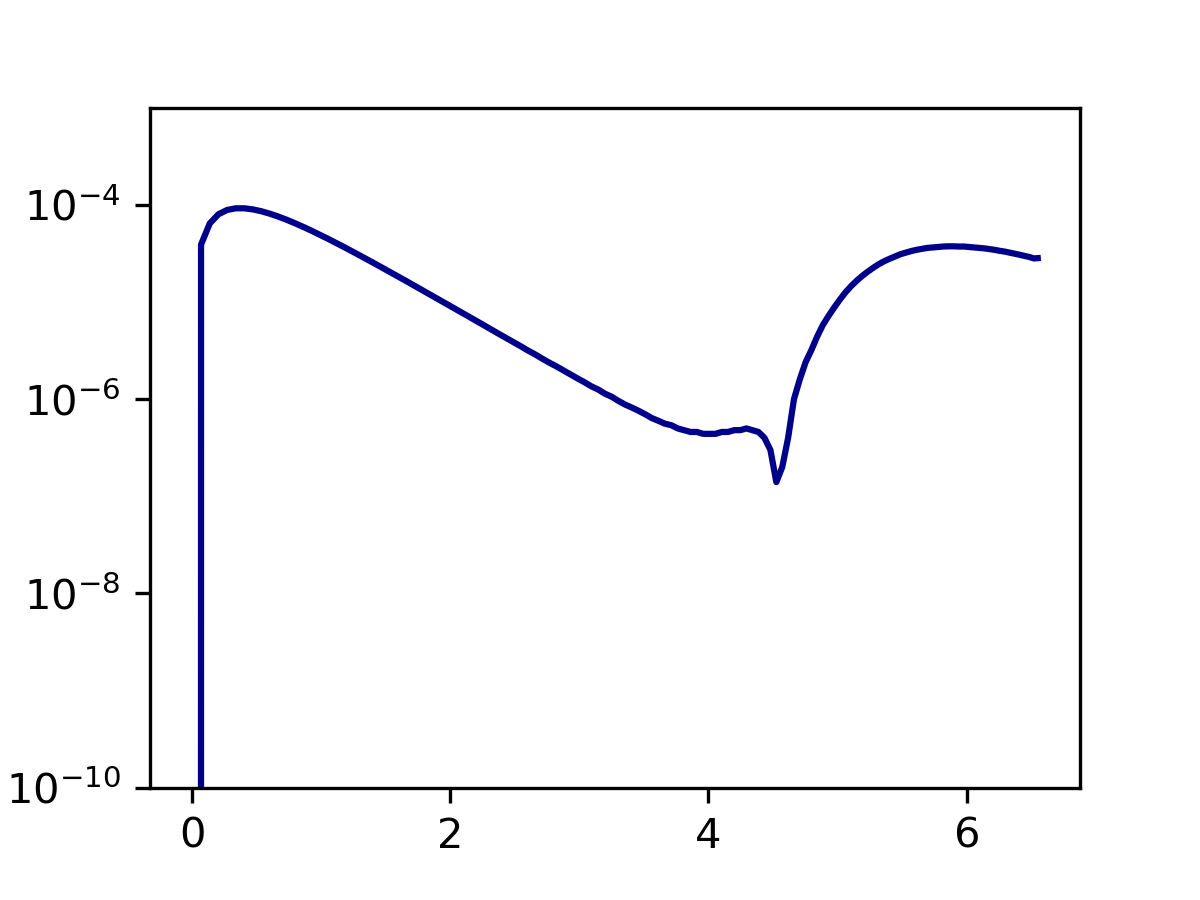}} ;
	
	\node [text width=0.01cm,align=center] at (+6.7,-2.7){$N_e$};

	\end{tikzpicture}

	\caption{Plot of energy violation during the numerical integration. The left panel shows the value in the case of negligible backreaction studied in \cref{sec:results_weak}, while the right panel shows the case of strong backreaction of \cref{sec:results_strong}.}
	\label{fig:energyA}
\end{figure}

\subsubsection*{Energy conservation}
We now discuss the accuracy of the numerical integrator in terms of energy conservation. In the left panel of \cref{fig:energy} we show the violation of the energy constraint defined in \cref{eq:cons} in the case of negligible backreaction, considered in \cref{sec:results_weak}. From this plot we can see that energy is well preserved during the evolution. Indeed, the violation of energy remains small throughout the evolution, and has a similar behavior to the single-field case shown in \cref{fig:energy}. As discussed in detail in \cref{sec:energy}, the peak in the violation of energy around the initial time is a consequence of the non-symplectic nature of the RK4 integrator employed to solve the differential equations, together with the rapidly oscillating and UV-peaked nature of the initial Bunch-Davies fluctuations.

In the right panel of \cref{fig:energy} we show the same plot in the case of strong backreaction. Also in this case, energy violation is under control throughout the evolution. This time, the initial peak is significantly higher than in the case of negligible backreaction. This is expected, as field fluctuations represent a larger fraction of the total energy of the system in this case: fluctuations are roughly of the same size, while the background potential energy $V(\phi)$ is much smaller in the case of strong backreaction. Indeed, we have $\phi^2 \simeq 210$ in the case of negligible backreaction and $\phi^2\simeq 30$ in the case of strong backreaction, making the background potential energy much smaller. 

Moreover, we can see that energy violation increases during the nonlinear phase in the last e-folds of evolution. This is also expected, as fluctuations are much higher during this phase and the dynamics is highly nonlinear. In order to independently check the accuracy of the time integration, we also performed the time-step convergence check described in \cref{sec:energy}. We increased \text{and} decreased the time-step of a factor of 10, and the simulation leads to the same results regarding power spectrum and all other observables considered in this chapter. This is an indication that the results of the simulation are physical and are not influenced by the small energy violations discussed in this section. 

The data presented in this section show that the RK4 is a good time integrator for the level of precision required in our analysis. Nevertheless, it would be valuable to implement an implicit symplectic integrator in our code. This could be particularly useful when simulating more complicated models of inflation, like the axion-SU(2) model described in \cref{sec:conclusions}. An example suitable for our case are high-order Gauss-Legendre integrators \cite{butcher1964implicit,Braden:2014cra,Pirvu:2021roq}, that preserve energy with great precision but are more challenging to implement. This is beyond the scope of the thesis, and will be subject of future work.

%
	

	
		


\section{Discussion and observational implications}
\label{sec:observations}
So far we presented the results of the lattice simulation. We now focus on the interpretation of these results and their implications on the observational constraint on the axion-U(1) model discussed in \cref{sec:obsA}. We start by discussing the suppression of non-Gaussianity in the strong backreaction regime, which is one of our main original findings. We give an interpretation of this suppression based on analytical arguments, and discuss the observational consequences of this result. Then, we also discuss the implications of our results on the statistics of scalar perturbation at large scales, such as the ones relevant for the CMB. All these topics will be further discussed in the conclusions.
\subsection{Suppression of non-Gaussianity and PBH production}
\subsubsection*{Prediction from perturbation theory}
 We now argue that the suppression of non-Gaussianity in the nonlinear regime can be already anticipated using perturbation theory in the $\xi$-constant approximation. In \cref{sec:obsA_bisp}, we saw that non-Gaussianity contained in the bispectrum can be parameterized by the following effective parameter:
\begin{equation}
f^{\rm(eff)}_{\rm NL}(\xi,x_1,x_2)=\frac{f_3(\xi,x_1,x_2)\mathcal{P}^3_{\rm vac}e^{6\pi\xi}}{\mathcal P^2_{\zeta}},
\end{equation}
where $\mathcal{P}_{\rm vac}$ is a constant. Using \cref{eq:ps_th}, and asuming that the sourced part dominates, one has $\mathcal{P}_{\zeta}\propto e^{4\pi\xi}$. Plugging this into the expression for $f_{\rm NL}^{\rm (eff)}$ we obtain:
\begin{equation}
\label{eq:supprfnl}
f^{\rm(eff)}_{\rm NL}(\xi,x_1,x_2)\propto e^{-2\pi\xi}.
\end{equation}
This shows that non-Gaussianity of $\zeta$ is suppressed when $\xi$ is large. Although perturbation theory is not reliable in this regime, and not all non-Gaussianity is contained in the bispectrum, the simulation confirms this intuition.
\subsubsection*{Central limit theorem interpretation}
The suppression of non-Gaussianity in the simulation and in \cref{eq:supprfnl} can be interpreted as natural consequence of the central limit theorem. To understand this, let us expand the term $F\tilde F$ in Fourier space as follows:
\begin{equation}
\left(	F_{\mu\nu}\tilde F^{\mu\nu}\right)(k)=\sum_{k^{\prime}} F_{\mu\nu}(k^{\prime})\,\,\tilde F^{\mu\nu}(k-k^{\prime}).
\end{equation}
This term acts as a source for the perturbations of the inflaton $\delta\phi$, as given by \cref{eq:gfparticle}. This expression shows that each Fourier mode of $F\tilde F$ is the sum of several non-Gaussian quantities. Each one is the product of two nearly Gaussian quantities $F$ and $\tilde F$. For small $\xi$, there is a small number of excited gauge field modes. Indeed, we saw in \cref{sec:gfparticles} that gauge modes are enhanced if $k<2\xi aH$, meaning that the range of excited gauge field modes grows linearly with $\xi$. Therefore, for small $\xi$, there is a small number of terms involved in this sum. If $\xi$ gets large, the number of excited gauge field modes grows and the sum converges to a Gaussian due to the central limit theorem. 

Our results show that the axion-U(1) model can lead to a large amount of non-Gaussianity only if $\xi\sim O(1)$. In this special case, a narrow window of gauge field modes is enhanced and can contribute to $\zeta$ in a non-Gaussian way. Due to the simplicity of this argument, we expect this suppression to be a general feature of models where matter fields are coupled linearly to the inflaton $\mathcal{L}\supset \phi f(X)$, with $f(X)$ being a nonlinear function of a generic matter field $X$, that could be for example a scalar $X=\psi$ or a gauge field $X=A^{a}_\mu$. If $X$ is copiously produced during inflation via some unspecified mechanism, its contribution to the statistics of $\zeta$ is expected to be nearly Gaussian for the same reason. An example where this also happens are models of axion inflation where $\phi$ is coupled to fermionic fields. Also in this case, non-Gaussianity has been shown to be suppressed when fermions are abundantly produced \cite{Adshead:2018oaa}.

\subsubsection*{Invalidating the PBH bound}
As we discussed in \cref{sec:obsA_PBH}, the axion-U(1) model has been constrained in the literature due to the efficient production of primordial black holes (PBH). Previous works, such as Refs. \cite{PhysRevD.87.103506, Garcia-Bellido:2016dkw}, obtained a bound of $\xi_{\rm CMB}\lesssim 1.66$ at CMB scales, corresponding to $\alpha/f<23$ for the quadratic potential. This bound, however, strongly depends on the assumption that $\zeta$ can be approximated by a $\chi^2$ distribution in the large $\xi$ regime:
$$\zeta(\vec{x})=\zeta_g^2(\vec{x})-\langle\zeta_g(\vec{x})\rangle^2,$$
where $\zeta_g$ is a Gaussian field. This corresponds to the $f_{\rm NL}\gg 1$ limit of the expansion in $\cref{eq:fnl}$, which is exactly the opposite of what we find. As a consequence, our results invalidate this bound. This allows for a much more interesting phenomenology at larger scales, such as CMB scales (discussed later) and intermediate scales corresponding to gravitational waves (GW) interferometers. In particular, it can be shown that values $\xi_{\rm CMB}<2.55$, compatible with current CMB bounds, can lead to a GW signal observable by LISA \cite{LISA1,Bartolo:2016ami}, advanced LIGO \cite{LIGOScientific:2016fpe}, and PTA-SKA \cite{1990pta,5136190,Kramer:2004hd}. Indeed, the gravitational wave signal of \cref{eq:GW} can be above the projected sensitivity of all these experiments in this parameter range \cite{Garcia-Bellido:2016dkw}.

Invalidating the existing PBH bound widens the available parameter space of the theory compatible with observations. However, a more careful investigation is needed to understand PBH production in this model. Indeed, as we can see in \cref{fig:ps_strong}, the power spectrum can be large during the nonlinear phase, potentially leading to a sizable production of PBH. Moreover, although the situation is simple for $\xi \gg 1$, $\xi$ does not grow too much due to the nature of the backreaction. This makes it important to quantify the small remaining non-Gaussianity. Indeed, we can see in \cref{fig:cum_strong} that the cumulants are not exactly zero at the end of the simulation due to the complicated dynamics, and PBH production is extremely sensitive to these effects. Therefore, one has to track the evolution of the power spectrum and the PDF of $\zeta$ during the last e-folds of inflation in order to precisely determine PBH abundance. Note that, at this level of precision, the phenomenology will also depend on the shape of inflationary potential during the last e-folds of in inflation. This careful study is beyond the scope of the thesis, although the lattice approach developed here fits this purpose.


%


\subsection{Non-Gaussianity at large scales}
\label{sec:CMB}
The results of last section imply that the current strongest constraints on the axion-U(1) model are the ones on the statistics of $\zeta$ at large scales corresponding to CMB experiments. At these scales, one has to be in the $\xi\sim O(1)$ regime in order to preserve the observed power spectrum of scalar perturbations. This allows for a large amount of non-Gaussianity, that is strongly constrained by CMB measurements. As we discussed in \cref{sec:obsA_bisp}, one can constrain $\xi_{CMB}\lesssim 2.55$ using the equilateral-shape bispectrum for this model. 

In \cref{sec:results_weak} we showed that, at CMB scales, the statistics of $\zeta$ is characterized by $$\kappa_5>\kappa_4>\kappa_3>1,$$ where $\kappa_i$ are the dimensionless cumulants defined in \cref{eq:momenta}. This means that most of non-Gaussianity of this model is \textit{not} contained in the bispectrum, but in higher order correlators. As a consequence, one should be able to constrain this model even further using high order statistics such as the trispectrum. This is possible with the lattice approach developed in this thesis, and is one of the possible prospects of the work of the thesis. Moreover, developing a trispectrum estimator for lattice simulations can be relevant in a number of applications, as we discuss in the conclusions.

Even more importantly, our results show that $n$-point functions can not fully describe the statistical properties of scalar perturbations from this model. Indeed, the fact that cumulants $\kappa_i$ grow with $i$ is an indication that $\zeta$ can not be efficiently described by a finite number of correlators $\langle \zeta^i\rangle$. In other words, the statistical information contained in the real space picture of $\zeta$ in \cref{fig:boxA}, resulting in a pronounced tail in the  one-point PDF of \cref{fig:hist_weak}, can hardly be anticipated by computing a small number $n$-point functions, such as bispectrum and trispectrum. 
This will be further discussed in the conclusions. 

\cleardoublepage
\newpage\null
\chapter{Conclusions}
\label{sec:conclusions}
\section{Summary}

In this thesis, we developed a numerical lattice simulation to study the inflationary epoch of the Universe. Lattice simulations are a well-established tool in primordial cosmology. Many codes have been developed in the last 20 years to study the end of inflation and the reheating phase after it \cite{Khlebnikov_1996,Prokopec_1997,latticeeasy,Frolov_2008,hlattice,Sainio_2012,Child_2013,Easther_2010,Lozanov_2020,figueroa2021cosmolattice}. We generalized this machinery to the deep inflationary era, and we used it to study both the simplest single-field scenario and the more complicated axion-U(1) model of inflation. Before providing an outlook and discussing future possible applications of the methodology developed in this thesis, we summarize the main original results.

\subsection{Part 1: single-field case}
The first part of the thesis was devoted to studying the most simple single-field scenario of inflation. This is the first time a lattice code is used to study the inflationary universe much before the end of inflation. Therefore, we first had to use the simulation to recover well-known results, and show how the predictions from the simulation compare to their analytical counterparts in terms of precision.
 In particular, we focused on recovering the power spectrum of primordial scalar fluctuations in two cases. In the first case, we looked at a simple slow-roll potential for the inflaton field, resulting in a nearly scale-invariant power spectrum. In the second, we considered a step potential, resulting in oscillations in the power spectrum.

 In both cases, the simulation allowed to recover the analytical results with great precision. To achieve this goal, we had to carefully analyze how the discretization affects the evolution of the small scalar fluctuations on the lattice. In this way, we were able to obtain a lattice prediction for the power spectrum that is independent of the numerical implementation. This allows comparing the output of the simulation directly with the analytical predictions, and potentially with data. This is well summarized by the right panel of \cref{fig:finalPS_stencils}, where we can see that different discretization schemes lead to the same result, independently of the particular lattice implementation. This issue has been neglected in all previous simulations in the context of reheating. In that case, the dynamics is highly nonlinear and much less constrained by observations. Therefore, a precise computation of the power spectrum is not relevant. This is different during the deep inflationary phase, where the power spectrum - and more generically the statistics of scalar perturbations - is much more constrained by experiments.
 
 
The lattice simulation developed in this first part was not needed to understand the physics of these simple models, which is well known and lies within the regime of validity of perturbation theory. However, this technical study was necessary to understand the main challenges of simulating inflation on a lattice and recovering the analytical results with good precision. This lays the groundwork for the second part of the thesis.

\subsection{Part 2: axion-U(1) inflation}
In the second part, we generalized the technique developed for the single-field case to the more complicated axion-U(1) model of inflation. We showed that the discretization has a much greater impact on the evolution of fields on the lattice, compared to the single-field case.
 Nevertheless, we were able to identify a discretization scheme for which the field dynamics on the lattice is equivalent to the one of continuous space. 

We presented the results of the simulation in two cases. In the first case, we studied the dynamics of the axion-U(1) system much before the end of inflation, which is relevant for large-scale observations such as the CMB. In this case, the dynamics lies within the regime of validity of perturbation theory, allowing for a comparison of some of our results with existing analytical computations in the literature. The power spectrum and bispectrum of $\zeta$ computed from the lattice simulation agree with previous analytical estimates, and improve upon them in terms of precision. 
Moreover, the simulation allowed us to discover new statistical properties of the curvature perturbation from the axion-U(1) system. We used the simulation to obtain a real-space picture of $\zeta$, from which we computed correlators such as $\langle \zeta^4(\vec{x}) \rangle$ and $\langle \zeta^5(\vec{x}) \rangle$. Our results show that the statistics of $\zeta$ within this model is rather unusual: most of non-Gaussianity is contained in high-order correlators, and the various dimensionless cumulants $\kappa_i$ of \cref{eq:momenta} grow with $i$. This suggests that $n$-point functions are not efficient in fully describing the statistics of $\zeta$, with potentially profound implications, as discussed later.

In the second case, we studied the phenomenology of the axion-U(1) system in the regime of strong backreaction, where standard perturbative computations do not apply.
 In realistic scenarios, this typically occurs at a later stage of the inflationary epoch, where the gauge field production is strong enough to affect the background evolution of the Universe.
Our main finding is that, as soon as the system enters this strong backreaction phase, the statistics of $\zeta$ rapidly converges to a Gaussian. This is a consequence of the large number of excited gauge field modes. In this regime, many uncorrelated gauge field modes contribute to the statistics of $\zeta$, which then becomes nearly Gaussian as a result of the central limit theorem. This result is in contradiction with several studies in the literature that explored the strong backreaction phase using perturbation theory outside its regime of validity. In particular, our study invalidates some constraints on this model coming from the overproduction of primordial black holes. These bounds were obtained assuming a highly non-Gaussian statistics for $\zeta$ \cite{PhysRevD.87.103506, Garcia-Bellido:2016dkw}, which is the opposite of what we find. Invalidating this bound widens the available parameter space of the theory. As a consequence, this model can potentially lead to a gravitational wave background signal within the range of upcoming experiments such as LISA \cite{Garcia-Bellido:2016dkw}. This will be further discussed in the next section.

\section{Outlook and future perspectives}
Our results demonstrate that lattice simulations can be a powerful tool to study the deep inflationary era and its theoretical predictions. On one hand, the simulation allowed us to study the axion-U(1) model beyond perturbation theory, giving new and insightful results. On the other hand, it allowed us to better understand the phenomenology of this model \textit{inside} the regime of validity of perturbation theory. In both cases, the simulation provided a full characterization of the statistics of primordial fluctuations beyond $n$-point functions, which can be hardly achieved with standard perturbative computations. This lays the groundwork for future research on the subject. In the remaining of this chapter, we are going to discuss a few of the possible applications of the methodology developed in this thesis, which constitute future directions of research.

\subsubsection*{Gravitational waves}
As mentioned in various parts of the thesis, gravitational waves (GW) are one of the most interesting predictions of the axion-U(1) model of inflation. Future experiments, such as LISA \cite{LISA1}, will open a new observational window of GW signals in the near future. 
Therefore, computing precise predictions for GW emission is particularly important. Similarly to scalar perturbations, the computation of inflationary GW emission from the axion-U(1) system presents several challenges related to the breakdown of perturbation theory \cite{Campeti:2022vom}. For this reason, extending the code to compute GW emission would be extremely valuable. In the axion-U(1) model, gravitational waves are sourced by the enhanced electromagnetic field \cite{Sorbo:2011rz}:
\begin{equation}
\hat	h^{\prime\prime}_{ij}+2\frac{a^\prime}{a}	\hat h^{\prime}_{ij}-\nabla^2	\hat h_{ij}=2 T^{GF,TT}_{ij},
\end{equation}
where $\hat h_{ij}$ is the transverse and traceless part of the tensor perturbation of the metric defined soon after \cref{eq:pertmet} and $T^{GF,TT}_{ij}$ is the transverse and traceless part of the gauge field stress-energy tensor \cref{eq:stress_GF}. This kind of sourced gravitational waves from an axion-U(1) system have been  already computed using a lattice code in the context of reheating simulations \cite{2020axiin,2020axiin2}. A similar methodology could be applied to the inflationary era, representing a natural extension of the work of this thesis.

\subsubsection*{Trispectrum and non-Gaussianity}
As we discussed in detail in \cref{sec:obsA}, our results imply that the strongest constraints on the axion-U(1) model are currently the ones from non-Gaussianity of $\zeta$ at CMB scales. Current bounds, discussed in \cref{sec:obsA_bisp}, are derived using the bispectrum on equilateral configurations. Our results show that most of non-Gaussianity is contained in higher-order correlators. For this reason, we expect the trispectrum $T$: $$\langle \zeta(\vec{k}_1)\zeta(\vec{k}_2)\zeta(\vec{k}_3)\zeta(\vec{k}_4) \rangle=T(\vec{k}_1,\vec{k}_2,\vec{k}_3,\vec{k}_4)\delta_D(\vec{k}_1+\vec{k}_2+\vec{k}_3+\vec{k}_4)$$ to be more relevant than the bispectrum in describing non-Gaussianity within this model and constraining the available parameter space of the theory. As no analytical estimate of the trispectrum is available for this model in the literature, a lattice computation would particularly useful. In this work, we only evaluated the bispectrum, and this is the first computation of a Fourier space 3-point function using a lattice simulation. Computing 4-point functions is much more challenging due to the high dimensionality of the integrals involved in the computation. Recently, some progress has been made on this problem, allowing to compute high-order statistics of this kind in an efficient way \cite{Philcox:2021eeh,Philcox:2021bwo}. Similar techniques could be applied to lattice codes, and would allow computing the trispectrum of $\zeta$ from the simulation.

Estimating the trispectrum can be relevant in several applications. First of all, the trispectrum is the first $n$-point Fourier correlator to be sensitive to parity-violating physics. Recent measurements point towards a parity-violating 4-point function of BOSS galaxies \cite{Philcox:2022hkh,Hou:2022wfj}. Although this detection is yet to be confirmed, the axion-U(1) model represents a natural candidate for this kind of signal due to the parity-violating nature of the Chern-Simons term $\phi F\tilde F$. Computing the trispectrum of $\zeta$ would allow establishing whether this model can lead to an observable amount of parity violation in the primordial fluctuations. 

Another application is the computation of scalar-induced GW emission from inflation. If scalar perturbations are large enough, they can source GW at second order in perturbation theory \cite{Mollerach:2003nq,Acquaviva:2002ud,Ananda:2006af,BaumannGw}. In this framework, the computation of the power spectrum of GW relies on computing the trispectrum of $\zeta$, and it is usually performed assuming that the 4-point function can be factorized in a product of 2-point functions using Wick's theorem \cite{Domenech:2021ztg}. If $\zeta$ is non-Gaussian this is not enough, and one has to compute the connected part of the trispectrum in order to evaluate this scalar-induced GW signal \cite{Domenech:2021ztg,Ragavendra:2020sop,Adshead:2021hnm,Cai:2018dig,Unal:2018yaa,Yuan:2020iwf,Atal:2021jyo}. Therefore, instrumenting a code for computing the trispectrum from lattice simulations can be crucial in studying this kind of GW signal from inflation.

Beyond the 4-point function, our work shows that non-Gaussianity from the axion-U(1) model is very particular. At large scales, the statistical distribution of $\zeta$ is characterized by a growing tower of cumulants $\kappa_5>\kappa_4>\kappa_3>1$, which are defined in \cref{eq:momenta}. This suggests that $n$-point functions, such as bispectrum and trispectrum, can not fully describe the statistics of $\zeta$ within this model. If we want to fully model non-Gaussianity, it would be relevant to come up with a statistical description that is independent of $n$-point functions. This poses a great challenge, as non-Gaussianity of scalar perturbations is usually described using $n$-point functions, such as power spectrum and bispectrum, both in theoretical modeling \textit{and} observations. The simulation presented in this work offers a unique opportunity in this direction: the output of our code is a real space picture of the pre-recombination universe, that could in principle be used to describe perturbations in a way that is independent of $n$-point functions. This can be hardly achieved by standard perturbative computations, and motivates further work in this direction.


\subsubsection*{Primordial black holes}
As discussed in \cref{sec:axionsim}, the lattice approach developed in this work can be very useful in studying the abundance of primordial black holes (PBH). Indeed, the simulation allowed us to obtain the full shape of the statistical distribution of $\zeta$ in real space, to which the production of PBH is known to be extremely sensitive \cite{Bullock:1996at,Seery:2006wk,Hidalgo:2007vk,Saito:2008em,Byrnes:2012yx,Young:2013oia,Bugaev:2013vba,Young:2014oea,Young:2015cyn,Franciolini:2018vbk,DeLuca:2019qsy,Atal:2018neu,Taoso:2021uvl,Ozsoy:2021qrg,Rezazadeh:2021clf,Davies:2021loj}. 

Although we focused on the axion-U(1) model, a similar lattice computation can be performed to study PBH production in other models of inflation. Typically, the large power spectrum needed for PBH production is achieved by a pronounced feature in the inflationary potential in single-field models of inflation, i.e. a localized deviation from the slow-roll trajectory \cite{Taoso:2021uvl,Rezazadeh:2021clf,Motohashi:2019rhu,Tasinato:2020vdk,Inomata:2021uqj}. In these cases, the large size of scalar fluctuations induces a nonlinear coupling between modes, and requires evaluating the full non-Gaussian shape of $\zeta$ to determine the production of PBH \cite{Pattison:2017mbe,Ezquiaga:2019ftu,Davies:2021loj}. 
The lattice approach developed in this thesis is well-suited for this purpose, and could be used to compute the full shape of $\zeta$ in analogy to what we did for the axion-U(1) model.


\subsubsection*{SU(2) gauge fields}

The methodology developed in this work can be generalized to more complicated models of inflation. A concrete example are axion-gauge models where the gauge field is a non-Abelian SU(2) field $A^{a}_{\mu}$, with $a\in\{1,2,3\}$. In this case, the part of the Lagrangian involving the gauge field reads:
\begin{equation}
\label{eq:nonab}
	\mathcal{L}\supset-\frac{1}{4}F^a_{\mu\nu}F^{a\mu\nu}-\frac{\lambda}{4f}\phi F^a_{\mu\nu}\tilde F^{a\mu\mu},
\end{equation}
where the gauge field strength tensor is
\begin{equation}
	F^{a}_{\mu\nu}=\partial_\mu A^a_{\nu} - \partial_\nu A^a_{\mu}+ig f^{abc}A^b_\mu A^c_\nu,
\end{equation}
and $f^{abc}$ are the structure constants of SU(2) and $g$ is the gauge coupling. $\tilde F^a_{\mu\nu}$ is the Hodge dual of $F^a_{\mu\nu}$, defined analogously to the U(1) case of \cref{eq:actionA}. This model has been extensively studied in the literature, both in cases where $\phi$ is the inflaton or a pseudo-scalar spectator field \cite{Maleknejad_2011,maleknejad2013gaugeflation,Adshead_2012,Adshead_2013,maleknejad2021su2r,Ishiwata:2021yne,Maleknejad:2016qjz,Papageorgiou:2018rfx,Maleknejad_2019,Lozanov_2019,Mirzagholi:2019jeb,Papageorgiou:2019ecb,Dimastrogiovanni:2016fuu,Agrawal:2017awz,Agrawal:2018mrg,Fujita:2022jkc}. The axion-SU(2) system leads to interesting observational signatures, such as non-Gaussian primordial gravitational waves and parity-violating physics in the early Universe. Similarly to the U(1) case, this model shows perturbativity issues that require going beyond perturbation theory to compute predictions in the parameter range relevant for observations \cite{Ferreira:2015omg,Peloso_2016,Papageorgiou:2018rfx,Maleknejad_2019,Lozanov_2019,Mirzagholi:2019jeb,Papageorgiou:2019ecb,Fumagalli:2020nvq, Inomata_2022}. Therefore, it would be important to develop a lattice simulation to study its inflationary dynamics.

Due to the non-Abelian nature of the gauge field, the nonlinear equations of motion are more complicated than in the U(1) case of \cref{eq:eomsa}. This makes the numerical implementation more challenging, both due to the complicated structure of the equations and to the increased number of variables to evolve. Nevertheless, the dynamics of this model shares several similarities with the U(1) case: there is a similar exponential production of gauge field particles caused by the background velocity of the axion field and the equation of motion for the axion field is analogous. Therefore, the aspects discussed in this thesis are relevant also for simulating this more complicated model. In particular, we expect the discretization scheme developed in \cref{sec:discA} to apply also to the SU(2) case. Hence, including a SU(2) gauge field in our simulation would be a natural extension of the work of the thesis.

\cleardoublepage\null
\appendix

\backmatter

\bibliographystyle{hunsrt}
\addcontentsline{toc}{chapter}{Bibliography}
\bibliography{Thesis.bib}


\end{document}